\pdfoutput=1

\documentclass[a4paper]{nusthesis}

\dsp 

\usepackage{indentfirst} 

\usepackage{silence} 
\usepackage{dcolumn}
\usepackage{bm}
\usepackage{hyperref}
\usepackage{physics}
\usepackage{amsmath,amsfonts,amsthm}
\usepackage{subcaption}
\usepackage[mathlines]{lineno}
\usepackage{graphicx}
\usepackage{bookmark}
\usepackage[
  backend=biber,
  style=ieee,
  citestyle=numeric-comp,
  giveninits=true,
  sorting=none,
  maxbibnames=99,
  dashed=false,
  doi=false]{biblatex}
\DeclareFieldFormat*{title}{``#1''\newunitpunct} 
\usepackage{microtype} 
\usepackage{tabulary} 
\usepackage{hhline}

\usepackage{enumitem}

\usepackage{hyperref}

\usepackage{listings}

\usepackage{amsthm}
\theoremstyle{definition}
 
\theoremstyle{plain}

\usepackage{mathtools}

\usepackage{interval}
\intervalconfig{soft open fences}

\usepackage[linesnumbered,ruled,vlined]{algorithm2e}
\SetAlgorithmName{Algorithm}{Algorithm}{Algorithm}
\IncMargin{0.5em}
\SetCommentSty{textnormal}
\SetNlSty{}{}{:}
\SetAlgoNlRelativeSize{0}
\SetKwInput{KwGlobal}{Global}
\SetKwInput{KwPrecondition}{Precondition}
\SetKwProg{Proc}{Procedure}{:}{}
\SetKwProg{Func}{Function}{:}{}
\SetKw{And}{and}
\SetKw{Or}{or}
\SetKw{To}{to}
\SetKw{DownTo}{downto}
\SetKw{Break}{break}
\SetKw{Continue}{continue}
\SetKw{SuchThat}{\textit{s.t.}}
\SetKw{WithRespectTo}{\textit{wrt}}
\SetKw{Iff}{\textit{iff.}}
\SetKw{MaxOf}{\textit{max of}}
\SetKw{MinOf}{\textit{min of}}
\SetKwBlock{Match}{match}{}{}

\usepackage{array}
\newcolumntype{L}[1]{>{\raggedright\let\newline\\\arraybackslash\hspace{0pt}}m{#1}}
\newcolumntype{C}[1]{>{\centering\let\newline\\\arraybackslash\hspace{0pt}}m{#1}}
\newcolumntype{R}[1]{>{\raggedleft\let\newline\\\arraybackslash\hspace{0pt}}m{#1}}

\usepackage{color}
\usepackage[usenames,dvipsnames]{xcolor}
\usepackage{soul}
\soulregister\cite7
\soulregister\ref7
\soulregister\pageref7
\soulregister\autoref7
\soulregister\eqref7

\renewcommand{\eqref}{Equation~\ref}

\newcommand*{\RootPicDir}{pic}
\newcommand*{\PicDir}{\RootPicDir}

\newcommand*{\SetPicSubDir}[1]{\renewcommand*{\PicDir}{\RootPicDir /#1}}
\newcommand*{\Pic}[2]{\PicDir /#2.#1}

\newcommand*{\RootExpDir}{exp}
\newcommand*{\ExpDir}{\RootExpDir}

\newcommand*{\SetExpSubDir}[1]{\renewcommand*{\ExpDir}{\RootExpDir /#1}}

\usepackage{lipsum} 

\begin{document}

\title{Spectral and entanglement transitions from non-Hermitian skin pumping}

\author{LI Qingya}
\prevdegrees{%
  B.Sci. (Hons.), NUS}
\degree{Doctor of Philosophy}
\field{Physics}
\degreeyear{2024}
\supervisor{Assistant Professor Lee Ching Hua}

\examiners{
Associate Professor Wang Qinghai, Department of Physics\\
Assistant Professor Yvonne Gao, Department of Physics
}

\maketitle

\declaredate{4 August 2024}
\declaresign{\RootPicDir /signature.png} 
\declarationpage
\begin{frontmatter}
  \begin{acknowledgments}

First and foremost, I would like to express my deepest gratitude to my supervisor, Prof. Lee Ching Hua, for his invaluable patience, understanding, guidance, and mentorship. From the very beginning, he introduced me to the intricate world of condensed matter physics, with a special focus on non-Hermitian physics. His precise advice and patient guidance allowed me to quickly grasp the fundamentals, avoiding many initial pitfalls. Prof. Lee is also brimming with innovative research ideas and inspirations. Conversations with him always spark new ideas and directions of investigation. He also had many fascinating solid works constructed, which have profoundly influenced my research interests and shaped the goals I have set for myself. Moreover, I have learned a lot about the communication and presentation skills of my research work, greatly improving my ability to craft research findings and effectively present results in papers, theses, and presentations. This has connected me to a broader world of the research community through more effective communication and exchange of ideas and knowledge. In the four years under Prof. Lee’s dedicated guidance and mentorship, I have transformed from an inexperienced student into an independent researcher capable of conducting scientific research autonomously and exploring research interests on my own. This is significant to me as it equips me with skills to pursue my personal interest in physics research throughout my life.

I would also like to thank my collaborators, Dr. Jiang Hui and Liu Sirui. Collaboration and discussion with these excellent researchers have been promising. My special thanks to Dr. Jiang Hui, a seasoned researcher, who has always provided invaluable insights whenever I encountered challenging or seemingly unsolvable problems. Discussions with her have consistently offered me new ideas and perspectives.

A big thanks to my family for their unwavering support and for keeping me grounded throughout this journey.

\end{acknowledgments}

  \tableofcontents 
  \begin{abstract}
Non-Hermitian physics has emerged as a rapidly evolving field in recent decades. Intriguing phenomena, including the non-Hermitian skin effect (NHSE) and various non-Hermitian phase transitions, lead to numerous unconventional properties elucidated by the theoretical framework of the generalized Brillouin Zone (GBZ), with promising applications. This thesis further advances our knowledge of non-Hermitian systems through a comprehensive analysis of band theories and critical phenomena exhibited by several proposed systems under NHSE alone or in conjunction with additional degrees of freedom.

The first project investigates the appearance of real spectra in non-Hermitian systems, enforced by the skin effect rather than the bulk symmetries. This study showcases simple ansatz models with robust open boundary condition (OBC) real spectra, complementing current efforts toward the stable design of non-Hermitian systems.

The second project designs a system of two oppositely oriented weakly coupled NHSE chains. Their competition in the skin effect leads to a new class of non-Hermitian critical transitions: the scaling-induced exceptional criticality (SIEC), which transcends existing non-Hermitian mechanisms such as NHSE-induced gap closures and exhibits dramatic divergent entanglement entropy dips, violating conventional logarithmic behavior. This study also advances the GBZ framework to describe and explain the system and the newly identified critical phenomenon.

The third project introduces spatial inhomogeneity as an additional degree of freedom to conventional non-Hermitian systems characterized by asymmetric but spatially constant hopping strengths. The scaled interplay between inhomogeneity and non-Hermiticity gives rise to unprecedented phenomena of GBZ bifurcation. This bifurcation mechanism not only alters the energy spectrum but also introduces an additional degree of freedom for stabilizing real spectra, giving rise to a new class of topological zero modes. We introduce the phase-space GBZ framework which unifies position and momentum space to address the challenges posed by spatial inhomogeneity and enables precise predictions of eigenstate distributions and energy spectra in inhomogeneous systems.

The final project explores the interplay between NHSE and the hyperbolic Bethe lattice. New forms of critical NHSE behavior of scale-induced spectral transition from real to complex were studied and further validated within the hyperbolic-like lattice geometry and a hierarchy of loop sizes. 

\end{abstract}

  \listoffigures
\end{frontmatter}

\SetPicSubDir{ch-intro}
\chapter{Introduction}
\label{ch-intro}
\vspace{2em}
\section{Historical development of non-Hermitian physics}
The mathematical formulation of modern quantum mechanics involves a few postulates, rigorously describing isolated quantum systems in terms of operators on Hilbert spaces~\cite{griffiths_introduction_2018,shankar_postulatesgeneral_1994,Albert,ClaudeQM,nottale2007derivation}. One of the postulates states that a physical observable is necessarily Hermitian, in order to enforce real eigenvalues as the results of measurements of corresponding observables and ensure probability-preserving time evolution for isolated systems, is of particular interest.\\
Let us introduce its details a bit. Given an operator $\hat{A}$, we consider the eigenfunction $\hat{A}\ket{\psi_n}=a_{n}\ket{\psi_n}$ and its Hermitian transpose $\bra{\psi_n}\hat{A}^{\dagger}=a_{n}^*\bra{\psi_n}$. If the operator is Hermitian $\hat{A}=\hat{A}^{\dagger}$,
\begin{align}
a_{n}\bra{\psi_n}\ket{\psi_n}=\bra{\psi_n}\underline{\hat{A}\ket{\psi_n}}=\underline{\bra{\psi_n}\hat{A}^{\dagger}}\ket{\psi_n}=a_{n}^*\bra{\psi_n}\ket{\psi_n},
\end{align}
its eigenvalues $a_{n}$ are enforced real. Furthermore, Hermiticity of the Hamiltonian $\hat{H}$ of a system preserves the inner product between states and hence the probability amplitude under time evolution~\cite{ballentine2014quantum,liboff2003kinetic,townsend2000modern,zettili2009quantum}:
\begin{align}
    \bra{\psi_n(t)}\ket{\psi_n(t)}= \bra{\psi_n(0)}e^{i\hat{H}^{\dagger}}e^{-i\hat{H}}\ket{\psi_n(0)}=\bra{\psi_n(0)}\ket{\psi_n(0)}.
\end{align}
The time-evolution operator $U=e^{-i\hat{H}}$ is unitary given the Hermitian Hamiltonian: $U^{-1}=U^{\dagger}$.

The above axiom was commonly believed until 1998 when Bender proposed the criteria of parity-time (PT) symmetry~\cite{bender_real_1998} to broaden the Hermiticity requirement. A quantum system is said to be symmetric under a transformation represented by an operator $\hat{A}$ if its Hamiltonian commutes with this operator. A PT symmetric system then has a Hamiltonian obeying $[PT,H]=0$ with a linear operator $P$ representing the space reflection and an antilinear operator $T$ representing the time reversal. Unbroken PT symmetric systems possess the physical properties of Hermitian systems such as the real spectra~\cite{bender_making_2007,carcassi2021four}, which states that non-Hermitian physics could also possibly be physically meaningful and bring non-Hermitian physics to the front~\cite{brody2013biorthogonal,ashida2020non,heiss2004exceptional,rotter2009non,HN1996prl,el2007theory, yao2018edge,heiss_physics_2012, Moiseyev_2011,kawabata2019symmetry,bergholtz2021exceptional}. 
Physically, non-Hermitian physics describes non-conservative systems, which can be constructed by introducing nonreciprocal hoppings, unbalanced gain and loss, or dissipative dynamics~\cite{lindblad1976generators,feynman2000theory}. In PT-symmetric non-Hermitian systems, the balance between gain and loss is restored and protected by the PT symmetry, leading to its conservative dynamics that is mathematically equivalent to Hermitian systems.  

PT symmetry has attracted intense interest since it effectively challenged the fundamental elements of QM. Research work on the properties and conditions of PT symmetric systems introduced a concept uniquely existing in non-Hermitian systems, known as the exceptional point (EP). EPs are found to be the transition point between broken and unbroken PT-symmetric phases, and they are singularities within the non-Hermitian Hilbert space. Mathematically, an EP is defined as a critical point where both eigenenergies and eigenstates coalesce. Being a special case of degenerate points, it does not have a Hermitian analogy with respect to its critical and defective nature. Extensive studies have been done on its intriguing critical properties and applications~\cite{heiss_physics_2012,zhou_observation_2018,kang_experimental_2017,kawabata_classification_2019,heiss_chirality_2001,hahn_observation_2016,ZOU2024,hahn_observation_2016,gonzalez_topological_2017,zhen_spawning_2015,San_Jose_2016}, and this phenomenon was experimentally verified in ~\cite{PhysRevLett123193901}.

On another front, in the 1980s the concept of topology appeared with the discovery of the quantum Hall effect (QHE). This phenomenon, observed in two-dimensional electron systems under a strong magnetic field, revealed that Hall conductance is quantized in integer multiples of $\frac{e^2}{h}$, and possessing conducting chiral edge states in one direction~\cite{PhysRevLett45494,Kane2005}. This quantization was then theoretically explained in 1982 through the concept of the Chern number~\cite{Thouless1982}, which is an integer defined in the momentum space (the Brillouin zone) of a system. We later named the Chern number a type of topological invariant since it characterizes the system's topological properties and remains constant under continuous transformations. The presence of conducting edge states was then understood by the bulk-edge correspondence principle, connecting the bulk topological invariants to the number of edge states.

In 2005, Kane and Mele extended the QHE to the quantum spin Hall effect (QSHE)~\cite{Bernevig2006,Kane2005}, which has edge states counter-propagating for different spin channels and characterized by a topological invariant similar $Z^2$ protected by time-reversal symmetry. This then led to the model of topological insulator in graphene, which was confirmed by the observation of the quantum spin Hall effect in HgTe/CdTe quantum wells in 2007~\cite{Koenig2007}.

The topological study holds significant importance because it broadened our understanding of the nature of electronic states and found a new phase of matter distinct from conventional materials that has potential in wide applications~\cite{Fu2007,Haldane2008,Khanikaev2013,Sun2016,Ozawa2019,Fleury2016,Brandenbourger2019}. An important property worth noting is the topologically protected edge states due to their non-trivial band topology. The topological edge states are robust against disorder, making them stable and resistant to localization, leading to numerous applications~\cite{Blanco-Redondo2016,Szameit2015}.

Then in the 2010s, the interplay of non-Hermitian physics and topology attracted great research interest, leading to the development of non-Hermitian topological phases~\cite{Kawabata2018nonHclass,Liu2019nonHclass,bergholtz2021exceptional,sayyad2021entanglement,wojcik2021eigenvalue,shiozaki2021symmetry,borgnia2020nonH,park2022nodal} exhibiting not only topologically protected edge states but two important phenomena that do not have Hermitian analogues, which are the exceptional points and the non-Hermitian Skin Effect.
The EPs, which have been introduced as the transitioning point between broken and unbroken PT symmetric phases, occur when the eigenvalues and eigenvectors of a non-Hermitian operator or Hamiltonian coalesce in the complex spectrum.
At EPs, the system's dynamics exhibits exceptional sensitivity to perturbations, with eigenvalues displaying square-root branch points in the complex plane.  EPs always induce critical behaviours in the system's response to external fields or parameters, leading to novel types of non-Hermitian phase transitions. Among different non-Hermitian phases or states, dramatic changes in the system's physical properties could occur, such as sudden shifts in the spatial distribution of eigenstates or the emergence of unidirectional transport. The discovery of these non-Hermitian phases of states expands the landscape of condensed matter physics and offers novel functionalities in photonic and electronic devices~\cite{lee2020ultrafast,cao2021non,zhou2021dual}.
The NHSE, which is a distinctive phenomenon observed in non-Hermitian systems where not only the edge state but the bulk states of the system localize at the boundaries~\cite{Martinez2018nonH,okuma2020topological,longhi2021non,peng2022manipulating,liang2022observation,zhu2020photonic,guo2021exact}, in contrast to the extended and spread Bloch bulk state in Hermitian systems. Typical NHSE arises in systems with asymmetric hopping amplitudes or gain and loss mechanisms, which break the symmetry of the Hamiltonian.
This asymmetry causes a nonreciprocal propagation of waves, leading to the exponential localization of states at the boundaries. The NHSE holding robust localized states opens new possibilities for developing robust waveguides, sensors, and devices and has potential applications in signal processing, quantum information, and other areas requiring precise control over wave dynamics~\cite{ghatak2020observation,helbig2020generalized,xiao2020non,Bouganne2020anomalous,song2019breakup,hou2020topo,longhi2018paritytime,ningyuan2015time,scheibner2020non,zeng2021real,long2021realeigenvalued,kawabata2020real,aharonov1996adiabatic,PhysRev99022111,Li2021protecting}.

Dozens of theoretical frameworks have been proposed to describe and explain the non-Hermitian topological phases and the above phenomena during these years. Until a breakthrough work in 2018 by Shunyu Yao and Zhong Wang~\cite{yao2018edge}, which is the construction of the generalized Brillouin zone (GBZ).

In Hermitian systems, Bloch band theory provides a framework for describing the energy bands of periodic systems, where wavefunctions are characterized by Bloch states effectively described within the Brillouin zone. However, in non-Hermitian systems, traditional Bloch band theory fails due to complex eigenvalues and the phenomenon of NHSE. To address these challenges, Yao and Wang proposed the GBZ as an extension of the traditional Brillouin zone. The GBZ framework allows for complex wavevectors $k$, which enables a proper description of the complex eigenstates of the system and their localization properties. This rigorously defined the phenomenon of NHSE and non-Hermitian topological transition, and restored the bulk-boundary correspondence by introducing new types of topological invariant specific to non-Hermitian systems. This work is particularly important because it links various aspects of non-Hermitian physics, providing a coherent theoretical foundation~\cite{yang2020non,yokomizo2019non,helbig2020generalized,Yao2018nonH2D,zhang2019correspondence,guo2021analysis,bartlett2021illuminating,zhang2022bulkbulk,cao2021non,Song2019BBC,kunst2018biorthogonal,koch2020bulk,imura2020generalized,qin2022non,jiang2022filling}.

The theoretical advances provided by the GBZ have spurred experimental efforts to realize and observe non-Hermitian topological phases. Systems such as photonic lattices, electronic circuits, and acoustic structures have been designed and realized to explore these phenomena. The GBZ framework has been instrumental in guiding these experimental studies, providing a robust theoretical basis for understanding observed behaviours.

With the GBZ, we can thoroughly understand conventional non-Hermitian physics and NHSE. This framework also facilitates investigations into more complex systems where NHSE interacts with other degrees of freedom, such as critical NHSE, exceptional non-Hermitian states, and new phases and phase transitions unique to the non-Hermitian regime.

\section{Critical phase transitions and entanglement entropy}
Within the realm of condensed matter physics, the study of phases and phase transitions is central to understanding the properties and behaviours of physical systems. A phase of matter is characterized by its physical properties or the arrangement of particles, and a phase transition is the transition between different phases of matter identified by their distinct physical properties. Some simple examples include ice melting and the ferromagnetic transition as the temperature varies.

Conventionally, phase transitions are divided into two categories based on the order parameter, which is a quantity that indicates the degree of order in physical systems undergoing a phase transition and captures changes in macroscopic properties: first-order phase transitions with a discontinuous order parameter and second-order phase transitions with a continuous order parameter exhibiting discontinuity in its first derivative. In the above examples, the order parameter is the density of matter and the net magnetization.

Second-order phase transitions are generally considered more interesting due to the existence of critical phenomena near the transition point and their implications in related areas. To understand this, we need to introduce the concept of the correlation length, which is fundamental in describing the scale over which fluctuations in a system are correlated. If two points within the system are separated by a distance smaller than the correlation length, their properties (such as spin orientation in a magnetic system) are likely to be strongly correlated. Conversely, if they are farther apart than the correlation length, their properties will be mostly uncorrelated. As a system approaches a critical point during a second-order phase transition, the correlation length diverges, with a critical exponent specific to the system. This infinitely large correlation length at the critical point indicates that fluctuations at one point in the system can influence distant parts of the system. For example, in a ferromagnet near the Curie temperature, the alignment of spins becomes correlated over large distances, leading to the formation of large domains of aligned spins and a global property change.

At the critical point, the system exhibits scale invariance, which means that it is invariant under transformations that change the length scale through the renormalization group (RG)~\cite{PhysRevLett110100402,Boer_2000,RevModPhys77259}. This self-similarity is a hallmark of critical phenomena and explains why universal behaviour can be described by scaling laws. Extreme susceptibility to disturbances is achieved at the critical point, where critical states are extremely sensitive to small changes in external factors, allowing wide applications in signal sensing~\cite{chen2017exceptional,wiersig2014enhancing,zhang2019quantumnoise,Zeng2019enhancedsensitivity,Budich2020NonHermitian,sahoo2022tailoring,hodaei2017enhanced}.

Moreover, scale invariance near a critical point can be generalized to conformal invariance, where critical systems are effectively described by conformal field theory (CFT). This leads to the conventional method of measuring and calculating entanglement entropy (EE) to study and characterize phase transitions, especially in the regime of quantum mechanics.
\\

Quantum phase transitions are of particular interest in condensed matter physics. Unlike classical systems where phase transitions are driven by varying temperature, quantum phase transitions occur at absolute zero, driven by quantum fluctuations and variations in external parameters. For instance, in Mott insulators, the energy band gap between the valence and conduction bands can close by altering electron-electron interactions, leading to a transition from an insulating state to a metallic state. Quantum fluctuations, influenced by multiple degrees of freedom, can result in complex and rich critical behaviours, giving rise to new phases of matter such as superconductivity, superfluidity, Anderson localization, and ordered magnetic phases.

Non-Hermitian quantum systems exhibit distinct critical behaviours compared to Hermitian quantum systems. They feature unique phenomena such as the emergence of exceptional points (EPs) as novel critical points and the non-Hermitian skin effect (NHSE), which influences state transport. These features challenge traditional concepts crucial to Hermitian critical transitions, including band gaps and state localization. Consequently, non-Hermitian systems display phenomena such as bulk-boundary correspondence (BBC) violations, robust directed amplifications, discontinuous Berry curvature, and anomalous transport behaviour. These distinctive characteristics suggest that non-Hermitian systems may exhibit novel critical behaviours not observed in traditional Hermitian systems.

The study of non-Hermitian quantum phase transitions and new phases of matter has advanced significantly since the development of the generalized Brillouin zone (GBZ) and the theoretical framework for non-Hermitian systems in 2018. For example, the emergent non-locality from NHSE allows topological transitions to occur even in the absence of gap closure, with the system size itself acting as a parameter to drive the model into different phases. Additionally, various line or point gap closures are made possible at non-Hermitian critical points.

Moreover, beyond NHSE, a new phase known as the exceptional boundary (EB) state has been identified, arising from the defective nature of non-Hermitian systems near EPs~\cite{cerf1997negative,salek2014negative,chang2020entanglement,lee2022exceptional,tu2022renyi,xue2024topologically,ZOU2024}. At EPs, geometric defectiveness blurs the distinction between occupied and unoccupied bands, and probability non-conservation can lead to fermionic occupancies effectively greater than one. This results in anomalously negative free-fermion entanglement entropy, as revealed through the 2-point function.

\section{Thesis Synopsis}

The rest of this thesis is organized as follows:

In \autoref{ch-prel}, we introduce technical details and mathematical preliminaries relevant to the theoretical description of non-Hermitian systems and critical phase transitions. These sections are designed to assist readers who may not be familiar with this field in understanding the subsequent content. For instance, we list essential tools required for analyzing non-Hermitian systems and explain how the generalized Brillouin zone is constructed and utilized as a successful framework for describing non-Hermitian inhomogeneous systems. Additionally, we approach critical phase transitions from the simplest Ising model to illustrate essential critical properties and how a critical system is characterized.

\autoref{ch-realSpec} investigates the appearance of real spectra in non-Hermitian systems, enforced by the skin effect rather than bulk symmetries. The realness of energy spectra has long been a fundamental focus in physics and holds the potential to inspire new research directions. Current efforts toward realizing stable real-energy non-Hermitian systems primarily rely on PT symmetry. In this work, we present simple ansatz models exhibiting robust real spectra under open boundary conditions (OBC), complementing ongoing efforts toward the stable design of non-Hermitian systems.

\autoref{ch-eedip} reports a new class of non-Hermitian critical transitions distinct from known Hermitian or non-Hermitian phase transitions. This critical phase transition, named the scaling-induced exceptional criticality (SIEC), exhibits dramatic divergent dips into the super-negativity in their entanglement entropy scaling, strongly violating current understanding of the EE in the CFT. This critical system is achieved through multiple differently-oriented weakly-coupled NHSE. Their competition in the skin effect leads to a combined scale-induced critical phase transition and the emergence of a novel type of scale-induced exceptional points (EPs), associated with an unexpected negative dip in the entanglement entropy. This study also advances the GBZ framework to describe and explain the system and the newly identified critical phenomenon.

\autoref{ch-inhomo} introduces spatial inhomogeneity as an additional degree of freedom to conventional non-Hermitian PBC systems, which are characterized by asymmetric but spatially constant hopping strengths. The scaled interplay between inhomogeneity and non-Hermiticity gives rise to a new phase of states in the spectral regimes that emerge from the non-zero difference between maximal and minimal hopping strengths. This phase exhibits numerous unprecedented phenomena, including discontinuous jumps between two branches of position-dependent GBZs and non-trivial topology in 2-component inhomogeneous systems. The phase-space GBZ, defined in a phase space encompassing both momentum and position, is completely constructed to thoroughly describe and analyze the inhomogeneous systems and to successfully explain the emergence and properties of this new phase.

\autoref{ch-bethe} explores the interplay between NHSE and the hyperbolic Bethe lattice, examining the band properties of critical NHSE within two distinct types of Bethe lattices. This study highlights phenomena unique to cNHSE states of Bethe lattices, such as alternate-layer localization acting as a bulk-boundary correspondence. These findings further confirm the potential applications of cNHSE in weak signal sensing and switching, advancing the understanding of the intrinsic nature of cNHSE and paving the way for investigations into cNHSE in higher dimensions.

We conclude the entire thesis, as well as discuss further directions for future research in \autoref{ch-concl}.

\SetPicSubDir{ch-prel}
\SetExpSubDir{ch-prel}

\chapter{Theoretical Preliminaries}
\label{ch-prel}
\vspace{2em}

\section{Mathematical formalism and properties of non-Hermitian systems}
As introduced above, Hermiticity, being one of the fundamental postulates in quantum mechanics (QM), is deeply related to the standard mathematical formalism in QM. The introduction of non-Hermiticity will lead to a necessary revision of the fundamental formalism and properties.

We start by considering a non-Hermitian operator $\hat{A}$ with $\hat{A} \neq \hat{A}^{\dagger}$. Its key difference from a Hermitian operator is that its (right) eigenstates are non-orthogonal but still independent of each other. To prove this, let us consider its $i^{\text{th}}$ (right) eigenstate $\psi_i^R$ from
\begin{gather}
\hat{A}\ket{\psi_{i}^{R}}=a_{i}\ket{\psi_{i}^{R}}\label{right},
\end{gather}
and the Hermitian adjoint of this eigenequation
\begin{gather}
\bra{\psi_{i}^{R}}\hat{A}^{\dagger}=a_{i}^{*}\bra{\psi_{i}^{R}}.
\end{gather}
Projecting the $i^{\text{th}}$ eigenstate onto the $j^{\text{th}}$ eigenstate, we observe that the difference between the conjugate transpose pair, $\hat{A} - \hat{A}^{\dagger}$, in general, does not yield a zero eigenvalue corresponding to $\psi_{i,j}$ and results in a non-zero projection:
\begin{align}
     \bra{\psi_{i}^{R}}\ket{\psi_{j}^{R}}&=\frac{a_{j}-a_i^{\dagger}}{a_{j}-a_i^{\dagger}}\bra{\psi_{i}^{R}}\ket{\psi_{j}^{R}}\notag\\
     &=\frac{\bra{\psi_{i}^{R}}a_j\ket{\psi_{j}^{R}}-\bra{\psi_{i}^{R}}a_{i}^{\dagger}\ket{\psi_{j}^{R}}}{a_{j}-a_i^{\dagger}}\notag\\
     &=\frac{\bra{\psi_{i}^{R}}A\ket{\psi_{j}^{R}}-\bra{\psi_{i}^{R}}A^{\dagger}\ket{\psi_{j}^{R}}}{a_{j}-a_i^{\dagger}}\notag\\
      &=\frac{\bra{\psi_{i}^{R}}(A-A^{\dagger})\ket{\psi_{j}^{R}}}{a_{j}-a_i^{\dagger}}\neq 0.
\end{align}

However, ${\ket{\psi_i^R}}$ are linearly independent and form a complete basis set spanning the Hilbert space, as the number of these basis elements equals the dimension of the space. To demonstrate this, we introduce a set of right eigenstates of $\hat{A}^{\dagger}$, or equivalently, a set of left eigenstates of $\hat{A}$
\begin{gather}
\hat{A}^{\dagger}\ket{\psi_{i}^{L}}=a_{i}^{*}\ket{\psi_{i}^{L}}
\Rightarrow \bra{\psi_{i}^{L}}\hat{A}=a_{i}\bra{\psi_{i}^{L}}, \label{left}
\end{gather}

which are necessary for establishing the essential bi-orthogonality relation with ${\ket{\psi^{R}}}$:
\begin{align}
&\bra{\psi_{i}^{L}}a_i\ket{\psi_{j}^{R}}=\bra{\psi_{i}^{L}}\hat{A}\ket{\psi_{j}^{R}}=\bra{\psi_{i}^{L}}a_j\ket{\psi_{j}^{R}}\notag\\
\Rightarrow & \bra{\psi_{i}^{L}}\ket{\psi_{j}^{R}}=\delta_{ij}\label{biortho}
\end{align}

considering both Eq.~\ref{right} and Eq.~\ref{left}.

To prove the linear independence of ${\ket{\psi^{R}}}$~\cite{brody_information_2013}, assume in contradiction that ${\ket{\psi^{R}}}$ are not linearly independent. Then there exists a set of ${c_i}$ with $\sum_i |c_i|^2=1$ such that $\sum_i c_i\ket{\psi_i^R}=0$. Applying ${\ket{\psi^L}}$ to the left, we get
\begin{align}
0=\bra{\psi_j^L}\left[\sum_i c_i\ket{\psi_i^R}\right]=c_i \bra{\psi_i^L}\ket{ \psi_i^R}.
\end{align}

Since $\psi_i^L \psi_i^R \neq 0$ from Eq.~\ref{biortho}, $c_i$ must be zero for all $i$, which contradicts the assumption. A similar argument applies to the set ${\ket{\psi^{L}}}$.

With the linear independence and bi-orthogonality of ${\ket{\psi^{L, R}}}$, the completeness relation for non-Hermitian operators is restored~\cite{brody_information_2013}:
\begin{align}
\sum_i\frac{\ket{\psi_i^R}\bra{\psi_i^L}}{\bra{\psi_i^L}\ket{\psi_i^R}}=I,
\end{align}

where $P_i=\frac{\ket{\psi_i^R}\bra{\psi_i^L}}{\bra{\psi_i^L}\ket{\psi_i^R}}$ is the non-Hermitian projection operator that satisfies $P_iP_j=\delta_{ij}P_j$.

Furthermore, the eigenvalues of non-Hermitian operators are generally complex:
\begin{align}
&\bra{\psi_{i}^{R}}(\hat{A}-\hat{A}^{\dagger})\ket{\psi_{i}^{R}}\notag\\
=&\bra{\psi_{i}^{R}}\hat{A}\ket{\psi_{i}^{R}}-\bra{\psi_{i}^{R}}\hat{A}^{\dagger}\ket{\psi_{i}^{R}}\notag\\
=& (a_i-a_i^{*})\bra{\psi_{i}^{R}}\ket{\psi_{i}^{R}}\neq 0,
\end{align}

where $a_i\neq a_i^{*}$ in general, unless special symmetries (e.g., PT symmetry) are present.

For a non-Hermitian quantum system with non-Hermitian Hamiltonian $H\neq H^{\dagger}$ having complex eigenenergies $E_n$, its time evolution operator $U=e^{iHt}$ is generally non-unitary: $U^{-1}\neq U^{\dagger}$, indicating that the system is not time-reversible anymore.

Based on the formalism described above, we are able to transition to the non-Hermitian bi-orthogonal Hilbert space~\cite{brody_information_2013} with the inner product defined as follows: For two states $\ket{\phi_1}=\sum_i c_i\ket{\psi_i^R}$ and $\ket{\phi_2}=\sum_i d_i\ket{\psi_i^R}$, their inner product is defined to be
\begin{align}
\bra{\phi_1}\ket{\phi_2}=\sum_{i,j}c_i^{}d_j\bra{\psi_i^L}\ket{\psi_j^R}=\sum_{i}c_i^{}d_i,
\end{align}

with the normalization $\bra{\psi_i^L}\ket{\psi_i^R}=1$.

In summary, the inner product convention defined in the non-Hermitian Hilbert space is expressed as
\begin{align}
\bra{\psi}G\ket{\phi},
\end{align}

where $G= \sum_i\frac{\ket{\psi_i^L}\bra{\psi_i^L}}{\bra{\psi_i^L}\ket{\psi_i^L}}$ represents one choice for the geometric matrix.

The expectation value of an arbitrary operator $\hat{O}$ in a pure state $\psi=\sum_i\psi_i^R$ with respect to the bi-orthogonal inner product is then defined as $\bra{\psi^L}\hat{O}\ket{\psi}/\bra{\psi^L}\ket{\psi}$ with $\ket{\psi^L}=\sum_i\psi_i^L$.

The bi-orthogonal basis is, in general, more convenient and physically meaningful in expressing non-Hermitian observables. However, it is challenging to implement experimentally due to the non-equal right and left eigenstates. Despite this, some experimental techniques, such as weak measurements and adiabatic measurements, have been developed to address the issue~\cite{PhysRevLett.77.983,PhysRev99022111,Matzkin_2012,PhysRev99022111}. Conventional orthogonal bases ${\ket{\psi^R}}$ remain widely used in the study of non-Hermitian physics.

\subsection{Non-Hermitian systems and the skin effect}
\begin{figure}
	\centering
	\includegraphics[width=.6\linewidth]{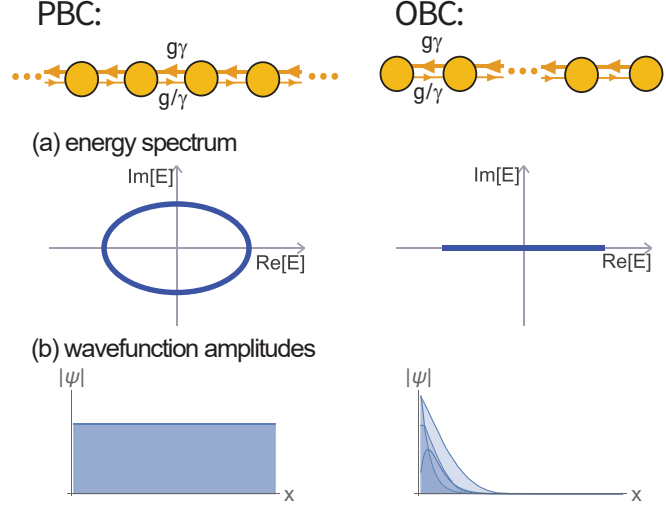}
	\caption{\textbf{Comparison between periodic boundary condition (PBC) and open boundary condition (OBC):} The one-dimensional non-Hermitian Hatano-Nelson chain consists of only nearest-neighbor hoppings, with right (left) hopping strength $t_R=\frac{g}{\gamma}$ and $t_L=g\gamma$. The non-unit Hermiticity factor $\gamma$ ($\gamma=0.8$ in the plot) introduces a directional bias in the hopping strength. (a) Under PBC, the eigenenergies form a complex spectral loop as the Bloch momentum becomes complex in non-Hermitian systems, $k\to k+i\log(\gamma)$ Eq.~\ref{PBCloop}. In contrast, the OBC chain has a purely real spectrum. (b) Random wavefunctions of the HN chains are plotted. For the closed PBC chain, Bloch wavefunctions are restored Eq.~\ref{PBCstate}; for OBC chains, boundary localization occurs for all states, which is the non-Hermitian Skin Effect.
}
	\label{obcpbclatticee}
\end{figure}
To elaborate on non-Hermitian quantum systems in explicit detail, we consider the Hatano-Nelson model, which is a one-dimensional non-Hermitian tight-binding chain with nearest-neighbor non-reciprocal hoppings. It does not have any symmetry and serves as the simplest example of a non-Hermitian lattice, with a Hamiltonian under the OBC or PBC given by
\begin{align}
H_{\text{HN,OBC}}&=\sum_{x=1}^{L-1}t_R\ket{x+1}\bra{x}+t_L\ket{x}\bra{x+1},\\
H_{\text{HN,PBC}}&= H_{\text{HN,OBC}}+t_R\ket{1}\bra{L}+t_L\ket{L}\bra{1},
\end{align}

where $x$ is the index of a site, $t_R=\frac{g}{\gamma}$ and $t_L=g\gamma$ are the right and left hopping terms, with $g$ being a constant hopping amplitude and $\gamma$ being the non-Hermiticity factor. The condition $\gamma \neq 1$ introduces non-reciprocal directional hopping strength to the chain, as shown in Fig. \ref{obcpbclatticee}.

The PBC system has an infinitely long chain or is effectively described by a chain of sites forming a loop. The introduced non-Hermiticity does not break the spatial periodicity, and we can still write the Hamiltonian in momentum space:
\begin{align}
H_{\text{HN,PBC}}(k)=t_R e^{-i k}+ t_L e^{i k},
\end{align}

where $k=2\pi n /L$ is the Bloch wave momentum. The eigenenergies, solved from the characteristic equation
\begin{align}
\text{Det}[H(k)-E\mathcal{I}]=0,
\end{align}

give
\begin{align}
E=t_R e^{i k}+ t_L e^{-i k}=2g\cos\left(k+i\log \gamma \right),\label{PBCloop}
\end{align}

which is equivalent to a Hermitian Bloch system with complex wave momentum $k\to k+i\log \gamma$, as shown by the spectrum in Fig. \ref{obcpbclatticee} (a). The wavefunctions are then solved as
\begin{align}
\psi_{x}=\gamma^x e^{i[k+i \log\gamma]x}=e^{ik x},\label{PBCstate}
\end{align}

which restores the Bloch states. This is intuitively expected through observing the tight-binding chain in Fig. \ref{obcpbclatticee}. Although electrons within the chain are more probable towards one of the directions, the directional pumping cancels out with itself through a circular chain under PBC.

However, under the OBC, the directional pumping results in all states accumulating at the boundary, as shown by the example states in Fig. \ref{obcpbclatticee} (b). This is the non-Hermitian skin effect (NHSE). It is interesting because a single change in the boundary site could result in a drastic change in the bulk of all states. The rigorous description of non-Hermitian systems under OBC, and the NHSE, requires the construction of the generalized Brillouin zone (GBZ), which we will introduce in the next section.

\subsection{Exceptional points}
\begin{figure}
	\centering
	\includegraphics[width=.75\linewidth]{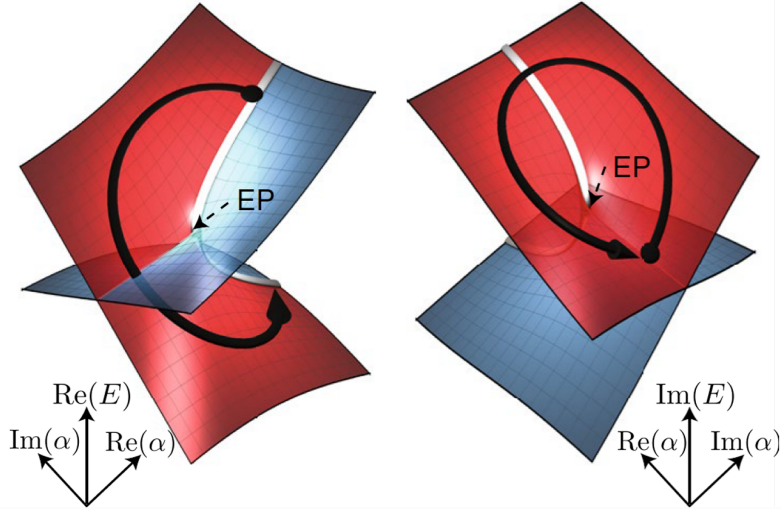}
	\caption{\textbf{Exceptional points on the two-sheeted Riemann surface of complex energy:}  Complex energy of a non-Hermitian system near the exceptional point is plotted. Tracing a loop to enclose the exceptional point, i.e., the black line in the figure, eigenenergy $E+$
 swaps with $E_-$. The image is adapted from Ref.~\cite{Özdemir2019}.
}
	\label{eps}
\end{figure}
As stated in \autoref{ch-intro}, exceptional points (EPs) are a unique phenomenon associated with non-Hermitian systems where both eigenvalues and eigenenergies coalesce, and the Hamiltonian becomes defective at the EPs.

To elaborate on its definition and properties, consider a minimal non-Hermitian Hamiltonian~\cite{heiss_physics_2012}
\begin{align}
H(\lambda)&=H_0+\lambda V\\
&=\begin{pmatrix}
\omega_1&0\\
0& \omega_2
\end{pmatrix} +\lambda \begin{pmatrix}
\epsilon_1 & \delta_1\\
\delta_2 & \epsilon_2
\end{pmatrix},
\end{align}

where $\lambda$ is a parameter changing in its complex parameter space, $\omega_{1,2}$ and $\epsilon_{1,2}$ determine the non-interacting energies $E_{1,2}=\omega_{1,2}+\lambda \epsilon_{1,2}$, and $\delta_{1,2}$ are the interactions repelling the two levels to avoid crossing. $H(\lambda)$ is only Hermitian if $\omega_{1,2}$, $\epsilon_{1,2}$, and $\lambda$ are real with $\delta_1=\delta_2^{*}$.

By assuming non-zero interacting terms $\delta_{1,2}\neq 0$, the system has two EPs where the two levels coalesce at these specific values of $\lambda=\lambda_{1,2}$:
\begin{align}
&\lambda_1=\frac{-i(\omega_1-\omega_2)}{i(\epsilon_1-\epsilon_2)-2\sqrt{\delta_1\delta_2}},\
&\lambda_2=\frac{-i(\omega_1-\omega_2)}{i(\epsilon_1-\epsilon_2)+2\sqrt{\delta_1\delta_2}},
\end{align}
calculated from the fact that EPs occur where the Hamiltonian is not diagonalizable and the energy levels have a square root singularity as a function of $\lambda$, with the energy levels solved from the eigenfunction of $H$ to be:
\begin{align}
E_{1,2}(\lambda)=\frac{1}{2}(\omega_1+\omega_2)+\lambda(\epsilon_1+\epsilon_2)\pm\sqrt{(\epsilon_1-\epsilon_2)^2+4\delta_1\delta_2}\sqrt{(\lambda-\lambda_1)(\lambda-\lambda_2)}.\label{sqrt}
\end{align}

We can confirm that $\lambda_{1,2}$ are indeed EPs by calculating their respective degenerate eigenenergy:
\begin{align}
E(\lambda_{1,2})=\frac{\epsilon_1\omega_2-\epsilon_2\omega_1\mp i\sqrt{\delta_1\delta_2}(\omega_1+\omega_2)}{\epsilon_1-\epsilon_2\mp 2 i\sqrt{\delta_1\delta_2}},
\end{align}
and the single eigenstate that exists:
\begin{align}
\ket{\psi_1}=\begin{pmatrix}
\frac{i\delta_1}{\sqrt{\delta_1\delta_2}}\\
1
\end{pmatrix} \text{ for }\lambda_1 \text{ and }\ket{\psi_2}=\begin{pmatrix}
-\frac{i\delta_1}{\sqrt{\delta_1\delta_2}}\\
1
\end{pmatrix} \text{ for }\lambda_2.
\end{align}

There are several main mathematical properties resulting from the square-root behaviour Eq.~\ref{sqrt}~\cite{heiss_physics_2012}.

Firstly, energy levels having an infinite derivative with respect to $\lambda$ at the EPs would be highly sensitive to the interaction parameters $\delta_{1,2}$ in the vicinity of EPs.

Secondly, the singular nature of EPs leads to unusual behaviours in the eigenenergies and eigenstates when encircling the EPs. As Fig. \ref{eps} suggests, encircling $\lambda_{1,2}$ will interchange the two energy levels. Moreover, consider the two linearly independent right eigenstates of $H$ near but not at the EPs $\ket{\phi_{1,2}^R}$ and their bi-orthogonal left counterpart $\bra{\phi_{1,2}^L}$. The normalization factor that enforces $\bra{\phi_i^L}\ket{\phi_i^R}=1$ tends to infinity as $\sim 1/(\lambda-\lambda_i)^{1/4}$ approaches the EPs. The fourth root behaviour leads to a necessary four-round pattern for a normalized eigenstate to be restored when encircling an EP (assuming counterclockwise encircling):
\begin{align}
\ket{\phi_1}\to -\ket{\phi_2} \to -\ket{\phi_1}\to \ket{\phi_2} \to \ket{\phi_1}.
\end{align}

An additional sign reversal is required if changing to clockwise encircling, suggesting a broken chiral symmetry.

Lastly, the defectiveness~\cite{lee_exceptional_2022} of EPs and the resulting single eigenstate blur the distinction between occupied and unoccupied states when considering the fermionic propagator $\langle c_{x_1,\alpha}^{\dagger},c_{x_2,z}\rangle=\bra{x_1,\alpha}P\ket{x_2,z}=\sum_k P^{\alpha z}(k)e^{ik(x_1-x_2)}$ with $P(k)$ being the biorthogonal projector of the occupied state in momentum space. Furthermore, when approaching EPs with singular projectors, the divergent behaviour of the projectors $P(k)$ and the propagators is not only scaled by the short-ranged distance difference $x_1-x_2$, but also scaled by the smallest lattice momentum point scales $k_0\sim \pi/L$~\cite{lee_exceptional_2022}, with $L$ being the finite number of sites in the system, which can be large. The large $L$ amplifies the asymmetry of the Hamiltonian and results in a highly asymmetric propagator, where a real-space cut (serving as a proposed boundary analogous to the OBC in NHSE) will truncate one of the elements of the propagator $\langle c_{x_1,+}^{\dagger},c_{x_2,-}\rangle$ much more than the other $\langle c_{x_1,-}^{\dagger},c_{x_2,+}\rangle$, leading to robust Exceptional Boundary (EB) states.

\section{Generalized Brillouin zone and non-Hermitian topology}
The generalized Brillouin zone (GBZ) is an extension of the conventional Brillouin zone, specifically designed for non-Hermitian systems under open boundary conditions (OBC). The GBZ framework is particularly important because it serves as a cohesive theoretical foundation for non-Hermitian physics, allowing us to thoroughly understand conventional non-Hermitian systems and the non-Hermitian skin effect (NHSE) and to investigate any unique physical phenomena associated with non-Hermitian systems. Specifically, the GBZ offers an effective framework for analyzing the eigenenergies and wavefunctions of a system's Hamiltonian, analogous to how Bloch theory describes Hermitian lattices under periodic boundary conditions, but within a non-Bloch context. The comprehensive knowledge about the eigenbands provided by the GBZ enables us to investigate various system properties, including predictions of observable measurements, entanglement entropy between sub-lattices, and topological features.

In non-Hermitian systems, the presence of complex eigenenergies and the NHSE leads to non-Bloch localization of all eigenstates to the boundary, making the traditional Brillouin zone insufficient. The GBZ is then constructed to analytically quantify the skin effects and formulate the band theory of non-Hermitian OBC systems, with the key idea of transforming the real momentum $k$ as the skin depth of the BZ, to a complex momentum $k+i\kappa$, as GBZ skin depths of varying amplitude. This complex skin depth effectively describes conventional non-Hermitian OBC systems.

\subsection{Example: Hatano-Nelson chain}
Before discussing the general GBZ formalism, we first explore the GBZ construction of the non-Hermitian Hatano-Nelson chain as an example. Again, consider the Hamiltonian
\begin{align}
H_{\text{OBC}}=&\sum_{x=1}^{L-1}g/\gamma\ket{x}\bra{x+1}+g\gamma\ket{x+1}\bra{x},\notag\\
H_{\text{PBC}}=&H_{\text{OBC}}+g/\gamma\ket{L}\bra{1}+g\gamma\ket{1}\bra{L},
\end{align}
where $g\gamma, g/\gamma\in \mathbb{R}$ are the asymmetric hopping amplitudes and $L$ is the system size. For simplicity, we assume $g$ and $\gamma$ to be positive.

Revisiting Hermitian systems where $\gamma=1$ under periodic boundary conditions, the Bloch Hamiltonian in momentum space with momentum $k$ is
\begin{align}
H_{\text{Hermitian,PBC}}(k)=g e^{-ik}+g e^{ik}.
\end{align}
The terms $e^{ik}$ for all $k$ form the Brillouin zone in the complex space, which is a unit circle. The energy spectrum of this system is determined as the eigenvalues via the characteristic equation
\begin{align}
\text{det}[H(k)-E I]=0,
\end{align}
giving
\begin{align}
E(k)=2g \cos k.
\end{align}
This provides all band information for the Hermitian Bloch system.

In non-Hermitian systems with $\gamma\neq 1$, we replace the BZ term $e^{ik}$ with a complex parameter $z$:
\begin{align}
H_{\text{PBC}}(z)=g \frac{1}{z}+g z.
\end{align}
Its characteristic equation $\text{det}(H(z)-E I)=0$ gives
\begin{align}
z^2-\frac{E}{g}z+1=0,\label{charac}
\end{align}
with the eigenenergies
\begin{align}
E(z)=g \frac{1}{z}+g z.
\end{align}
From Eq.~\ref{charac}, there are two solutions $z_1$ and $z_2$ as roots. For each eigenenergy $E(z)$, the GBZ $z$ is determined from the GBZ condition $|z_1|=|z_2|$. We derive this condition by examining the details of the wavefunctions.

Consider the $n^{\text{th}}$ eigenstate of the Hamiltonian $\Psi_n$ obeying $H\Psi_n=E_n\Psi_n$, expressed as $\Psi_n=(\psi_1, \psi_2,..., \psi_L)^{\text{T}}$. The Schrödinger equation $H\Psi_n=E_n\Psi_n$ is written as
\begin{align}
\frac{g}{\gamma} \psi_{x-1}+g\gamma\psi_{x+1}=E\psi_{x},\label{rec}
\end{align}
for $x=1, 2,..., L$. Based on the theory of linear difference equations, the general solution of the recursion equation Eq.~\ref{rec} is
\begin{align}
\psi_x=z_1^x\phi^{(1)}+z_2^x\phi^{(2)},
\end{align}
where $z_{j}$ with $j=1,2$ fulfills
\begin{align}
\frac{g}{\gamma z_j} +g\gamma z_j=E.\label{Espec}
\end{align}

Now, if we introduce the open boundary condition (OBC) $\psi_{0}=0$ and $\psi_{L+1}=0$, the former gives
\begin{align}
&\phi^{(1)}=-\phi^{(2)},
\end{align}
and the latter results in
\begin{align}
z_1^{L+1}=z_2^{L+1}.
\end{align}
Rewriting the above equations gives
\begin{align}
\frac{z_1}{z_2}=e^{2i\theta_{n}}\text{, }\left( \theta_n=\frac{n\pi}{L+1}\text{, }n=1,...,L \right),
\end{align}
which specifies that
\begin{align}
|z_1|=|z_2|.\label{GBZcond}
\end{align}
Moreover, Eq.~\ref{Espec} leads to
\begin{align}
&\frac{g}{\gamma z_1} +g\gamma z_1=\frac{g}{\gamma z_2}+g\gamma z_2\notag\\
\Rightarrow & z_1 z_2=\frac{1}{\gamma^2},
\end{align}
and hence $|z_{1,2}|=\frac{1}{\gamma}$. We can then solve the GBZ of this Hatano-Nelson chain to be
\begin{align}
&z_{n,1}=\frac{1}{\gamma}e^{i\theta_{n}}\text{, }z_{n,2}=\frac{1}{\gamma}e^{-i\theta_{n}}.
\end{align}
The band structure, including the eigenenergies and wavefunctions of the system, can be analytically solved as
\begin{align}
&\psi_{n,x}=(\frac{1}{\gamma} e^{i\theta_n})^x-(\frac{1}{\gamma} e^{-i\theta_n})^x,\\
&E_n=2g \cos \theta_n.
\end{align}
These $E_n$ form discrete energy levels in a finite open chain. If the system size $L$ becomes larger, these energy levels become dense and eventually form a spectrum in the thermodynamic limit $L\to \infty$. We can see that the continuum bands at the thermodynamic limit can be directly obtained from Eq.~\ref{GBZcond} where Eq.~\ref{Espec} gives
\begin{align}
&z_1=\frac{1}{\gamma} e^{i\theta} \text{, }z_2=\frac{1}{\gamma} e^{-i\theta},\\
&E=2g \cos \theta.
\end{align}
which are the continuum bands with $\theta\in\mathbb{R}$. $|z_1|=|z_2|$ is said to be the condition for the GBZ of the continuum bands.
\subsection{Formalism of generalized Brillouin zone}

\begin{figure}
	\centering
	\includegraphics[width=.9\linewidth]{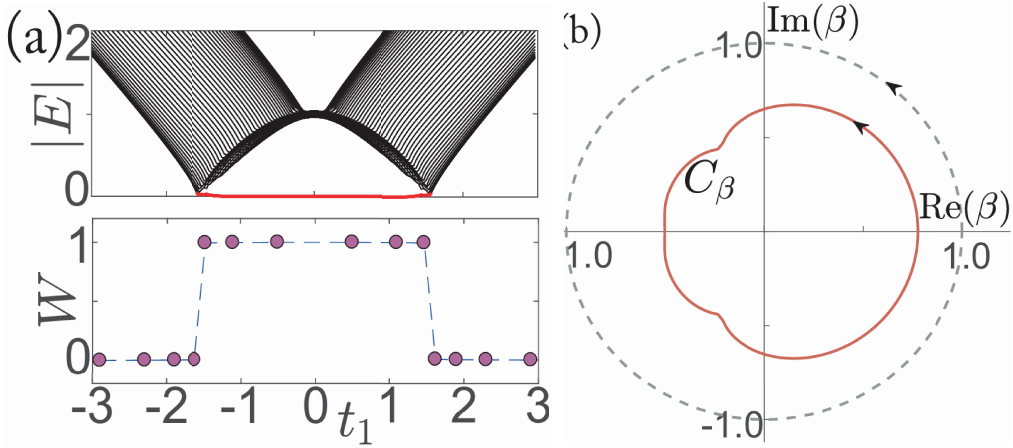}
	\caption{\textbf{The generalized Brillouin zone (GBZ):} (a) (Top) The existence of robust zero modes (red line) in the diagram of the OBC energy |E| as a function of the hopping parameter $t_1$  (bottom) is effectively described by the GBZ and the topological invariant $W$ developed from the GBZ. (b) The complex GBZ term $\beta$ (denoted as $z$ in this thesis)  forms the non-unitary GBZ loop in the thermodynamic limit. The dashed line represents the conventional unit BZ for comparison. The figure is adapted from Ref.~\cite{yao_edge_2018}, Yao's 2018 work, which introduced the GBZ for the first time. Note that the hopping parameters in the figure are defined differently than those in this thesis.
}
	\label{zeromode}
\end{figure}

After the simple example of the non-Hermitian HN chain, we look at the general formalism of GBZ~\cite{jin_bulk-boundary_2019}. Consider the non-Hermitian Hamiltonian of a one-dimensional tight-binding model with a $q$ number of degrees of freedom within each unit cell and a range of hopping of $N$:
\begin{align*}
H=\sum_{x}^{L}\sum_{i=-N}^{N}\sum_{\mu,\nu=1}^{q}t_{i,\mu\nu}c^{\dagger}_{n+i,\mu}c_{n,\nu},
\end{align*}
where $x$ ($L$) is the index (total number) of unit cells, $i$ ($N$) is the (largest) hopping length in terms of the number of unit cells, $\mu,\nu$ ($1$) are the index (total number) of degrees of freedom inside a unit cell. Similarly, we consider its $n^{th}$ eigenstate $\Psi=(\psi_{1,1},...,\psi_{1,q},\psi_{2,1},...,\psi_{2,q},...,\psi_{L,q})^{T}$. Note that we eliminate the index $n$ for simpler notation. Substituting this eigenstate into the eigen-equation $H\Psi=E\Psi$ leads to the bulk equation
\begin{align*}
&\sum_{i=-N}^{N}\sum_{\nu=1}^{q}t_{i,\mu\nu}\psi_{x-i,\nu}=E\psi_{x,\mu}.
\end{align*}
The recursion equation leads to the linear combination of each element number $\psi_{n,\mu}$:
\begin{align*}
&\psi_{x,\mu}=\sum_{j=1}^{2qN}\phi_{x,\mu}^{(j)},\\
&\phi_{x,\mu}^{(j)}=z_{j}^{x}\phi_{\mu}^{(j)},
\end{align*}
where $2qN$ is the total number of solutions from the degrees of freedom involved in this set of recursion equations. By imposing that $\phi_{x,\mu}^{(j)}$ is an eigenstate, substituting the above expression into the bulk equation leads to
\begin{align}
&\sum_{\nu=1}^{q}\mathcal{H}_{\mu\nu}^{(j)}(z_{j})\phi_{\nu}^{(j)}=E\phi_{\mu}^{(j)}\label{gbz11},\\
\text{with } &\mathcal{H}_{\mu\nu}^{(j)}(z_{j})=\sum_{i=-N}^{N}t_{i,\mu\nu}(z_{j})^{-i}.\notag
\end{align}
Eq.~\ref{gbz11} has $q*[N-(-N)]=2qN$ equations corresponding to $2qN$ number of $z_j$ for each $\psi_{x,\mu}$ as expected.

Rewrite Eq.~\ref{gbz11} as
\begin{align*}
&\phi_{\mu}^{(j)}=f_{\mu}(z_{j}, E, \mathcal{S})\phi_{1}^{(j)}.
\end{align*}
Consider the boundary condition
\begin{align}
&\psi_{x,\mu}=0 \text{ for } L+1\leq x\leq L+N\text{ and } 1\leq \mu\leq q,\\
&\psi_{x,\mu}=0 \text{ for } 0\leq x\leq 1-N\text{ and } 1\leq \mu\leq q,\label{gbz31}
\end{align}
we have $2qN$ equations for each boundary equation
\begin{align*}
\sum_{j=1}^{2qN}(z_{j})^{1-n}f_{\mu}(z_{j}, E, \mathcal{S})\phi_{1}^{(j)}=0,\\
\sum_{j=1}^{2qN}(z_{j})^{n}f_{\mu}(z_{j}, E, \mathcal{S})(z_{j})^{L}\phi_{1}^{(j)}=0,
\end{align*}
and they form an eigen-equation with eigenstate $(\phi_{1}^{(1)},\phi_{1}^{(2)},...,\phi_{1}^{(2qN)})^{T}$ and eigenvalue $0$. The determinant of this eigen-equation is an algebraic equation for $z$ with an order of $2qN$ and solving it gives the $2qN$ amount of solutions $z_{j}$, in which we number them as
\begin{align*}
|z_{1}|\leq |z_{2}| \leq ... \leq |z_{2qN}|.
\end{align*}

In the $L\to\infty$ limit, the determinant has only one asymptotic leading term that is proportional to $\prod_{j=Nq+1}^{2Nq}(z_{j})^{L}$ if $|z_{qN}|\neq |z_{qN+1}|$; or two asymptotic leading terms proportional to $\prod_{j=Nq+1}^{2Nq}(z_{j})^{L}$ and $(z_{Nq})^{L}\prod_{j=Nq+2}^{2Nq}(z_{j})^{L}$ respectively if $|z_{qN}|\neq |z_{qN+1}|$. The former would have the boundary requirement being independent of $L$, which is not physical and eliminated.
\subsection{Non-Hermitian topological phase transition and the bulk-boundary correspondence }

\begin{figure}
	\centering
	\includegraphics[width=.75\linewidth]{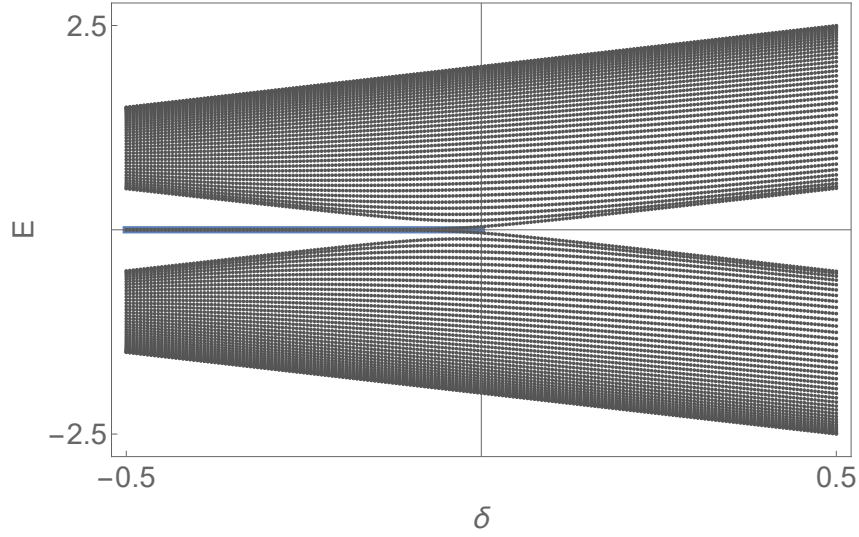}
	\caption{\textbf{The existence of topological zero modes in the topological phase of the Hermitian SSH model:} Setting $g_1=1+\delta$ and $g_2=1$ results in the OBC energy as a function of $\delta$. The zero edge state appears when $\delta\leq1$.
}
	\label{zeromodeH}
\end{figure}

In topological band theory, quantum physical states are classified into distinct topological phases with discrete topological invariants associated with their band structure and parameter space. Any topological phase transition necessarily involves a change in the corresponding topological invariant. To be more detailed, states that are adiabatically disconnected from each other are considered to belong to different topological phases, or are topologically distinct, where a phase transition necessarily occurs when changing from one to the other.

Topological physics is deeply studied due to the newly found unique properties of the new topological phases of matter and the associated phase transition points, which are distinct from the conventional critical behaviour of all phases of matter we have known. Examples of applications of such new phenomena include topological insulators and signal measurements. Hence, investigating proper topological invariants that correctly describe new topological phases with unique properties is of deep interest.

An example topological invariant for Hermitian systems is the winding number $\nu$ of a double-band SSH model, with a Hamiltonian
\begin{align}
H_{\text{SSH,OBC}}&=\sum_n^L(g_1\hat{c}_{A,x}^{\dagger}\hat{c}_{B,x}+\text{h.c.})+\sum_n^{L-1}(g_2\hat{c}_{B,x}^{\dagger}\hat{c}_{A,x+1}+\text{h.c.}),\\
H_{\text{SSH,PBC}}&=H_{\text{SSH,OBC}}+(g_2\hat{c}_{B,L}^{\dagger}\hat{c}_{A,1}+\text{h.c.}),
\end{align}
where $\hat{c}_{j,x}$ ($\hat{c}_{j,x}^{\dagger}$) is the annihilation (creation) operator of a particle on the site $j=A, B$ of the $x^{th}$ unit cell, $L$ is the total number of unit cells, and h.c. stands for the Hermitian conjugate pair of the preceding term.

The PBC SSH model, having the Bloch Hamiltonian
\begin{align}
H_{\text{SSH,PBC}}(k)\begin{pmatrix}
0 & g_1+g_2e^{-ik}\\
g_1+g_2e^{ik} &0
\end{pmatrix},
\end{align}
has the eigenenergies $E_{\pm}(k)$ and eigenstates $\ket{\Psi_{\pm}(k)}$ given by
\begin{align}
E_{\pm}(k)&=\pm |h|,\\
\ket{\Psi_{\pm}(k)}&=\frac{1}{\sqrt{2}}\left(\pm e^{i\varphi},1\right)^{T},
\end{align}
with $h=g_1+g_2 e^{-ik}$, $|h|=\sqrt{g_1^2+g_2^2+2g_1g_2 \cos k}$, and $e^{i\varphi}=\frac{h}{|h|}$. The eigenenergies form two gapped continuum bands of the spectrum with a band gap $2\Delta=2(E_+-E_-)$ that closes at $|g_1|=|g_2|$. It was found that by varying the hopping parameter, e.g., $g_1$, the properties of systems before and after the band-closing point demonstrate drastically different topological properties [see Fig. \ref{zeromodeH}]. This topological phase transition at $|g_1|=|g_2|$ is characterized by the winding number, which is defined as
\begin{align}
\nu_{\pm}=&\frac{i}{\pi}\int_{-\pi}^{\pi}\bra{\Psi_{\pm}}\frac{d}{dk}\ket{\Psi_{\pm}(k)}dk,
\end{align}
and calculated to be
\begin{align}
\nu_{\pm}=&-\frac{1}{2\pi}\int_{-\pi}^{\pi}\frac{d\varphi(k)}{dk} dk,
\end{align}
which is essentially how $\varphi(k)$ accumulates along the BZ, or equivalently, the number of loops $e^{i\varphi}$ makes around the origin of the complex plane along the BZ. This indeed has distinct values of one for $g_1<g_2$ and zero for $g_2<g_1$. We say that the system is topologically non-trivial for the parameters $g_1<g_2$ and topologically trivial for $g_2<g_1$. The winding number, which is calculated from the PBC systems, is a purely bulk property, but effectively characterizes the number of edge states, or the zero modes, in the corresponding OBC systems, as shown in Fig. \ref{zeromodeH}. This is the well-known bulk-boundary correspondence, which states that a property evaluated in the bulk could imply a property at the edge as a topological result. It indicates the important nature of zero edge states, which is their robustness and topological protection against local disorder or perturbation as long as the bulk band gap does not close, and this stability leads to numerous applications and new research directions~\cite{KANE20133,Yao2018nonH2D,zhang2019correspondence,guo2021analysis,bartlett2021illuminating,zhang2022bulkbulk,cao2021non,Song2019BBC,kunst2018biorthogonal}.

For non-Hermitian systems, as we have discussed, the BZ framework fails, and the non-Hermitian GBZ is introduced to describe the band structure, as shown in Fig. \ref{zeromode}. It is found that by replacing the BZ with the GBZ in the winding number, the BBC is effectively recovered with the winding number integrating along the GBZ correctly characterizing the presence of edge states. The updated GBZ winding number is
\begin{align}
\nu_{\pm}=&\frac{i}{\pi}\oint_{\text{GBZ}}\bra{\Psi_{\pm}(z)}\frac{d}{dz}\ket{\Psi_{\pm}(z)}dz,
\end{align}
where $\ket{\Psi_{\pm}(z)}$ are the eigenstates of the PBC non-Bloch Hamiltonian $H(z)$, and $z$ is solved from the OBC procedure in the previous section and fulfills the condition $z_{qN}=z_{qN+1}$.

\section{Critical scaling and entanglement entropy}
\subsection{Critical phase transition and scale invariance}

\begin{figure}
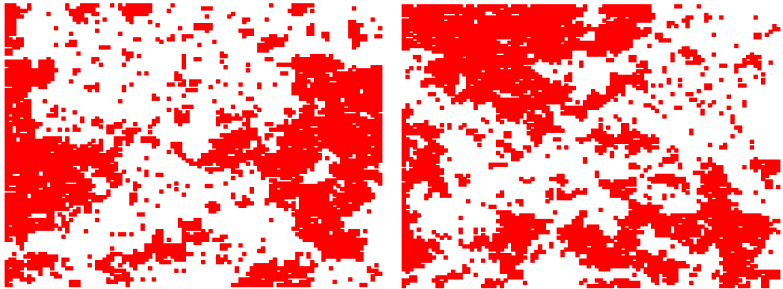

	\centering
	\includegraphics[width=.35\linewidth]{\Pic{pdf}{scaleinv1}}\includegraphics[width=.35\linewidth]{\Pic{pdf}{scaleinv2}}
	\caption{\textbf{Demonstration of scale invariance in the Critical Ising model:} Example spin configuration for the Ising model calculated from a Monte Carlo analysis, with all ``up'' spins colored in red. Left: Direct output of the spins. Right: Output of spins under a scale-factor 3 spin-block transformation. They are effectively indistinguishable from each other macroscopically. The figure is adapted from Ref.~\cite{Dstanford}.
}
	\label{scaleinv}
\end{figure}
As discussed in Sec. \ref{ch-intro}, critical behaviours at transition points between distinct phases of matter are of great interest to condensed matter physicists. The physics within the critical regime is interesting due to unique features such as scale invariance.

We take the Ising model as an example. The Ising model consists of spins $s$ interacting with a nearest-neighbouring hopping $J$. The Hamiltonian is
\begin{align}
H=-J\sum_{<ij>}s_is_j,
\end{align}
where the sum is taken over nearest-neighbor lattice sites and $J$ is the reduced hopping constant, which is inversely proportional to the temperature. The system will have all spins pointing in the same direction at the zero-temperature ground state. However, raising the temperature introduces non-zero thermal fluctuations, forcing the configuration of spins to depart from uniformity. For very low temperatures, the majority of the spins will still be aligned, and the system will have a net magnetization as a macroscopic property. At high temperatures, though, the system's order is completely destroyed, and the spins will be evenly distributed up and down, with zero net magnetization. At some critical temperature $T_c$, the Ising model undergoes a phase transition.

Precisely at the critical temperature, the Ising model exhibits several interesting behaviours, including the property of scale invariance. Scale invariance means that a configuration of spins looks the same at all levels of magnification. More formally, it means that if we trace out the short length degrees of freedom in the partition function, the Hamiltonian for the remaining degrees of freedom will be precisely the same as the original $H$, with an identical coupling. Scale invariance can be rigorously explained by the renormalization group (RG) transformation, where critical systems define a fixed point of the renormalization group. We will not delve deeply into the mathematical details, but will explain scale invariance through a simple demonstration in Fig. \ref{scaleinv}. The left figure represents a typical configuration of spins in the critical Ising model, and the right figure groups each block of 9 spins together as one unit, according to the majority of the individuals within the block, to form a new configuration of spins. The distribution of cluster sizes in the two systems looks strikingly similar. Scale invariance ensures that both figures are statistically identical, in terms of the probability of occupation.

The scale invariance in the Ising model has the following result: The correlation function, which is defined as the expectation value of the product of two spins separated by distance $r$, is written as:
\begin{align}
G(r)=\langle s(r)s(0)\rangle=\text{Tr} ,s(r)s(0)e^{-H}.
\end{align}

If we perform a renormalization group transformation with a scale factor $b$, which is equivalent to grouping together a block of spins, with $b$ spins on each side, into a single unit. Consider a correlation function between two of these units, $r$ blocks apart. This blocked system would give the same two-point correlation function $G(r)$, under scale invariance. On the other hand, we can also write the correlator of the blocks in terms of correlations of individual sites in the original system that make up the blocks. This correlator will consist of correlation functions between the sites in block $A$ and the sites in block $B$, multiplied by a factor accounting for the number of spins in each block, and the interactions between them. Since this factor is insensitive to the distance between the blocks, it can only depend on the scaling parameter $b$. We then write the scale invariance statement as
\begin{align}
G(r) = f(b)G(br).
\end{align}
If we perform one more spin-blocking transformation with another scale factor $a$ on the above system, the scale invariance is further written as $G(r) = f(a)f(b)G(abr)$. This is equivalent to performing the same shift in a single transformation, with factor $ab$: $G(r) = f(ab)G(abr)$, leading to the result $f(ab) = f(a)f(b)$, which implies
\begin{align}
f(b)=b^y,
\end{align}
where $y$ is some constant. Above is usually expressed by defining the scaling dimension $x$
\begin{align}
G(r)=b^{-2x}G(r/b).
\end{align}
Motivated by the scale invariance of the critical Ising model, we have heuristically derived that the correlation function will be a simple power law with no dimensional parameters to form exponential. In other words, it has an infinite correlation length, the system exhibits long-range correlations extending over arbitrarily large distances, and any local change affects the entire system.

The above 1D example is special and simple given its constant term in the Hamiltonian. In higher dimensions or other more complicated critical systems, RG transformations typically give rise to more complicated interactions and cannot be exactly computed in most cases. However, an infinite correlation length and the scale invariance at the second-order phase transition point are always assured. This implies invariance under a more general group of transformations, known as conformal field theory (CFT), and the CFT is commonly used to study and characterize second-order phase transitions, or more commonly, quantum phase transitions.

Scale invariance and conformal invariance, along with the associated conformal field theory, are the key methods of investigating and characterizing quantum phase transitions. We apply methods derived from these theories when studying critical phenomena in non-Hermitian systems in this work.
\begin{figure}
	\centering
	\includegraphics[width=.75\linewidth]{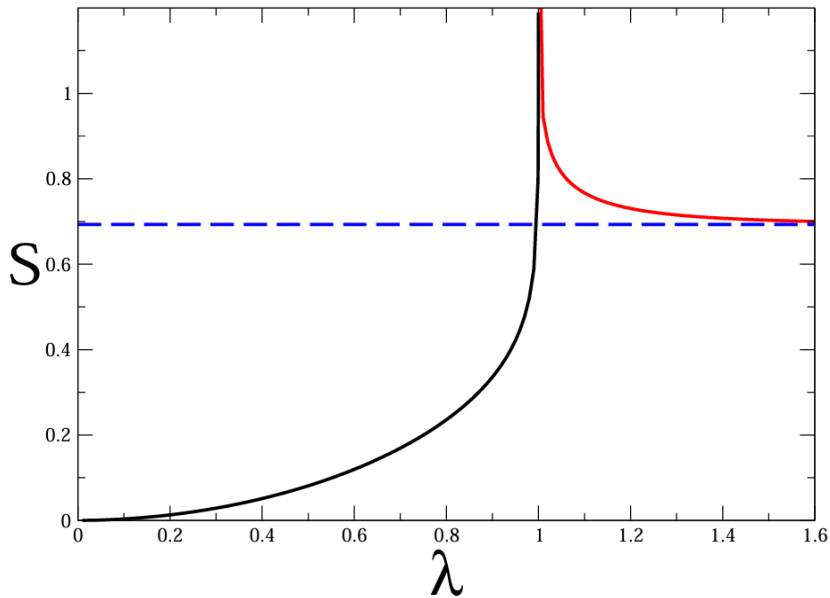}
	\caption{\textbf{Entanglement entropy for the 1D Ising chain in a transverse field:} entanglement entropy is plotted as a function of $\lambda$, where a critical phase transition occurs at $\lambda=1$. The dashed line represents the limit for $\lambda\to\infty$, yielding a pure ferromagnetic ground state with $S=\log(2)$, as expected. The figure is adapted from Ref.~\cite{Calabrese_2004}.
}
	\label{isingEE}
\end{figure}

\subsection{Conformal field theory and entanglement entropy}

The conformal field theory (CFT) is a field theory that is invariant under conformal transformations, which locally include scaling, rotations, and translations. This makes CFT particularly powerful for describing critical points where scale invariance is a hallmark. This CFT description is especially powerful in (1+1) dimensions, where many critical points are exactly solvable using CFT techniques.

Again, we will not delve into the details of CFT but will focus on how CFT characterizes critical phase transitions, introducing the concept of entanglement entropy.

Entanglement entropy is essentially a measure of how much a given quantum state is quantum mechanically entangled. For example, for a pure state $\ket{\Psi}$ with density $\rho_\text{tot}=\ket{\psi}\bra{\Psi}$, or a mixed state at finite temperature $\rho_\text{tot}=\frac{e^{-\beta H}}{\text{Tr}\left[e^{-\beta H}\right]}$ where $\beta=1/(k_B T)$, the system is divided into two parts $A$ and $B$ with the factorized Hilbert space
\begin{align}
H_{\text{tot}}=H_A\otimes H_B,
\end{align}
resulting in a reduced density matrix $\rho_A$ for the subsystem $A$:
\begin{align}
\rho_A=\text{Tr}_B\rho_{\text{tot}}.
\end{align}
The entanglement entropy $S_A$ is defined by
\begin{align}
S_A=-\text{Tr}_A\rho_A\log\rho_A.
\end{align}
The entanglement entropy (EE) obeys the well-known Area Law, which states that the leading divergent term of EE in a ($d+1$) dimensional Quantum Field Theory is proportional to the area of the ($d-1$) dimensional boundary $\partial A$. Equivalently, the EE scales with the boundary between two partitions.

Exceptions occur within ($1+1$) dimensional CFT, where the EE $S_A$ of a subsystem $A$ of length $l$ in an infinite system under CFT is given by~\cite{Calabrese_2004,Vidal_2003}
\begin{align}
S_A=\frac{c}{3}\log(l),
\end{align}
where $c$ is the central charge. It measures the number of degrees of freedom or the ``strength'' of the theory, and different CFTs are characterized by different values of $c$. In other words, we can measure the EE to obtain the central charge and effectively characterize different classes of quantum phase transitions. [See Fig. \ref{isingEE}].
\SetPicSubDir{ch-realSpec}
\SetExpSubDir{ch-realSpec}

\chapter{Real non-Hermitian energy spectra without any symmetry}
\label{ch-realSpec}
\vspace{2em}
The realness of energy spectra is a crucial property in physics research, as it signifies stable systems that are practically realizable in the laboratory. The study of energy spectra's realness has always been fundamental to the development of new fields in physics and can potentially open up new directions of study. For example, as introduced in \ref{ch-intro}, Hermiticity has been a long-standing postulate in quantum systems since it ensures real spectra of the system. Later, after the discovery of non-Hermitian systems protected by PT-symmetry also possessing real eigenenergies, non-Hermitian systems soon came into focus, leading to a surge of new discoveries of novel phenomena, properties, and applications in non-Hermitian physics. Interestingly, the study of non-Hermitian physics, initially motivated by real-energy PT-symmetric non-Hermitian systems, has evolved far beyond the constraints of PT-symmetry. Many such systems, despite their exotic theoretical results, are still limited by their complex energy spectra and the challenges of experimental realization beyond the PT-symmetric phase.

In this work, we investigate how the non-Hermitian skin effect (NHSE) offers an alternative route to achieving real spectra and system stability. We present various ansatz models with different classes of energy dispersions that exhibit large regions in parameter space with real spectra, even in the absence of obvious symmetries. These minimal local models can be readily implemented in non-reciprocal experimental setups, such as electrical circuits with operational amplifiers.

\section{Introduction}

Non-Hermitian systems has recently inspired intense research efforts for their unconventional mathematical properties and physical robustness, such as enlarged topological symmetries~\cite{Kawabata2018nonHclass,Liu2019nonHclass,bergholtz2021exceptional,sayyad2021entanglement,wojcik2021eigenvalue,shiozaki2021symmetry,borgnia2020nonH,park2022nodal}, exceptional point sensing~\cite{chen2017exceptional,wiersig2014enhancing,zhang2019quantumnoise,Zeng2019enhancedsensitivity,Budich2020NonHermitian,sahoo2022tailoring,hodaei2017enhanced}, quantized classical responses~\cite{li2021quantized}, modified bulk-boundary correspondences~\cite{yao2018edge,yang2020non,yokomizo2019non,helbig2020generalized,Yao2018nonH2D,zhang2019correspondence,guo2021analysis,bartlett2021illuminating,zhang2022bulkbulk,cao2021non,Song2019BBC,kunst2018biorthogonal,koch2020bulk,imura2020generalized,qin2022non,jiang2022filling}, unconventional entanglement entropy scaling~\cite{li2021critical,lee2022exceptional,sayyad2021entanglement,chen2022characterizing,chang2020entanglement,zhou2020renormalization,yokomizo2021scaling,pan2021point}, enhanced Rabi oscillations~\cite{lee2020ultrafast,cao2021non,zhou2021dual}, effective non-Hermitian curved space~\cite{lv2021curving,marcus2022pt}. Yet, many of these exciting phenomena are often difficult to probe experimentally due to their intrinsically unstable nature from complex eigenenergies. While real eigenenergies can be symmetry-enforced, i.e., through PT-symmetry~\cite{bender_real_1998,el2018non,stegmaier2021topological,fring2022introduction,xiao2021observation,Schomerus2013topo,Weimann2016topo,zdemir2019ParitytimeSA}, doing so is incompatible with realizing many of the most exotic non-Hermitian phenomena~\cite{ghatak2020observation,helbig2020generalized,xiao2020non,Bouganne2020anomalous,song2019breakup,hou2020topo,longhi2018paritytime,ningyuan2015time,scheibner2020non,zeng2021real,long2021realeigenvalued,kawabata2020real,aharonov1996adiabatic,PhysRev99022111,Li2021protecting}.

In this work, we carefully investigate how the non-Hermitian skin effect (NHSE)~\cite{Martinez2018nonH,okuma2020topological,longhi2021non,peng2022manipulating,liang2022observation,zhu2020photonic,guo2021exact} can also enforce real non-Hermitian spectra, even for lattices whose couplings and momentum-space descriptions do not admit any obvious symmetry. NHSE has been heavily associated with modified bulk-boundary correspondences~\cite{yao2018edge,pan2021point} and, in our context, implies that a system can robustly possess a real spectrum in the presence of a boundary, even though its bulk is unstable with complex eigenenergies. Physically, this is because the directed amplification in a NHSE lattice can be stabilized by the interfering wavepackets from a boundary or spatial inhomogeneity, a mechanism that is unrelated to conventional symmetry protection.

We elucidate this route towards real eigenspectra in terms of the inverse skin depth $\kappa$~\cite{yao2018edge,lee2020unraveling}, which is an additional degree of freedom that mathematically takes the role of imaginary momentum. It physically controls the accumulation and interference of skin states at a boundary, and mathematically replaces the Bloch description of lattice by an effective \emph{surrogate}~\cite{lee2020unraveling} model that can look completely different. Specifically, we shall show that in a bounded lattice, the reality of the spectrum is only destroyed by a spontaneous symmetry breaking process that can occur much later than the explicit symmetry breaking at the Bloch level.

\section{OBC vs PBC spectra}

We first review and distinguish the approaches for computing the eigenenergy spectrum under open vs. periodic boundary conditions (OBCs vs PBCs). Given a generic non-Hermitian Hamiltonian $H(k)$, PBC eigenenergies $\tilde E$ and OBC eigenenergies $\bar E$ are obtained very differently. To find the set of $E\in\tilde E$, we simply solve for eigenenergies $E$ such that characteristic polynomial $P(E,k) = \text{Det}[H(k)-E\,\mathbb{I}]=0$, where $k\in [0,2\pi)$. However, under OBCs, translation invariance is broken, and in general the spectrum is \emph{not} indexed by real momentum $k$. Instead, the OBC spectrum $E\in \bar E$ is given by eigenenergies $E$ that solve
\begin{equation}
    P(p)=P(E,k+i\kappa) = \text{Det}[H(k+i\kappa)-E\,\mathbb{I}]=0,
\end{equation}
and are \emph{degenerate in both} $E$ and $\kappa$. Here $\kappa$ is the imaginary part of the complexified momentum $p=k+i\kappa$ that also represents the inverse decay length (skin depth) of eigenstates viz. $e^{ipx}\sim e^{-\kappa x}$, i.e., taking the role of a length scale~\cite{qi2013exact,gu2016holographic} not present in Hermitian systems. This $\kappa$ degeneracy is required because OBC skin eigenstates have exponential spatial profiles, and we need a superposition of two of them with identical $E$ and $\kappa$ to satisfy OBCs at two arbitrarily separated boundaries~\footnote{As we interpolate between OBCs and PBCs, we observe a peculiar scaling behaviour of the corresponding effective $\kappa$~\cite{li2021impurity}.}. In general, we write $\kappa=\kappa(k)$ to emphasize its $k$-dependence, and $p=k+i\kappa(k)$ is known as the generalized Brillouin zone (GBZ)~\cite{yao2018edge, yang2020non, yokomizo2019non, Yao2018nonH2D, zhang2019correspondence, guo2021analysis,guo2021non,yang2020non}. $H(p)=H(k+i\kappa(k))$ is also referred to as the surrogate Hamiltonian, which is used instead of the original Bloch Hamiltonian $H(k)$ in computing topological invariants~\footnote{However, the topological eigenenergies themselves fall outside of the purview of our prescription, because they are isolated solutions that are not adiabatically connected to any Bloch solution.}~\cite{ghatak2019new,song2019realspace,liu2021real} and spectral properties under OBCs.

Note that the above prescriptions for the OBC and PBC eigenenergies cocide in the case of Hermitian lattices, since as $p$ cycles through real values $[0,2\pi)$, every value of $E$ lies on the real line and will be visited at least twice, both with $\kappa =\text{Im}(p)=0$. 

\subsection{Minimal model with different OBC vs PBC spectra}

\begin{figure}
    \includegraphics[width=\linewidth]{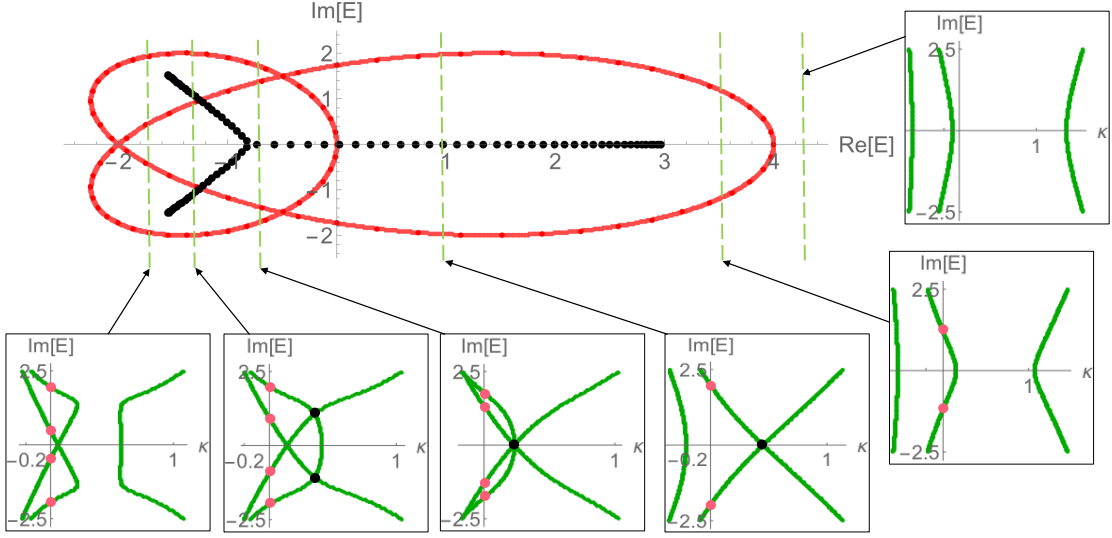}
    \caption{\textbf{Real OBC vs. real PBC spectra in terms of the symmetry of $\kappa(E)$ solutions.} Shown are the OBC (black) and PBC (red) spectra of the model $H_\text{min}$ with dispersion $E(z)=z+z^{-1}+2z^2$; the inset plots shows $\text{Im}[E(z)]$ as a function of $\kappa=-\log|z|$ (green $\kappa$ curves) at various fixed $\text{Re}[E(z)]$ slices. PBC eigenenergies (red dots) correspond to $\{\text{Re}[E(z)],\text{Im}[E(z)]\}$ consistent with $\kappa=0$, and OBC eigenenergies (black dots) correspond degeneracies in both $\kappa$ and $E(z)$ (green curve intersections). In particular, we have real OBC eigenenergies when the green curves intersect at $\text{Im}(E)=0$, which often occurs even when the PBC eigenenergies are non-real. }
    \label{fig:1} 
\end{figure}

As a concrete demonstration, we consider a minimal model with $H_\text{min}(z)=z+\frac1{z}+Az^2$, where $z=e^{ik}$. In real space, it contains two symmetric nearest neighbor (NN) hoppings and another uncompensated next-nearest-neighbor (NNN) hopping: $H_\text{min}=\sum_x |x+1\rangle\langle x| + |x\rangle\langle x+1|+A|x+2\rangle\langle x|$. Clearly, its PBC spectrum is given by $\tilde E = 2\cos k +Ae^{2ik}$, and is entirely complex unless $A=0$, as plotted as the thick red curve in Fig.~\ref{fig:1} for $A=2$. However, large segments of its OBC spectrum lie on the real line, as shown by the black dots. 

Below, we explain how one can visually derive the OBC and PBC spectra. We turn to the plots of $\kappa$ solutions vs. $\text{Im}(E)$, for fixed $\text{Re}(E)$ slices. Going from large to small $\text{Re}(E)$, we find that PBC eigenenergies (red) first appear when we pass $\text{Re}(E)=4$, followed by OBC eigenenergies (black) after $\text{Re}(E)=3$. In the green $\text{Im}(E)$ vs. $\kappa$ plots, it is apparent that PBC eigenenergies appear when the green $\kappa$ solution curves cross $\kappa=0$, while OBC eigenenergies only appear when the $\kappa$ curves intersect. This is exactly what was prescribed earlier - $\kappa=0$ gives the PBC spectrum, while $\kappa$ degeneracies give the OBC spectrum. As $\text{Re}(E)$ decreases further, additional PBC winding loops appear, and they correspond to additional $0$ crossings of $\kappa$ from the green curve that is emerging from small $\kappa$. At $\text{Re}(E)\approx -0.79$, that green curve goes over the original black $\kappa$ intersection, thereby splitting it into two black intersections. That corresponds to  the two black OBC branches away from the real $E$ line. Finally, when these two $\kappa$ intersections gap out at sufficiently negative $\text{Re}(E)$, the OBC eigenenergies disappear.  

All in all, we see that real OBC eigenenergies correspond to intersections of the $\kappa$ curves at $\text{Im}(E)=0$, which can exist even if PBC eigenenergies are already complex, i.e., if the $\kappa$ curves cross $\kappa=0$ at $\text{Im}(E)\neq 0$. As such, the breakdown of the reality of the spectra can be understood as the breaking of the symmetry of the $\kappa$ curves. While a non-real PBC spectrum only requires the $\kappa$ curves to have asymmetric zero crossings, which are almost guaranteed in a system without PT symmetry, a non-real OBC spectrum requires that symmetry to be \emph{spontaneously broken}, i.e., broken at the level of ``extrema'' corresponding to the intersection points. Tellingly, it is often much harder to have asymmetric $\kappa$ intersections rather than $\kappa$ zero crossings, and that explains the relative robustness of real OBC spectra compared to real PBC spectra.

\section{Parameter spaces for real eigenenergies}

We next present the parameter space for real OBC spectra for several paradigmatic models. It has to be emphasized that almost all eigenenergies (under both OBCs and PBCs) fundamentally depends on the form of the dispersion $P(E,p)$, and only indirectly on the form of the Hamiltonian. The exceptions are the eigenenergies of isolated topological modes, which are protected by bulk eigenstate topology, but we will not focus on them here. 

An important simplifying symmetry for OBC (but not PBC) spectra is its invariance under constant translations of $\kappa$, i.e., $H(p)$ and $H(p+i\kappa_0)$ always possess identical OBC spectra for any fixed $\kappa_0$. This is because the OBC spectra is determined by $\kappa$ crossings, which are unaffected by overall translations. As such, each model belongs to an equivalence class of models related by hopping rescalings, all possessing the same OBC spectrum. As an illustration, $H_\text{min}(z)=z+\frac1{z}+Az^2$ has identical OBC spectrum as $H'_\text{min}(z)=e^{-\kappa_0}z+\frac{e^{\kappa_0}}{z}+Ae^{-2\kappa_0}z^2$, for which the $A=0$ case reduces to the famed Hatano-Nelson model with unequal NN couplings~\cite{koch2020bulk,gong2018topological,zhang2022symmetry}.

\subsection{Separable energy dispersions \texorpdfstring{$P(E,p)$}{Lg}}
We first discuss separable dispersions, namely those with $P(E,p)=F(E) + G(p)$, where $F(E)$ is a function of $E$ and $G(p)$ is a Laurent polynomial of $z=e^{ip}$. As long as $G(p)$ gives an OBC spectral curve that does not contain branches, we can in principle achieve a real spectrum by modifying the model such that $F(E)$ conformally~\cite{tai2022zoology} maps the curve onto the real line. For this reason, the non-Hermitian SSH model and its variants can all possess real spectra~\cite{yao2018edge,Lee2019anatomy,tai2022zoology}.

\subsubsection{Single-component Hamiltonians}
We start with the single-component Hamiltonians, whose characteristic polynomials are simply given by $P(E,p)=H(p)-E$. As discussed above, cases with only two NN hoppings are trivial, since they are always reducible to the equivalence class of $H(z)=z+z^{-1}$. As such, the minimal nontrivial case is the 3-hoppings model $H_\text{min}(z)$, which we just examined. There are two ways to generalize to the next level of sophistication through a fourth hopping term, namely
\begin{equation}
    H_\text{1-band}^1(z)=z+\frac1{z}+Az^2+Bz^3  ,
\end{equation}
\begin{equation}
    H_\text{1-band}^2(z)=z+\frac1{z}+Az^2+\frac1{z^2}{B} ,
\end{equation}
where $z=e^{ip}$. These two models capture all the possibilities for Hamiltonians with hoppings spanning four sites, up to reflection and translation symmmetry. Note that for single-component models, the onsite term is just a trivial constant. Also, all meaningful models must possess both left and right hoppings, since otherwise the OBC spectrum will collapse onto a single point~\cite{longhi2020non,martinez2018non}.

\begin{figure}
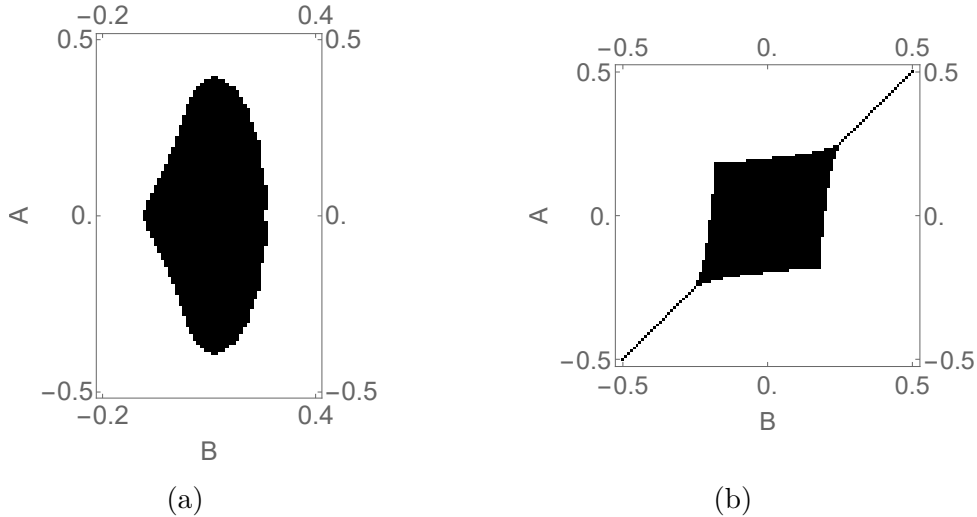

\centering
    \subfloat[]{\includegraphics[width=.33\linewidth]{\Pic{pdf}{fig2a}}}
    \hspace{0.1\linewidth}
   \subfloat[]{ \raisebox{0.4cm}{ \includegraphics[width=.4\linewidth]{\Pic{pdf}{fig2b}}}}
    \caption{
        \textbf{Parameter space of real OBC spectra (black) for 1-band 2-component models}: (a) $H_\text{1-band}^1$ and (b) $H_\text{1-band}^2$. In (b), trivial real spectra appear with $A=B$. The numerical threshold is  $\text{Max}|\text{Im}(E)|<\epsilon=10^{-6}$.}
    \label{fig:2} 
\end{figure}

As we can see from Figs.~\ref{fig:2}a and b, there exists a rather large (black) region in the $(A,B)$ parameter space where the spectra still remain real, despite $A$ and $B$ manifestly breaking any possible symmetry. Indeed, the $Az^2+Bz^3$ term of $H_\text{1-band}^1(z)$ gives rise to robustly complex eigenenergies under PBCs, even though it can still give a real spectrum for $A$ as large as $0.35$ (Fig.~\ref{fig:2}a). For $H_\text{1-band}^2(z)$ with dispersion $2\cos p +Ae^{2ip}+Be^{-2ip}$, the OBC spectrum is trivially real for $A=B$, but still remains real for a large parameter region away from that (Fig.~\ref{fig:2}b). Physically, that is so because interference from waves reflected off a boundary are sufficient in preventing a wavepacket from being amplified indefinitely. 

\subsubsection{Two-component Hamiltonians}
In 2-band models, we have 
\begin{equation}
    P_2(E,p)=E^2 - [\text{Tr}H_\text{2}(p)]E + \text{Det}H_\text{2}(p),
    \label{P2Ep}
\end{equation}
such that separable dispersions correspond to Hamiltonians with $p$-independent traces, which can occur when the diagonal terms are either zero or constant. Physically, this corresponds to the absence of homogeneous same-sublattice net hoppings across different unit cells.

\begin{figure*}
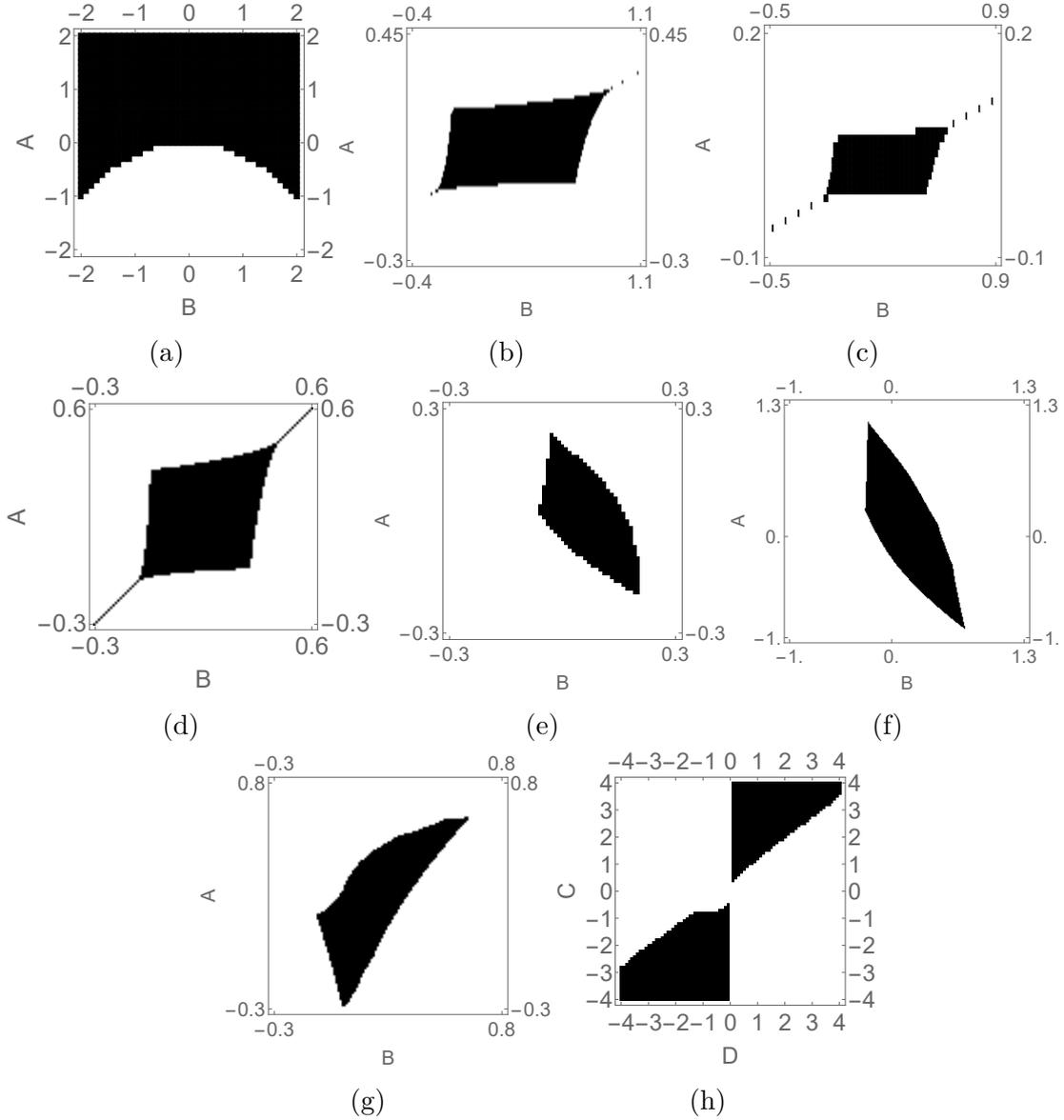

    \subfloat[]{\raisebox{-0.5\height}{\includegraphics[width=.3\linewidth]{\Pic{pdf}{fig3a}}}}
    \subfloat[]{\raisebox{-0.5\height}{\includegraphics[width=.33\linewidth]{\Pic{pdf}{fig3b}}}}
    \subfloat[]{\raisebox{-0.5\height}{\includegraphics[width=.33\linewidth]{\Pic{pdf}{fig3c}}}}\\
    \subfloat[]{\raisebox{-0.5\height}{\includegraphics[width=.34\linewidth]{\Pic{pdf}{fig3d}}}}
    \subfloat[]{\raisebox{-0.5\height}{\includegraphics[width=.33\linewidth]{\Pic{pdf}{fig3e}}}}
    \subfloat[]{\raisebox{-0.5\height}{\includegraphics[width=.31\linewidth]{\Pic{pdf}{fig3f}}}}\\
    \subfloat[]{\raisebox{-0.5\height}{\includegraphics[width=.33\linewidth]{\Pic{pdf}{fig3g}}}}
    \subfloat[]{\raisebox{-0.5\height}{\includegraphics[width=.3\linewidth]{\Pic{pdf}{fig3h}}}}
    \centering
    \caption{
        \textbf{Parameter space of real OBC spectra (black) for 2-bands 2-component separable models:} (a) $H_\text{2-band}^0$; (b) $H_\text{2-band}^1$ with $C=2,D=3$. (c) $H_\text{2-band}^1$ with $C=1/2,D=1$; (d) $H_\text{2-band}^1$ with $C=2,D=2$; (e) $H_\text{2-band}^2$ with $C=2,D=3$;	(f) $H_\text{2-band}^2$ with $C=2,D=1$; (g) $H_\text{2-band}^3$ with $C=2,D=2$; (h) $H_\text{2-band}^3$ with $A=0.1,B=0.1$. The numerical threshold is  $\text{Max}|\text{Im}(E)|<\epsilon=10^{-6}$. Note that the PBC spectra of all of these models are complex. }
    \label{fig:3} 
\end{figure*}

In this case of constant trace, the dispersion is essentially determined by $F(E)=-G(p)=\text{Det}H_2(p)$, directly generalizing the case of 1-component models upon the conformal transform $E\rightarrow E^2-[\text{Tr}H_2]E$. The only difference is that $G(p)$ now contains products of matrix elements of $H_2(p)$, and as such is usually a higher-order polynomial describing OBC spectra with no analytic solution. While a higher-order $G(p)$ can result in a more branched and hence complex OBC spectrum, with appropriate model design, there can still be larger parameter regions with real OBC spectra, see Fig.~\ref{fig:3}. 

For instructive purposes, we first study the 2-component model with almost trivial NHSE:
\begin{equation}
    H_\text{2-band}^0(z)=\left(z+Az^{-1}\right)\sigma_x+B\sigma_y,
\end{equation}
$\sigma_x,\sigma_y$ the Pauli matrices. Upon performing the $\kappa_0$ translation $p\rightarrow p+i\kappa_0=p+i\log\sqrt{A}$, we find the dispersion $E^2=-G(p)=B^2+2A\cos 2p$. As such, the spectrum is real whenever $B^2>-2A$, as numerically verified in Fig.~\ref{fig:3}a.

Next, inspired by the 1-component cases, we present the following models with nontrivial real spectral parameter regions:
\begin{equation}
    H_\text{2-band}^1(z)=\left(Az^2+\frac1{z}+C\right)\sigma_++\left(z+\frac{B}{z^2}+D\right)\sigma_-,
\end{equation}
\begin{equation}
    H_\text{2-band}^2(z)=\left(Az^2+\frac1{z}+C\right)\sigma_++\left(z+Bz^3+D\right)\sigma_-,
\end{equation}
\begin{equation}
    H_\text{2-band}^3(z)=\left(\frac{A}{z^2}+\frac1{z}+C\right)\sigma_++\left(z+\frac{B}{z^3}+D\right)\sigma_-,
\end{equation}
where $\sigma_\pm=(\sigma_x\pm i\sigma_y)/2$ and $z=e^{ip}$. In these models, the hoppings are different in either direction, and it is surprising from Fig.~\ref{fig:3}b-h that real OBC spectra not just exists, but in fact over large parameter space regions. For instance, in $H^2_\text{2-band}$ with $C=2,D=1$ (Fig.~\ref{fig:3}h), the OBC spectrum can be real even for $A>1$, which corresponds to a very unbalanced set of physical couplings.

\subsection{Inseparable energy dispersions \texorpdfstring{$P(E,p)$}{Lg}}

We next consider more sophisticated dispersions which contain products of $E$ and $z=e^{ik}$. In general, they are not analytically tractable, although their spectral graph structure can often still be heuristically predicted~\cite{tai2022zoology}. In this work, we shall limit ourselves to 2-component models. From Eq.~\ref{P2Ep}, those with inseparable dispersions correspond to those with $p$-dependent traces, i.e., those with same-sublattice hoppings across unit cells.
	
\subsubsection{Analytically tractable examples}
	
\begin{figure*}
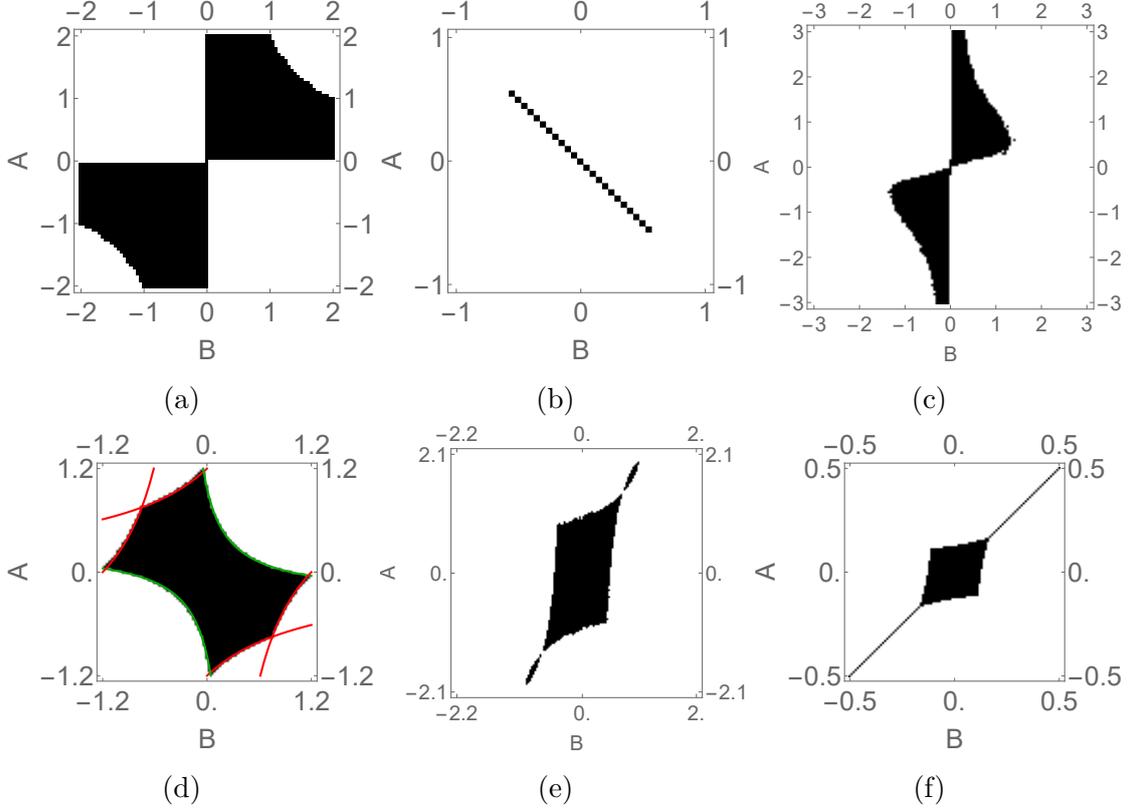

    \subfloat[]{\raisebox{-0.5\height}{\includegraphics[width=.33\linewidth]{\Pic{pdf}{fig4a}}}}
    \subfloat[]{\raisebox{-0.5\height}{\includegraphics[width=.33\linewidth]{\Pic{pdf}{fig4b}}}}
    \subfloat[]{\raisebox{-0.5\height}{\includegraphics[width=.33\linewidth]{\Pic{pdf}{fig4c}}}}\\
    \subfloat[]{\raisebox{-0.5\height}{\includegraphics[width=.33\linewidth]{\Pic{pdf}{fig4d_withbound}}}}
    \subfloat[]{\raisebox{-0.5\height}{\includegraphics[width=.33\linewidth]{\Pic{pdf}{fig4e}}}}
    \subfloat[]{\raisebox{-0.5\height}{\includegraphics[width=.33\linewidth]{\Pic{pdf}{fig4f}}}}
    \caption{\textbf{Parameter space of real OBC spectra (black) for 2-component inseparable models:} (a) $H_\text{in}^1(z)$; (b) $H_\text{in}^2(z)$; (c) $H_\text{in}^3(z)$; (d) $H_\text{in}^4(z)$ with the regime boundary approximately depicted by curve $(0.28+A)(0.28+B)=0.35$ and its reflection across $A=-B$ (green), together with curve $-0.8AB+A=1.2$ and its reflections across $A=-B$ and $A=B$ (red); (e) $H_\text{in}^5(z)$; (f) $H_\text{in}^6(z)$. The system length is $L=30$ for all and the numerical threshold is  $\text{Max}|\text{Im}(E)|<\epsilon$ where $\epsilon=10^{-5}$ for (c), and $\epsilon=10^{-6}$ for the other cases. }
    \label{fig:4} 
\end{figure*}

\begin{figure} 
    \includegraphics[width=.7\linewidth]{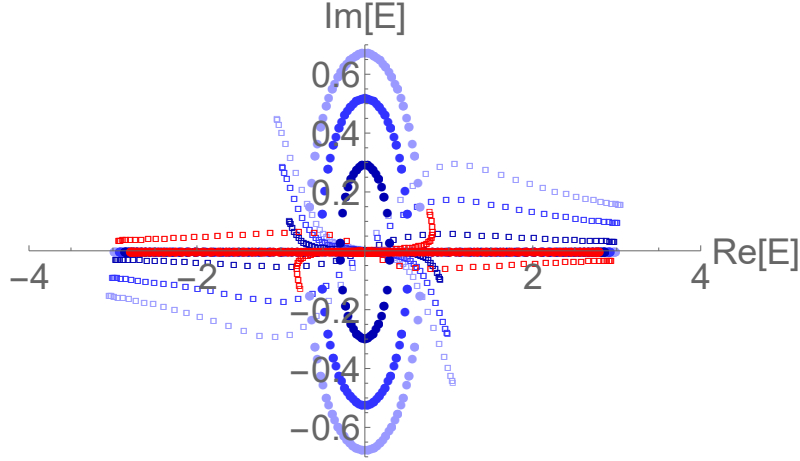}
    \centering
    \caption{
        Spectra of model $H_\text{in}^1$ (Eq.~\ref{Hin1}) with fixed $B=0.9$ and varying $A$: $A=1$ (red, showing a real spectrum), $A=1.1, 1.2, 1.3$ (blue, showing decaying complex spectra). Dots/square represents OBC/PBC eigenenergies.  The OBC spectrum becomes real as the PBC spectrum crosses the real axis.
    }
    \label{fig:5}
\end{figure}

 First, we introduce an inseparable case whose condition for real spectrum can still be analytically derived. Consider
\begin{equation}
    H_\text{in}^1(z)=\left(\begin{matrix}
        Az^3 & \sqrt{C} \\ \sqrt{C} & B/z^3
    \end{matrix}\right),
    \label{Hin1}
\end{equation}
with dispersion given by the characteristic polynomial
\begin{equation}
    P_\text{in}^1(E,z)=E^2-\left(Az^3+\frac{B}{z^3}\right)E+AB-C,
\end{equation}
where $z=e^{ip}$. To make analytic headway, we perform the hopping rescaling $p\rightarrow p+i\log \sqrt[6]{A/B}$ so as to symmetrize the $Az^3+Bz^{-3}$ term to $\sqrt{AB}(z^3+z^{-3})$. Doing so does not change the OBC spectrum; yet, due to the symmetric occurrence of $z$ and $z^{-1}$, we also know that $\kappa=0$, i.e., real $p$ constitutes a potential solution for the OBC spectrum. Whether real $p$ indeed generates \emph{the correct solution} depends on whether a competing degenerate $\kappa$ solution exists. For this model, it is not hard to show that, with $p\rightarrow i\log \sqrt[6]{B/A}$, 
\begin{eqnarray}
    P_\text{in}^1(E,z)&=&E^2-2\sqrt{AB}E\cos 3p+AB-C\notag\\
    &=& \left(E-\sqrt{AB}\cos 3p\right )^2+AB\sin^23p-C=0,\qquad
\end{eqnarray}
admits real $E$ solutions generated by real $p$ as long as $0\leq AB\leq C$. Outside of this regime, complex pairs of $E$ solutions appear (Fig.~\ref{fig:4}a), leading to non-real spectra. Fig.~\ref{fig:5} shows the boundary behaviour of spectra from real to non-real as the parameter varies. This criterion $0\leq AB\leq C$ entails that the real spectrum hinges on the presence of nonzero coupling, and is numerically verified in Fig.~\ref{fig:4}a. 

Next, we introduce another Hamiltonian which admits real spectrum for parameters that make it separable:
\begin{equation}
    H_\text{in}^2(z)=\left(\begin{matrix}
        Az & 1+z^2 \\ 2+1/z^2 & Bz
    \end{matrix}\right),
    \label{Hin2}
\end{equation}
with 
\begin{equation}
    P_\text{in}^2(E,z)=E^2-(A+B)Ez+(AB-2)z^2-3-z^{-2}.
\end{equation}
Although its GBZ and thus OBC spectrum is not analytically solvable for general parameters, for the special case when $A=-B$, the trace term disappears, and we simply obtain $E=\pm \sqrt{(2+A^2)z^2+z^{-2}+3}$. Again, this is of Hatano-Nelson form, and setting $p\rightarrow p+i\log\sqrt[4]{2+A^2}$, we obtain $\bar E =\pm\sqrt{3+2\sqrt{2+A^2}\cos 2p}$. This is real for $A=-B$ with $|A|<1/2$, as reflected in its numerical parameter space diagram (Fig.~\ref{fig:4}b).

\subsubsection{More general examples}

Next, we introduce a few more models whose propensity for real OBC spectra cannot be predicted through any simple way:
\begin{equation}
    H_\text{in}^3(z)=\left(\begin{matrix}
        2+Az & 2 \\ 1 & 1+B/z
    \end{matrix}\right),
    \label{Hin3}
\end{equation}
\begin{equation}
    H_\text{in}^4(z)=\left(\begin{matrix}
        1+Az & 1+z \\ 2+1/z & 1+Bz
    \end{matrix}\right),
    \label{Hin4}
\end{equation}
\begin{equation}
    H_\text{in}^5(z)=\left(\begin{matrix}
        Az & 1+z \\ 2+1/z & B/z
    \end{matrix}\right),
    \label{Hin5}
\end{equation}
\begin{equation}
    H_\text{in}^6(z)=\left(\begin{matrix}
        z+Az^2 & 2 \\ 1 & 1/z+B/z^2
    \end{matrix}\right).
    \label{Hin6}
\end{equation}
Like $H_\text{in}^1$ of Eq.~\ref{Hin1}, the model $H_\text{in}^3$ contain constant off-diagonal couplings. However, despite its simple algebraic form, it is not analytically tractable, and in fact behaviours completely differently from $H_\text{in}^1$, and in fact the other models too. 

\begin{figure}
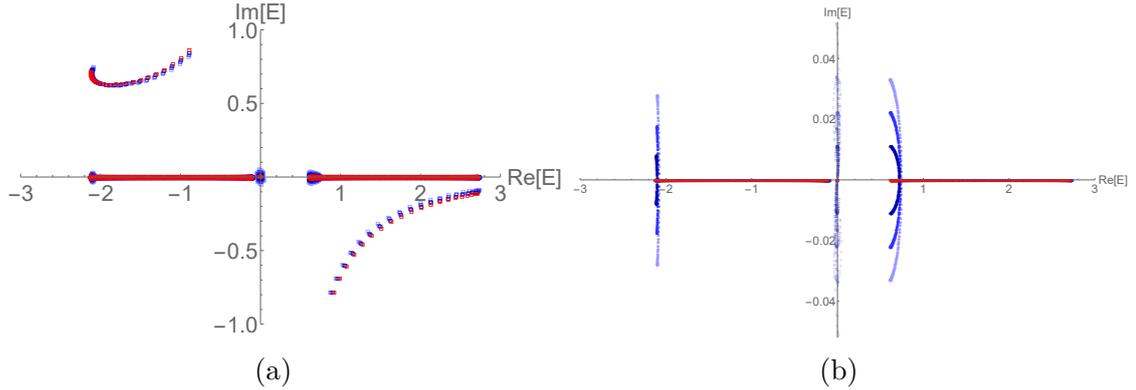
 
\centering
    \subfloat[]{	\includegraphics[width=.49\linewidth]{\Pic{pdf}{Hin6SpecVsA}}}
    \subfloat[]{	\includegraphics[width=.49\linewidth]{\Pic{pdf}{Hin6SpecVsAzoomin}}}
    \caption{
        (a) Spectra and (b) its zoomed-in view for model $H_\text{in}^6$ (Eq.~\ref{Hin6}) with fixed $B=0.3$ and varying $A$ values: $A=0.3$ (red, showing a real spectrum) and $A=0.32, 0.34, 0.36$ (blue, showing decaying complex spectra). Dots/square represents OBC/PBC eigenenergies.}
    \label{fig:6}
\end{figure}

The real spectrum parameter region of $H_\text{in}^6$ is an interesting intersection of a diagonal $A=B$ line segment, and a smaller extended region. These two parameter subregions have different origins: For the former, the OBC spectrum is real because the PBC spectrum is also real; larger $A=B$ closes the band gap and releases complex eigenenergies. For the latter, the PBC spectrum is always complex (example in Fig.~\ref{fig:6}), but an open boundary creates sufficient interference to stop indefinite amplification, leading to a completely real spectrum.

\section{Discussion}

In this work, we have seen that real non-Hermitian OBC spectra are generically more robust against small variations in system parameters compared to real PBC spectra. This can be explained in terms of the inverse skin depth $\kappa(E)$ solutions curves - while the PBC spectrum becomes complex once $\kappa(E)\neq 0$ at $\text{Im}E \neq 0$, a complex OBC spectrum requires $\kappa(E)$ curves to \emph{intersect} at $\text{Im}E \neq 0$, which is a more demanding condition. While there exists a very general electrostatics approach~\cite{yang2022designing} that returns possible parent Hamiltonians for any desired real OBC spectrum and skin localization, this work showcases particularly simple ansatz models that have the benefit of being as local as possible.

The discovery of these Hamiltonians with robust real spectra marks a significant advancement in the design of stable non-Hermitian systems, providing a pathway to spectral stability that operates independently of symmetry constraints, including the widely studied PT-symmetry framework. Whereas PT-symmetric systems achieve real eigenvalues through balanced gain and loss, these new Hamiltonians demonstrate that spectral stability can arise through entirely different mechanisms. This development is crucial for engineering stable non-Hermitian systems across a broader range of conditions and applications, addressing the recent surge in theoretical work on general, symmetry-free non-Hermitian models that, until now, have remained experimentally elusive due to stability challenges, i.e., non-Hermitian skin clusters and pseudogaps~\cite{li2022non,shen2021nonhermitian,tahir2011PG} can be realized not just with enhanced stability, but also with generically non-reciprocal platforms such as circuits with MOSFETs or operational amplifiers~\cite{budich2020sensor,yang2022observation,hofmann2019chiral,helbig2020generalized,zou2021observation,ningyuan2015time,stegmaier2021topological,blais2021circuit}. Yet, the NHSE is not guaranteed to yield real spectra, and in Appendix A, we have also listed down models which failed, to aid further search efforts.

\section{Appendix: Inseparable energy dispersions not giving real spectra}

While the main text had discussed various models with real spectra despite having no favorable symmetries, it is also important to record models where this is does not occur. As discussed, NHSE-induced real spectra depends on the algebraic properties of the $\kappa(E)$ curves, and it is instructive to list the models where they do not behave favorably as such.

For instance, the following deformations of $H_\text{in}^1(z)$ have no real spectra:
\begin{itemize}
    \item $	H_\text{in,m1}^1(z)=\left(\begin{matrix}
        Az^{n} & 0 \\ c_{1} & Bz^{-n}
    \end{matrix}\right)$ and
    \item $ H_\text{in,m2}^1(z)=\left(\begin{matrix}
        Az^{n} & -c_{1} \\ c_{1} & Bz^{-n}
    \end{matrix}\right),$
    for sufficiently small $c_1$, which is expected as they do not fulfil the condition of $0\leq AB\leq C$, where $C=0$ for $H_\text{in,m1}^1$ and $C=c_{1}^2$ for $H_\text{in,m2}^1$.
    
    \item	$ H_\text{in,m3}^1(z)=\left(\begin{matrix}
        Az^{n} & 0 \\ c_{1} & Bz^{n}
    \end{matrix}\right).$
\end{itemize}
Modifications of $H_\text{in}^2(z)$ and  $H_\text{in}^4(z)$ with diagonal terms having all positive powers of $z$ only:
\begin{itemize}
    \item $H_\text{in,m1}^4(z)=\left(\begin{matrix}
        1+Az & 1+z \\ 2+z & 1+Bz
    \end{matrix}\right).$
\end{itemize}

Modifications of $H_\text{in}^3(z)$ with diagonal terms having same sign of power of $z$:
\begin{itemize}
    \item By having same sign of power of $z$ at the non-diagonal terms\\
    
    $H_\text{in,m1}^3(z)=\left(\begin{matrix}
        2+Az & c_{2} \\ 1 & 1+Bz
    \end{matrix}\right),$
    with $c_{2}=1, -1, 0$.
\end{itemize}
Modifications of $H_\text{in}^5(z)$ with diagonal terms having positive and negative powers of $z$:
\begin{itemize}
    \item 	$H_\text{in,m1}^5(z)=\left(\begin{matrix}
        Az & 1+z^2 \\ 2+z^{-2} & Bz^{-1}
    \end{matrix}\right),$
    \item By having same sign of power of $z$ at the non-diagonal terms\\
    
    $H_\text{in,m2}^5(z)=\left(\begin{matrix}
        Az & 1+z \\ 2+z & Bz^{-1}
    \end{matrix}\right),$
\end{itemize}
Modifications of $H_\text{in}^6(z)$ with diagonal terms having same sign of power of $z$:
\begin{itemize}
    \item 	
    $H_\text{in,m1}^6(z)=\left(\begin{matrix}
        1+z+Az^2 & 1 \\ 2 & 1+z+Bz^{-2}
    \end{matrix}\right).$\\
    $H_\text{in,m2}^6(z)=\left(\begin{matrix}
        1+z+Az^2 & 1 \\ 2 & 1+z^{-1}+Bz^{2}
    \end{matrix}\right).$
    
\end{itemize}
Models with non-symmetric hopping systems:
\begin{itemize}
    \item 
    $H_\text{c1}(z)=\left(\begin{matrix}
        Az^2 & z \\ 2 & Bz^{-2}
    \end{matrix}\right).$
    \item 	$H_\text{c2}(z)=\left(\begin{matrix}
        Az^2 & z \\ 2 & Bz^{2}
    \end{matrix}\right).$
    \item 
    $H_\text{c3}(z)=\left(\begin{matrix}
        Az^2 & z+2 \\ 2 & Bz^{-2}
    \end{matrix}\right).$
    \item 
    $H_\text{c4}(z)=\left(\begin{matrix}
        Az^2 & z+2 \\ 2 & z^{-1}+Bz^{-2}
    \end{matrix}\right).$
    \item 
    $H_\text{c5}(z)=\left(\begin{matrix}
        1+z+Az^2 & 2 \\ 2 & Bz^{-2}
    \end{matrix}\right).$
\end{itemize}

\section{Appendix: Spectra of illustrative models}

Here, we display the spectra of selected models discussed, such as to contrast the OBC vs. PBC spectra and their propensities for being real.

\begin{figure} 
\centering
    \includegraphics[width=.7\linewidth]{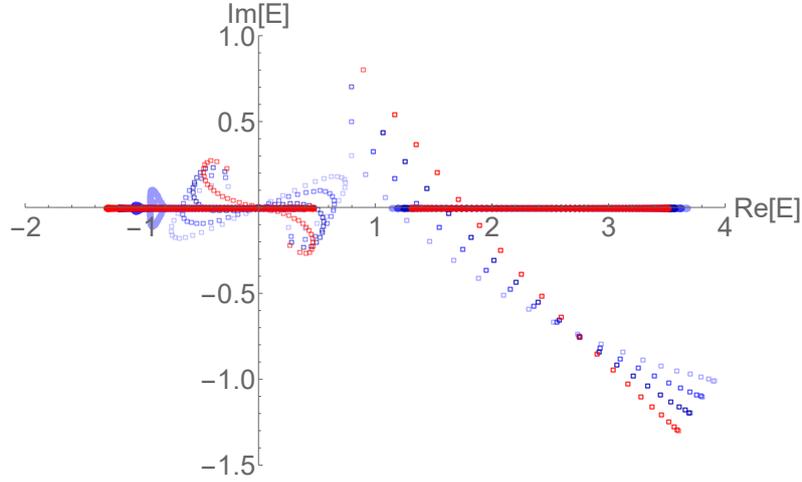}
    \caption{
        Spectra of model $H_\text{in}^4$ (Eq.~\ref{Hin4}) with fixed $B=0.2$ and varying $A$ values: $A=0.1$ (red) and $A=0.3, 0.5, 0.7$ (blue, showing decaying complex spectra). Dots/square represents OBC/PBC eigenenergies. The OBC spectrum becomes real as the near-zero PBC spectrum shifts across the real axis.}
    \label{fig:7}
\end{figure}

\begin{figure} 
\centering
    \includegraphics[width=.7\linewidth]{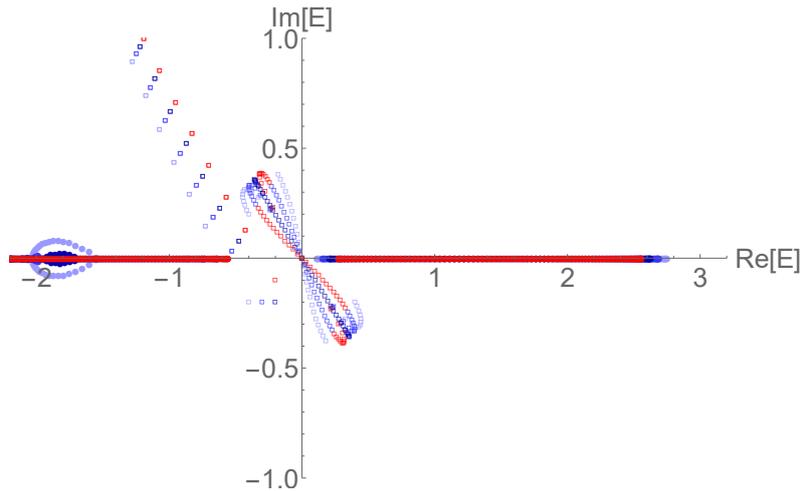}
    \caption{
        Spectra of model $H_\text{in}^5$ (Eq.~\ref{Hin5}) with fixed $B=0.2$ and varying $A$ values: $A=0.1$ (red) and $A=0.2, 0.3, 0.4$ (blue, showing decaying complex spectra). Dots/square represents OBC/PBC eigenenergies.}
    \label{fig:8}
\end{figure}

\begin{figure}
\centering
    \includegraphics[width=.7\linewidth]{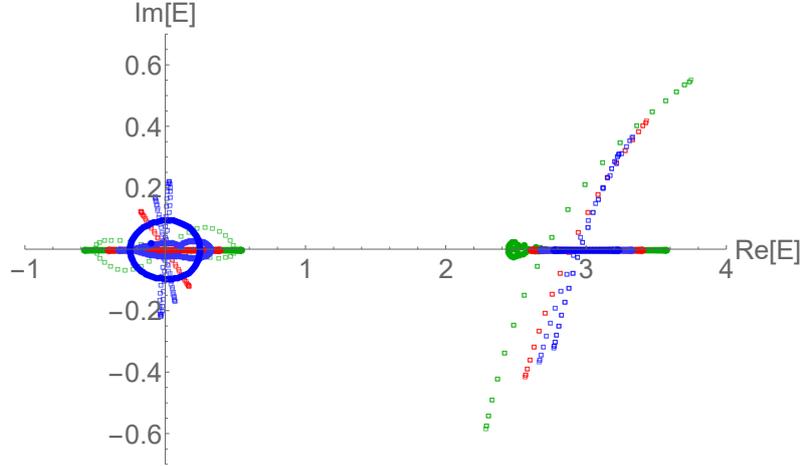}
    \caption{
        Spectra of model $H_\text{in}^3$ (Eq.~\ref{Hin3})  with fixed $B=0.4$ and varying $A$ values: $A=0.15, 0.3$ (blue, showing decaying complex spectra), $A=0.45$ (red, real OBC spectrum) and $A=0.9$ (green). Dots/square represents OBC/PBC eigenenergies. As A increases, the complex ``bubble'' on the left side of the spectrum vanishes and re-emerges on the right side. A real spectrum shortly appears during these states.}
    \label{fig:9}
\end{figure}

\begin{figure}
    \includegraphics[width=\linewidth]{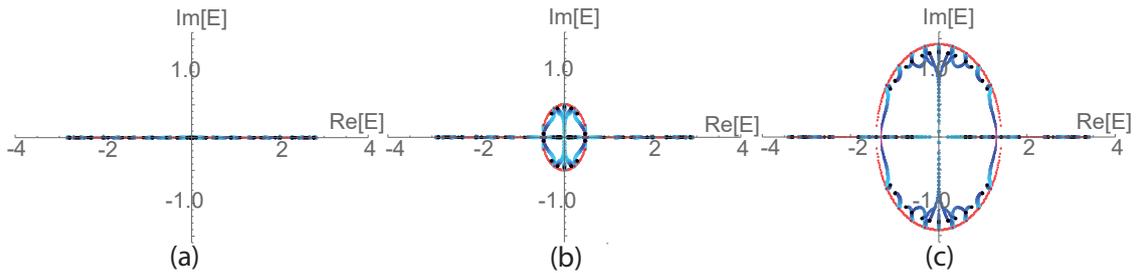}
    \caption{The PBC-OBC spectral interpolation of the model $H_\text{in}^1$ (Eq.~\ref{Hin1}) with parameters $A=B=\sqrt{2},1.5$ and $2$ for (a-c) respectively. Red and black dots represent PBC and OBC eigenenergies, respectively. The blue-purple curves trace the evolution of selected eigenenergies as PBCs are smoothly deformed into OBCs by gradually disabling the boundary hoppings. For this model, once a complex ``bubble'' forms in the PBC spectrum, the OBC spectrum also ceases to remain real.}
    \label{ABspects}
\end{figure}

\section{Appendix: Error tolerance in determining the parameter region of real spectra}
In general, the real spectra of our models were determined numerically. In most cases, it is clear whether the collapse onto the real energy line occurs, and the parameter region does not depend on the numerical error tolerance $\epsilon$ for $\text{Max}\text{Im}(E)$. However, in some cases that are potentially afflicted with the critical skin effect~\cite{li2021critical,liu2020helical,rafi2021critical}, there is strong sensitivity to system size, and different $\epsilon$ also gives rise to different parameter regions for real spectra. Shown below are a few illustrative cases.

\begin{figure}
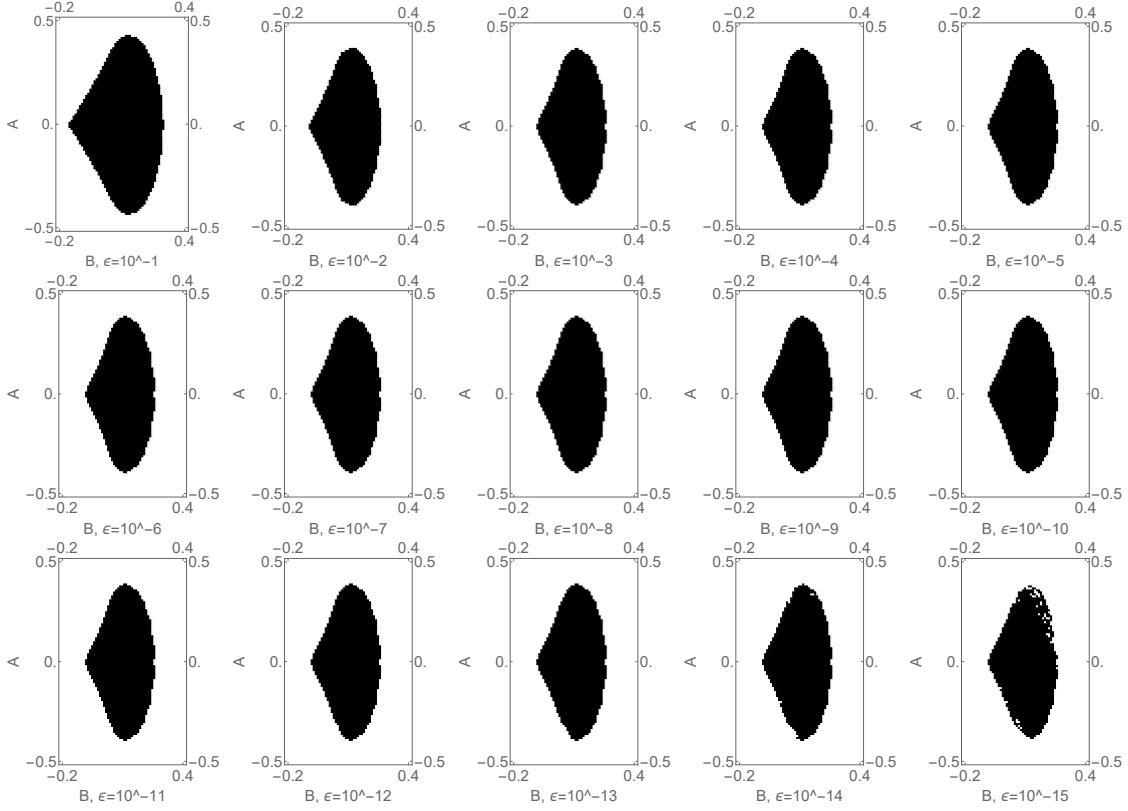

\centering
    \includegraphics[width=.19\linewidth]{\Pic{pdf}{fig11_1}}
    \includegraphics[width=.19\linewidth]{\Pic{pdf}{fig11_2}}
    \includegraphics[width=.19\linewidth]{\Pic{pdf}{fig11_3}}
    \includegraphics[width=.19\linewidth]{\Pic{pdf}{fig11_4}}
    \includegraphics[width=.19\linewidth]{\Pic{pdf}{fig11_5}}\\
    \includegraphics[width=.19\linewidth]{\Pic{pdf}{fig11_6}}
    \includegraphics[width=.19\linewidth]{\Pic{pdf}{fig11_7}}
    \includegraphics[width=.19\linewidth]{\Pic{pdf}{fig11_8}}
    \includegraphics[width=.19\linewidth]{\Pic{pdf}{fig11_9}}
    \includegraphics[width=.19\linewidth]{\Pic{pdf}{fig11_10}}\\
    \includegraphics[width=.19\linewidth]{\Pic{pdf}{fig11_11}}
    \includegraphics[width=.19\linewidth]{\Pic{pdf}{fig11_12}}
    \includegraphics[width=.19\linewidth]{\Pic{pdf}{fig11_13}}
    \includegraphics[width=.19\linewidth]{\Pic{pdf}{fig11_14}}
    \includegraphics[width=.19\linewidth]{\Pic{pdf}{fig11_15}}
    \caption{ Dependence of the real spectrum parameter region for the model $H_\text{1-band}^1$ on the threshold $\epsilon$. From top left to bottom right, $\epsilon$ is set $10^{-1}$, $10^{-2}$, $10^{-3}$, ... ,$10^{-15}$ respectively. $\epsilon=10^{-3}$ to $\epsilon=10^{-9}$ give exactly same parameter region, which only exhibits some degeneration beyond $\epsilon=10^{-14}$. This shows a clear independence on the threshold $\epsilon$. 
    }
\end{figure}

\begin{figure}
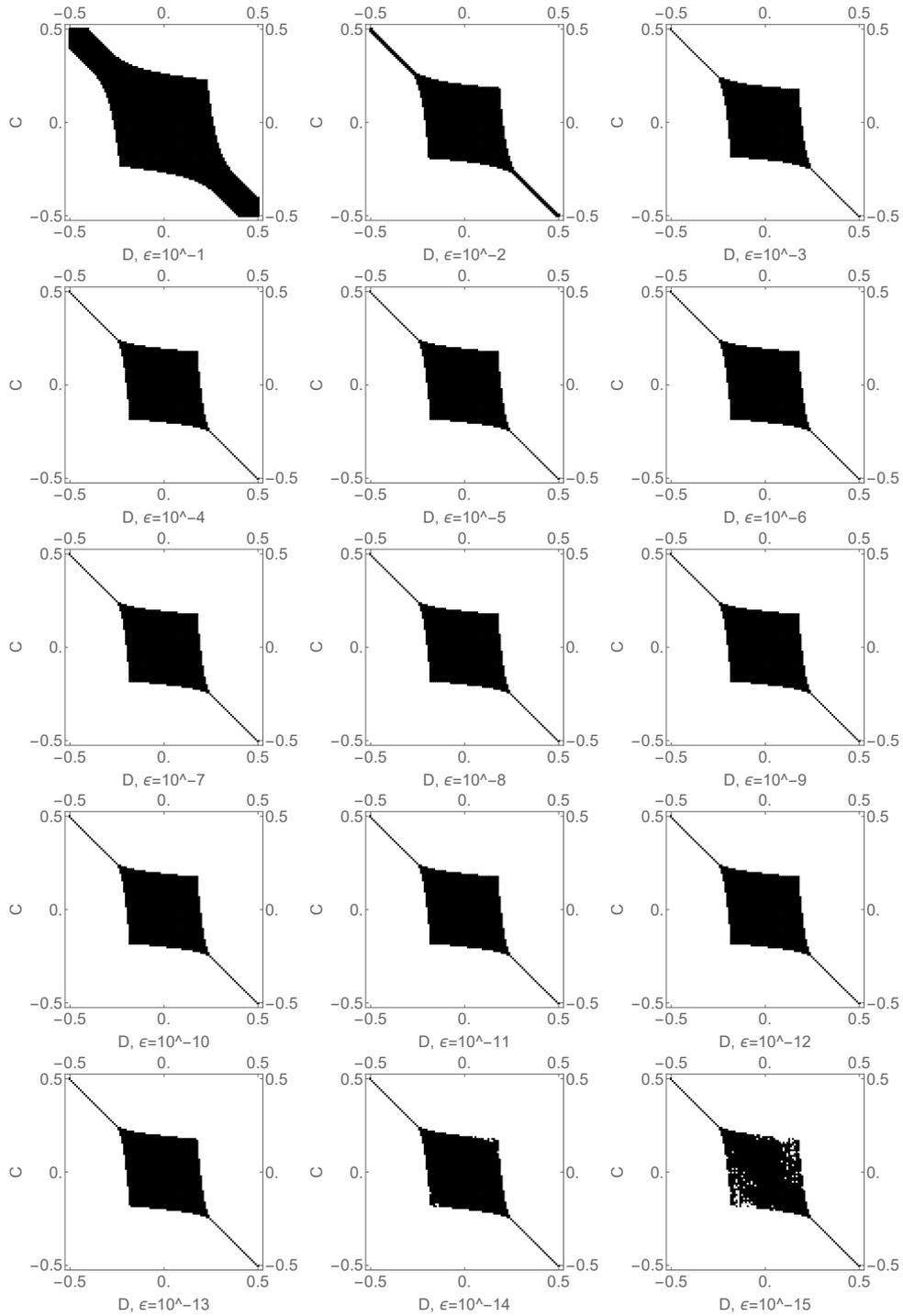

\centering
    \includegraphics[width=.28\linewidth]{\Pic{pdf}{CD1}}
    \includegraphics[width=.28\linewidth]{\Pic{pdf}{CD2}}
    \includegraphics[width=.28\linewidth]{\Pic{pdf}{CD3}}\\
    \includegraphics[width=.28\linewidth]{\Pic{pdf}{CD4}}
    \includegraphics[width=.28\linewidth]{\Pic{pdf}{CD5}}
    \includegraphics[width=.28\linewidth]{\Pic{pdf}{CD6}}\\
    \includegraphics[width=.28\linewidth]{\Pic{pdf}{CD7}}
    \includegraphics[width=.28\linewidth]{\Pic{pdf}{CD8}}
    \includegraphics[width=.28\linewidth]{\Pic{pdf}{CD9}}\\
    \includegraphics[width=.28\linewidth]{\Pic{pdf}{CD10}}
    \includegraphics[width=.28\linewidth]{\Pic{pdf}{CD11}}
    \includegraphics[width=.28\linewidth]{\Pic{pdf}{CD12}}\\
    \includegraphics[width=.28\linewidth]{\Pic{pdf}{CD13}}
    \includegraphics[width=.28\linewidth]{\Pic{pdf}{CD14}}
    \includegraphics[width=.28\linewidth]{\Pic{pdf}{CD15}}
    \caption{	 Dependence of the real spectrum parameter region for the model $H_\text{1-band}^1$ on the threshold $\epsilon$. From top left to bottom right, $\epsilon$ is set $10^{-1}$, $10^{-2}$, $10^{-3}$, ... ,$10^{-15}$ respectively. $\epsilon=10^{-3}$ to $\epsilon=10^{-12}$ give exactly same parameter region, which only exhibits some degeneration beyond $\epsilon=10^{-14}$. This also shows a clear independence on the threshold $\epsilon$. 
    }
\end{figure}

\SetPicSubDir{ch-inhomo}
\SetExpSubDir{ch-inhomo}

\chapter{Phase-space generalized Brillouin zone for spatially inhomogeneous systems}
\label{ch-inhomo}
\vspace{2em}

As illustrated in Chapter~\ref{ch-prel}, the non-Hermitian skin effect (NHSE) is conventionally characterized on non-Hermitian tight-binding lattices with translationally invariant but asymmetric hopping amplitudes under open boundary conditions (OBC). The asymmetric, translationally invariant hopping leads to the constant directed amplification of eigenstates, where the ``pumped'' eigenstates are confined by the system's hard open boundary.

The emergence of the NHSE has fundamentally transformed the understanding of band structure characterization through the construction of the generalized Brillouin zone (GBZ). This introduces a constant GBZ factor $z$ from the consistent degree of asymmetry between hopping amplitudes in opposite directions. The GBZ framework built around this factor provides a comprehensive mathematical model that accurately describes the properties of these non-Hermitian systems.

The NHSE has attracted significant research interest over the years, leading to the discovery of novel phenomena, properties, and applications facilitated by the GBZ framework.

However, as observed in the intrinsic mechanism of the NHSE, the localization and accumulation of eigenstates do not require a hard boundary or constant amplification. Any underlying directed pumping can generally lead to localized state accumulations as long as translational symmetry is broken.

Moreover, there have been intense research on the distinct properties and relations of systems under periodic boundary conditions (PBC) and open boundary conditions (OBC). PBC and OBC are typically considered as two discontinuous cases and are often studied separately.While there are notable relationships between their properties, such as the bulk-boundary correspondence protected by topology, and the critical non-Hermitian skin effect where spectra gradually transition between PBC and OBC spectra as the system size varies, little attention has been given to systems with boundary conditions that lie between OBC and PBC. These ``soft boundary'' lattices lack a hard open cut-off but have their translational invariance broken by variations between non-zero neighboring hoppings. As systems that lie between OBC and PBC, they offer a deeper understanding and further development in the study of localization and transport, exhibiting unique features not present in strictly periodic or open systems.

Motivated by the reasons above, we propose an additional degree of freedom to the hopping amplitude of non-Hermitian systems, termed spatial inhomogeneity. The hopping amplitudes are defined by an arbitrary function that varies with real-space position. These spatially inhomogeneous hopping amplitudes break the translational symmetry in PBC non-Hermitian lattices, leading to effective state accumulation within the bulk region, characterized by the arbitrary hopping function.

The inhomogeneous asymmetry within non-Hermitian systems introduces spatially inhomogeneity to the conventional GBZ factor $z$, which adds an additional dimension to the generalized Brillouin zone. We refer to this as the phase-space GBZ, where the complex momentum deformation, used to describe the state accumulation profile, depends on both the spatial position and the Bloch momentum.

In this work, we successfully constructed the phase-space GBZ, which serves as a unified framework providing comprehensive descriptions of the energies and eigenstates of one-dimensional inhomogeneous systems. Additionally, our phase-space GBZ offers a solid theoretical foundation for several new phenomena, including the categorization of spectral regions within which states exhibit distinct behaviours and the emergence of new spectral branches. Most importantly, we discovered a new phase of states with real energies and locally non-smooth eigenstates specifically associated with spatial inhomogeneity. This new phase of inhomogeneous states exhibits topologically non-trivial properties characterized by its inhomogeneous phase.

\section{Introduction}

The non-Hermitian skin effect (NHSE) represents a very robust form of localization caused by the directed amplification from asymmetric hoppings. 
Such non-Bloch behavior has led to various paradigm shifts in the way band structures are characterized, as epitomized by modified non-Bloch topological invariants defined on the generalized Brillouin zone (GBZ)~\cite{yao2018edge,yao2018non,yokomizo2019non,lee2019anatomy,zhang2020correspondence,xiong2018does,longhi2019probing,kawabata2020non,lee2020ultrafast,song2019realspace,jiang2023dimensional,wang2024amoeba}. Competition between more than one type of asymmetric hoppings can furthermore modify critical scaling properties~\cite{arouca2020unconventional,liu2024non,li2020critical,qin2023universal,liu2020helical,rafi2022critical,yang2024percolation,yokomizo2021scaling,lin2023topological}and induce non-Hermitian pseudo-gaps~\cite{li2022non} that break the one-to-one correspondence between the open and periodic boundary conditions (OBC and PBC) spectra, challenging existing notions of non-Bloch topological bulk-boundary correspondences~\cite{martinez2018topological,kunst2018biorthogonal,deng2019non,zhu2020photonic,imura2020generalized,yang2020non,guo2021non,yang2022non,shen2024non,yang2024non,ye2025observing,okuma2023non}.

While most existing investigations have focused on NHSE localization against hard open boundaries~\cite{schindler2021dislocation,bhargava2021non,zhang2022universal,lee2019hybrid,kawabata2020higher,okuma2020topological,li2020topological,li2021quantized,li2020topological,sun2021geometric,zhang2023electrical,lee2021many,shen2022non,zhang2022review,shang2022experimental,fang2022geometry,guo2023anomalous,wang2023non,liu2023reentrant,shen2023observation,qin2024dynamical,shen2024enhanced,qin2024kinked,yang2024anatomy,Lei2024,qin2024kinked,xiong2024non}, i.e., physical edges, the underlying directed pumping mechanism gives rise to localized state accumulations as long as translation symmetry is broken. Recently, it has been recognized that spatial inhomogeneities that are not abrupt cut-offs can support other less-known but interesting phenomena such as scale-free eigenstates~\cite{li2021impurity,guo2023accumulation,li2023scale,wang2023scale,xie2024observation,liu2024emergent} and non-Hermitian Anderson localization~\cite{jiang2019interplay,longhi2019topological,longhi2023inhibition,zhai2020many,wang2021anderson,zeng2022real,wang2023observation,acharya2024localization,longhi2024robust}.
These enigmatic phenomena arise because of the nontrivial interplay between the emergent nonlocality from directed non-Hermitian pumping, and the momentum non-conserving spatial inhomogeneities associated with Stark localization, quantum confinement, and band flattening~\cite{liu2020generalized,longhi2020non,longhi2022non,qi2023localization,li2024fate}. 
However, a unified framework encompassing the breadth of these exciting interplays does not yet exist.

To provide this unified framework, we propose in this work the new concept of phase-space GBZs, where the complex momentum deformation for describing the state accumulation profile depends on the spatial position $x$ in addition to the Bloch momentum $k$. By considering the subtle but important skin contributions from spatial hopping gradients, we obtained an analytic ansatz that accurately predicts the energy spectrum and most skin eigenstate profiles in various one-dimensional spatial inhomogeneity profiles, from soft boundaries to impurities to physical edges. This phase-space GBZ construction is also generalized to 2-component systems, where it is shown to accurately predict a new topological phase transition arising from the tuning of the soft boundary width.

Our phase-space GBZ gives a firm theoretical basis for a number of new phenomena. Most salient is the bifurcation of the GBZ into various branches due to spatial inhomogeneity. OBC eigensolutions can exhibit robust discontinuous jumps between them, and as such, acquire non-exponential spatial profiles, distinct from conventional NHSE states. Such ``inhomogeneous skin'' regions also give rise to unconventional but universal spectral branching in the complex plane, beyond what is allowed by spatially homogeneous GBZs. Additionally, in multi-component scenarios, varying the spatial hopping inhomogeneity can also drive topological phase transitions unique to NHSE-pumped states.

\section{Results}

\subsection{1D monoatomic chain with spatially inhomogeneous asymmetric hoppings}

Conventionally, the non-Hermitian skin effect (NHSE) has been rigorously characterized in translationally invariant tight-binding lattices with \emph{constant} but asymmetric hopping amplitudes and hard OBC boundaries~\cite{yao2018edge,kunst2018biorthogonal,yao2018non,lee2019anatomy,song2019non,lee2020unraveling,yokomizo2019non}. Due to this translational invariance, the concept of momentum-space lattice (BZ) remains intact, except that the momenta acquire imaginary contributions, i.e., the GBZ formalism.

In this work, we relax the requirement for translation invariance by modulating the hopping strengths between neighboring sites with a spatially inhomogeneous profile $g(x)$. We first study the minimal (nearest-neighbor) 1D model with a monoatomic unit cell (see Sect. \ref{sec_2b} for more complicated unit cells) under PBCs:
\begin{align}
	H&=\sum_{x=1}^{L-1}g(x)\left(\frac{1}{\gamma}\ket{x}\bra{x+1}+\gamma \ket{x+1}\bra{x}\right)\notag\\
 &+g(L)\left(\frac{1}{\gamma}\ket{L}\bra{1}+ \gamma \ket{1}\bra{L}\right),
 \label{1bH}
\end{align}

on a ring with $L$ sites, where $g(x)/\gamma$ and $g(x)\gamma$ are the left and right hoppings between sites $x$ and $x+1$, respectively.
Here, we have fixed the local hopping asymmetry $\gamma$ to a constant value, since any desired spatial profile of the hopping asymmetry $\gamma(x)$ can be easily obtained from Eq.~\ref{1bH} via a local basis transformation $\ket{x}\rightarrow \gamma^{-x}\Pi_{x'=1}^{x-1}\gamma(x')\ket{x}$, $\bra{x}\rightarrow \bra{x}\gamma^{x}/\Pi_{x'=1}^{x-1}\gamma(x')$, with $\gamma$ set to $\left(\Pi_{x'=1}^{L}\gamma(x')\right)^{1/L}$. 
PBCs are used instead of OBCs, such that the only source of spatial inhomogeneity is from $g(x)$.

\begin{figure}
    \includegraphics[width=0.8\linewidth]{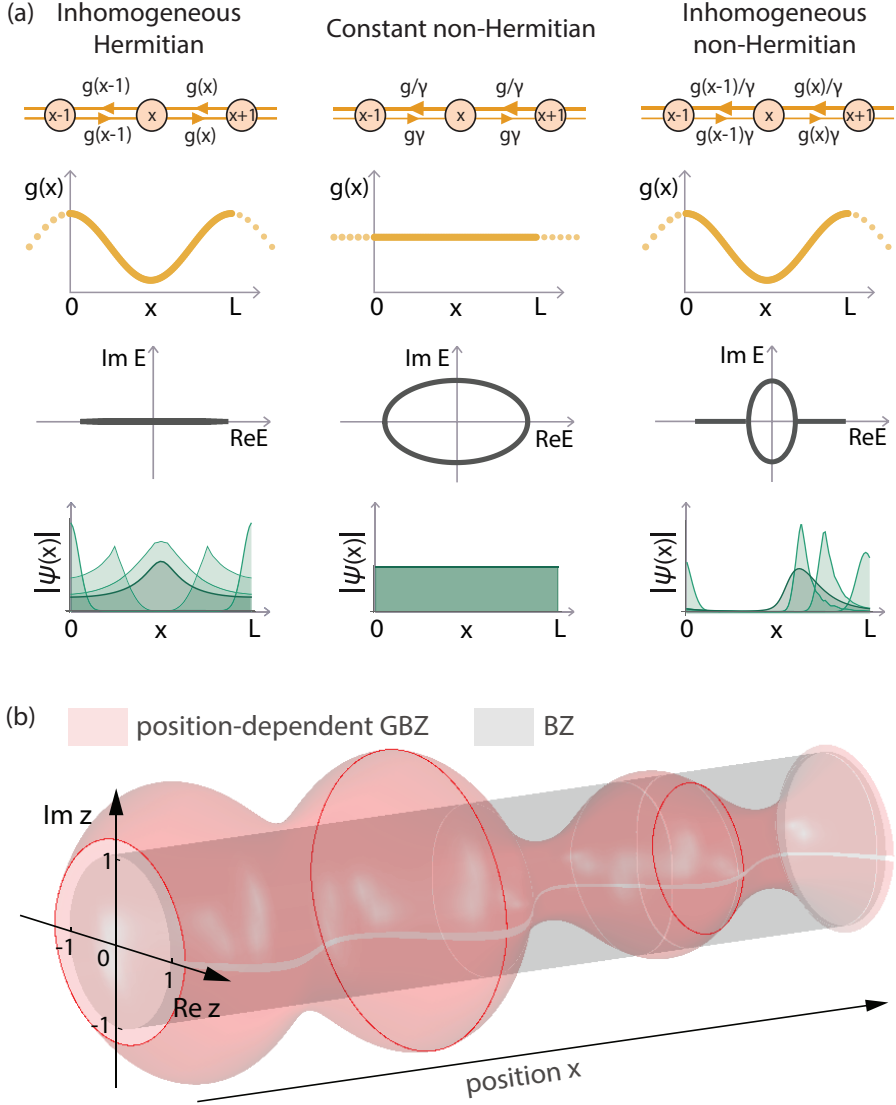}
    \caption{\textbf{Non-trivial Interplay of spatial inhomogeneity and anisotropy in the lattice hoppings. } 
    (a) Left: Spatial inhomogeneity in the lattice hoppings (with profile given by $g(x)$) lead to inhomogeneities in the eigenstates $\psi(x)$ that are locally proportional to $1/\sqrt{g(x)}$. Center: Without spatial hopping inhomogeneity, i.e., PBCs with constant $g(x)$, the eigenstates exhibit constant magnitudes despite non-Hermitian hopping anisotropy $\gamma$ (asymmetric hoppings). Right: With the simultaneous presence of spatial inhomogeneity and asymmetry in the hoppings, new tails branch out from the eigenspectrum and the wavefunctions accumulate asymmetrically against $g(x)$ troughs, which behave like ``partial'' boundaries. 
    (b) An inhomogeneous non-Hermitian hopping strength profile $g(x)$ results in a position-dependent generalized Brillouin zone (GBZ) $ z_{n}(x) = \psi_n(x+1)/\psi_n(x) $ defined in phase space, as given in Eq.~\ref{zdefine}, such that the wavefunctions exhibit different spatial decay or growth rates at different positions. Shown is the GBZ for an illustrative $g(x) = \left(\sin(4\pi x / L) \cdot \cos(2\pi x / L) + 1.2\right)^{-1}$ with $ \gamma = 0.2 $.
    }
    \label{cartoon2}
    \end{figure}

The PBC ansatz Hamiltonian Eq.~\ref{1bH} encompasses the usual well-studied limits as special cases, as we first schematically describe in the following. For $\gamma=1$ [Fig.~\ref{cartoon2}a (Left)], it reduces to a Hermitian nearest-neighbour tight-binding chain with real spectrum and eigenstates $\psi(x)$ that depends only \emph{locally} on the texture $g(x)$. For constant $g(x)=g$ with non-Hermitian $\gamma\neq 1$ [Fig.~\ref{cartoon2}a (Center)], it reduces to the usual Hatano-Nelson model~\cite{gong2018topological,schindler2021dislocation,HN1996prl,HN1997prb,HN1998prb,PhysRevB.92.094204,claes2020skin}  with a complex elliptical spectrum. Even though the eigenstates $\psi(x)$ are pumped leftwards by the hopping asymmetry, they do not have anywhere to accumulate against due to the PBCs and uniform $g(x)$, and thus exhibit spatially uniform amplitudes $|\psi(x)|$.

But with the simultaneous presence of nontrivial hopping asymmetry $\gamma \neq 1$ and non-constant spatial hopping profile $g(x)$ [Fig.~\ref{cartoon2}a (Right)], the non-Hermitian pumped eigenstates are able to accumulate partially against the $g(x)$ inhomogeneities. As schematically shown, illustrative eigenstates generally accumulate towards the right of the $g(x)$ minimum, which is at the center. Notably, due to the directed pumping from right to left, the $g(x)$ minimum behaves like a partial ``boundary'' that non-locally prevents most of the state from occupying the region on its left. However, unlike skin states under genuine OBCs, the spatial state accumulation is manifestly non-exponential, thereby precluding any direct characterization through complex non-Bloch momentum. Also, as compared to the spectrum in Fig.~\ref{cartoon2}a (Center), the inhomogeneity in $g(x)$ has given rise to additional spectral branches at the side, reminiscent of (but distinct from) the OBC spectra of uniform models with asymmetric hoppings~\cite{lee2020unraveling,yang2022designing,tai2023zoology,zhang2024observation,li2020critical,liu2020helical,rafi2022critical,qin2023universal}.

\subsection{Phase-space GBZ for inhomogeneous single-component chains}\label{Result_A}

\subsubsection{General formalism setup}

\noindent To rigorously characterize the anomalous non-local consequences from spatially non-uniform $g(x)$, we examine the Schr\"{o}dinger eigen-equation $H\ket{\psi_{n}}=E_{n}\ket{\psi_{n}}$ of Eq.~\ref{1bH}:
\begin{align}
   \frac{ g(x)}{\gamma }\psi_{n}(x+1)+g(x-1)\gamma \psi_{n}(x-1)=E_{n} \psi_{n}(x),\label{bulk}
\end{align}
where $E_n$ is the eigenenergy of the eigenstate $\ket{\psi_n}=\sum_x \psi_n(x)| x\rangle$. For later convenience, we define the lower and upper bounds of $g(x)$ by $g_\text{min}=\text{Min}[g(x)]$ and $g_\text{max}=\text{Max}[g(x)]$, such that $g_\text{min}=g_\text{max}$ only if the hoppings are completely homogeneous in space.

Since $g(x)$ acts like an energy rescaling factor in the Hermitian continuum limit, local energy conservation requires that $g(x)|\psi_n(x)|^2$ remains invariant.
Extending this to generic non-Hermitian cases where non-local pumping arises from $\gamma\neq 1$, we propose an ansatz for the state amplitude $\psi_{n}(x)$ at site $x$ to be 
\begin{gather}
\psi_{n}(x)=\frac{1}{\sqrt{g(x)}}\gamma^x\prod_{x'=1}^{x}\beta_n(x')\label{ansatz},
\end{gather}
where the $\gamma^x$ keeps track of the pure exponential state accumulation, and $\prod_{x'=1}^{x}\beta_n(x')$ denotes the part of the state accumulation that depends specifically on the $g(x)$ profile, which is the key quantity that we will focus on in this work.

Together, $\gamma$ and $\beta_n(x)$ define the \emph{phase-space} GBZ, as schematically illustrated in Fig.~\ref{cartoon2}b:
\begin{align}
    z_n(x)=\frac{\psi_n(x)}{\psi_n(x-1)}={\sqrt{\frac{g(x-1)}{g(x)}}}\gamma\beta_n(x),\label{zdefine}
\end{align}

which is defined in position-momentum phase space, depending \emph{both} on the position $x$, and the state momentum~\footnote{In more complicated spatially-inhomogeneous lattices with further hoppings, which we will not consider in this work, not only will there be more $\beta_n(x)$ branches, $\gamma$ would also depend on the momentum~\cite{tai2023zoology,lee2020unraveling,lee2022exceptional,zhang2023electrical,lee2019hybrid,lee2019anatomy,li2020topological,kawabata2020higher,okuma2020topological,borgnia2020non,sun2021geometric,li2021quantized,lee2021many,zhang2022review,shen2022non,jiang2023dimensional,qin2023universal}.} indexed through $n$. Different from the usual spatially homogeneous (i.e., constant $g(x)=g$) \emph{OBC} case, where we simply have $z(x)=\gamma$, we have the new $\beta_n(x)$ factor that we call the phase-space GBZ factor, which would compensate for the usual $\gamma$ rescaling factor in a homogeneous PBC system. Its spatial periodicity $\beta_n(x+L)=\beta_n(x)$ is inherited from that of the hopping amplitude profile $g(x)$ and the state amplitude $\psi_n(x)$.

To determine the form of the phase-space GBZ factor $\beta_n(x)$, we substitute this ansatz [Eq.~\ref{ansatz}] into the bulk equation Eq.~\ref{bulk} and arrive at
\begin{gather}
    \frac{\sqrt{g(x)}g(x)}{\sqrt{g(x+1)}}\beta_{n}(x+1)+\frac{{\sqrt{g(x)g(x-1)}}}{\beta_{n}(x)}=E_{n}.\label{Bulk}
\end{gather}
This expression Eq.~\ref{Bulk} applies to generic hopping inhomogeneity $g(x)$, even discontinuous ones to a good approximation (see Sect. \ref{sec_discont_gx}). However, for most cases that we shall consider, we will further make the key assumption of local spatial continuity: If we consider a sufficiently smooth hopping function $g(x)$ such that the spatial gradients satisfy
\begin{align}
    &g'(x)\ll g(x),\label{SmthApprox}\\
    &\beta'_n(x)\ll\beta_n(x),\label{SPT}
\end{align}
in the thermodynamic limit of large $L$, Eq.~\ref{Bulk} simplifies to
\begin{flalign}
    &\beta_{n}(x+1)+\frac{1}{\beta_{n}(x)}=\frac{E_{n}}{ g(x)}&&\label{BulkApprox}\\
     \xRightarrow{\text{continuous GBZ}}\quad &\beta_{n}(x)+\frac{1}{\beta_{n}(x)}=\frac{E_{n}}{g(x)}.&&\label{BulkApproxSmooth}
\end{flalign}
The second line, obtained by assuming local continuity of the phase-space GBZ, decouples the inter-dependency between neighbouring $\beta_n(x)$ and $\beta_n(x+1)$, such that $\beta_n(x)$ can be solved solely from the local hopping $g(x)$ and the eigenenergy $E_n$.
In practice, the locally continuous GBZ assumption can be justified a posteriori by comparing its analytic predictions with numerical diagonalization results. From our results presented later, it turns out that even the rapidly oscillating phases in the wavefunctions do not compromise this assumption, at least for the majority of the reasonably smooth eigenstates.

From Eq.~\ref{BulkApproxSmooth}, the phase-space GBZ factor $\beta_{n}(x)$ can be directed solved in terms of the eigenenergy $E_n$ and the spatial hopping amplitude profile $g(x)$, with a pair of solutions $\beta_{n,\pm}(x)$ given by
\begin{align}
    \beta_{n,\pm}(x)=&\exp\left(\pm \cosh^{-1}\left(\frac{E_{n}}{2g(x)}\right)\right)\notag\\
    =&\frac{E_{n}}{2g(x)}\pm i \sqrt{1-\frac{E_{n}^2}{4g^2(x)}}\ ,\label{PTSol1}
\end{align}
with $|\beta_{n,+}|\geq 1$. To keep track of the extent of spatial state accumulation, we decompose $\beta_{n,\pm}$ as 
\begin{align}
    \beta_{n,\pm}(x)&=\exp\{i[k_{n,\pm}(x)+i\kappa_{n,\pm}(x)]\}\notag\\
    &=e^{ik_{n,\pm}(x)}e^{-\kappa_{n,\pm}(x)},
    \label{skinDepth}
\end{align}
where 
{\small
\begin{align}
    \kappa_{n,\pm}(x)=-\log|\beta_{n,\pm}(x)|=\mp\text{Re}\left[ \cosh^{-1}\left(\frac{E_{n}}{2g(x)}\right)\right]\ 
    \label{skinDepth2}
\end{align}}
represents the local contribution to the \emph{inverse inhomogeneous} skin depth at position $x$, and the phase $k_{n,\pm}(x)$ describes the effective Bloch-like phase oscillations with spatially varying wavenumber $dk_{n,\pm}(x)/dx$.

\subsubsection{Spatially inhomogeneous GBZ branches}

\noindent Although Eq.~\ref{PTSol1} or Eq.~\ref{skinDepth2} may look superficially similar to that of the usual Hatano-Nelson model~\cite{gong2018topological,schindler2021dislocation,HN1996prl,HN1997prb,HN1998prb,PhysRevB.92.094204,claes2020skin}, where $g(x)$ is constant, the inhomogeneity of $g(x)$ brings about various new levels of subtleties. First, labeling the $\beta_{n,\pm}(x)$ solutions such that
\begin{gather}
    |\beta_{n,+}(x)|\geq 1 \geq |\beta_{n,-}(x)|.\label{absBeta}
\end{gather}
We identify the following distinct regions in real space $x$:

\begin{itemize}

\item \textbf{Pure skin region:} $ |\beta_{n,+}(x)|= |\beta_{n,-}(x)|=1$, i.e., $\kappa_{n,\pm}=0$, such that the spatial state profile in these positions is purely exponential (just like usual non-Hermitian skin modes), arising only from the $\gamma^x$ term in Eq.~\ref{ansatz}. It occurs in the region $|\text{Re}(E_n)|\leq 2g(x)$ and $\text{Im}(E_n)=0$. 
\\

\item \textbf{Inhomogeneous skin region:} nonconstant $|\beta_{n,+}(x)|>1 > |\beta_{n,-}(x)|$, such that the spatial state profile is manifestly non-exponential. 
It occurs when $\text{Im}(E_n)\neq0$ or $|\text{Re}(E_n)| >2g(x)$, which represents pockets of weak hopping with no spatially homogeneous analog.

\end{itemize}

Of course, the same physical eigenstate $\psi_n(x)$ can exhibit both pure skin and inhomogeneous skin behaviors at different locations $x$. 
But, whether exhibiting pure or inhomogeneous skin, $\psi_n(x)$ can only incorporate one of the two possible $\beta_{n,\pm}(x)$ solution branches at any particular point $x$. 

We define the choice function $\sigma(x)$ that takes values of $\pm 1$ depending on which branch is chosen at position $x$; exactly how $\sigma(x)$ can be determined will be detailed in the next subsection. Notating the chosen branch as $\beta_n(x)=\beta_{n,\sigma(x)}(x)$ with 
{\small
\begin{align}
    \kappa_{n}(x)=-\log|\beta_{n}(x)|=-\sigma(x)\text{Re}\left[ \cosh^{-1}\left(\frac{E_{n}}{2g(x)}\right)\right],
    \label{skinDepth3}
\end{align}}
we distinguish between two different scenarios for the phase-space GBZ:
\begin{itemize}
\item \textbf{Continuous phase-space GBZ:} Either $\beta_n(x)=\beta_{n,+}(x)$ or $\beta_n(x)=\beta_{n,-}(x)$ for all $x$, such that only one branch is ever realized, i.e., $\sigma(x)=\pm 1$ for all $x$. 
\item \textbf{Discontinuous phase-space GBZ:} $\beta_n(x)$ switches (jumps) between the $\beta_{n,+}(x)$ and $\beta_{n,-}(x)$ branches at the so-called GBZ inversion points $x_\text{jump}$, which exist due to the spatial inhomogeneity from $g(x)$. 

\end{itemize}

\noindent In a nutshell, the phase-space GBZ connectivity can be classified by the number of $x_\text{jump}$ points where $\sigma(x)$ alternates between $+1$ and $-1$. 
An even number of alternations must occur since $\sigma(x)$ is periodic in $x$ and has to switch an even number of times. A continuous/discontinuous phase-space GBZ corresponds to a zero/nonzero number of $x_\text{jump}$ points -- in this work, we shall explicitly examine only cases with at most two $x_\text{jump}$ points, since more complicated cases can be broken down into multiple discontinuous GBZs in real space and analyzed separately.

\begin{figure}
    \centering
    \includegraphics[width=\linewidth]{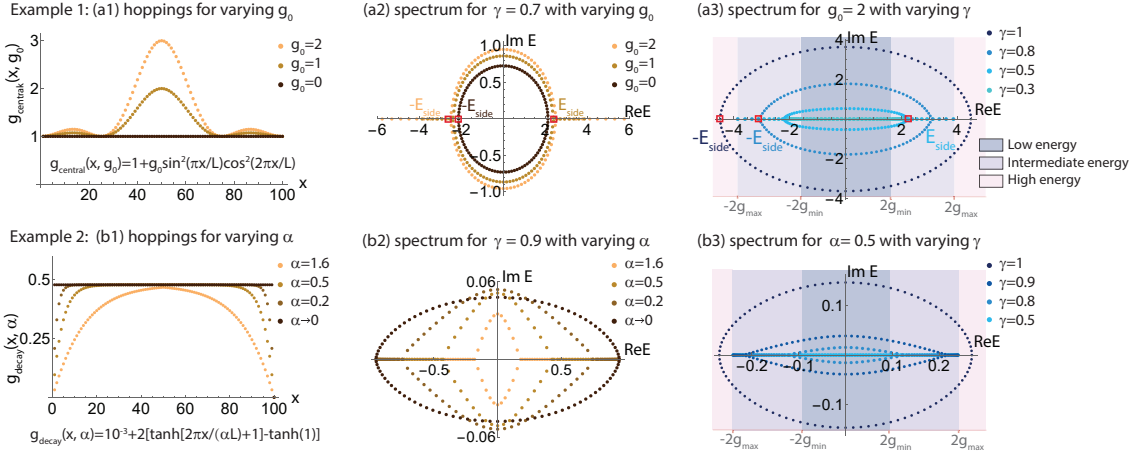}
    \caption{
    \textbf{How the spatial profile $g(x)$ and asymmetry $\gamma$ of the non-Hermitian hoppings affect their energy spectra. } 
    Shown are two illustrative example models with hopping profiles (a1-a3) $g(x)=g_{\text{bump}}(x,g_0)=1 + g_0 \sin^2(\pi x/L)\cos^2(2 \pi x/L)$ and (b1-b3) $g(x)=g_{\text{dip}}(x,\alpha)=10^{-3}+2\left(\tanh(\frac{2 \pi x}{\alpha L} + 1)-\tanh(1)\right)$ for $x\leq L/2$ and symmetric $g(x)$ about $x=\frac{L}{2}$ for $x>L/2$, where $g_0$ and $\alpha$ respectively control the extent of spatial inhomogeneity. (a1,b1) $g(x)$ profiles for these two models. (a2,b2) The corresponding energy spectra at fixed $\gamma$, which branch out into real spectral ``tails'' at branch points $\pm E_\text{side}$ for sufficiently strong inhomogeneity $g_0$ or $\alpha$. (a3,b3) Their corresponding energy spectra at fixed $g_0$ or $\alpha$. As the non-Hermiticity departs from the Hermitian ($\gamma=1$) limit where $E_{\text{side}}=2g_{\text{min}}$, the spectral loops expands and eventually engulfs the real ``tails'' when $E_{\text{side}}$ exceeds $2g_{\text{max}}$. As elaborated in the text, real tails can only exist in the so-called intermediate energy regime where $2g_\text{min}<\text{Re}[E]\leq 2g_\text{max}$.
}
\label{inhomo_figure2}
\end{figure}

Note that $\kappa_n(x)$ controls the eigenstate amplitude not just locally at $x$, but, in fact, non-locally:
\begin{equation}
|\psi_n(x)|=\frac{\gamma^{x}}{\sqrt{g(x)}} \prod\limits_{x'=1}^{x}e^{-\kappa_{n}(x')},
\label{magpsi}
\end{equation}
which, in terms of the spatial gradient of the state amplitude, takes the form:
\begin{equation}
\frac{\text{d}}{\text{d}x}\left(\log{(}\sqrt{g(x)} \left|\psi_n(x)\right|{)}\right)=\log\gamma -\kappa_n(x).
\label{magpsi2}
\end{equation}

As such, $\kappa_n(x)$ can be interpreted as the spatially dependent correction factor that compensates for the exponential ``pure skin'' accumulation from $\gamma^x$, such that the wavefunction satisfies PBCs. It represents a \emph{position-dependent} reduction of the inverse skin localization length $-\log\gamma$. As the simplest example, consider the spatially homogeneous Hatano-Nelson model where $g(x)=g$ and PBC energy can be
\begin{equation}
E_n=g(\gamma e^{ip_n} + \gamma^{-1}e^{-ip_n})=2g\cosh\left(\log \gamma + i p_n\right),
\label{EnHN}
\end{equation}
where real momentum $p_n\in [0,2\pi)$. We have
\begin{equation}
\kappa_n(x)=\text{Re}\left(\cosh^{-1}(E_n/2g)\right)=\log \gamma,
\end{equation}
which serves to exactly cancel off the $\gamma^x$ accumulation [Eq.~\ref{magpsi}] in the case of the Hatano-Nelson model.

\subsubsection{Determining the allowed energy spectrum}

\noindent The PBC condition $\psi_{n}(x+L)=\psi_{n}(x)$ of the system imposes an important constraint on $\beta_n(x)$ that enables its spectrum $E_n$ and eigenstates $\psi_n(x)$ to be uniquely solved. Substituting the PBC condition into the ansatz Eq.~\ref{ansatz}, we obtain
\begin{align}
   &\gamma^{-L}=\prod_{x=1}^{L}\beta_n(x),\label{GBZBC0}
\end{align}
which, from Eq.~\ref{skinDepth2}, is equivalent to the following handy constraint on the skin depth and eigenenergies:
\begin{align}
    \log(\gamma)=& \frac{1}{L}\sum_{x=1}^{L} \kappa_n(x)\notag\\
    =&-\frac{1}{L}\sum_{x=1}^{L} \sigma(x)\,\text{Re}\left(\cosh^{-1}\left(\frac{ E_n}{2g(x)}\right)\right).
    \label{GBZBC}
\end{align}
In particular, Eq.~\ref{GBZBC} allows the spectrum $E_n$ to be mapped out once $\sigma(x)$ is determined, as will be explained in the following pages. By restricting the 2D complex energy plane to satisfy the constraint equation, the spectrum takes the form of 1D curves or branches, as will be demonstrated later in Figs.~\ref{inhomo_figure2}-\ref{0908fig3}. 

As a corollary to Eq.~\ref{GBZBC}, it is not possible for an entire non-Hermitian system to consist only of pure skin regions (with $\kappa_n(x)=0$), since $\log \gamma\neq 0$. This is just another way of saying that any net exponential NHSE accumulation in a periodic system must be smoothed out and compensated in a $g(x)$-dependent way, as given by Eqs.~\ref{skinDepth3} and \ref{magpsi}.

\subsection{Anatomy of non-Hermitian spectra and eigenstates for spatially inhomogeneous hoppings between monoatomic unit cells}

Fig.~\ref{inhomo_figure2} presents two illustrative models of how the non-Hermitian spectrum behaves as the hopping asymmetry $\gamma$ and spatial inhomogeneity are varied. They are $g(x)=g_\text{bump}(x,g_0)$ and $g(x)=g_\text{dip}(x,\alpha)$, as shown in Figs.~\ref{inhomo_figure2}a1 and~\ref{inhomo_figure2}b1. In the first model, $g_0$ controls the height of the bump at $x=L/2$ [Fig.~\ref{inhomo_figure2}a1]. In the second model, $\alpha$ controls the depth of the dip in the hopping amplitude near the ends $x=1,L$ [Fig.~\ref{inhomo_figure2}b1].  

Even though both hopping profiles result in rather different spectra, they universally behave in qualitatively similar ways [Figs.~\ref{inhomo_figure2}a2,~\ref{inhomo_figure2}b2]. In both cases, the spectrum initially assumes the form of the Hatano-Nelson PBC spectral ellipse [Eq.~\ref{EnHN}] (black) when there is no spatial hopping inhomogeneity, i.e., $g_0=0$ or $\alpha=0$. However, when the spatial inhomogeneity $g_0$ or $\alpha$ is introduced, it generically deforms the spectral loop and saliently introduces real spectral branches or ``tails'' at its sides. We call the eigenenergy branch points where the branches join the loop as $\pm E_\text{side}$ [red square markers in (a2) and (a3)]. These real spectral branches exist only when the non-Hermiticity is not excessively strong, emerging in Figs.~\ref{inhomo_figure2}a3,~\ref{inhomo_figure2}b3 as the hopping asymmetry $\gamma$  decreases towards the Hermitian limit ($\gamma=1$). As will be proven later in this section, these branches can only exist for $2g_\text{min}<|\text{Re}(E)|\leq 2g_\text{max}$ (dubbed the intermediate energy regime), implying that they can only appear when the spectral loop is contained within $|\text{Re}(E)|< 2g_\text{max}$.

Since directed amplification is supposed to continue indefinitely around a homogeneous PBC loop, the appearance of these real eigenenergy branches reveals how spatial inhomogeneity physically behaves like ``partial boundaries'' that can nevertheless completely stop the amplification. For $g_\text{dip}(x,\alpha)$ [Fig.~\ref{inhomo_figure2}b2], it is also interesting that $\text{Max}(\text{Im}(E))$ peaks slightly at moderate values of $\alpha\approx 0.2$, indicative of a slight enhancement of NHSE feedback gain due to hopping inhomogeneity, even though nonzero $\alpha$ corresponds to regions of weak hoppings that should have reduced the overall hopping energies.

\subsubsection{Real spectral branches from GBZ discontinuities}

\noindent To derive the real eigenenergy tail segments that would appear given a generic $g(x)$ and $\gamma\neq 1$, we turn to the boundary condition (Eq.~\ref{GBZBC}), which constrains the set of possible $E_n$ eigensolutions for a given GBZ choice function $\sigma(x)$. Since the full solution set of Eq.~\ref{GBZBC} consists of 1D spectral curves in the complex energy plane, additionally restricting to the real line ($\text{Im}(E)=0$) reduces the solutions to one or more isolated points. In particular, for constant $\sigma(x)=-\text{Sgn}[\log \gamma]$, i.e., continuous phase-space GBZs, the real eigenenergies are found to be $E_n=\pm E_\text{side}$, as defined by
\begin{align}
\qquad\text{Re}\sum_{x=1}^{L} \cosh^{-1}\left(\frac{E_\text{side}}{2g(x)}\right)=L|\log\gamma|.
   \label{sideReE2}
\end{align}

To show that $\pm E_\text{side}$ indeed bounds the two sides of a complex spectral loop [Figs.~\ref{inhomo_figure2}a2,~\ref{inhomo_figure2}a3,~\ref{inhomo_figure2}b2,~\ref{inhomo_figure2}b3], we invoke the following relations between a generic eigenenergy $E_n$ and its $k_n(x)$ and $\kappa_n(x)$ [Eq.~\ref{skinDepth}]:
\begin{align}
    &\frac{\text{Re}(E_{n})}{2g(x)}=\cos(k_{n}(x))\cosh(\kappa_{n}(x)),\label{smoothRe}\\
    &\frac{\text{Im}(E_{n})}{2g(x)}=-\sin(k_{n}(x))\sinh(\kappa_{n}(x)),\label{smoothIm}
\end{align}

which can be obtained by separating the real and imaginary parts of Eq.~\ref{BulkApproxSmooth}. Since $\kappa_n(x)$ cannot identically vanish due to the boundary condition [Eq.~\ref{GBZBC}], $\sin( k_n(x))$ must vanish identically for the real eigenenergy $E_\text{side}$ [Eq.~\ref{smoothIm}]. To continue satisfying Eq.~\ref{GBZBC} for $\text{Re}(E)<E_\text{side}$, $k_n(x)$ in Eq.~\ref{smoothRe} can simply be tuned up; however, doing so inevitably also introduces non-zero $\text{Im}(E)$, as required by Eq.~\ref{smoothIm}. As such, for $\text{Re}(E)<E_\text{side}$, $k_n(x)$ generically generates continuous spectral curves extending into the complex plane.

Most interestingly, if the GBZ choice function $\sigma(x)$ were to vary with $x$, exhibiting jumps between $\pm 1$ values, it is possible to realize a whole continuum of real eigenenergies $E_n$, i.e., the real spectral ``tails'' that all satisfy Eq.~\ref{GBZBC}, as illustrated in Fig.~\ref{0908fig3}a. Previously, with constant $\sigma(x)$, real energies with $E_n>E_\text{side}$ cannot satisfy Eq.~\ref{GBZBC} because the inverse cosh function is monotonically increasing, such that their $\text{Re}\left(\cosh^{-1}\left(E_n/2g(x)\right)\right)$ contributions must exceed that from $E_\text{side}$. However, if $\sigma(x)$ is non-uniform (i.e., exhibits phase-space GBZ discontinuities), jumping at $x=1$ and $x=x_\text{jump}$:
\begin{align}
       \qquad\sigma(x) =\begin{cases}
            1 & 1\leq x< x_\text{jump}\\
            -1 & x_\text{jump}\leq x \leq L ,
        \end{cases}\ ,\label{sigmacases}
\end{align}
the constraint Eq.~\ref{GBZBC} becomes
{\small \begin{align}
\left({\sum_{x=x_\text{jump}}^{L} - \sum_{x=1}^{x_\text{jump}-1}} \right)\text{Re}\left(\cosh^{-1}\left(\frac{E_n}{2g(x)}\right)\right)=L|\log\gamma|,
   \label{GBZBCinv}
\end{align}}
which can be satisfied by a continuum of real $E_n>E_\text{side}$ as the GBZ discontinuity position $x_\text{jump}$ is decreased continuously from $L$.
As shown for an illustrative $g(x)$ [Fig.~\ref{0908fig3}b], such discontinuities are indeed numerically observed [Fig.~\ref{0908fig3}c1] in a typical real-energy state whose absolute energy lies between $|E_{\text{side}}|$ and $|2g_{\text{max}}|$.

In the above, we have established that, due to the freedom in toggling between two different GBZ solutions $\pm \kappa_n(x)$ in a position-dependent manner, as encoded by $\sigma(x)$ jumps, extensively many real energy eigenstates can exist in a PBC system with spatially inhomogeneous hopping amplitudes. This is of profound physical significance because these real energy states do not grow with time, unlike almost every eigenstate in a clean PBC NHSE system~\footnote{For instance, the spectrum of a spatially homogeneous Hatano Nelson model is an ellipse in the complex energy plane, and only two isolated $\pm E_\text{side}=\pm g(\gamma+\gamma^{-1})$ energies are real.}, which has complex energy due to unfettered directional amplification. The juxtaposition of two different GBZs $\kappa_n(x)$ at a spatial GBZ discontinuity $x=x_\text{jump}$ effectively gives rise to an effective spatial ``barrier'' that curtails directional state growth, at least for eigenstates lying in the real spectral branches.

\subsubsection{Spectral behavior in low, high and intermediate energy regimes}

Having discussed how the hopping inhomogeneity leads to the real spectral segments, here we discuss how it affects the full non-Hermitian ($\gamma\neq 1$) spectral behavior in the whole complex plane. To showcase the universality of our arguments, we introduce an additional model with a spatially inhomogeneous hopping profile $g(x)$ that contains a smooth bump at $x=L/2$ and a sharp bump at $x=1$ or $L$ [Fig.~\ref{0908fig3}]. 

Below, we classify the eigenenergies $E_n$ into 3 regimes by comparing $\text{Re}(E_n)$ against the lower and upper bounds of $g(x)$, as colored in Figs.~\ref{inhomo_figure2}a3, \ref{inhomo_figure2}b3 and Fig.~\ref{0908fig3}a: 

\begin{itemize}[leftmargin =.5cm]
\item \textbf{Low energy regime with $|\text{Re}(E_n)|\leq 2g_{\text{min}}$:} 
Only complex $E_n$ allowed, with $\kappa_n(x)\neq 0$ (inhomogeneous skin) across all $x$.

To see why the eigenenergies $E_n$ must be complex, first note that $k_n(x)\neq 0$, because $|\text{Re}(E_n)/2g(x)|\leq 1$ in the LHS of Eq.~\ref{smoothRe}, but $|\cosh(\kappa_{n}(x))|\geq 1$ on the RHS. This is the only possibility because the special case $|\text{Re}(E_n)/2g_{\text{min}}(x)|=1=|\cosh(\kappa_{n}(x))|$ with $|\text{Re}(E_n)|=2g_{\text{min}}(x)$ cannot hold, as $\kappa_n(x)$ cannot identically vanish due to the PBC condition [Eq.~\ref{GBZBC}]. Eq.~\ref{smoothIm} then forces $\text{Im}(E_n)$ to be nonzero.

Note that this low energy regime also exists even when $g(x)$ is spatially uniform, since the above arguments do not involve the details of the $g(x)$ profile, only that $|\text{Re}(E_n)|< 2g_{\text{min}}$. This is already evident from Fig.~\ref{inhomo_figure2}a2, where the low energies form the top of the spectral loops for any $g(x)$ profile. In particular, similar to the spatially homogeneous case, the phase-space GBZ is also continuous with $\kappa_n(x)\neq 0$ everywhere [Fig.~\ref{0908fig3}c3], as required for complex $E_n$ with $k_n(x)\neq 0$ [Eq.~\ref{smoothIm}].

\begin{figure}
    \centering
    \includegraphics[width=0.8\linewidth]{\Pic{pdf}{fig3_v2}}
    \caption{ 
    \textbf{GBZ bifurcations and jumps in the inhomogenous skin regions. }
    (a) Numerically obtained PBC energy spectrum of an illustrative model with $g(x)=1/(\sin(2 \pi x/L) \cos(\pi x/L) + 0.3)$, $L=100$ and $\gamma=0.7$, which satisfies Eq.~\ref{GBZBC}. The branch points $\pm E_\text{side}$ in the intermediate energy regime connect the complex spectral segments with the real ``tails'', whose existence is constrained by Eq.~\ref{sideReE3}. Here, no states exist in the high energy regime. 
    (b) The doubled spatial hopping profile $2g(x)$ is shown against three three illustrative chosen eigenenergy values, which determine the pure and inhomogenous skin regions for their respective eigenstates $\psi$ and GBZs $\kappa(x)$ shown below.
    (c1-c3) The GBZ bifurcates into two branches $\pm \kappa_{n}(x)=\sigma(x)\text{Re}\left(\cosh^{-1}(E_n/2g(x))\right)\neq 0$ at inhomogeneous skin regions (pale yellow) where the hoppings are locally weak, i.e., $\text{Re}(E_n)>2g(x)$ or  $\text{Im}(E_n)\neq0$. The GBZ branch chosen by the numerical eigenstates (green), as computed from Eq.~\ref{magpsi2}, can jump abruptly in the inhomogeneous skin region ($\kappa(x)\neq 0$), as for $E_1$. The jump position $x_{\text{jump}}=27$ is consistent with Eq.~\ref{GBZBCinv}. However, no jump may occur even if GBZ bifurcation occurs, as for $E_2=E_\text{side}$ where the numerically-obtained GBZ adheres to one GBZ solution throughout. No jump can possibly occur when no pure skin region (pale yellow) exists and the GBZ solutions never get to meet, as for $E_3$ not real in the low energy regime.
    }
    \label{0908fig3}
\end{figure}

\item \textbf{High energy regime with $|\text{Re}(E_n)|>2g_{\text{max}}$:}
Either no states at all, or two branches of complex $E_n$ with $\kappa_n(x)\neq 0$ (inhomogeneous skin) meeting at an isolated real eigenenergy point $E_\text{side}$. 

These two possibilities correspond to $\gamma>0.3$ ($\gamma = 0.3$) in Fig.~\ref{inhomo_figure2}a3 and $\gamma>0.5$ ($\gamma = 0.5$) in Fig.~\ref{inhomo_figure2}b3. The argument partially mirrors that of the low-energy regime. Here, $\kappa_n(x)\neq 0$ because $|\text{Re}(E_n)/2g(x)|>1$ in the LHS of Eq.~\ref{smoothRe}, but $|\cos(k_{n}(x))|\leq 1$ on the RHS. With $\kappa_n(x)\neq 0$ for all $x$, Eq.~\ref{smoothIm} forces $\text{Im}(E_n)$ to also not disappear, except for the possible special real solution, which we call $E_n=E_\text{side}$, where $k_n(x)=0$ for all $x$.

\item \textbf{Intermediate energy regime with $2g_{\text{min}}< |\text{Re}(E_n)|\,\leq\, 2g_{\text{max}}$:}

Most interesting is that this intermediate energy regime is characterized by $E_n$ lying between the lower and upper bounds $2(g_\text{min},g_\text{max}]$ of the hopping energy $g(x)$ [Figs.~\ref{inhomo_figure2}a3,~\ref{inhomo_figure2}b3 and Fig.~\ref{0908fig3}a], which has no analogue in the spatially homogeneous limit (where $g_\text{min}=g_\text{max}$). This regime is most intricate because, for a fixed value of $E_n$, there \emph{simultaneously} exist spatial intervals with Re $ E_n>2g(x)$ as well as Re $ E_n\leq 2g(x)$, giving rise to coexisting locally high and low energy regions where $\kappa_n(x)=0$ and $\kappa_n(x)\neq 0$, respectively. 

In particular, it is only in this intermediate energy regime that a continuum of real $E_n$ can exist. 
As shown in Figs.~\ref{0908fig3}a,b, if $2g_\text{min}< E_\text{side}\leq 2g_\text{max}$, the two complex spectral branches from the low-energy regime would meet at $E_n=E_\text{side}$ and continue as a real energy tail that extends till $E_n=2g_\text{max}$, the upper limit of the intermediate energy regime. These real energies in the tail correspond to phase-space GBZ discontinuities at $x=x_\text{jump}$, as given in Eq.~\ref{GBZBCinv}, although nonzero contributions to the sum only come from $\kappa_n(x)\neq 0$ (inhomogeneous skin) regions where $g(x)<E_n/2$.

To understand the role of $E_\text{side}$, note that it is the ``extremal'' solution of the boundary equation [Eq.~\ref{GBZBC}] that still keeps $\sigma(x)$ constant, i.e., $\log(\gamma)= \frac{1}{L}\sum_{x=1}^{L} \kappa_{\text{side}}(x)$. For other real energies $E_n$ with $|E_n|>E_\text{side}$, the constraint $|\kappa_{n}|>|\kappa_{\text{side}}|$ would then require position-dependent $\sigma(x)$ such as to satisfy the boundary equation [Eq.~\ref{GBZBC}]. The upper limit of this continuum of real energies is given by $E_n=2g_\text{max}$, since  $\cosh \kappa_n(x)\leq 1$ and $\sin k_{n}(x)=0$ in Eq.~\ref{smoothRe} and Eq.~\ref{smoothIm}.

As shown in Figs.~\ref{0908fig3}a,b, if $2g_\text{min}< E_\text{side}\leq 2g_\text{max}$, the two complex spectral branches from the low-energy regime would meet at $E_n=E_\text{side}$ and continue as a real energy tail that extends until $E_n=2g_\text{max}$, the upper limit of the intermediate energy regime. These real energies in the tail correspond to phase-space GBZ discontinuities at $x=x_\text{jump}$, as given in Eq.~\ref{GBZBCinv}, although non-zero contributions to the sum only come from $\kappa_n(x)\neq 0$ (inhomogeneous skin) regions where $g(x)<E_n/2$. As shown in Figs.~\ref{0908fig3}c1,c2 for eigensolutions $E_1,E_2$ that lie in the intermediate energy regime [Fig.~\ref{0908fig3}a purple], the jump/s can be numerically extracted from the spatial wavefunction profile via Eq.~\ref{magpsi2}. In Fig.~\ref{0908fig3}c1, the numerical fit to either $\pm\kappa(x)$ solution is good, except at the rather abrupt jump. In Fig.~\ref{0908fig3}c2, there is no jump as the numerical wavefunction adheres to only one GBZ solution branch throughout; in Fig.~\ref{0908fig3}c3, no jump is possible because the two $\kappa(x)$ solutions do not even touch.

However, if $E_\text{side}$ exists in the high-energy regime, i.e., $E_\text{side}>2g_\text{max}$ or
\begin{equation}
\frac{1}{L}\sum_{x=1}^L\cosh^{-1}\left(g_\text{max}/g(x)\right)<|\log\gamma|,
\label{sideReE3}
\end{equation}
there will be no real spectral tail, and the complex spectral branches simply meet at the real point $E_\text{side}$ and terminate there [Figs.~\ref{inhomo_figure2}a3,~\ref{inhomo_figure2}b3]. This would definitely be the case when $g(x)$ is uniform, since $g(x)=g_\text{max}$. Hence Eq.~\ref{sideReE3} gives the threshold for the absence of real eigenenergies: as the hopping asymmetry $\log \gamma$ is increased, directed amplification becomes stronger, and greater spatial inhomogeneity $g(x)$ in the hoppings is needed to stop the amplification and produce asymptotically dynamically stable eigensolutions.

\end{itemize}

\begin{figure}
    \includegraphics[width=0.9\linewidth]{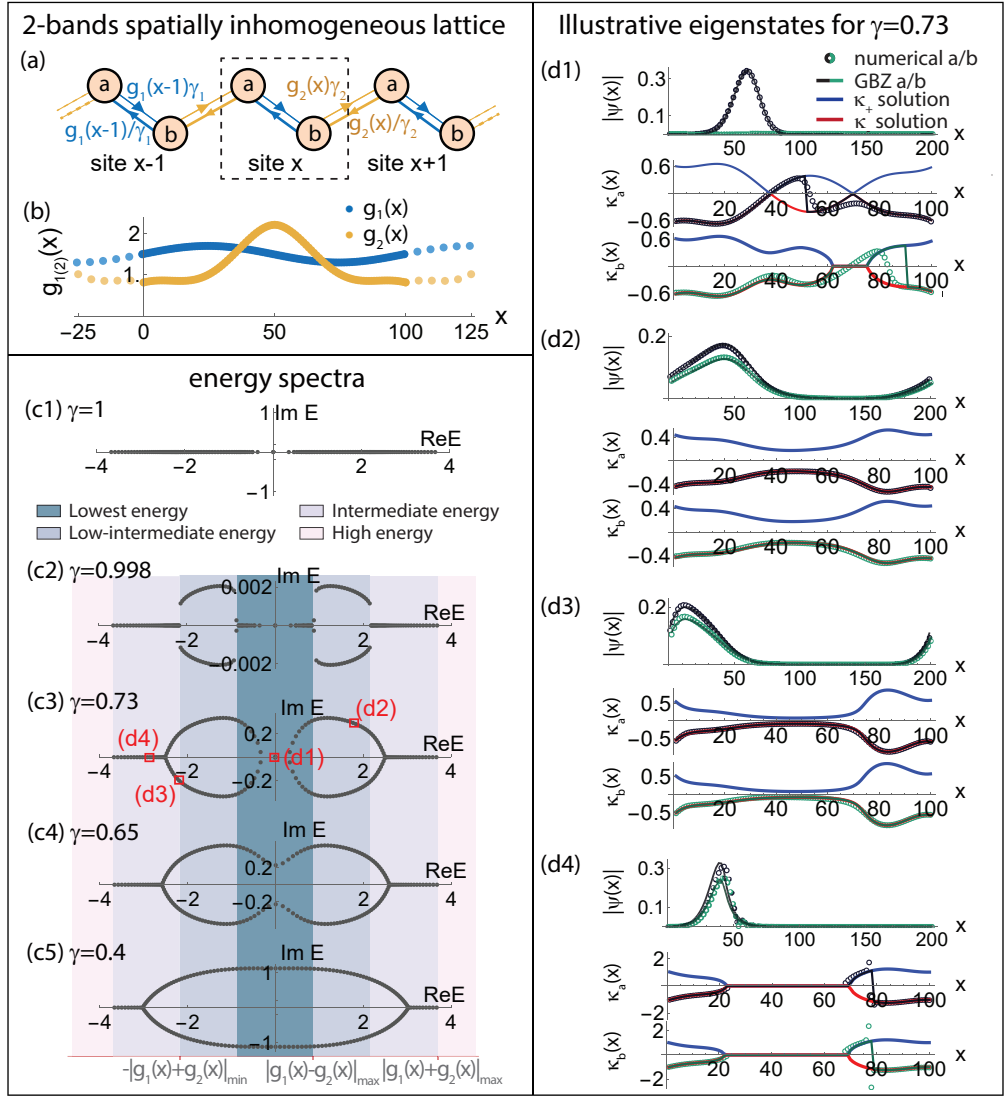}
	\caption{\textbf{The energy regimes and bifurcated GBZs of 2-component spatially inhomogeneous non-Hermitian lattices. }
    (a) Schematic of our spatially inhomogeneous PBC lattice with two atoms per unit cell (dashed), such that the inhomogeneous intra-cell hoppings and inter-cell hoppings have independent profiles $g_1(x)$ and $g_2(x)$.  
    (b) The illustrative profiles used, with $g_1(x)= 1.5 + 0.2 \sin(2 \pi x/L)$ and $g_2(x)=(0.001|(x - L/2)^{1.8}|+ 0.5)^{-1} + 0.2 \cos(4 \pi x/L) + 0.01$, with $L=200$.
    (c1-c5) Energy spectra as the combined hopping asymmetry $\gamma=\gamma_1\gamma_2$ is tuned from the Hermitian $\gamma=1$ limit to $\gamma=0.4$. Different from the 1-component case, there are four distinct energy regimes and two spectral loops that terminate at real ``tails'' in the lowest and intermediate energy regimes. A spectral transition occurs as $\gamma$ is lowered from $\gamma=0.73$ to $0.65$, when the two complex loops coalesce into one, destroying the zero mode. 
    (d1-d4) Wavefunctions of 4 representative numerical eigenstates (black/green for component $a/b$) from (c3), and their approximations by the phase-space GBZs $\pm\kappa_a(x),\pm\kappa_b(x)$. Analogous to 1-component cases, states on the real ``tails'' possess discontinuous phase-space GBZs and exhibit $x_{j,\text{jump}}$ jumps such as to satisfy the PBC constraint Eq.~\ref{2BspecGBZ}. Interestingly, the zero mode of (d1) exhibits completely distinct $\kappa_{a,b}(x)$ profiles both analytically and numerically, a spectacular consequence of the spatial inhomogeneity of the hoppings. 
    }
	\label{Fig4} 
\end{figure}

\subsection{Phase-space GBZ for inhomogeneous two-component chains}\label{sec_2b}

Here, we generalize previous results on monoatomic unit cells to spatially inhomogeneous lattices with diatomic unit cells, such that the odd and even hoppings have independent spatial profiles $g_1(x)$ and $g_2(x)$. Having a non-trivial unit cell leads to the appearance of multiple spectral bands that not only enrich the phase-space GBZ and complex spectral graphs~\cite{yang2020non,song2019realspace,esaki2011edge,shen2018topological,PhysRevA.97.052115,manna2023inner}, but also host special topological ``soft-interface'' zero modes that have no analogue in spatially homogeneous PBC or OBC systems.
   
A 1D two-component (diatomic unit cell) lattice with spatially inhomogeneous hoppings is defined by

{\small \begin{align}
	H_\text{2-comp}&=\sum_{x=1}^{L}g_1(x)\left(\frac{1}{\gamma_{1}}\ket{x,a}\bra{x,b}+\gamma_{1}\ket{x,b}\bra{x,a}\right)\notag\\&+\sum_{x=1}^{L-1}g_{2}(x)\left(\frac{1}{\gamma_{2}}\ket{x,b}\bra{x+1,a}+\gamma_{2}\ket{x+1,a}\bra{x,b}\right)\notag\\
 &+g_2(L)\left(\frac{1}{\gamma_2}\ket{L,b}\bra{1,a}+\gamma_2\ket{1,b}\bra{L,a}\right),
\end{align}}

as illustrated in Fig.~\ref{Fig4}a, generalizing the well-known (spatially uniform) SSH model. Each diatomic unit cell is indexed by its position $x$, and hoppings across atoms $a,b$ within each unit cell have amplitudes $g_1(x)\gamma_1$ and $g_1(x)/\gamma_1$ in either direction, where $\gamma_1$ is the intra-cell hopping asymmetry. Likewise, hoppings connecting atoms $b,a$ across adjacent unit cells have amplitudes $g_2(x)\gamma_2$ and $g_2(x)/\gamma_2$ in either direction.

The construction of its eigensolutions and phase-space GBZ proceeds analogously to the previous single-component case, albeit with important new features. Writing the $n$-th eigenstate as $\Psi_{n}=\sum_{x}\sum_{ j =a,b}\psi_{n, j }(x)\ket{x, j }$ and substituting it into the time-independent Schr\"{o}dinger eigenequation $H_\text{2-comp}\Psi_{n}=E_{n}\Psi_{n}$ yields the bulk equations 
\begin{align}
    &\frac{g_{1}(x)}{\gamma_{1}}\psi_{n,b}(x)+g_{2}(x-1)\gamma_{2}\psi_{n,b}(x-1)=E_{n} \psi_{n,a}(x),\notag\\
&g_{1}(x)\gamma_{1}\psi_{n,a}(x)+\frac{g_{2}(x)}{\gamma_{2}}\psi_{n,a}(x+1)=E_{n} \psi_{n,b}(x),
\label{2bBulk1}
\end{align}
and is subject to the PBC boundary constraint
 \begin{align}
     \psi_{n, j }(x+L)=\psi_{n, j }(x),\label{2bBound1}
 \end{align}
where $ j =a,b$ labels the sublattice component. Inspired by the success of the single-component ansatz Eq.~\ref{ansatz}, we write down the two-component ansatz for the state amplitude as
\begin{align}
\psi_{n, j }(x)&=\frac{\gamma_1^{x}\gamma_2^x}{\sqrt{g_{1}(x)g_{2}(x)}}\prod_{x'=1}^{x}\beta_{n, j }(x').\label{2bansatz}
\end{align}
Assuming sufficiently smooth odd/even hopping amplitude profiles and local phase-space GBZ continuity:
\begin{align}
    &g'_{1,2}(x)\ll g_{1,2}(x),\\
    &\beta'_{n, j }(x)\ll\beta_{n, j }(x),
\end{align}
the bulk equations Eq.~\ref{2bBulk1} for a state with eigenenergy $E_n$ reduce to
\begin{align}
\frac{1}{2}\left(\beta_{n, j }(x)+\frac{1}{\beta_{n, j }(x)}\right)=\omega_{ j }(E_n,x),\label{2bBulk2}
\end{align}
with
\begin{align}
     &\omega_{a}(E_n,x)=\frac{E_n^2-g_{1}^2(x)-g_{2}^2(x-1)}{2\sqrt{g_{1}(x-1)g_{1}(x)g_{2}(x-1)g_{2}(x)}},\notag\\
    &\omega_{b}(E_n,x)=\frac{E_n^2-g_{1}^2(x)-g_{2}^2(x) }{2\sqrt{g_{1}(x)g_{1}(x+1)g_{2}(x-1)g_{2}(x)}}.\label{omega}
\end{align}
Importantly, the equations governing $\beta_{n,a}(x)$ and $\beta_{n,b}(x)$ are completely decoupled, with information of the inter-sublattice couplings entering only through $\omega_{ j =a,b}(E_n,x)$.

In fact, by substituting
\begin{align}\label{replacement}
    &\omega_{ j }(E_n,x)\to\frac{E_{n}}{2g(x)},\qquad \gamma_1\gamma_2\to \gamma,
\end{align}
Eqs.~\ref{omega} for each $ j =a,b$ can be solved in a way identical to the single-component case, even though additional subtleties emerge (as will be discussed later). The intra- and inter-cell hopping asymmetries $\gamma_1,\gamma_2$ appear only through their product $\gamma_1\gamma_2$ because they contribute to non-Hermitian skin pumping successively in a symmetric manner. Thus, the two phase-space GBZ branches can be defined exactly analogously through $\beta_{n, j, \pm}(x)=\exp(ik_{n, j,\pm}(x) -\kappa_{n, j,\pm}(x))$, with $\kappa_{n, j ,\pm}(x)$ and phase $k_{n, j ,\pm}(x)$ being the local contributions to the inverse skin depth and phase:
\begin{align}
&|\beta_{n, j ,\pm}(x)|=\exp( \pm  \cosh^{-1}(\omega_{ j }(E_n,x))),\label{2bSol1}\\
&\kappa_{n, j ,\pm}(x)=\mp \text{Re}\left( \cosh^{-1}(\omega_{ j }(E_n,x))\right).\label{2bSol2}
\end{align} 
As in the 1-component case, the phase-space GBZ may potentially jump discontinuously between the $\kappa_{n, j ,\pm}(x)$ solutions at certain positions $x_\text{jump}$, as encoded in the GBZ choice functions [Eq.~\ref{sigmacases}] $\sigma_{ j }(x)=\pm 1$, depending on whether the eigenwavefunction $ \psi_{n, j }(x)$ assumes the $+1$ or $-1$ branch at $x$. For a given eigenenergy $E_n$, the exact locations of the jumps in $\sigma_{ j }(x)$ can be determined by enforcing the PBC condition
\begin{align}
    \log(\gamma_1\gamma_2)={-}\frac{1}{L}\sum_{x=1}^{L}\sigma_{ j }(x)\text{Re}\left( \cosh^{-1}(\omega_{ j }(E_n,x))\right),
    \label{2BspecGBZ}
\end{align}
such that the eigenstate profile is explicitly
{\small
\begin{align}
&|\psi_{n, j }(x)|\notag\\=&\frac{\gamma_1^{x}\gamma_2^x}{\sqrt{g_{1}(x)g_{2}(x)}}\prod_{x'=1}^{x}\exp\left(\sigma_{ j }(x) \cosh^{-1}(\omega_{ j }(E_n,x))\right).\label{2bandprofile}
\end{align}}

\begin{figure}
\centering
    \includegraphics[width=\linewidth]{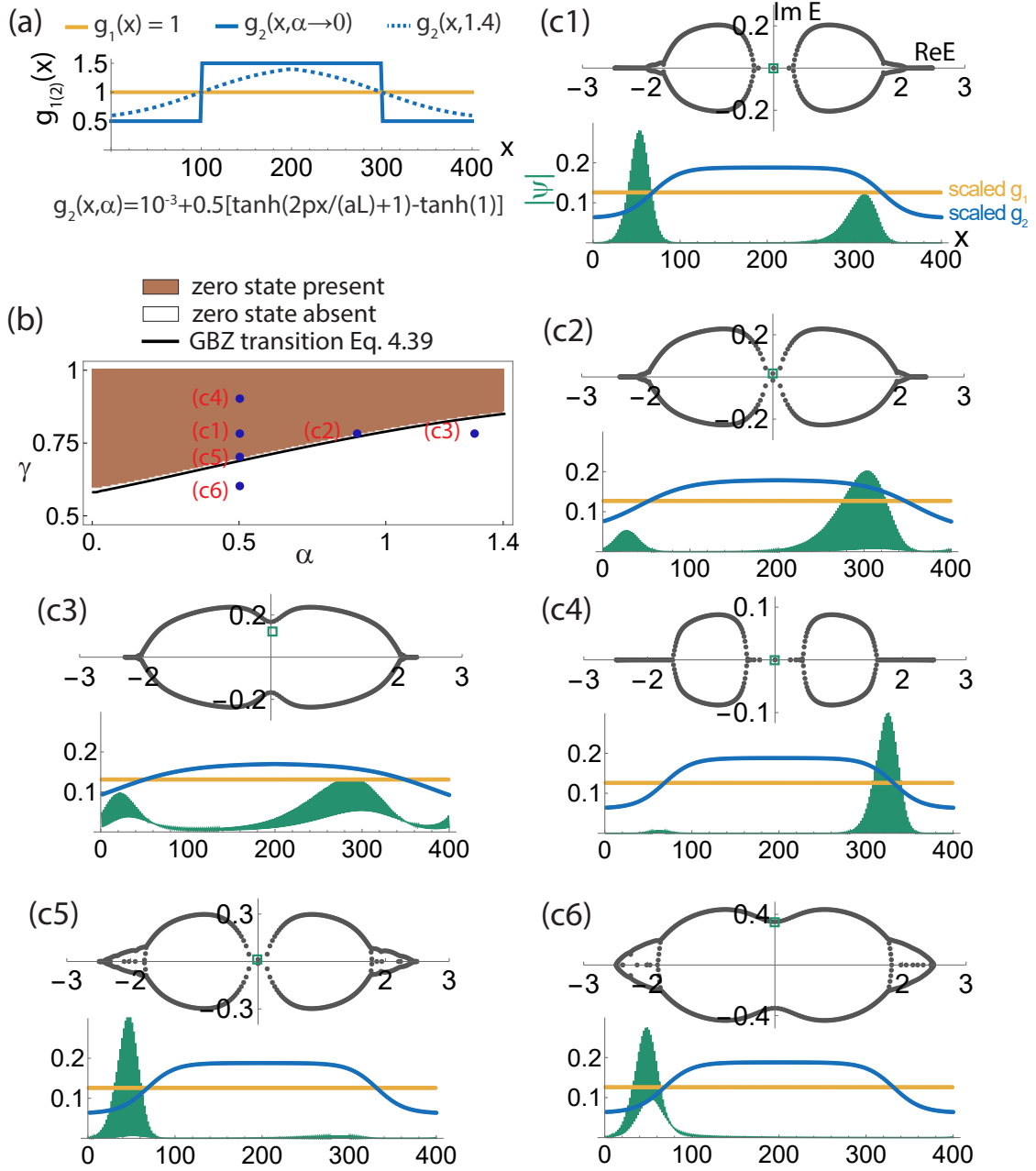}
    \caption{\textbf{Topological phase transition from varying the strength and smoothness of the spatial hopping profile. }
    (a) PBC hopping profile, with uniform intra-unit cell hoppings $g_1(x)=1$ and non-uniform inter-unit cell hopping strengths $g_{2}(x,\alpha)=1+0.5\tanh( 2 \pi \frac{x - \frac{L}{4}}{\alpha L})$, $-\frac{L}{2}<x\leq \frac{L}{2}$, with $L=200$. $g_1(x)$ and $g_2(x)$ intersect to form domain walls at $x=\frac{L}{4}$ and ${3L}{4}$, with wall steepness diverging to form hard boundaries as $\alpha\rightarrow 0$.
    (b) The numerically-determined region (brown) with robust zero modes (numerical tolerance $|E|<0.015$) in the parameter space of $\alpha$ and non-Hermitian asymmetry $\gamma$. It is accurately bounded by the transition curve (black) from Eq.~\ref{PDD2}. 
    (c1-c6) Spectra and corresponding zero mode (or minimal $|E_n|$ state) [Eq.~\ref{2bandprofile}] at illustrative parameter values given in (b). 
    Topological zero modes always occupy only one sublattice (dashed or solid green) per unit cell at the same $x$, just like familiar SSH zero modes, and accumulate against the domain wall intersections $g_{1}(x)=g_2(x)$.
    }
    \label{fig_topo_smoothness}
\end{figure}

\subsubsection{Real spectra and phase-space GBZ discontinuities for two-component chains}

On first impression, the two-component spatially inhomogeneous problem may seem no more complicated than the 1-component case, since its spectrum can be obtained by replacing $E_n/2g(x)$ with $\omega_{ j }(E_n,x)$, $ j =a,b$ [Fig.~\ref{Fig4}]. However, since $\omega_{ j }(E_n,x)$ has a more complicated spatial dependence, it, in fact, sets more sophisticated and subtle constraints on the spectrum.

Previously, it was established that a real energy continuum requires the presence of phase-space GBZ discontinuities $x_\text{jump}$ [Eq.~\ref{GBZBCinv}], which occur when $|E_n/2g(x)|\leq 1$ for some but not all $x$, i.e., $2g_\text{min}\leq\text{Re}(E_n)\leq 2g_\text{max}$ (the intermediate energy regime). For this 2-component case, a real spectral continuum thus requires that $|\omega_{ j }(E_n,x)|\leq 1$ for some but not all $x$, such that Eq.~\ref{2BspecGBZ} can be satisfied for a continuum of real $E_n$ by continuously adjusting $x_\text{jump}$ in $\sigma_{ j }(x)$. Substituting this condition into Eqs.~\ref{omega} 
and assuming sufficiently smooth spatial inhomogeneities such that $g_{1,2}(x)\approx g_{1,2}(x\pm 1)$,
{\small
\begin{align}
         |g_1(x)+g_2(x)|_{\text{min}}&<|\text{Re}(E_n)|<|g_1(x)+g_2(x)|_{\text{max}}\label{Eside2b2}\\
          \text{or}\qquad\qquad 0& <|\text{Re}(E_n)|\leq |g_1(x)-g_2(x)|_{\text{max}}.
    \label{Eside2b}
\end{align}}
Unlike the 1-component case where $g_1(x)=g_2(x)=g(x)$, real energies can also exist within an additional low(est) energy regime $0 <|\text{Re}(E_n)|< |g_1(x)-g_2(x)|_{\text{max}}$ [Eq.~\ref{Eside2b}], where $|\text{Re}(E_n)|$ is not larger than the hopping amplitude difference between the odd and even bonds. This new regime obviously does not exist in the 1-component case where Eq.~\ref{Eside2b2} simply reduces to the intermediate energy regime. As such, the spectral plane is divided into up to 4 different energy regimes:

\begin{itemize} [leftmargin = 0.5cm]
\item \textbf{Lowest energy regime with $|\text{Re}(E_n)|< |g_1(x)-g_2(x)|_{\text{max}}$: } Real energies are possible.
\item \textbf{Low-intermediate energy regime with $|g_1(x)-g_2(x)|_{\text{max}}\leq |\text{Re}(E_n)|\leq |g_1(x)+g_2(x)|_{\text{min}}$:} Only complex energies are possible; however, this regime may not exist for sufficiently dissimilar $g_1(x),g_2(x)$.
\item \textbf{Intermediate energy regime with $|g_1(x)+g_2(x)|_{\text{min}}<|\text{Re}(E_n)| \leq |g_1(x)+g_2(x)|_{\text{max}}$:} Real energies are possible.
\item \textbf{High energy regime with $|g_1(x)+g_2(x)|_{\text{max}}<|\text{Re}(E_n)|$:} Complex energy branches (if any) join up along the real line.
\end{itemize}
In particular, continua of real energies can exist in two distinct scenarios: in the intermediate energy regime, which exists only for spatially inhomogeneous systems (similar to the 1-component case), or in the lowest energy regime, where the contrast between adjacent $g_1(x),g_2(x)$ hoppings is sufficient to block directed amplification.

Shown in Fig.~\ref{Fig4} is an illustrative two-component model [Fig.~\ref{Fig4}a] with intra- and inter-component hopping inhomogeneity profiles $g_1(x),g_2(x)$ that intersect at two different locations [Fig.~\ref{Fig4}b]. As non-Hermiticity is introduced and $\gamma$ departs from unity [Fig.~\ref{Fig4}c], spectral loops appear across the lowest to intermediate energy regimes (cyan to light purple). Even though real energies are allowed in the lowest energy regime (cyan), the complex loops join up to form one loop with sufficiently strong non-Hermitian asymmetry $\gamma$, a phenomenon with no single-component analog.

It is instructive to examine the GBZ profiles of the representative eigenstates indicated in Fig.~\ref{Fig4}c3 by red hollow squares. As presented in Fig.~\ref{Fig4}d, only the real energy eigenstates (d1) and (d4) possess pure skin regions where at least one GBZ component is degenerate, i.e., $\kappa_a(x)=0$ or $\kappa_b(x)=0$. These degeneracies allow for ``hidden'' switching of the actual GBZ branches adopted by the numerical eigenstates $\psi$, which in turn gives rise to spectacular GBZ jumps, i.e., discontinuities in $\sigma_j(x)=\pm 1$ necessary for satisfying Eq.~\ref{2BspecGBZ}. While both GBZ components $\kappa_{a,b}(x)$ look identical in (d2)-(d4), they can become completely distinct when the spatial gradient $g_2(x)-g_2(x-1)$ dominates over the eigenenergy $E_n$ in $\omega_{a/b}(E=0,x)$ of Eq.~\ref{omega}, as for the zero mode of (d1).

\subsection{Topological transitions from spatial inhomogeneity}

Most interestingly, spatial hopping inhomogeneity in a 2-component lattice can also drive topological phase transitions. Ordinarily, in a non-Hermitian SSH model with uniform $g_1(x)=g_1$ and $g_2(x)=g_2$ hopping amplitudes, it is well-known that a topological phase boundary occurs at ``domain walls'' where they swap. Here, with spatially inhomogeneous $g_1(x)$ and $g_2(x)$, it is reasonable to expect that  $g_1(x)=g_2(x)$ intersections still function as topological interfaces, since they demarcate the regions $g_1(x)<g_2(x)$ and $g_2(x)>g_1(x)$ that are supposed to represent different phases.

\begin{figure}
\centering
    \includegraphics[width=\linewidth]{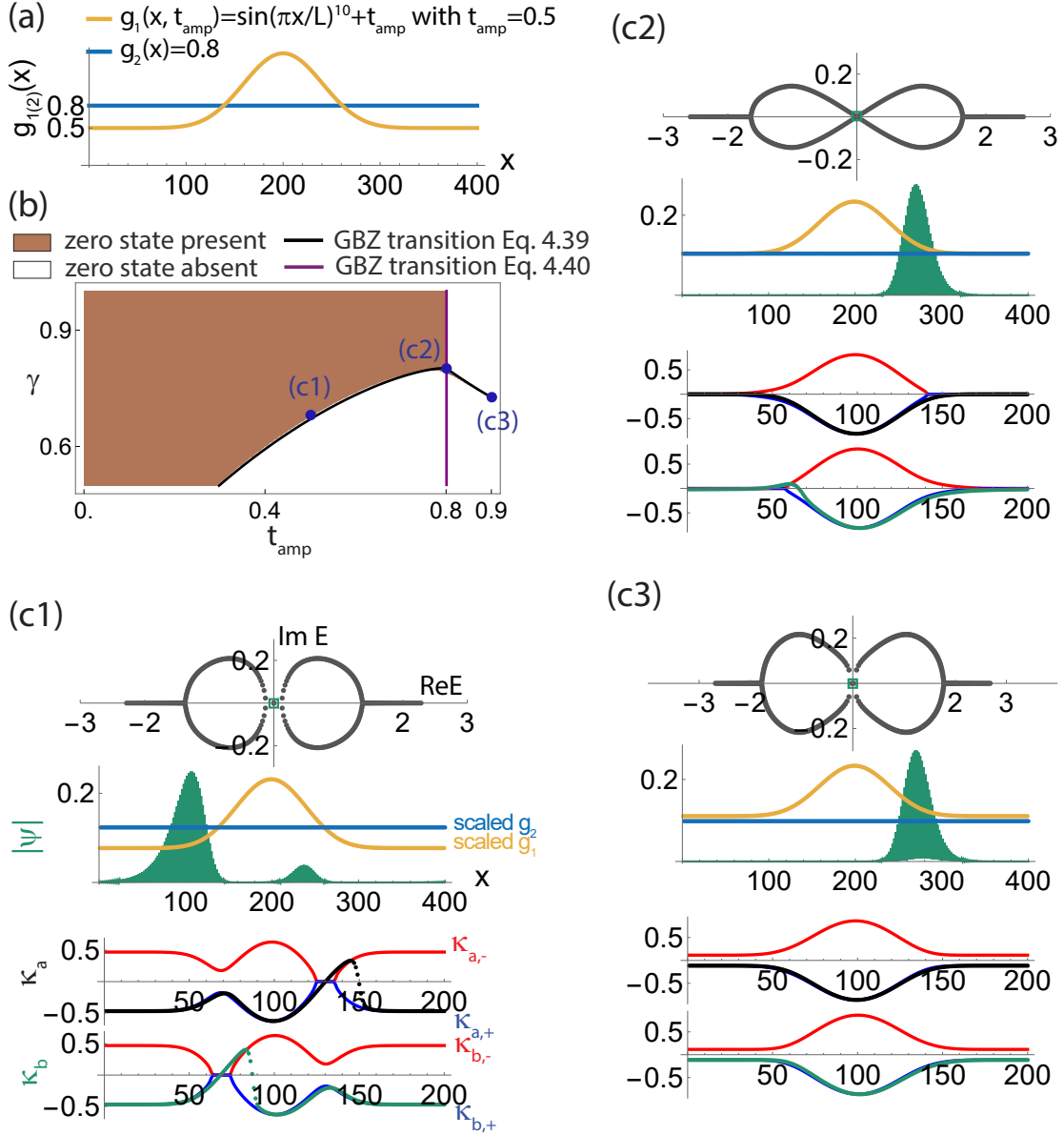}
    \caption{\textbf{Topological phase transition and the crucial role of GBZ jumps. } 
    (a) The PBC hopping profile, as given by constant $g_2(x)$ and inhomogeneous $g_1(x,t_{\text{amp}}) =\sin(\pi x/L)^{10}+t_{\text{amp}}$, $L=200$, with the offset $t_{\text{amp}}$ controlling whether $g_1(x)$ and $g_2(x)$ intersect to form topological domain walls.    
    (b) The numerically determined region hosting topological zero modes (brown, with numerical tolerance $|E|<0.02$) in the parameter space of $\gamma$ and $t_\text{amp}$. It is accurately demarcated by Eq.~\ref{PDD22} (purple) and Eq.~\ref{PDD2} (black), which gives the phase-space GBZ threshold boundaries between continuous and discontinuous GBZs. 
    (c1-c4) Numerically-obtained spectra, state profiles [Eq.~\ref{2bandprofile}] and GBZ occupancies of illustrative $E=0$ zero modes in (b). Deeper into the topological regime (c1,c4), the zero modes are always localized near the ``soft boundary'' defined by $g_1(x) = g_2(x)$ and occupy only one sublattice (dashed or solid green) per unit cell, even though the other sublattice may be occupied away from the soft boundary. However, unlike the bulk zero mode (c3), only the topological states in (c1) and (c4) have discontinuous $\kappa_{j}(x)$ jumping between two branches of $\kappa_{j,\pm}(x)$. 
    }
    \label{fig_topo_amplitude}
    \end{figure}

However, PBC spatial inhomogeneity complicates the stability of zero modes in various ways. Firstly, among two non-empty regions separated by a domain wall, one of them must already possess non-trivial bulk topology in a bipartite system. Secondly, the bulk is now spatially inhomogeneous, such that its GBZ description rightly lives in phase space and not just momentum space. Thirdly, since phase space encompasses position coordinates, the shape of the domain wall itself affects the topology. Indeed, as will be shown in Figs.~\ref{fig_topo_smoothness} and \ref{fig_topo_amplitude}, the topological modes are not fully confined to $g_1(x)=g_2(x)$ interfaces but, in fact, penetrate nonlocally and nonexponentially into other parts of the system.

\subsubsection{Topological criteria}

Here, we describe a new type of topological robustness protected by GBZ bifurcations. Unlike topological modes at open boundaries, which are simply protected by bulk topological invariants, the inhomogeneous PBC isolated zero modes (i.e., in Fig.~\ref{Fig4}d1) turn out to be crucially protected by GBZ jumps. These jumps facilitate the realization of real spectra (which includes $E_n=0$), as discussed in the paragraphs surrounding Eqs.~\ref{GBZBCinv} and \ref{2BspecGBZ}, and are also further elaborated in Sect.~\ref{sec:single_state}.

As such, to have robust isolated zero modes (and not just trivial $E=0$ crossings), the following two criteria on $\gamma$ and $g_1(x),g_2(x)$ must be satisfied:
\begin{align}
& |g_1(x)-g_2(x)|_{\text{min}}=0,\label{PDD22}\\
 \& \quad   &|\log(\gamma)|< |\log(\Tilde{\gamma}_{\text{topo}})|,\label{PDD2}  
\end{align}
where 
{\small
\begin{align}
    &\Tilde{\gamma}_{\text{topo}}=\exp\left( \frac{1}{L}\sum_{1<|\omega_{ j }(E=0,x)|}\cosh^{-1}|\omega_{ j }(E=0,x)|\right)\label{sideReE2zero},
\end{align}}

is the $\gamma$ threshold that separates continuous and discontinuous GBZ scenarios,with $j=a$ or $b$ yielding negligible differences in the phase diagram (see Sect.~\ref{sec:single_state}). 

The criterion $|g_1(x)-g_2(x)|_{\text{min}}=0$ [Eq.~\ref{PDD22}] is simply that the intra- and inter-unit cell hopping profiles  $g_1(x)$ and $g_2(x)$ intersect to form spatial ``domain walls'' that, by construction, have one side with nontrivial bulk topology. Its implications for the bifurcated GBZ are as follows: when Eq.~\ref{PDD22} is satisfied and $g_1(x)$ and $g_2(x)$ are sufficiently smooth, $\kappa_{n=0,j,\pm}(x)\approx 0$ [Eqs.~\ref{omega},~\ref{2bSol2}] for some $x$ and $j=a,b$. Since $\kappa_{n=0,j,\pm}\neq 0$ for most other $x$, Eq.~\ref{PDD22} can hence be understood as the requirement for the simultaneous existence of both inhomogeneous and pure skin regions in the zero eigenmode. 

It is also insightful to recast the criterion $|\log(\gamma)|< |\log(\Tilde{\gamma}_{\text{topo}})|$ [Eq.~\ref{PDD2}] into a much more intuitive approximate form when $g_1(x),g_2(x)$ are sufficiently smooth. Neglecting spatial gradients in Eq.~\ref{omega}, 
\begin{align}
\kappa_{n, j ,\pm}(x)&={\mp}\text{Re}\left(\cosh^{-1}|\omega_{ j }(E=0,x)|\right)\notag\\
&\approx{\mp} \text{Sgn}[g_1(x)-g_2(x)]\log\frac{g_1(x)}{g_2(x)},\label{PDD4}
\end{align}
which is just the inverse decay length of SSH-type topological modes without additional skin localization, valid when the inhomogeneities in $g_1(x),g_2(x)$ contribute only ``adiabatically''. Eq.~\ref{PDD4} simplifies $\Tilde{\gamma}_{\text{topo}}$ (for $\Tilde{\gamma}_{\text{topo}}<1$) to
\begin{eqnarray}
\Tilde{\gamma}_{\text{topo}}&=&e^{-\frac{1}{L}\sum_{x=1}^L\kappa_{n, j ,-}(x)}\notag\\
&\approx &\left(\prod_{g_2(x)>g_1(x)}^L\frac{g_2(x)}{g_1(x)}\prod_{g_2(x)<g_1(x)}^L\frac{g_1(x)}{g_2(x)}\right)^{\frac{1}{L}}\notag\\
&=&e^{\left\langle\left|\log\frac{g_2(x)}{g_1(x)}\right|\right\rangle},
\label{PDD3}
\end{eqnarray}
where the average $\left\langle\left|\log\frac{g_2(x)}{g_1(x)}\right|\right\rangle$ is taken over all $x\in[1,L]$. As such, for sufficiently smooth $g_1(x),g_2(x)$, the criterion Eq.~\ref{PDD2} reduces to
\begin{equation}
|\log\gamma|<\left\langle\left|\log\frac{g_2(x)}{g_1(x)}\right|\right\rangle,
\label{PDD5}
\end{equation}
which places an upper bound on the non-Hermitian hopping asymmetry $|\log\gamma|$, above which it pumps all states across the inhomogeneous PBC chain and destroys the zero modes. This result applies exclusively for ``soft spatial boundaries'' with no OBC analog: it gives the upper bound for the eigenfunction to be ``patched up'' to satisfy the PBC condition [Eq.~\ref{2BspecGBZ}] through appropriately placed GBZ jumps in $\sigma_j(x)$. Note that the criterion Eq.~\ref{PDD2} (or Eq.~\ref{PDD5}) is trivially satisfied in the Hermitian limit of $\log\gamma=0$, in which only the criterion Eq.~\ref{PDD22} exists.

\subsubsection{Illustrative examples for the topological criteria}

In Figs.~\ref{fig_topo_smoothness} and Fig.~\ref{fig_topo_amplitude}, we showcase numerical topological phase diagrams of two illustrative models and how their phase boundaries are accurately determined by the criteria given in Eqs.~\ref{PDD22} and \ref{PDD2}. We also present some illustrative eigenstates [Eq.~\ref{2bandprofile}] and highlight the characteristic amplitude profiles and GBZ jumps of isolated zero eigenmodes.

Fig.~\ref{fig_topo_smoothness} presents a spatial hopping profile [Fig.~\ref{fig_topo_smoothness}a] whose topological phase boundary [Fig.~\ref{fig_topo_smoothness}b] is well approximated by Eq.~\ref{PDD2} alone. It features a uniform $g_1(x)=1$ (yellow) and a $g_2(x)$ (blue) which cuts $g_1(x)$ at domain walls $x=\frac{L}{4}$ and $x=\frac{3L}{4}$. The wall steepness is controlled by a parameter $\alpha$; as $\alpha\rightarrow 0$, the steepness diverges, leading to an OBC-like hard boundary. The phase boundary curve matches excellently with results from numerical diagonalization (brown) and indicates that, as the domain wall becomes softer (with larger $\alpha$), the zero mode becomes more fragile, being more easily destroyed as $\gamma$ moves away from the Hermitian limit $\gamma=1$. 

Even though all eigenstates (green) are localized to some extent in a spatially inhomogeneous setting, topological zero modes characteristically occupy only one sublattice, as plotted in Figs.~\ref{fig_topo_smoothness}c1,c4. As the topological line gap closes and forms a point gap [Figs.~\ref{fig_topo_smoothness}c2,c3,c5,c6], the eigenstates start to disperse away from the domain walls, and both sublattices assume nonzero occupancy.

In Fig.~\ref{fig_topo_amplitude}, we present a different spatial hopping profile [Fig.~\ref{fig_topo_amplitude}a] such that its topological phase boundary [Fig.~\ref{fig_topo_amplitude}b] is demarcated by both Eqs.~\ref{PDD22} and \ref{PDD2}. As the parameter $t_\text{amp}$ increases, the region where $g_1(x)$ (yellow) is larger than $g_2(x)$ (blue) broadens until it finally occupies the whole system and no domain wall exists. This scenario is exactly demarcated by the purple line, to the right of which $|g_1(x)-g_2(x)|_\text{min}=0$ [Eq.~\ref{PDD22}] no longer holds. 

That the zero modes are fundamentally protected by GBZ jumps can be seen in Fig.~\ref{fig_topo_amplitude}c, which showcases three illustrative eigenstates on the threshold boundary given by $|\log \gamma|=|\log\Tilde\gamma_\text{topo}|$ [Eq.~\ref{PDD2}]: (c1) within the topological phase, (c2) at the boundary given by Eq.~\ref{PDD22}, (c3) outside the topological phase, as well as a reference (c4) deep within the topological phase.
Evidently, numerical GBZ jumps (green) in their GBZ branches of blue and red curves confirm that the topological nature of the $E=0$ state corresponds to the presence of a discontinuous jump between $\kappa_{j,\pm}(x)$ and the simultaneous presence of both inhomogeneous and pure skin regions.

\subsection{Discussion}

In this work, we have formulated a new theoretical framework for generically treating the interplay of the NHSE and spatial lattice hopping inhomogeneity, as encoded by $\gamma$ and $g(x)$, respectively. This is a subtle scenario because the spatially non-uniform energy scale not only competes with NHSE accumulation through Wannier-Stark localization but also distorts the skin accumulation and deforms the effective lattice momentum in a position-dependent manner. 

Central to our formalism is the phase-space generalized Brillouin zone (GBZ), which captures the effective non-Bloch deformation in both position and momentum space. For any PBC eigensolution $E_n$, the phase-space GBZ bifurcates into two possible solution branches within regions of relatively weak hoppings $2g(x)<E_n$, leading to non-exponential ``inhomogeneous skin'' state profiles. Crucially, discontinuous jumps in the adopted GBZ branch give rise to an emergent degree of freedom that results in real ``tails'' in the energy spectrum. Physically, these real eigensolutions represent states that are prevented from indefinite growth by spatial inhomogeneity.

Two-component settings encompass new forms of topological robustness as different $g(x)$ components intersect to form spatial domain walls. The real spectral solutions from GBZ jumps can also exist at very low energies, including the topological zero modes in particular. Unlike the well-known topological edge modes, these isolated zero modes are protected by the phase-space GBZ bifurcations, lending their robustness from the emergent freedom in the GBZ jump positions. As shown both theoretically and numerically, such topological phase boundaries can be accurately predicted through our criteria given by Eqs.~\ref{PDD22} and \ref{PDD2}.

By generalizing Eqs.~\ref{2bansatz} and \ref{omega}, our phase-space GBZ framework can be extended to inhomogeneous systems with arbitrarily many components, such that the GBZ of each component depends non-linearly on the inter-component hoppings. Additionally, generalizing the spatially inhomogeneous hoppings beyond nearest neighbors replaces Eq.~\ref{2bBulk2} with a higher-degree Laurent polynomial that splits the GBZ solutions into multiple branches. In all, these are expected to generate far more intricate GBZ jumps, leading to many more hidden degrees of freedom that stabilize new emergent spectral branches, some possibly containing topological modes with higher symmetry~\cite{Chen2013, Chiu2016, Wen2017, Chen2011}. Further generalization to higher dimensions and multiple interacting boundary conditions could lead to significant new subtleties in the already fragmented GBZ structure, opening up a vast playground for future research into non-Hermitian localization.

Experimentally, spatially inhomogeneous non-Hermitian systems are as accessible as their usual uniform lattice counterparts. Non-Hermitian lattice models have already been realized in electrical circuits~\cite{lee2018topolectrical,helbig2019band,Wang2020,hofmann2020reciprocal,zou2021observation,su2023simulation,hohmann2023observation,zhang2023electrical,zhang2024observation,zou2024experimental,guo2024scale},
 cold atoms~\cite{li2019, ren2022,liang2022dynamic,zhou2022}, 
 photonics~\cite{song2020two,  yu2021nonhermitian, song2023nonhermitian, parto2020non,lin2021optics, zhong2021nonhermitian}, 
 programmable quantum simulators~\cite{kamakari2022,koh2023measurement,peng2020simulating,chertkov2023characterizing,Shen2023,Shen2024,koh2024realization,koh2022prl,okuma2022nonnormal}, 
 and mechanical/acoustic systems~\cite{yang2022nonhermitian, braghini2021nonhermitian, jin2022exceptional}. In most metamaterial platforms, the effective hopping strengths can be spatially tuned in a versatile manner -- for instance, the individual components of an electrical circuit array can be tuned at will, with effectively asymmetric couplings simulated using operational amplifiers~\cite{lee2018topolectrical,Ezawa2019c,Ezawa2019d}. The phase space GBZ can be reconstructed from the eigenstate profiles, which can for instance be obtained in electrical circuits through impedance measurements~\cite{Imhof2018,Lu2023,Zhang2022nonHermitian,Franca2024} alongside the resonance spectrum.

\section{Methods}

\begin{figure}
    \includegraphics[width=\linewidth]{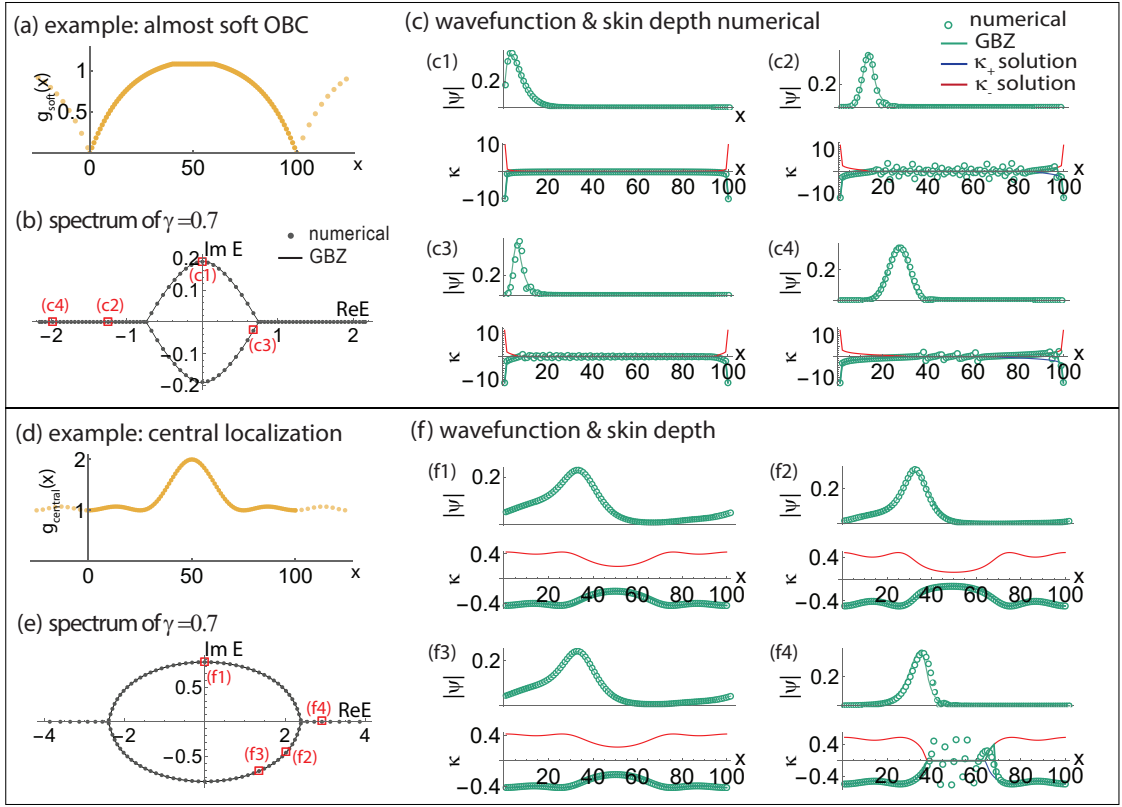}
	\caption{\textbf{Phase-space GBZ: further examples. } Our phase-space GBZ approach well-approximates both energy spectra [Eq.~\ref{sideReE2}] and wavefunctions [Eq.~\ref{ansatz} and Eq.~\ref{PTSol1}] in smooth 1D single-band inhomogeneous lattices. 
    Two different inhomogeneous hopping profiles are chosen, $g(x)=g_{\text{soft}}(x)$ (a-c) from Eq.~\ref{osci_exp} and $g(x) =g_{\text{bump}}(x)$ (d-f)
 from Eq.~\ref{osci_exp}, with $\gamma=0.7$. 
 The numerical results agree almost perfectly with the GBZ solutions, not just in the spectra (b,e), but also the wavefunctions (green) as well as their fits with both continuous and discontinuous GBZ regions (blue and red in c,f). Some fluctuations in the numerically reconstructed GBZ are unavoidable in the pure skin region of finite inhomogeneous systems, as in the discussions in Sec.~\ref{pure_fluc}, but they effectively average to zero in the phase-space GBZ.
        }
	\label{fig7A} 
\end{figure}

\begin{figure}
    \includegraphics[width=\linewidth]{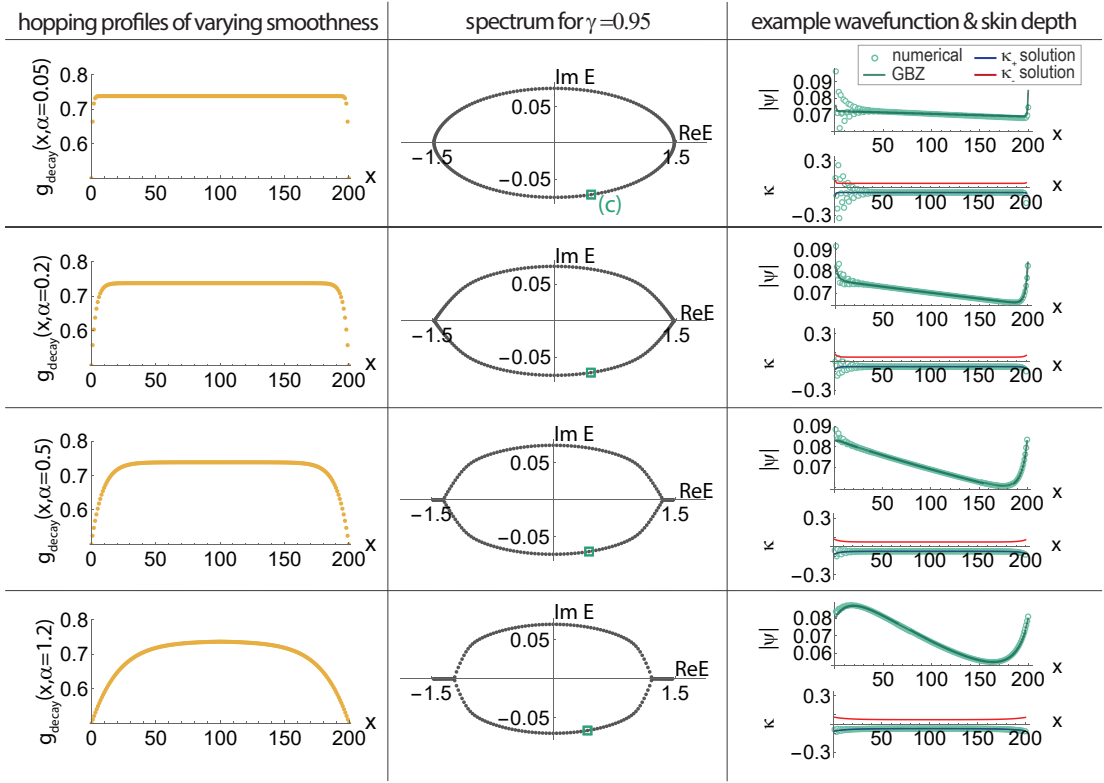}
	\caption{\textbf{Phase-space GBZ: effect of discontinuities in the hopping profile $g(x)$. }
 Numerical reconstruction of the phase-space GBZ becomes more noisy with increasing local non-smoothness in $g(x)$. The cases plotted are based on $g_{\text{dip}}(x,\alpha)=10^{-3}+(\tanh(\frac{2 \pi x}{\alpha L} + 1)-\tanh(1))$ for $-L/2<x\leq L/2$ [Eq.~\ref{NS_exp}], with varying smoothness parameter $\alpha=0.05, 0.2,0.5,1.2$ and fixed $L=200$, $\gamma=0.95$. A small $\alpha$, corresponding to a sharp discontinuity in $g(x)$, gives rise to significant Gibbs-like wavefunction fluctuations and hence fluctuating numerically reconstructed GBZ (green). Increasing $\alpha$ gradually suppresses these fluctuations until a smooth hopping profile and illustrative wavefunction are achieved. The phase-space GBZ, which is always a smooth curve, does not capture local fluctuations but consistently represents the best averaged trend, providing an optimal smoothed-fit for the wavefunctions.
 }
	\label{fig7B} 
\end{figure}

\subsection{Limitations on the phase-space GBZ approach}

Although the phase-space GBZ has been shown to be effective in describing the spectra of inhomogeneous single- and double-component systems, local deviations, fluctuations, or discrepancies may still be observed for certain states, as exemplified in Fig.~\ref{0908fig3}c1 and c2, particularly near non-smooth regions of $g(x)$ or at the transition of $\kappa(x)$ into the pure skin region. In this section, additional examples of phase-space GBZs in single-component systems are presented, and the limitations of the phase-space GBZ method in various finite-size physical systems are examined.

We first showcase additional examples of two contrasting inhomogeneous $g(x)$ hopping profiles and their phase-space GBZ approximations in Fig.~\ref{fig7A}, such as to validate the accuracy of our approach. The two illustrative systems are:
{\small
\begin{align}
        &g_{\text{soft}}(x)= \begin{cases}
            5\tanh(\pi \frac{x}{L}+1)+10&\text{ for }x< \frac{2L}{5}\\
            5\tanh(\pi \frac{2}{5}+1)+10&\text{ for }\frac{2L}{5}\leq x\leq \frac{3L}{5}\\
            5\tanh(\pi-\frac{\pi x}{L}+1)+10&\text{ for }x> \frac{3L}{5}
        \end{cases},\label{decay1_exp}\\
        &g_{\text{bump}}(x)=1 + \sin(\pi x/L)^2 \cos(2 \pi x/L)^2,\label{osci_exp}
    \end{align}}
both with $\gamma=0.7$.

Their full spectra and illustrative states are well approximated by the phase-space GBZ compared to numerical results (Figs.~\ref{fig7A}b,~\ref{fig7A}c,~\ref{fig7A}e,~\ref{fig7A}f), with minor but observable localized fluctuations: slight deviations appear near the non-smoothness at $x=0$ and $x=L$ in the hopping profile $g(x)=g_{\text{soft}}(x)$ (Fig.~\ref{fig7A}a) as shown in Fig.~\ref{fig7A}c2; fluctuations also occur in the pure skin regions of states with coexisting pure and inhomogeneous skins (Figs.~\ref{fig7A}c2,~\ref{fig7A}c4,~\ref{fig7A}f4), where these GBZ $\kappa(x)$ fluctuations lead to localized deviations in the corresponding state profiles. We will elaborate on the reasons and the degree of effect of these two sources of inaccuracy later in this section.

\subsubsection{Non-smoothness in the hopping function}

In introducing the phase-space GBZ method, we assume sufficiently smooth hopping functions in Eq.~\ref{SmthApprox} such that $g(x\pm 1)$ is approximated as $g(x)$, leading to the local continuity of the continuous phase-space GBZ.

While Eq.~\ref{SmthApprox} remains valid in the thermodynamic limit $L\to\infty$ for any continuous hopping function $g(x)$ with $g(x+L)=g(x)$, the approximation precision diminishes for finite $L$ due to the local non-smoothness of $g(x)$. To illustrate the extent and characteristics of this diminishing accuracy, consider a hopping function with strongly decaying boundary hoppings:
{\small
\begin{align}
    g_{\text{dip}}(x,\alpha)&=\frac{1}{2}-\tanh(1)\notag\\
    &+\begin{cases}
        \tanh(\frac{2\pi x}{\alpha L}+1) &\text{ for }0<x\leq \frac{L}{2}\\
        \tanh(\frac{2\pi(L-x)}{\alpha L}+1) &\text{ for }\frac{L}{2}<x\leq L
    \end{cases},\label{NS_exp}
\end{align}}

as shown in Fig.~\ref{fig7B}. With $L=200$, the smoothness $\alpha$ is varied from $0.05$ to $1.2$, and the phase-space GBZ approximation is compared to numerical results.

With fixed $g_{\text{min}}$ and $g_{\text{max}}$, a reduction in $\alpha$ makes the violation of Eq.~\ref{SmthApprox} more pronounced, amplifying fluctuations in the wavefunction near the non-smooth regions of $g(x)$. Although the phase-space GBZ assumes perfectly smooth hopping functions and cannot fully capture these fluctuations, it effectively approximates the overall trend of the wavefunction, as demonstrated in Fig.~\ref{fig7B}.

When local fluctuations in the wavefunctions are of interest, it becomes necessary to enhance the smoothness of $g(x)$, most easily done by just increasing the number of sites within a fixed real-space length. In the special case of discontinuous $g(x)$ with sharp jumps over a small range, our phase-space GBZ can be modified to accommodate sharp boundaries, as discussed later in Sect.~\ref{sec_discont_gx}.

\subsubsection{Pure skin fluctuation in finite-size systems}\label{pure_fluc}

In the phase-space GBZ approach, inhomogeneous systems are approximated by assuming the thermodynamic limit $L\to \infty$, where the bulk relation becomes fully decoupled between neighbouring sites:
\begin{align}
    &\beta_{n}(x)+\frac{1}{\beta_{n}(x)}=\frac{E_{n}}{g(x)}.
\end{align}
Within this framework, discontinuous phase-space GBZ states, which have real eigenenergies and consist of both the pure skin and inhomogeneous skin regions, have $\kappa_{n,\pm}(x)$ that goes from being zero to non-zero at the boundary between pure skin and inhomogeneous skin regions.

In practice, with finite $L$, neighbouring $\beta_n(x)$ does differ slightly, necessitating the use of the original bulk equation [Eq.~\ref{BulkApprox}] without the approximation $\beta_n(x+1)\approx \beta_n(x)$.

This then brings in problems at the pure-inhomogeneous skin boundaries, where neighbouring phase-space GBZ factors $\beta_{n,\pm}(x)$ and $\beta_{n,\pm}(x+1)$ are located within different skin regions, and only one of them is complex. This is mathematically prohibited in Eq.~\ref{BulkApprox} for real $E_n$ and $g(x)$. A necessarily non-zero imaginary component of $\beta_{n}(x)$ of pure skin, or equivalently $\kappa_{n}(x)\neq 0$ in the pure skin regions, is required to balance the imaginary equation of Eq.~\ref{BulkApprox} in finite-size systems.

Nevertheless, the phase-space GBZ solutions $\kappa_{n,\pm}(x)=0$ still provide an effective best-fit approximation for the fluctuating $\kappa_{n}$. Moreover, despite the fluctuations in $\kappa_{n}(x)$, their impact on the wavefunctions remains minimal, as demonstrated in examples of Figs.~\ref{fig7A}b,~\ref{fig7A}c,~\ref{fig7A}e,~\ref{fig7A}f, where the effect of GBZ fluctuations is restricted locally by the effectively zero skin in the pure skin regions and is compressed by the inhomogeneous skin, which amplifies the wavefunctions at one of the sides.

\begin{figure}
    \includegraphics[width=\linewidth]{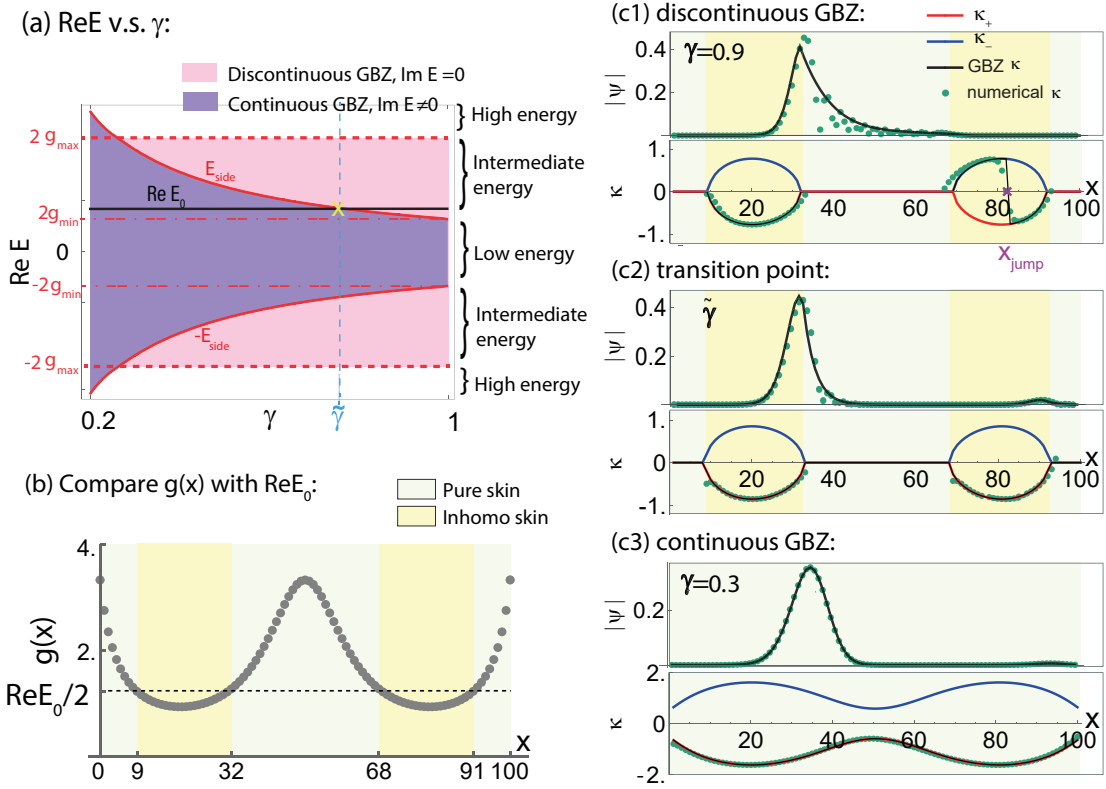}
	\caption{\textbf{$\gamma$-driven transition between continuous and discontinuous phase-space GBZs. }
(a) The phase diagram of a 1-component lattice with $g(x)=1/(\sin(2 \pi x/L) \cos(\pi x/L) + 0.3)$ and $L=100$, in the Re$E$-$\gamma$ parameter space. The curves $\pm E_{\text{side}}$, as calculated from Eq.~\ref{sideReE2}, separate the continuous and discontinuous phase-space GBZ phases.
For a fixed chosen value of Re$E_0$ (black), one can always find the threshold value $\gamma=\Tilde{\gamma}$ [Eq.~\ref{sideReE}] where the transition occurs. Here, $E_0=E_{\text{side}}=2.46$ when $\gamma$ is tuned to the value $\Tilde{\gamma}=0.76$.
(b) For a eigenstate corresponding to $E_0$, its pure/inhomogeneous skin regions correspond to positions where $2g(x)$ is above/below the value of Re$E_0$.
(c) Illustrative states at various $\gamma$, correspond to (c1) the discontinuous GBZ phase, (c2) at the transition point and (c3) in the continuous GBZ phase. Clear-cut regions of pure and inhomogeneous skins exist in the discontinuous GBZ and the transition point states. Within the inhomogeneous skin region, the local continuity of the GBZ ensures numerical $|\kappa(x)|$ consistently adheres to one of $|\kappa_{\pm}(x)|$, even near the sign-inversion site. This enables the calculation of the single $x_{\text{jump}}$ according to Eq.~\ref{GBZBCinv}. 
 }
	\label{fig3old} 
\end{figure}

\subsection{Phase-space GBZ transitions for a fixed state}
\label{sec:single_state}
The energy eigenvalue $E_{\text{side}}$ [Eq.~\ref{sideReE2}] not only bounds the spectral loop on both sides but also serves a crucial role in confining the real ``tails'' within the intermediate energy regime, which is bounded by $\pm 2g_{\text{max}}$. In essence, $E_{\text{side}}$ acts as a transition point between continuous and discontinuous phase-space GBZ, and is calculated from the boundary requirement Eq.~\ref{sideReE2} (Eq.~\ref{2BspecGBZ}) in single-component (double-component) chains, given known non-Hermiticity $\gamma$ ($\gamma_1 \gamma_2$) and hopping function $g(x)$ ($g_{1,2}(x)$).

Here, we elaborate on how one can alternatively investigate the phase-space GBZ by examining how a specific $E_0$ state undergoes transitions as a certain parameter is varied. For instance, in the topological investigation of the zero-energy state $E_{0}=0$, the non-Hermiticity $\gamma$ is tuned such that the state undergoes a transition between topologically trivial and non-trivial phases. We hereby propose the concept of threshold value $\Tilde{\gamma}$ of $\gamma$, defined for a fixed specified $E_0$. In a single-component system, it is given by
\begin{align}
   \Tilde{\gamma}(E_0)=&\exp\left(-\frac{1}{L}\sum_{x=1}^{L} \cosh^{-1}\left(\frac{\text{Re }E_{0}}{2g(x)}\right)\right).\label{sideReE}
\end{align}
It gives the value of $\gamma$ across which a phase transition between continuous and discontinuous phase-space GBZ occurs, as demonstrated in Fig.~\ref{fig3old}. The state $E_0$ is real and resides within the discontinuous GBZ phase if $|\log(\gamma)|\leq|\log(\Tilde{\gamma})|$, but has to be complex otherwise.

Fig.~\ref{fig3old}a shows the phase diagram of state $E_0$ in the Re$E$-$\gamma$ space, where $\Tilde{\gamma}(E_0)$ is determined as the point where Re$E_0$ becomes the $E_\text{side}$. For a fixed value of Re$E_0$, $\Tilde{\gamma}(E_0)$ separates the (c3) continuous (complex $E_0$) and (c1,c2) discontinuous (real $E_0$) phases.  Crucially, the real state $E_0$ within the discontinuous GBZ phase, and consequently its phase-space GBZ solutions $\kappa_{n,\pm}$, remain robust against slight variation in $\gamma$, as verified by comparing (c1) and (c2). Adjusting $\gamma$ solely shifts the discontinuity position $x_{\text{jump}}$ along the real-space position.

A similar threshold non-Hermiticity $\Tilde{\gamma}$ is defined for 2-component states with energy $E_0$ within the lowest energy regime or the intermediate energy regime:
{\small
\begin{align}
    &\Tilde{\gamma}(E_0)=\exp\left(-\frac{1}{L}\sum_{1<|\omega_{ j }(\text{Re} (E_0),x)|}\cosh^{-1}|\omega_{ j }(\text{Re}(E_0),x)|\right),
\end{align}}
where $\Tilde{\gamma}_{\text{topo}}=\Tilde{\gamma}(0)$ sets the threshold non-Hermiticity for the topological zero mode.

\begin{figure}
    \centering
    \includegraphics[width=\linewidth]{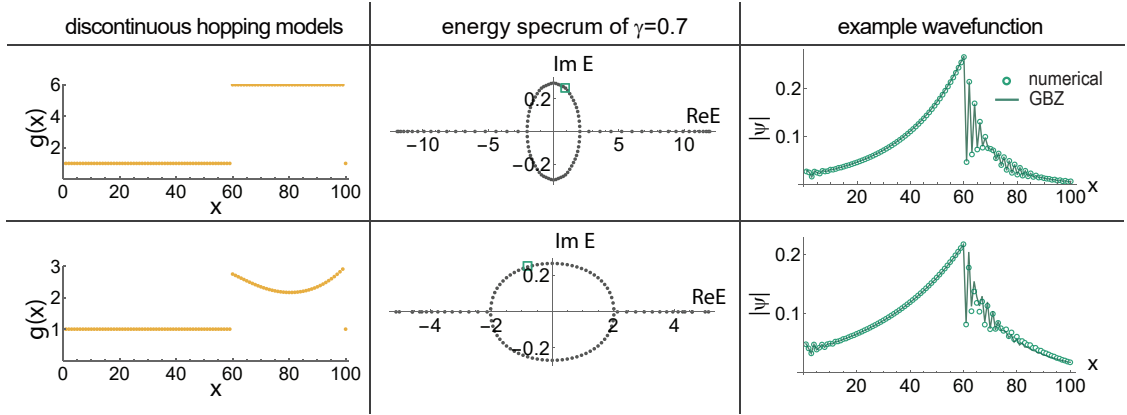}
    \caption{\textbf{Incorporating isolated discontinuities in the hopping profiles into the phase-space GBZ. } Although the accuracy of the phase-space GBZ approach may be reduced when non-smoothness of $g(x)$ is introduced, isolated non-smoothness or discontinuities can be effectively addressed by matching superposed solutions at the discontinuity sites, as given by Eq.~\ref{NS3}. (a) Two illustrative hopping profiles with discontinuities at $x=0$ and $\frac{3L}{5}$, both with $g(x)=1$ for $0\leq x < \frac{3L}{5}$. For the remaining lattice, the upper example has $g(x)=6$ and the lower has $g(x)=1/(\sin(2 \pi x/L) \cos(\pi x/L) + 2)$. (b) Their spectra at $\gamma=0.7$, both which exhibit characteristic real tails. (c) By applying Eq.~\ref{NS3} at $x=0$ and $\frac{3L}{5}$ for two illustrative states (indicated by green hollow squares in (b)), the resultant wavefunction prediction (solid line) is seen to agree well with the numerical wavefunctions (circles) for both $g(x)$ profiles.} 
    \label{fig10} 
\end{figure}

\subsection{Handling isolated hopping discontinuities}\label{sec_discont_gx}
Here we extend our phase-space GBZ formalism to handle systems with local isolated non-smoothness or discontinuity in the hoppings, such that they are treated as additional boundaries. 
Consider a non-Hermitian inhomogeneous lattice with a discontinuous hopping function
\begin{align*}
    &g(x)=
    \begin{cases}
      f_{1}(x) & \text{ for } x_{0}\leq x< x_{1}\\
      f_{2}(x) & \text{ for } x_{1} \leq x< L+x_{0}
    \end{cases},
\end{align*}
where $f_{1,2}(x)$ are smooth functions with $f_{1}(x_1-1)\neq f_2(x_1)$. Its eigenstate solution $\psi(x)$ is solved by separately considering $\psi^{(1,2)}(x)$:
\begin{align}
&\psi(x)=
\begin{cases}
     \psi^{(1)}(x) & \text{ for } x_{0}\leq x\leq x_{1}\\
     \psi^{(2)}(x) & \text{ for } x_{1} \leq x\leq L+x_{0}
\end{cases},
\end{align}
with two boundary conditions at each discontinuous point:
 \begin{align}
    &\psi^{(1)}(x_{0})=\psi^{(2)}(x_{0}),\\
    &f_{1}(x_{0})\psi^{(1)}(x_{0}+1)=f_{2}(x_{0})\psi^{(2)}(x_{0}+1).\label{NSdeal}
\end{align}
$\psi^{(1,2)}(x)$ could be written as a superposition of phase-space GBZ state solutions of \emph{both} smooth-hopping sub-lattices simultaneously:
\begin{align}
    &\psi^{(1,2)}(x)=c_{+}^{(1,2)}\phi_{+}^{(1,2)}(x)+c_{-}^{(1,2)}\phi_{-}^{(1,2)}(x),\label{NS1}
\end{align}
with
\begin{align}
     &\phi_{\pm}^{(1,2)}(x)= \frac{1}{\sqrt{f_{1,2}(x)}}\gamma^x\prod_{x'=1}^{x}\beta_{n,\pm}^{(1,2)}(x).
\end{align}
The superposed state solutions fulfill the bulk relation:
{\small
\begin{align}
f_{1,2}(x)&/\gamma\psi^{(1,2)}(x+1)+f_{1,2}(x-1) \gamma\psi^{(1,2)}(x-1)=E_{n},\label{jumpSubBulk}
\end{align}}
for all $x$ defined on $f_{1,2}(x)$. In contrast to smooth-hopping states that have only one of $\beta_{n,\pm}(x)$ contributing to the wavefunction, each non-smooth jump in $g(x)$ requires simultaneous contributions from both solutions, where one governs the overall trend and the other accounts for the fluctuation near the jump. Generally, two sub-lattices with two non-smooth jumps involve four non-trivial coefficients $c_{\pm}^{(1,2)}$ to be solved: 
\begin{align}
    \text{Det}\begin{bmatrix}
c_{+}^{(1)} & c_{-}^{(1)} & -c_{+}^{(2)} & -c_{-}^{(2)} \\
F_{1,+,+}& F_{1,-,-}& -F_{2,+,+}&-F_{2,-,-}\\
U_{1,+,+}& U_{1,-,-}& -U_{2,+,+}& -U_{2,-,-}\\
W_{1,+,+}&W_{1,-,-} &-W_{2,+,+}& -W_{2,-,-}
    \end{bmatrix}=0,\label{NS2}
\end{align}
where 
\begin{align}
     F_{a,\mu,\nu}&=f_{a}(x_{0})c_{\mu}^{(a)}\gamma\beta_{n,\nu}^{(a)}(x_{0}+1),\\
U_{a,\mu,\nu}&=c_{\mu}^{(a)}\prod_{x=x_{0}+1}^{x_{1}}\gamma\beta_{n,\nu}^{(a)}(x),\\
W_{a,\mu,\nu}&=f_{a}(x_{1})c_{\mu}^{(a)}\prod_{x=x_{0}+1}^{x_{1}+1}\gamma\beta_{n,\nu}^{(a)}(x).
\end{align}
In practice, with large $L$, strong NHSE accumulation occurs in one of the sub-lattices due to the existence of inhomogeneous skin regions:
\begin{align}
    &\prod_{x=x_{0}+1}^{x_{1}}\gamma\beta_{n,-}^{(j)}(x)\ll\prod_{x=x_{0}+1}^{x_{1}}\gamma\beta_{n,+}^{(j)}(x),\label{NS4}
\end{align}

and Eq.~\ref{NS2} can be simplified by the approximation $c_{-}^{(j)}\to 0$, which is justified by Eq.~\ref{NS4} . For $j=1$, Eq.~\ref{NS2} reduces to
\begin{align}
    \text{Det}\begin{bmatrix}
        U_{2,+,+}+U_{1,+,+}  & U_{2,-,-}\\
            W_{2,+,+}+W_{1,+,+} &W_{2,-,-}
    \end{bmatrix}=0,\label{NS3}
\end{align}

whose solution gives all $c_{-}^{(j)}=0$ and completes the phase-space GBZ construction for systems with isolated non-smoothness. Fig.~\ref{fig10} demonstrates examples of $g(x)$ with a pair of discontinuities and how their states are well approximated by the GBZ approach with an additional discontinuity [Eq.~\ref{NS3}].

\SetPicSubDir{ch-eedip}
\SetExpSubDir{ch-eedip}

\chapter{Non-Hermitian entanglement dip from scaling-induced exceptional criticality}
\label{ch-eedip}
\vspace{2em}

As discussed in Chapters \ref{ch-intro} and \ref{ch-prel}, critical phase transition is a central focus in condensed matter physics. Intense research has been conducted on both Hermitian and non-Hermitian critical behaviours, leading to the discovery of novel phases of matter and profound implications in conformal and statistical field theory, entanglement entropy, and thermodynamics. Within the study of critical phenomena, entanglement entropy (EE) serves as a key factor characterizing phase transitions. Typically, in both Hermitian and non-Hermitian critical systems, 1D criticality is characterized by entanglement entropy scaled logarithmically with system size of the system as a direct consequence predicted by the conformal field theory (CFT).

If we go beyond 1D systems, for example, to two-dimensional topological insulators and superconductors, the EE scaling is influenced by the presence of edge states and Majorana modes. Similar logarithmic scaling behaviour is observed, though usually accompanied by secondary corrections that reveal the additional degrees of freedom and more complex topological properties. These corrections to the EE scaling offer a deeper understanding of the mechanisms of phase transitions and the properties of criticality and are effectively determined by measurable parameters corresponding to the increased complexity of the systems contributing to the EE scaling.

In this chapter, we report a new class of non-Hermitian critical transitions that diverge dramatically from this conventional scaling behaviour. This phenomenon, which we term scaling-induced exceptional criticality (SIEC), is marked by dips into super-negative EE that contradict established predictions and current understandings in CFT. SIEC arises from non-local antagonism between multiple amplification channels, resulting in scale-dependent spectral properties and exceptional crossings that occur only at specific system sizes.

We present a general approach for constructing SIEC systems and elucidate the emergence of this unconventional scaling-induced critical behaviour through a rigorously developed, scale-dependent generalized Brillouin zone (GBZ) framework. The GBZ, introduced in Chapter \ref{ch-prel}, provides a complete mathematical description of conventional non-Hermitian systems through the analytic eigenenergies and eigenstates of the Hamiltonian. Here, the scaled-dependent generalized GBZ is adapted to explain the emergence of the SIEC-induced dips in EE and the breakdown of scale invariance at exceptional points.

The divergent EE dips observed near the scaling-induced exceptional points signify a breakdown of scale invariance, a characteristic usually associated with second-order critical points. This scale-dependent GBZ offers a comprehensive understanding of the breakdown of scale invariance and the emergence of scale-dependent exceptional points (EPs). The study of SIEC thus broadens our understanding of the interplay between non-Hermitian amplification, system size, and entanglement entropy.

This chapter focuses on 1D free fermion systems, though we expect similar SIEC phenomena to occur in higher-dimensional systems.

\section{Introduction}

As established in Chap. \ref{ch-intro} and Chap. \ref{ch-prel}, entanglement entropy (EE) is a crucial factor in characterizing critical Hermitian quantum phase transitions. It universally associates logarithmic entanglement entropy (EE) scaling $S\sim \log L$ with 1D criticality~\cite{PhysRevLett.113.250401, herviou2019entanglement,fring2019eternal,guo2021entanglement,ding2022information,tu2022general, PhysRevB.105.L241114,itable2023entanglement,fossati2023symmetry,kawabata2023entanglement,PhysRevB.109.024204, e26030272, Wei2017, PhysRevB.109.024306, PhysRevResearch.6.023081}, where $L$ is the system size, according to the boundary conformal field theory~\cite{calabrese2004entanglement, Cardy:2004hm, PhysRevLett.116.026402, Taddia_2016, PhysRevB.95.115122, PhysRevB.96.045140, PhysRevLett.130.250403, PhysRevLett.132.086503, Jones2022}.
The established results in critical entanglement scaling have been challenged in non-Hermitian contexts recently~\cite{PhysRevLett.113.250401, herviou2019entanglement,fring2019eternal,guo2021entanglement,ding2022information,tu2022general, PhysRevB.105.L241114,itable2023entanglement,fossati2023symmetry,kawabata2023entanglement,PhysRevB.109.024204, e26030272, Wei2017, PhysRevB.109.024306, PhysRevResearch.6.023081}, 
where the exceptional points (EPs)~\cite{bender_real_1998,Bender1999, Berry:2004ypy,  Heiss_2004, PhysRevE.69.056216, bender_making_2007,Rotter_2009, PhysRevA.80.052107,Moiseyev_2011, PhysRevX.8.031079,kawabata2019symmetry, PhysRevLett.123.066405,PhysRevB.106.L041302,mandal2021symmetry} introduces geometric defectiveness that blurs the distinction between occupied and unoccupied bands, and the probability non-conservation leads to fermionic occupancies effectively greater than one, together giving rise to negative free-fermion EE~\cite{cerf1997negative,salek2014negative,chang2020entanglement,lee2022exceptional,tu2022renyi,xue2024topologically,ZOU2024}. This is further exacerbated by the non-Hermitian skin effect (NHSE)~\cite{PhysRevB.97.121401, PhysRevLett.123.016805, PhysRevB.99.201103, PhysRevLett.124.250402,  PhysRevB.102.205118, PhysRevLett.124.086801,PhysRevLett.125.126402, okuma2021quantum, PhysRevLett.127.066401, PhysRevB.104.L241402, PhysRevB.104.L161106, Li2021, PhysRevB.104.195102, Shen2022, zhang2022review, Shang_2022, Zhang2022, FangHuZhouDing, PhysRevB.107.085426, Zhou2023, Wan_2023, PhysRevLett.131.207201, Okuma_2023, Lin2023, le2023volume, jana2023emerging,shen2023observation, Lei2024}, which produces macroscopically many highly defective eigenstates indexed by complex momenta~\cite{bender_making_2007, El-Ganainy2018, yao2018edge,yin2018geometrical,jiang2018topological,PhysRevLett.123.066404, jiang2019interplay,PhysRevLett.125.226402,li2020critical,li2020topological,PhysRevLett.125.126402,bergholtz2021exceptional,yokomizo2021scaling,guo2021exact,zou2021observation,li2021impurity,li2022non,li2022direction,yang2022designing,yang2023percolation,jiang2022filling,liu2023reentrant,tai2023zoology,jiang2023dimensional,manna2023inner,qin2023universal,shen2024enhanced,guo2023anomalous,zhang2024observation,qin2024dynamical,yang2024anatomy,hamanaka2024multifractality,ekman2024liouvillian,qin2024kinked,liu2024localization}. Despite these complications, all known non-Hermitian free-fermion critical systems still exhibit the celebrated $S\sim \log L$ scaling, even if c is not the central charge.

In this work, we report a new class of non-Hermitian critical transitions that exhibit dramatic divergent dips into the super-negativity in their entanglement entropy scaling, strongly violating conventional logarithmic behaviour and current understanding of the EE in the CFT. We named it the scaling-induced exceptional criticality (SIEC), since this criticality comes from the non-local competition between multiple directed amplification channels, such as two reversely-oriented weakly-interacted NHSE used in this work. This results in strongly scale-dependent spectra, where the eigenbands exhibit an exceptional crossing only at a particular system size. This unconventionally scaling-induced exceptional point leads to the new class of scale-induced critical phase transition, which is dominated by how the generalized Brillouin zone (GBZ) sweeps through the exceptional crossing with increasing system size.
We provide a general approach for constructing SIEC systems, and elaborate how such dips can be traced to the breaking of scaling invariance as well as the emergence of supposedly scale-free gapless points, through a rigorous scale-dependent generalized Brillouin zone (GBZ) construction.

The study of SIEC in this work focus on 1D free fermions model, whereas SIEC is expected to occur more prevalently in higher-dimensional or even interacting systems, where antagonistic NHSE channels generically proliferate. 
Moreover, SIEC-induced entanglement dips generalize straightforwardly to kinks in other entanglement measures such as Renyi entropy, and serve as spectacular demonstrations of how algebraic and geometric singularities in complex band structures manifest in quantum information.

\begin{figure*}
	\centering
	\includegraphics[width=\linewidth]{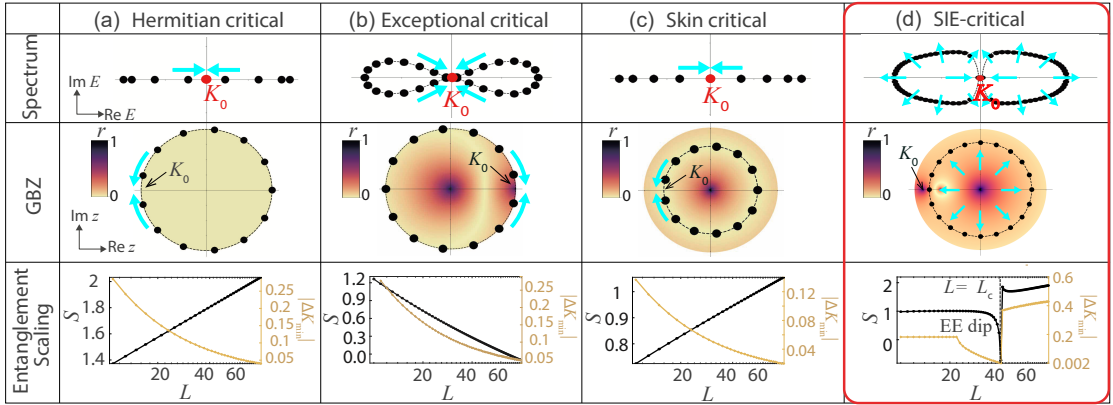}
\caption{\label{fig:fig1} \textbf{Uniqueness of scaling-induced exceptional critical (SIEC) transitions in violating logarithmic entanglement scaling.} (a) For a Hermitian critical point $K_0$, eigenmomentum states (black dots) simply become denser and approach $K_0$ (red) as the system size $L$ increases (cyan arrows), such that the closest approach $\Delta K_\text{min}\sim \pi/L$ (brown curve), giving rise to $S\sim \log L$ (black curve). 
(b) Conventional logarithmic EE scaling from usual $\Delta K_\text{min}\sim \pi/L$ eigenmomentum convergence still governs an exceptional non-Hermitian critical point $K_0$, albeit with a negative coefficient ($S\sim -\log L$) due to divergent 2-point functions at exceptional criticality where $r\rightarrow 0$ (dark purple).  
(c) Logarithmic EE scaling also persists under the NHSE, since we still have $\Delta K_\text{min} \sim \pi/L$  in the generalized Brillouin zone (GBZ)  $z=\text{e}^{iK (k)}$. 
(d) By contrast, in our SIE-critical (SIEC) systems, the spectrum and GBZ themselves change with $L$ (expanding cyan arrows), 
exhibiting critical touching $K_0$ (dark purple) only at a particular $L=\lfloor L_c\rfloor$. This leads to peculiar $\Delta K_\text{min}$ scaling and a characteristic entanglement dip in $S$.
}
\end{figure*}

\section{Conventional non-Hermitian criticality: Log-Entanglement scaling} 
Before going to the SIEC entanglement dips, we take some time to review the properties and entanglement entropy (EE) scaling of well-known types of criticality. This review will clarify the significance of SIEC and highlight its unique features.

Consider an arbitrary non-interacting 1D lattice Hamiltonian 
\begin{equation}
\mathcal{H}=\sum_{k,\mu} E_\mu(k)|\psi_{k,\mu}^R\rangle\langle \psi^L_{k,\mu}|=\sum_{k,\mu} E_\mu(k)\psi_{k,\mu}^R[\psi^L_{k,\mu}]^*|k\rangle\langle k|,
\end{equation}
where $|k\rangle$ is the quasi-momentum basis state and $|\psi_{k,\mu}^{R}\rangle,\langle \psi^L_{k,\mu}|$ are respectively the right and left eigenbands~\cite{Brody_2014} of $\mathcal{H}$ with band index $\mu$. We are interested to the entanglement entropy between two truncated sub-lattices on the occupied band. We assume that fermions occupy an arbitrary set of bands $\mu\in occ.$, such that state occupancy is described by the occupied band projector 
$P=\sum_{k,\mu\in occ.} |\psi_{k,\mu}^R\rangle\langle \psi^L_{k,\mu}|=\sum_kP_k|k\rangle\langle k|$, where $P_k=\sum_{\mu\in occ.}\psi_{k,\mu}^R[\psi^L_{k,\mu}]^*$. 
For an entanglement cut where the region $[x_L,x_R]$ is to be truncated, the biorthogonal~\footnote{We used the biorthogonal $\bar P$ such as to directly connect to the band occupancy interpretation; some other works~\cite{fring2019eternal,ding2022information,PhysRevB.105.L241114,itable2023entanglement,kawabata2023entanglement,PhysRevB.109.024204,e26030272,PhysRevB.109.024306,PhysRevResearch.6.023081} used the right eigenstate for both the bra and the ket, with very different results.} free-fermion EE is given by~\cite{PhysRevB.109.024306, kawabata2023entanglement}
\begin{equation}
S= -\text{Tr}[\bar P\log\bar P +(\mathbb{I}-\bar P)\log(\mathbb{I}-\bar P)],
\label{S}
\end{equation}
where $\bar P = \bar{\mathcal{R}}P\bar{\mathcal{R}}$ is the occupied band projector that is restricted to $x\notin [x_L,x_R]$ by the real-space projector $\bar{\mathcal{R}}=\sum_{x\notin [x_L,x_R]}|x\rangle\langle x|$. In real-space, $\bar P$ is a block Toeplitz matrix with nonvanishing blocks $\bar P_{xx'}=\langle x | \bar P|x'\rangle$ (for $x,x'\notin [x_L,x_R]$) given by the Fourier transform 
\begin{eqnarray}
\bar P_{xx'}
&=& \sum_{k} \text{e}^{ik(x'-x)}\sum_{\mu\in occ.}\psi_{k,\mu}^R[\psi^L_{k,\mu}]^*=\sum_{k} \text{e}^{ik(x'-x)}P_k.\notag\\
\label{barP}
\end{eqnarray}

\noindent Based on Eq.~\ref{barP}, we first review why the EE scales like $\log L$ in Hermitian critical points, as illustrated in Fig.~\ref{fig:fig1}(a). Consider a  critical point $K_0$ (red) where the gap closes but the eigenspace remains full-rank, i.e., non-defective. Due to the possible mixing of the occupied and unoccupied eigenstates at gap closure, the occupied band projector $P_k$ generically become discontinuous, i.e., singular at the critical point. However, due to the discrete momentum spacing (black dots) which scales like $2\pi/L$ (cyan arrows) in a finite lattice, the momentum sum in Eq.~\ref{barP} only samples $P_k$ down to a distance $\Delta k =k-K_0$ bounded below by $\Delta K_\text{min}\sim\pi/L$ (brown curve) from the critical point, i.e.,  
$\bar P_{xx'}\sim\int_{\pi/L} \text{e}^{i\Delta k(x'-x)}P_{\Delta k} \,d(\Delta k)$ 
with a UV cutoff $\pi/L$. 
Analytic singularities result in power-law decay of their Fourier coefficients~\cite{PhysRevD.100.016016, aloisio2022fourier}~\footnote{In general, the exponent of the power-law decay of the Fourier coefficients is directly dependent on the singularity branching order~\cite{PhysRevD.100.016016, PhysRevA.95.013604}.}. Consequently, we expect $P_{xx'}$, which are also the 2-point functions, to become long-ranged and cause substantial entanglement across the cut. Mathematically, this entanglement can be shown to diverge logarithmically with the inverse of the UV cutoff~\cite{CARDY1988377, PhysRevLett.97.050404}, thereby yielding $S\sim \log L$ entanglement scaling~\footnote{It can be proven that the $\sim 1/L$ cutoff away from a branching singularity in the symbol $P_k$ causes the eigenvalues of $\bar P$ to deviate from $0$ and $1$ with $\log L$ spacing~\cite{CARDY1988377}.}.

Such $\log L$ entanglement scaling persists even if the critical point is geometrically defective (with completely non-orthogonal eigenstates), as is possible in non-Hermitian systems, as illustrated in Fig.~\ref{fig:fig1}(b). For a particular band, the overlap between its left and right eigenstates $\langle\psi^L_k|$, $|\psi^R_k\rangle$ can be quantified through the phase rigidity~\cite{Ding2022, PhysRevResearch.5.033042, Zou2023, Wang2020}
$r(k)= \frac{\langle\psi^L_k|\psi^R_k\rangle}{\sqrt{\langle\psi^L_k|\psi^L_k\rangle\langle\psi^R_k|\psi^R_k\rangle}}, $
which is always unity in Hermitian systems, but ranging between $0\leq r(k)\leq 1$ in non-Hermitian systems. In particular, at an exceptional critical point (EP) $K_0$ where $r(K_0)=0$~\cite{PhysRevResearch.5.033042} (dark purple), the occupied and unoccupied eigenstates coalesce and the occupied band projector $P_k$ is not just discontinuous, but in fact divergent. Specifically, one or more matrix elements of $P_k$ can be shown to diverge with $\Delta k=k-K_0$ as $(\Delta k)^{-B}$, which scales with $\Delta K_\text{min}\sim L^{B}$, the model-dependent exponent $B$ related to the order of the EP~\cite{lee2022exceptional,ZOU2024}. 
Such divergences in $P_k$ have also been recently shown to lead to $S\sim \log L$ entanglement scaling, albeit with a negative coefficient determined by $B$ ~\cite{lee2022exceptional,ZOU2024}.

Another key non-Hermitian phenomenon is the non-Hermitian skin effect (NHSE), but below we explain why the critical $\log L$ EE scaling still generically holds, as shown for the prototypical gapless non-Hermitian SSH [Fig.~\ref{fig:fig1}(c)]. In general, NHSE ``skin'' states can be modeled with exponentially decaying state profiles $\text{e}^{ikx}\text{e}^{-\kappa(k)x}$ for each wavenumber $k$, the inverse decay length $\kappa(k)$ 
determined through detailed boundary condition analysis~\cite{yao2018edge, PhysRevLett.123.066404}\footnote{This can involve various subtleties when there is more than one competing NHSE channel, as in our entanglement dip model discussed later.}. 
As such, the periodic and open boundary condition (PBC and OBC) eigenstates are approximately related through a complex quasi-momentum deformation $k\rightarrow K(k)=k+i\kappa(k)$, such that $\psi^R_{k,\mu}\rightarrow \psi^R_{K,\mu}$.  
This path $K(k)$, $k\in [0,2\pi)$ is known as the generalized Brillouin zone (GBZ), and can be visualized as a closed loop $z=\text{e}^{iK(k)}$ that deviates from the unit circle $\text{e}^{ik}$ due to the NHSE, as illustrated in Fig.~\ref{fig:fig1}(c).

Importantly, the modified entanglement entropy due to the NHSE can be directly computed through this GBZ deformation, since the eigen-equations for the left and right eigenstates depend on $K(k)$ according to~\footnote{In Eq.~\ref{biortho2}, $\mathcal{H}$ and $E_\mu$ are regarded as functions of $K$, such that the conjugation operator only acts on the coefficients of $K$, and not $K$ itself.}
\begin{align}
\mathcal{H}(K)|\psi^R_{K,\mu}\rangle&=E_\mu(K)|\psi^R_{K,\mu}\rangle,\notag\\
\mathcal{H^\dagger}(K^*)|\psi^L_{K^*,\mu}\rangle&=E^*_\mu(K^*)|\psi^L_{ K^*,\mu}\rangle,
\label{biortho2}
\end{align}
such that in the presence of skin state accumulation, $P_k\rightarrow \sum\limits_{\mu\in occ.}|\psi^R_{K,\mu}\rangle[|\psi^L_{ K^*,\mu}\rangle]^\dagger=\sum\limits_{\mu\in occ.}|\psi^R_{K,\mu}\rangle\langle \psi^L_{K_\mu}|= P_K$. 
As such, replacing $P_k$ in Eq.~\ref{barP} by $P_K$, we obtain the 2-point functions under the NHSE as
\begin{eqnarray}
\bar P_{xx'} \xrightarrow[\,\text{NHSE}\,] \quad\sum_{k} \text{e}^{ik(x'-x)}P_{K(k)=k+i\kappa(k)}.
\label{barP2}
\end{eqnarray}
Note that the Fourier sum for $\bar P_{xx'}$ is still with respect to $k$ (and not $K$) because the NHSE does not alter the definition of the basis transform between real- and momentum-space.
If $P_{K}$ encounters a singularity $K_0$ along the GBZ path $K(k)$, then, like the above cases, the system would be critical and the EE would scale like 
\begin{equation}
|S|\sim |\log\Delta K_\text{min}|,
\label{SKmin}
\end{equation}
where $\Delta K_\text{min}$ is determined~\footnote{To satisfy OBCs, the discrete effective momenta on the GBZ are obtained by subdividing $2\pi$ into $L+1$ instead of $L$ intervals, and discarding the endpoints.} by the closest approach of $K(2\pi n/(L+1))$, $n=1,...,L$ to the singularity. Since $\Delta K_\text{min}$ would still scale like $L^{-1}$ (brown curve) except in rare cases with sufficiently badly behaved $K(k)$~\footnote{Even when the GBZ $K(k)=k+i\kappa(k)$ contains cusps, the singularities encountered by $P_K(k)$ at most acquire different branching numbers, and that can only modify the coefficient of the $\log L$ EE scaling.}, the key takeaway is that the NHSE itself cannot give rise to critical entanglement scaling other than $S\sim \log L$.

\section{Unconventional SIEC from scale-dependent generalized Brillouin zone }
The prominent non-Hermitian phenomena, exceptional point (EP) defectiveness and the NHSE, are shown not able to bring in by themselves violations of $\log L$ entanglement scaling. This is because the closest approach to the critical point $K_0$ obeys $\Delta K_\text{min}\sim L^{-1}$ for all but the most pathological GBZ paths $K(k)$.  

In this work, a key insight is that qualitatively new entanglement scaling \emph{can} however occur if the GBZ trajectory $K(k)$ also depends on $L$, i.e., is scale-dependent. As $L$ increases, $\Delta K_\text{min}$ may vary non-monotonically with $L$, reaching a minimum when the $K(k)$ cuts across a singularity $K_0$, as highlighted in Fig.~\ref{fig:fig1}(d) in red. The GBZ loop $\text{e}^{iK(2\pi n/(L+1))}$, $n=1,...,L$ expands with increasing $L$ (outward cyan arrows on loop of black dots), such that it intersects $K_0$ (dark purple) only at a special value of $L=L_c$. As such, we have rapid unconventional $\Delta K_\text{min}\propto |L-L_c|$ scaling around $L\approx L_c$ (kink on brown curve). Since $S\sim -\log\Delta K_\text{min}$, we thus also expect $\log L$-violating EE scaling.

\begin{figure}
	\centering
	\includegraphics[width=\linewidth]{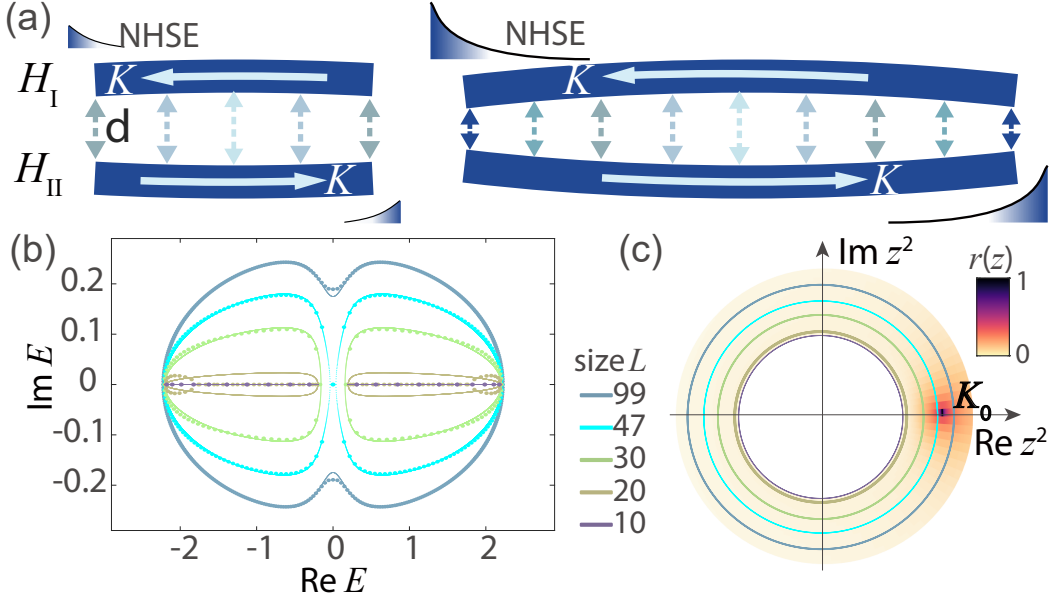}
	\caption{
\textbf{Scaling-dependent critical spectrum and GBZ from antagonistic NHSE.} 
(a) Our construction [Eq.~\ref{Hcoupled}] involves weakly coupling two chains $H$ with opposite NHSE and momentum label $k$. Only for sufficiently long chains (Right) is the NHSE strong enough to effectively close the two chains into a loop with PBC-like dynamics. 
(b)	Strongly $L$-dependent spectrum which undergoes a gapped to gapless exceptional critical transition at $L=47\approx L_c$ (cyan). (c) The corresponding $L$-dependent GBZ $z=\text{e}^{iK(k)}$ (colored loops), which pass through the critical point $K_0$ (where phase rigidity $r(z)=0$) when $L$ sweeps pass $L_c$. Parameters are $\delta = 8.4 \times 10^{-4}$ and $t_L=1.62$, $t_R=0.89$ in Eq.~\ref{SSH}.
	}
	\label{fig:fig2}
\end{figure}
For such unconventional entanglement scaling, the system must meet the following conditions: (i) it must support qualitatively distinct dynamics at small $L$ and large $L$, such as to have very different spectra and hence GBZs in these limits; (ii) it must possess at least two bands (components) to exhibit an exceptional critical transition between these limits. 

Condition (i) can be generically met by weakly coupling two finite 1D NHSE chains $\nu=\text{I}$, $\text{II}$ with oppositely directed NHSE pumping to induce inter-chain tunneling [Fig.~\ref{fig:fig2}(a)]. As an important departure from usual literature, we label their basis $|x\rangle_\nu,|k\rangle_\nu$ in opposite directions [Fig.~\ref{fig:fig2}(a)], such that both chains are described by the same Hamiltonian $H$. This labeling, which assigns the momentum $k$ based on the NHSE pumping, would turn out instrumental in our GBZ construction. Our system, with weak coupling $\delta$, takes the form
\begin{eqnarray}
\label{Hcoupled}
\mathcal{H}
&=& \sum_{x,x'=1}^LH_{xx'}|x\rangle_\text{I}\langle x'|_\text{I}+\sum_{x,x'=1}^LH_{xx'}|x\rangle_{\text{II}}\langle x'|_{\text{II}}\notag\\
&& + \delta \sum_{x=1}^L \Big[|x\rangle_{\text{I}}\langle L+1-x|_{\text{II}}+|L+1-x\rangle_{\text{II}}\langle x|_{\text{I}}\Big]\notag\\
&=& \sum\limits_{k;\nu}H_{k}|k\rangle_\nu\langle k|_\nu+ \delta \sum_k \Big[|k\rangle_{\text{I}}\langle -k|_{\text{II}}+|-k\rangle_{\text{II}}\langle k|_{\text{I}}\Big],\qquad
\end{eqnarray}
where the last line describes only the bulk structure. Crucially, skin states grow exponentially with distance $L$ in each chain, concomitantly enhancing the effective inter-chain tunneling probability. At small $L$, we have two effectively uncoupled OBC chains (faint double-arrows in Fig.~\ref{fig:fig2}(a)). But at large $L$, the exponentially larger tunneling probabilities (dark blue double-arrows) effectively close the two antagonistic chains into a loop, forming a PBC-like configuration. 
This construction indeed produces a distinctively scale-dependent spectrum, as shown in Fig.~\ref{fig:fig2}(b) for a minimal illustrative model with coupled
\begin{equation}
H_\text{SSH}(k)=(t_L+\text{e}^{-ik})\sigma_++(t_R+\text{e}^{ik})\sigma_-
\label{SSH}
\end{equation}
chains, where $\sigma_\pm = (\sigma_x\pm i \sigma_y)/2$ are the Pauli matrices (see Fig.~\ref{fig:fig4}).

While no exact analytic solution exists for the GBZ of generic $\mathcal{H}$ [Eq.~\ref{Hcoupled}], below we outline a general procedure for deriving an approximate GBZ that nevertheless captures its scaling properties accurately:
\begin{enumerate} [leftmargin =.5cm]
\item First, for an isolated chain $H$, solve $\text{Det}(H_k-E\,\mathbb{I})=0$ to obtain the bulk dispersion of $H$, i.e., relation between $z=\text{e}^{ik}$ and energy $E$. 

\item Solve the above dispersion polynomial and obtain all possible solutions $z_i$ for any given $E$. For each $z_i$, obtain the corresponding eigenstates $\ket{\zeta(z_i)}_\text{I}$ and $\ket{\zeta(z_i)}_\text{II}$ based on $H$ and $H^T$. While values of $E$ that satisfy the conventional GBZ condition $|z_i|=|z_j|$, $i\neq j$ correspond to the OBC spectrum of an \emph{isolated} chain, this condition will be severely violated for the coupled system $\mathcal{H}$, even with tiny $\delta \neq 0$.

\item Introduce the inter-chain coupling $\delta$ and solve the OBC eigenequation $\mathcal{H}\ket{\Psi_\text{GBZ}}=E\ket{\Psi_\text{GBZ}}$ for both coupled chains using the ansatz $\ket{\Psi_\text{GBZ}}=\sum_{i}c_i\sum\limits_{x=1}^L z_i^x\ket{\zeta(z_i)}_\text{I}\oplus z_i^{L+1-x}\ket{\zeta(z_i)}_\text{II}$, where $c_i$ are linear combination coefficients to be eliminated. The idea is to express the $\delta$ coupling terms entirely in terms of the OBC boundary contributions, both of which are what did not exist in the single-chain bulk solution in (2) above. 

\item From the above, solve for the full GBZ solutions $z_i$ in terms of $L$, $\delta$ and $H$ model parameters, eliminating $E$ using the bulk dispersion from (1) if necessary. Due to the $z_i^{L+1-x}$ factor in our ansatz, which arises from the antagonistic momenta in Eq.~\ref{Hcoupled}, the GBZs $z_i$ typically satisfy a $L$-degree polynomial involving the model parameters, leading to generic $\sim 1/L$-th power scaling of $z_i$ with $\delta$. 

\end{enumerate}  

\noindent This GBZ effectively encapsulates the non-perturbative effects of the coupling $\delta$ through the scale-dependent complex deformation $K(k)$, henceforth allowing properties of the full system $\mathcal{H}$ to be computed just through a single chain $H$. As detailed in Supplement sect.~\ref{suppmat} and plotted in [Fig.~\ref{fig:fig2}(c)], the dominant scale-dependent GBZ for our minimal example $H_\text{SSH}$ [Eq.~\ref{SSH}]~\footnote{The subdominant GBZ solutions $|z_i|<|z_1|$ can control the lower bound of the decaying eigenstates, as detailed in Supplement Sect.~\ref{suppmat}} scales like
\begin{equation}
z_1=\text{e}^{iK(k)}\sim (\alpha_0\delta)^{1/(L+1)}\text{e}^{ik},
\label{z1}
\end{equation}
$\alpha_0$ a model-dependent constant. Eq.~\ref{z1} accurately predicts the spectrum of $\mathcal{H}_k$  upon substituting $k\rightarrow -i\log z_1$ [Fig.~\ref{fig:fig2}(b)]. Correctly associating the antagonistic directions with opposite momenta [Eq.~\ref{Hcoupled}] was crucial in deriving this scaling-dependent GBZ and spectrum, which cannot be analytically derived as a function of $L$ using conventional approaches~\cite{li2021critical, yokomizo2021scaling,rafi2022critical,qin2023universal,yang2023percolation,wang2023non}.

\section{SIEC and Entanglement Dip} 
The system size-dependent GBZ can give rise to a new phenomenon dubbed ``scaling-induced exceptional criticality'' (SEIC) if the GBZ loop $\text{e}^{iK(k)}$ sweeps through an EP, as shown by $r(K)=0$ in dark purple in Fig.~\ref{fig:fig2}(c), when $L$ is varied across a special value $L_c$.
For $H_\text{SSH}$ and related models~\ref{suppmat}, substituting Eq.~\ref{z1} into $H_\text{SSH}(K(k))$ and demanding that it reduces to the Jordan form $\sigma_-$ at the EP yields 
\begin{eqnarray}
\label{eqLc}
L_c&=&-\alpha_0\delta/\log t_R -1,\\ 
K_c&=&K|_{L=L_c} = \pi+\frac{i\alpha_0\delta}{L+1},
\label{eqKc}
\end{eqnarray}

which accurately predicts when the exceptional transition occurs, corroborated by numerical results in Fig.~\ref{fig:fig2}(b). As $L_c$ is typically non-integer, actual lattice $\mathcal{H}$ models with integer $L$ experience dramatic SEIC divergences when $L\rightarrow L_c$, as controlled by~\footnote{$\Delta K_\text{min}$ measures the departure from a critical point at the same $L$, while $\Delta_K$ measures that from a theoretical critical point limit at another, possibly non-integer, $L$.}
\begin{eqnarray}
\Delta_K=K(k)-K_c
&\approx & (k-\pi)+i\delta\frac{\alpha_0 (L_c-L)}{L_c^2},
\label{DeltaK}
\end{eqnarray}

\begin{figure}
	\centering
	\includegraphics[width=1\linewidth]{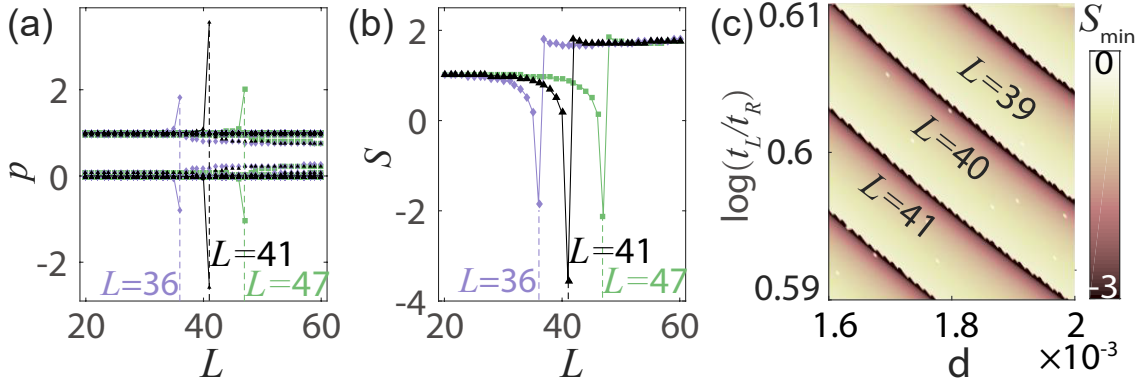}
	\caption{
	\textbf{Entanglement dip from SIEC.} (a) Spikes in the $\bar P$ eigenvalues $p$ occur near $L\approx L_c$ [Eq.~\ref{eqLc}] for different inter-chain couplings $\delta=0.83\times10^{-3}, 1.6843\times 10^{-3}$ and $3.03\times 10^{-3}$ (green, black and purple respectively) of Eq.~\ref{SSH}, with hopping asymmetries $t_L,t_R=1.2e^{\pm 0.3}$. 
	(b) Corresponding entanglement dips at these $L\approx L_c$, with characteristic $S\sim (L_c-L)^{-1/2}$ behaviour for $L<L_c$.
	(c) Depth of entanglement dips in the parameter space of $t_L/t_R, \delta$ and the system sizes $L$ realizing them, with large dips of $S_\text{min}\leq -3$ along narrow bands separating regions of different $L$.}
	\label{fig:fig3}
\end{figure}

such that~\ref{suppmat} $H_\text{SSH}(K)\approx (t_L-t_R^{-1})\sigma_+-it_R\Delta_K\sigma_-$ for $L\approx L_c$, with the expected square root cusp in its eigenenergies $\pm\mathcal{E}_\text{SSH}(K)=\pm \sqrt{i(1-t_Lt_R)\Delta_K}$.
In particular, its occupied band projector, which assumes the form $P(K)=(\mathbb{I}-H(K)/\mathcal{E}(K))/2$ for 2-component models, has divergent off-diagonal term $-\sqrt{i(t_L-t_R^{-1})/4t_R\Delta_K}\,\sigma_+$. Hence, with $k=\pi$
, we obtain the unique scaling divergence
\begin{equation}
P\sim -\sqrt{\frac{t_L-1/t_R}{4t_R\alpha_0\delta }}\frac{L_c}{\sqrt{L_c-L}}\,\sigma_+,
\end{equation} 
which, unlike in conventional critical scenarios of Figs.~\ref{fig:fig1}(b,c), diverges due to $L\rightarrow L_c$ rather than vanishing momentum spacings. This square-root divergence shows up in the $\bar P$ spectrum near $L\approx L_c$ [Fig.~\ref{fig:fig3}(a), also see \ref{suppmat}] which in turn translates to a characteristic $S\sim 1/\sqrt{|L_c-L|}$ divergence in the EE which we call the \emph{entanglement dip} [Fig.~\ref{fig:fig3}(b)]~\footnote{Since the eigenvalues of the projector $P$ are strictly $0$ or $1$, entanglement can be regarded as a measure of how much the entanglement cut, i.e., $\bar{\mathcal{R}}$ causes $\bar P=\bar{\mathcal{R}}P\bar{\mathcal{R}}$ to depart from being a true projector onto the occupied bands (which extend across the cut), and spectacularly so for negative entanglement}. 

While such entanglement dips can in principle be infinitely deep, in practice its depth is limited by the how closely $L_c$ approaches an integer. Shown in Fig.~\ref{fig:fig3}(c) is the minimal entanglement $S_\text{min}$ as the hopping asymmetry $t_L/t_R$ and inter-chain coupling $\delta$ are varied. For each lattice size $L$, the EE dip becomes particularly deep along curves  in parameter space (dark pink), even reaching $S_\text{min}\approx -3$ in the region shown.

\begin{figure}
	\centering
	\includegraphics[width=1\linewidth]{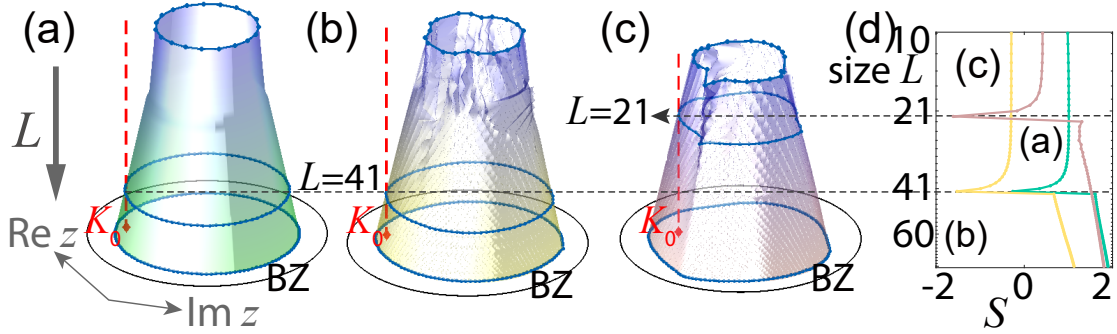}
	\caption{
\textbf{Entanglement dips and scale-dependent GBZ in more general models satisfying Eq.~\ref{Hcoupled}.} Shown are the $L$-dependent GBZs for illustrative (a) $H=(t_L+1/z)\sigma_+ + (t_R+z)\sigma_-$, (b)	$H=(t_L+1/3z^{-2})\sigma_+ + (t_R+z)\sigma_-$ (middle) and (c)	$H=t_L\sigma_+ + (0.6+0.12i+z+0.1z^{-2})\sigma_-$ (right), all with weak inter-chain coupling $\delta=1.6\times 10^{-3}$ and $t_L,t_R=1.2\text{e}^{\pm 0.3}$ (here, $L$ increases from top to bottom). 
(d) While their GBZs can be irregularly-shaped, entanglement dips in $S$ consistently emerge at values of $L$ where the GBZ loop intersects the critical point $K_0$ (red).
}
\label{fig:fig4}
\end{figure}

SIEC and entanglement dips are not limited to the coupled $H_\text{SSH}$ chains explicitly computed so far. They are expected to show up whenever the spectrum depends on $L$, and becomes gapless at an EP at a special $L\approx L_c$. This requires a scale-dependent GBZ that can be generically designed by coupling subsystems with competing NHSE pumping.  
Presented in Fig.~\ref{fig:fig4} are other models containing oppositely-directed NHSE channels, albeit with lower symmetry. Evidently, they all exhibit scale-dependent GBZs, and importantly exhibit entanglement dips at $L\approx L_c$ whenever the GBZ encounters a critical point $K_0$ (red).

\section{Conclusion} 	
We have uncovered a new class of critical transitions marked by characteristic $S\sim (L_c-L)^{-1/2}$ EE dips, departing from the almost-universal $S\sim\log L$ critical scaling for free fermions, Hermitian or otherwise.  
Its peculiar scaling behaviour originates from the scale-dependence of the GBZ itself, which fundamentally alters how a critical point can be approached. 

Physically, such dramatically suppressed EE represent the unique non-conservation of probability from antagonistic non-Hermitian pumping, which never occurs in ordinary NHSE processes where the gain/loss can be ``gauged away'' with a basis redefinition. Beyond the EE, entanglement dips also translate to kinks in the Renyi entropy, whose measurement prospects we discuss in \ref{suppmat}.

\section{Appendix: system size-dependent generalized Brillouin zone  and the energy spectrum} \label{suppmat}
Recent work has revealed the crucial importance of the generalized Brillouin zones (GBZ) in describing non-Hermitian lattice Hamiltonians under open boundary conditions (OBCs)~\cite{PhysRevLett.125.126402,Mong2011winding,PhysRevB.102.205118}. For a given Hamiltonian $H(z)$ where 
 $z=\exp(ik)$, the two roots $z$ of the characteristic (dispersion) equation $\text{det}(H(z)-E)=0$ with equal absolute values $|z|$ represents two ``standing wave'' solutions that can be superposed to satisfy OBCs at both ends. This is the typical way the GBZ is defined~\cite{yao2018edge,xiong2018does,PhysRevB.99.201103,song2019realspace,longhi2019probing,PhysRevLett.125.126402,PhysRevLett.123.066404,PhysRevLett.125.226402,lee2020ultrafast,longhi2020non,lee2020unraveling}.

However, there exist many systems where the usual GBZ approach gives a poor approximation to the actual numerical spectrum. Consider two weakly coupled chains exhibiting the non-Hermitian skin effect (NHSE) in opposite directions. In the weak coupling limit, we expect the presence of localized skin states at the ends of the individual chains. Due to the oppositely directed skin effect, these state accumulations must also be oppositely directed. Consequently, it is not tenable to exploit the same ``standing waves''  to satisfy the system's boundary conditions.

In other words, for such systems with competitive skin effect channels, it is not possible to accurately model the the net skin effect using the usual GBZ approach. Below, we shall present a new alternative approach using \emph{different} momentum wavevector directions in different chains, and show that how we can derive a \emph{size-dependent GBZ} which accurately considers the effect of the coupling different skin effect channels.

\subsection{ Coupled 1D chains with opposite non-Hermitian skin effect}\label{suppmat1a}

\begin{figure}
	\centering
	\includegraphics[width=\linewidth]{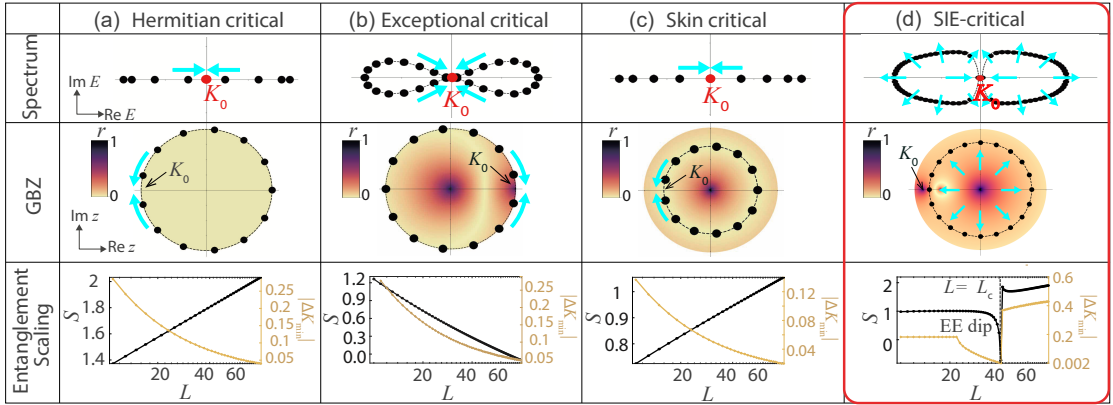}
	\caption{\textbf{ Scaling-dependent GBZ, spectrum and eigenfunctions of the 4-component model $\mathcal{H}$ [Eq.~\ref{EqS1}].}
(a) The two chain 4-component model [Eq.~\ref{EqS1}] incorporates equal but opposite hopping asymmetries $t_L$ and $t_R$ in both chains I and II. Even a small inter-chain coupling $\delta$ can lead to significant scaling dependence when $t_L \neq t_R$. 
(b) Agreement of the scale-dependent skin depth $\log |z|$ obtained via two different ways: numerically, by solving $\det(\mathcal{H}(z)-E_0)=0$ from the numerical eigenspectrum $E_0$ and choosing slowest decaying $z$ solution (blue crosses); analytically, via the scale-dependent generalized Brillouin zone (GBZ) we derived [Eq.~\ref{EqGBZ4}]  (red dots) . $L'$ [Eq.~\ref{Lp}] represents the system size at which the skin depth (GBZ) transitions from being scale-independent to scale-dependent. 
(c-f, c'-f') Energy spectra at various system sizes $L$ and illustrative eigenfunctions at energies $E_0$ (green diamonds in (c-f)) of smallest absolute value. 
Excellent agreement is observed between the numerical (blue) and analytically-derived GBZ (red) spectra [Eq.~\ref{EqGBZ4}] and eigenfunctions [Eq.~\ref{z1z2},Eq.~\ref{Eqpsi}]. The lower bound for the eigenfunctions (lower red curves) results from the superposition of $z_1$ and $z_2$ solutions in Eq.~\ref{Eqpsi}, and is generally non-monotonic except at the critical scale $L\approx L_c$ [Eq.~\ref{Eqcritical1}], where it exhibits a perfectly exponential profile since both $z$ solutions coalesce. Parameters are $t_L = 1.2\text{e}^{0.3}, t_R = 1.2\text{e}^{-0.3}$, and $\delta = 1.6\times 10^{-3}$ for all panels.}
	\label{picS1}
\end{figure}

We consider a model featuring non-Hermitian 1D chains with opposing NHSE effects, represented by $H$ and $H^{\dagger}$, , which are weakly coupled via a small parameter $\delta$. In the simplest scenario, this coupling occurs exclusively between corresponding pairs of sites in a ladder-like arrangement, resulting in the coupled Hamiltonian having the following form:
\begin{equation}\label{EqS1}
	\begin{split}
		&	\mathcal{H} =
		\begin{pmatrix} 
			H & \delta \mathbb{I}  \\
			\delta  \mathbb{I} & H^{T}
		\end{pmatrix}\ ,\\
	\end{split}
\end{equation}
where $\delta \mathbb{I}$ denotes the identity matrix with a multiplier operating in the basis of $H$ or $H^T$, as shown in Fig.~\ref{picS1}(a).  We specialize $H$ to the well-known non-Hermitian SSH model~\cite{yao2018edge,yin2018geometrical}, which is simple enough for analytic solutions and yet topologically nontrivial:
\begin{equation}\label{EqS2}
	\begin{split}
		H =\sum_n \begin{pmatrix}
			0 & 1\\
			0& 0\\
		\end{pmatrix}|n\rangle\langle n-1|+\begin{pmatrix}
			0 & t_L\\
			t_R& 0\\
		\end{pmatrix}|n\rangle\langle n|+ \begin{pmatrix}
			0 & 0\\
			1& 0\\
		\end{pmatrix}|n\rangle\langle n+1|\ .
	\end{split}
\end{equation} 
Diagonalizing $H$ in the Fourier basis gives eigenenergies
\begin{equation}
E=\pm \sqrt{(t_L + 1/z )(t_R + z )}\ .
\label{Ez}
\end{equation}

Under the periodic boundary condition (PBC), we simply have $z=\text{e}^{ik}$ where $k$ is the momentum.
In the absence of any NHSE, this should also approximately represent ``bulk'' spectrum under open boundary conditions (OBCs), since boundary effects should only affect a subdominant proportional of the eigenstates.

However, it is well-known that the existence of NHSE leads to the great sensitivity of the entire spectrum to the boundaries, such that we must have~\cite{yao2018edge} $z = \text{e}^{ik - \kappa(k)}$ for $E(z)$ to approximate the OBC spectrum. Here, $\kappa(k)$ represents the complex deformation of the momentum $k \rightarrow k + i\kappa(k,\kappa \in \mathbb{R})$, which is what defines the GBZ.

Here, we first show the procedure computing the GBZ of fully uncoupled $H$ chains. We note that to satisfy OBC at both ends of a 1D chain, an eigenfunction$|\Psi\rangle$ must be a linear combination~\cite{yao2018edge}  of at least two independent non-Bloch wave functions {$\sum_n z_1^n( |n,A\rangle+\phi_{z_1}|n,B\rangle)$ and $\sum_nz_2^n ( |n,A\rangle+\phi_{z_2}|n,B\rangle)$, where $A,B$ and $n$ label the sublattices and unit-cells, $z_1,z_2$ are the two solutions of Eq.~\ref{Ez} which satisfy $z_1z_2=t_R/t_L$, and $\phi_{z_{1,2}}=(t_R+z_{1,2})/E$. Crucially, the OBCs $\langle 0,B|\Psi\rangle=\langle L+1,A|\Psi\rangle=0$ enforce the constraint $z_1^{L+1}\phi_{z_2}=z_2^{L+1}\phi_{z_1}$ such that $(\phi_{z_1}/\phi_{z_2})^{1/L}\approx1$.} Consequently, for our uncoupled chain $H$, the GBZ satisfies $|z_1|=|z_2|$ and is simply given by $\left\{z=\sqrt{{t_R}/{t_L}}\exp(ik)|\,k=\frac{2m\pi}{L+1}, m=1,2,\cdots, L\right\}$, i.e., with skin depth  $\kappa_{H}=\frac1{2}\log |z|=\frac{1}{2}\log(t_R/t_L)$. Its transpose conjugate is an oppositely directed chain $H^{T}$ that possesses skin depth $\kappa_{H^{\dagger}}=-\kappa_{H}=-\frac{1}{2}\log(t_R/t_L)$.

Note that the important condition $|z_1|=|z_2|$, which is commonly used as the GBZ constraint, will no longer be valid if we consider coupled chains with different NHSE directions.
\\

Before discussing the coupled ($\delta\neq 0$) case, we note that our system possesses spatial inversion symmetry $\mathcal{I}$ represented by 
\begin{eqnarray}
	\mathcal{I} \mathcal{H}\mathcal{I}=\mathcal{H},\qquad \mathcal{I}=\sum_n(\tau_x\otimes \sigma_x)\ket{L+1-n}\bra{n},
\end{eqnarray}

such that the wave function accumulates similarly at both ends of the chain. This $\mathcal{I}$ symmetry is a consequence of the inversion symmetry between $H$ and $H^\dagger$. It is not a fundamental requirement for the general construction of size-dependent effective GBZ. However, it offers to simplify the system's boundary conditions by half, and is thus helpful to obtaining the analytic results.

\subsection{The new approach: Scale-dependent energies and generalized Brillouin zone }\label{suppmat1b}

We next derive the size-dependent effective GBZ for two antagonistic NHSE chains coupled by a non-zero but small $\delta$ coupling. Since $\delta$ is small, the effect of the coupling term is insufficient to alter the orientations of the NHSE within each of the two chains, which are in opposite directions. The usual GBZ approach, which assigns the \emph{same} decay parameter magnitude $|z|$ to all subsystems (i.e., both chains), is then doomed to be inadequate. 

In our new approach, we assign the decay parameter $z$ to one chain and $1/z$ to the other chain, leading to the OBC eigenstate in general consisting of a linear combination of two or more eigensolutions, which we will first solve for, separately. Writing each $z$ eigensolution as $ \ket{\varphi}=\sum_{n,\alpha}\varphi_{\alpha,n}\ket{n,\alpha}$ with sublattice $\alpha= A, B$(chain I), $A', B'$(chain II), we have $\varphi_{A(B),n}=\varphi_{B'(A'), L+1-n}$. Concretely,
\begin{eqnarray}
&\varphi_{A,n}=z^n,&\ \varphi_{B,n}=\phi_z z^n\ ;\notag\\
&\varphi_{A',n}=\phi_z z^{L+1-n},&\  \varphi_{B',n}=z^{L+1-n}\ .
\label{psiAB}
\end{eqnarray}
Due to spatial inversion symmetry $\mathcal{I}$, the decay in both chains are exactly equal but opposite -- in more generic cases where the two coupled chains are not related by $\mathcal{I}$, they would need to be separately solved for.

In the above, the assignment of $z$ and $1/z$ decay parameters to chains I and II only addresses their relative decay directions. To solve for what value/s the decay parameter $z$ should take, we need to incorporate the weak couplings and the open boundary conditions. In our approach, we will (i) first solve for the bulk relation between $z$ and the corresponding eigenenergy $E$ in the uncoupled limit, and then (ii) invoke the weak couplings and boundary conditions to obtain a size-dependent effective GBZ.

In the given Hamiltonian $\mathcal{H}$, the full bulk equations take the form
\begin{eqnarray}
	&&\left\{\begin{array}{ll}
		t_L\varphi_{B,n} +     \varphi_{B,n-1} + \delta  \varphi_{A',n} &= E \varphi_{A,n}\\
		t_R\varphi_{A,n} +      \varphi_{A,n+1} + \delta  \varphi_{B',n} &= E \varphi_{B,n}\\
		t_R\varphi_{B',n} +     \varphi_{B',n-1} + \delta  \varphi_{A,n}& = E \varphi_{A',n}\\
		t_L\varphi_{A',n} +      \varphi_{A',n+1} + \delta  \varphi_{B,n} &= E \varphi_{B',n}
	\end{array}\right. \\ &&\quad \stackrel{\mathcal{I}}{\longrightarrow} \left\{\begin{array}{ll}
		t_L\varphi_{B,n}+     \varphi_{B,n-1}  + \delta  \varphi_{B,L+1-n} &= E \varphi_{A,n}\\
		t_R\varphi_{A,n} +      \varphi_{A,n+1} + \delta  \varphi_{A,L+1-n} &= E \varphi_{B,n}\\
	\end{array}\right. ,
\end{eqnarray}
with site $n=2,3,\cdots,L-1$ and eigenvalue $E$. Due to $\mathcal{I}$ symmetry, the four equations (from two chains with two sublattices each) are simplified to two equations. Substituting Eq.~\ref{psiAB} into the above, 
\begin{eqnarray}
	\left\{\begin{array}{rl}
		(-E+(t_L+     1/z)\phi_z)z^n+ \phi_z z^{L+1-n}\delta&=0\\
		((t_R+     z) -E\phi_z)z^n+  z^{L+1-n}\delta &=0
	\end{array}\right.  \ .
\end{eqnarray}
For step (i), we work in the $\delta\ll 1$ limit and neglect the inter-chain couplings $\delta$ to obtain 
\begin{eqnarray}
	\left\{\begin{array}{rl}
		(-E+(t_L+     1/z)\phi_z)z^n&\approx 0\\
		((t_R+     z) -E\phi_z)z^n&\approx 0
	\end{array}\right. \\ \Rightarrow\begin{pmatrix} 
		-E& t_L+    1 /z\\
		t_R+     z& -E
	\end{pmatrix}\begin{pmatrix} 
		1 \\
		\phi_z
	\end{pmatrix}=(H(z)-E)\begin{pmatrix} 
		1 \\
		\phi_z
	\end{pmatrix} =0 \  .
\end{eqnarray}
Via the same steps as in the previous subsection on a single chain, these bulk equations give (Eq.~\ref{Ez})
\begin{eqnarray}
	E=\pm \sqrt{(t_L+     1/z)(t_R+     z)}\ ,\quad \phi_z=\frac{t_R+     z}{E} =\frac{E}{t_L+  1/ z}\ ,
	\label{Eqenergy}
\end{eqnarray}
which, for a particular eigenvenergy $E$, corresponds to two 2 solutions $z=z_1,z_2$ which satisfy 
\begin{align}
	&\qquad E^2 = (t_L+ 1/z)(t_R+ z)\  , \\
	&t_Lz^2 + (t_Rt_L + 1 - E^2)z + t_R = 0\ .
\end{align}
Without loss of generality, we label them in the order $|z_2|\leq \sqrt{|t_R/t_L|}\leq |z_1|$ since, according to Vieta's formulas,
\begin{eqnarray}
	z_1z_2= t_R/t_L\ .
	\label{z1z2}
\end{eqnarray}
If the 2-component subsystem $H(z)$ existed independently, the GBZ solution would just have been given by $|z_1|=|z_2|$, with the OBC skin spectrum given by the set of $E$ satisfying this equality constraint. However, we will see later that even in the presence of very weak inter-chain couplings, the incorporation of the full boundary conditions will give rise to a very different constraint from $|z_1|=|z_2|$.

We next proceed to step (ii) where we consider the boundaries at unit-cells $1$ and $L$ without neglecting the weak couplings $\delta$. To satisfy the boundary conditions, we now consider an eigensolution that superposes all the (two) different single-chain GBZ solutions $z_1,z_2$ from step (i), we write the ansatz OBC eigenfunction as
\begin{eqnarray}\label{Eqpsi}
	\ket{\Psi_{\text{GBZ}}}=|\varphi_{z_1}\rangle-c|\varphi_{z_2}\rangle=\sum^L_{n=1}\sum_{\alpha=A,B,A',B'}\psi_{n,\alpha} |n,\alpha\rangle\\
    =\sum_n \left(\begin{pmatrix} 
		z_1^n\\
		z_1^n\phi_{z_1}\\
		z_1^{L+1-n}\phi_{z_1}\\
		z_1^{L+1-n}
	\end{pmatrix}-c \begin{pmatrix} 
		z_2^n\\
		z_2^n\phi_{z_2}\\
		z_2^{L+1-n}\phi_{z_2}\\
		z_2^{L+1-n}
	\end{pmatrix}\right)\ket{n},
\end{eqnarray}
where the coefficient $c$ controls the relative amplitude of the $z_1,z_2$ non-Bloch contributions. We next substitute this ansatz Eq.~\ref{Eqpsi} into the open boundary conditions, where $\psi_{n,\alpha}$ disappear at $n=0$ and $L+1$ (and beyond):
\begin{eqnarray}
\label{eqS13}
	&&\left\{\begin{array}{ll}
		t_L\psi_{B,1} + \delta  \psi_{A',1}  &= E \psi_{A,1}\\
		t_R\psi_{A,L} + \delta  \psi_{B',L}  &= E \psi_{B,L}\\
		t_R\psi_{B',1}  + \delta  \psi_{A,1} & = E \psi_{A',1}\\
		t_L\psi_{A',L}  + \delta  \psi_{B,L} &= E \psi_{B',L}
	\end{array}\right. \stackrel{\mathcal{I}}{\longrightarrow}\left\{\begin{array}{ll}
	\delta\psi_{B,L}= E\psi_{A,1}-t_L\psi_{B,1} &\\
	\delta  \psi_{A,1}=E\psi_{B,L}-t_R\psi_{A,L} &\\
			\end{array}\right..\end{eqnarray}
The extra terms involving the small inter-chain coupling $\delta$, which were not taken into account in deriving the bulk $z_1,z_2$ solutions, must exactly compensate the missing terms due to the open boundaries. By explicitly substituting $E$ in terms of $z_{1,2}$, $\phi_{z_{1,2}}$ and the hoppings using Eq.~\ref{Eqenergy}, we obtain
\begin{eqnarray}
	&&\left\{\begin{array}{rl}
		(  1 - \delta z_1^L)\phi_{z_1}&= c(    1 - \delta z_2^L)\phi_{z_2}\\
		(  z_1^{L}- \delta)z_1&= c( z_2^{L}- \delta)z_2
	\end{array}\right.\ , 
\end{eqnarray}

which yields
\begin{eqnarray}
	(      1- \delta z_1^L)(       z_2^{L}- \delta)z_2\phi_{z_1}=(  1   - \delta z_2^L) (       z_1^{L}- \delta)z_1 \phi_{z_2}\ .
\end{eqnarray}
We comment that $z_1,z_2$, which were previously derived under the uncoupled ($\delta=0$) approximation, has now become dependent on the coupling $\delta$ through $E$ (even though this dependence is hidden behind Eq.~\ref{Eqenergy}, which was derived under the uncoupled approximation).

Since the inter-chain couplings are weak, we are safe to assume that they cannot fundamentally reverse the NHSE direction within each chain. That is, in the case of $|t_R/t_L|<1$, 
we have $ |z_2| \leq \sqrt{|t_R/t_L|} \leq |z_1|\leq 1$. Together with $\delta\ll 1$ and $|z_{j}|<1$, we can thus approximate $1 - \delta z_{j}^L\approx 1$ ($j=1,2$), thereby simplifying the above to
\begin{equation}
	\begin{split}
		( z_2^{L+1}- \delta z_2)\phi_{z_1}&\approx (z_1^{L+1}- \delta z_1)\phi_{z_2}\ .
	\end{split}
\end{equation}
Moving all the $\delta$ terms to the right-hand side of the equation,
\begin{equation}
	\begin{split}
		z_1^{L+1}\phi_{z_2}-  z_2^{L+1}\phi_{z_1}&\approx \delta (z_1\phi_{z_2}-z_2\phi_{z_1})\ .\\
	\end{split}
\end{equation}
Recalling that $ z_1z_2=t_R/t_L$, we can further simplify the above into 
\begin{equation}
	\begin{split}
		z_1^{L+1}\phi_{z_2}- \frac{1}{z_1^{L+1}}\left( \frac{t_R}{t_L}\right)^{L+1}\phi_{z_1}&\approx\delta (z_1\phi_{z_2}-z_2\phi_{z_1})\ ,\\
	\end{split}
\end{equation}
that is, 
\begin{equation}\label{EqGBZ1}
	z_1^{L+1}\approx \frac{\delta (z_1\phi_{z_2}-z_2\phi_{z_1})}{2\phi_{z_2}}+ \sqrt{\left(\frac{\delta (z_1\phi_{z_2}-z_2\phi_{z_1})}{2\phi_{z_2}}\right)^2 +\left(\frac{t_R}{t_L}\right)^{L+1}\frac{\phi_{z_1}}{\phi_{z_2}}}\ ,
\end{equation}
with the larger root chosen since $|z_2| \leq \sqrt{|t_R/t_L|} \leq |z_1|\leq 1$.

Below, we highlight two extreme but highly relevant cases:
\begin{itemize}
	\item  $ \delta \ll  {t_R^L}/{t_L^L}$, which is the regime where the coupling $\delta$ is so weak that it remains negligible despite the presence of NHSE competition. In this case, only the second term under the square root in Eq.~\ref{EqGBZ1} survives: 
	\begin{equation}\begin{split}
			z_1^{L+1}&\approx \frac{\delta (z_1\phi_{z_2}-z_2\phi_{z_1})}{2\phi_{z_2}}+\sqrt{\frac{t_R^{L+1}}{t_L^{L+1}}\frac{\phi_{z_1}}{\phi_{z_2}}} + \frac{1}{2}\left(	\sqrt{\frac{t_R^{L+1}}{t_L^{L+1}}\frac{\phi_{z_1}}{\phi_{z_2}}}\right)^{-1}\left(\frac{\delta (z_1\phi_{z_2}-z_2\phi_{z_1})}{2\phi_{z_2}}\right)^2\\
			&\approx\sqrt{\frac{t_R^{L+1}}{t_L^{L+1}}\frac{\phi_{z_1}}{\phi_{z_2}}}\quad.\end{split}\end{equation}
	As $\sqrt[L+1]{{\phi_{z_1}}/{\phi_{z_2}}}\approx 1$ for moderately large $L$, where $\phi_z=(t_R+z)/E$, we recover
	\begin{equation}
		z_1,z_2\approx \sqrt{\frac{t_R}{t_L}}\text{e}^{\pm ik} \ ,
	\end{equation}
	with $k=n\pi/(L+1)$, $n=1,2,\cdots, L$. This is just the usual GBZ expression for satisfying OBCs $\psi(x=0) = \psi(x=L+1) = 0$ in the uncoupled ($\delta=0$) case.\\
	
	\item  $ \delta \gg {t_R^L}/{t_L^L}$, which defines the strongly coupled regime. Here, the first term under the square root in Eq.~\ref{EqGBZ1} dominates, such that	
	\begin{equation}\label{eqGBZ2}
		\begin{split}
			z_1^{L+1}&\approx \frac{\delta (z_1\phi_{z_2}-z_2\phi_{z_1})}{\phi_{z_2}}+\left(\frac{\delta (z_1\phi_{z_2}-z_2\phi_{z_1})}{2\phi_{z_2}}\right)^{-1}\frac{t_R^{L+1}}{t_L^{L+1}}\frac{\phi_{z_1}}{\phi_{z_2}}\\	
			&\approx  \frac{\delta (z_1\phi_{z_2}-z_2\phi_{z_1})}{\phi_{z_2}} \ .
	\end{split}\end{equation}

In principle, the above expression can be exactly solved to yield the GBZ (i.e., $z_1$), which we see must depend on the system size $L$. However, because $\phi_{z_1},\phi_{z_2}$ depend on $E$, which further depends on the system parameters in a complicated manner, it is useful to perform some approximations such that the right-hand side does not depend explicitly on $E$. One convenient series of approximations is 
	\begin{equation}\label{eqGBZ3}
		\begin{split}
			& z_1=\left(\delta\times \frac{(z_1\phi_{z_2}-z_2\phi_{z_1})}{\phi_{z_2}}\right)^{\frac{1}{L+1}}\text{e}^{ik} \\
			&\xrightarrow{\sqrt[L+1]{{\phi_{z_2}}/{\phi_{z_1}}}\approx 1} \left(\delta\times \frac{(z_1\phi_{z_2}-z_2\phi_{z_1})}{\phi_{z_1}}\right)^{\frac{1}{L+1}}\text{e}^{ik} \\ 
			&\xrightarrow{\phi_z=(t_R+z)/E} \left(\delta \times \frac{t_R( t_L z_1- t_R/z_1)}{t_L(t_R+z_1)}\right)^{\frac{1}{L+1}}\text{e}^{ik} \ .
	\end{split}\end{equation}
While this expression technically needs to be solved self-consistently for $z_1$, noting that the right-hand side depends very slowly due to the $1/(L+1)$ exponent, we can simply approximate $z_1$ in it with suitable constants, i.e., $z_1\approx -1$ such as to obtain
	\begin{equation}
		z_1\approx\left( \delta \times\frac{t_R( t_R- t_L)}{t_L(t_R-1)}\right)^{\frac{1}{L+1}}\text{e}^{ik}  \ ,
	\end{equation}
	with $k=2n\pi/(L+1)$, $n=1,2,\cdots, L$. 
	
	This approximate expression for the GBZ is dependent on the system size $L$ in the form of the exponent $1/(L+1)$, which gives weak dependence on the inter-chain coupling $\delta$ as well as a constant factor containing the other system parameters. While it is not the only possible approximation, it is relatively compact and agrees well with the numerical GBZ values  [Fig.~\ref{picS1}(b)] obtained by substituting the exact OBC spectrum $E_0$ of our 4-component physical system into $\text{det}(\mathcal{H}(z)-E_0\mathbb{I})=0$, and solving for $z$. Also, it predicts a spectrum $E_\text{GBZ}=\pm \sqrt{(t_L+1/z_1)(t_R+z_1)}$ that agrees very well with that numerically-obtained exact OBC spectrum $E_0$, as shown in Fig.~\ref{picS1}(c-f). Regardless of the exact choice of approximation, the $\delta^{1/(L+1)}$ dependence holds universally, and suggests a slow but $\delta$-dependent convergence of the GBZ onto the unit circle in the $L\rightarrow \infty$ limit. In Fig.~\ref{picS1}(c'-f'), the $|z_{1}|^{x},|z_{2}|^{x}$ spatial decay profiles predicted by the GBZ (brown) near $E_0=0$ also exhibit excellent agreement with the upper and lower envelope limits of the numerical eigenstate $\Psi$ closest to $E_0=0$ in Fig.~\ref{picS1}(c-f).
\end{itemize}

In all, our approximate size-dependent GBZ solution for our coupled HN model system is given by
\begin{equation}\label{EqGBZ4}
	\begin{split}
		&\  z =\exp(iK)=\left \{\begin{array}{ll}  
			\sqrt{t_R/t_L} \text{e}^{ik}, \qquad  &\qquad L \leq L' \\
			\\
			\text{e}^{-\frac{\alpha}{L+1}} \text{e}^{ik}, &\qquad L \geq L'\\
		\end{array}\right.\ ,\qquad \alpha = -\log \left|\delta \times \displaystyle\frac{ t_R( t_R- t_L)}{t_L(t_R-1)}\right|\ ,\\
	\end{split}
\end{equation} 
where 
\begin{equation}
L'=-\alpha/\log\sqrt{t_R/t_L}-1\ ,
\label{Lp}
\end{equation}
is the transition length in which the energy spectrum or skin depth $-\log|z|$ switches from being size-independent to size-dependent [Fig.~\ref{picS1}(b)]. The value of $L'$ is determined by setting both expressions for the GBZ to be equal, i.e., $|z_1|=\sqrt{t_R/t_L} = \exp\left(-{\alpha}/({L'+1})\right)$. The very good agreement between our analytic GBZ solution (Eq.~\ref{EqGBZ4}) and exact numerics is presented in Fig.~\ref{picS1} for the skin depth $-\log|z|$, spectrum $E$ and eigenfunction decay.

After solving $z_1$ and $z_2=t_R/(t_Lz_1)$[Eq.~\ref{EqGBZ4}],
substitute into the bulk function that $\det (H(z)-E)=0$, we have the effective energy satisfying  $$E = \pm \sqrt{(t_L+1/z_1)(t_R+z_1)}= \pm \sqrt{(t_L+1/z_2)(t_R+z_2)}.$$

In summary, for the model involving weakly coupled chains with opposing skin depths, the spatial inversion symmetry is maintained by the arrangement of wave functions across the two chains. This symmetry indicates that the skin directions of the two chains are oriented in opposite directions.
This is reflected in the wave function as  $ \ket{\psi}= \sum_{n}z^n\ket{n,\text{I}} + \tilde z^n\ket{n,\text{II}} $ with $\tilde z^n=z^{L+1-n}$.

In a nutshell, we have demonstrated  that as the system size $L$ increases, the GBZ and spectrum always switches from being size-independent to size-dependent as $L$ crosses $L'$. This behaviour is captured by our approximate GBZ expression in Eq.~\ref{EqGBZ4}, which shows excellent agreement with numerical results obtained from exact diagonalization. Our method involved neglecting the influence of weak coupling terms in the bulk equations while retaining their effect in enforcing boundary conditions. This approach facilitated an analytical treatment of the energy expression derived from the characteristic equation of a single chain, $\det(H(z)-E)=0$. This approximation is valid because, although extremely weak couplings minimally impact the bulk properties, they significantly affect the boundary conditions due to the heightened sensitivity of non-Hermitian systems to spatial inhomogeneities.
To conclude, with nonzero coupling between chains with antagonistic NHSE, the GBZ is found to be no longer given by the usual $|z_1|=|z_2|$ condition, but instead takes on a $L$-dependent behaviour that crucially involves the coupling strength $\delta$.
\subsection{ Effective size-dependent two-component model }\label{suppmat1c}

The key objective of constructing a GBZ is to effectively ``remove'' the NHSE, allowing the Hamiltonian evaluated on the GBZ—referred to as the ``surrogate Hamiltonian''~\cite{lee2020unraveling}—to accurately reproduce the spectrum under open boundary conditions (OBCs) without requiring an explicit diagonalization of the Hamiltonian under OBCs. In our case, the size-dependent GBZ derived in the previous subsection has successfully incorporated the effects of both coupling and the antagonistic NHSE.

As such, working in our effect size-dependent GBZ $z$, we can accurately study $\mathcal{H}$ (Eq.~\ref{EqS1}) as two uncoupled copies of 2-component chains $H$ i.e.,$\mathcal{H} = H(K) \oplus H^{T}(-K)$, where $K=-i\log z$. The effects of the coupling and NHSE are implicit in the imaginary part of $K$. From Eq.~\ref{EqS2},
\begin{equation}\label{Eqeffectivemodel} 
H(K) = \begin{pmatrix} 
		0 & t_L + \text{e}^{-iK}\\
		t_R+ \text{e}^{iK}& 0
	\end{pmatrix}, 
\end{equation}
even though our approach would work well for generic $H(k)$ of similar levels of complexity. To recall, 
\begin{equation}\label{Eqeffectivemodel1} 
K=-i\log z=k+i\frac{\alpha}{L+1}\ ,
\end{equation}
from Eq.~\ref{EqGBZ4}, where the real part $ k=\text{Re} K=2m\pi/(L+1)$ with $m=1,2,...,L$, and the imaginary part $\text{Im} K=\displaystyle\frac{\alpha}{L+1}$ is controlled by 
\begin{equation}\label{Eqeffectivemodel2} 
\alpha =   \left \{\begin{array}{ll}-\log \left|\delta \times \displaystyle\frac{t_R(t_R-t_L)}{t_L(t_R-1)}\right|, &\qquad L\geq L'\\
	-(L+1)\log \sqrt{t_R/t_L} \ ,&\qquad L\leq L'
\end{array}\right. \ ,
\end{equation}
from Eq.~\ref{EqGBZ4} and \ref{Lp}. From direct diagonalization of $H(K)$, the eigenenergies are given by
\begin{equation}
	\begin{split}
		&\mathcal{E}=\pm \sqrt{(t_L+\exp(-iK))(t_R+\exp(iK))}\ ,\\
	\end{split}
	\label{2bandenergy}
\end{equation} 
which has a similar form as Eq.~\ref{Ez}; but when evaluated on the GBZ (Eq.~\ref{EqGBZ4}), it should reproduce the OBC spectrum of the entire original 4-component model $\mathcal{H}$ (Eq.~\ref{EqS1}). Note that we have used the notation $\mathcal{E}$ to refer to the eigenenergies predicted by the GBZ, which are supposed to closely approximate the true eigenenergies $E$.

Importantly, because of the size dependence of the GBZ, $H(K)$ can possess exceptional points (EPs) at special sizes $L=L_c$ where the corresponding $K=K_c$ gives rise to $t_R+\exp(iK_c)=0$ and $t_L+\exp(-iK_c)\neq 0$. When that occurs, $H(K)$ is of the Jordan form $\sigma_+$ and is not of full rank (defective). We will soon investigate how this scaling-induced EP lead to unconventional entanglement behaviour.
Substituting $K_c=k_c +i\alpha/(L_c+1)$,
\begin{equation}
	\begin{split}\label{Eqcritical1}
		&t_R+\exp(iK_c)=t_R+\exp(-\alpha/(L_c+1)+ik_c)=0 \\
		&\Rightarrow k_c=\pi ,\qquad  {\alpha}/{(L_c+1)}=-\log t_R\\
		& \Rightarrow L_c=-\frac{\alpha}{\log t_R}-1\ ,
	\end{split}
\end{equation} 
with $\alpha = -\log \left|\delta \times \displaystyle\frac{t_R(t_R-t_L)}{t_L(t_R-1)}\right| $.
Here, $L_c$ represents the critical length in which an EP appears, and should not be confused with $L'=-\alpha/\log\sqrt{t_R/t_L}-1$ (Eq.~\ref{Lp}), which is the length above which the GBZ becomes size dependent. Evidently, such a scaling-induced EP exists only if $L_c>L'$.

\begin{figure}
	\centering
	\includegraphics[width=\linewidth]{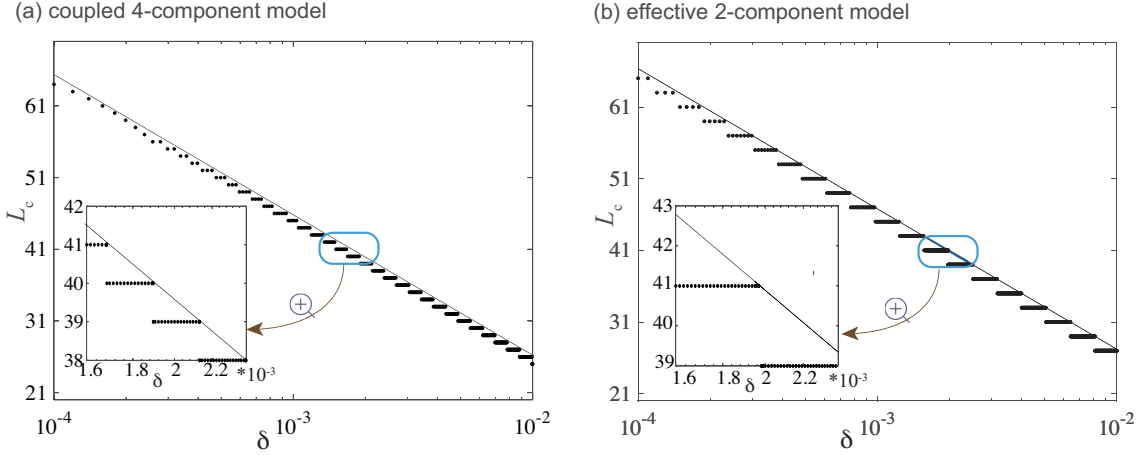}	
	\caption{\textbf{ The critical system size $L_c$ dependence on $\delta$ for the coupled 4-component model [Eq.~\ref{EqS1}] and its effective two-component approximation [Eq.~\ref{Eqeffectivemodel}].}
The spectral gap of our model closes at an EP	at the critical system size $L_c$, which is analytically predicted to depend logarithmically on the small inter-chain coupling parameter $\delta$. Excellent agreement is observed between the numerically determined integer $L_c$ (black dots) and its analytical prediction [Eq.~\ref{Eqcritical1}]. Since $\delta$ is built into the GBZ for the two-component, single-chain approximation [Eq.~\ref{Eqeffectivemodel}], its excellent numerical agreement is testimony to the validity of our scale-dependent GBZ approach. 
}
	\label{picS2}
\end{figure}

Even though $L_c$ was derived from the approximately-obtained GBZ, it already predicts the scaling-induced EP transition to excellent accuracy, when compared to the critical $L_c$ obtained from exact diagonalization spectra (which must be an integer), as shown in Fig.~\ref{picS2}. 

\subsubsection{ Critical behaviour near the scaling-induced exceptional point}\label{EEP}
To investigate the entanglement properties around the scaling-induced exceptional points (EPs) mentioned earlier, we first analyze the behaviour of the Hamiltonian close to these points. We first expand the size-dependent quasi-momentum $K$, both its real and imaginary parts, as $\Delta_K=K-K_c$ where $K=k+i\alpha/(L+1)$ and $K_c=\pi+ i\alpha/(L_c+1)$ [See Eqs.~\ref{EqGBZ4} and \ref{Eqcritical1}], such that $\Delta_K=\text{Re}\Delta_K+i\text{Im}\Delta_K$ is given by
\begin{align}
\text{Re}\Delta_K&= k-\pi\notag\ ,\\
\text{Im}\Delta_K&= \alpha\left( \frac{1}{L+1}-\frac{1}{L_c+1}\right)\ .
\end{align}
With these, the Hamiltonian (Eq.~\ref{Eqeffectivemodel}) can be expanded near $K=K_c$ as
\begin{equation}
	\begin{split}
		H(K) &\xrightarrow{\Delta_K=K- K_c } \begin{pmatrix} 
			0 & t_L + \text{e}^{-iK_c}\text{e}^{-i\Delta_K}\\
			t_R+ \text{e}^{iK_c}\text{e}^{i\Delta_K}& 0
		\end{pmatrix}\\
		&\xrightarrow[t_R + \exp(iK_c) = 0]{\text{e}^{i\Delta_K}\approx 1+i\Delta_K} \begin{pmatrix} 
			0 & t_L - 1/t_R \\
			-it_R\Delta_K& 0
		\end{pmatrix}\ .\\
	\end{split}
\end{equation}
This yields the eigenenergies of $H(K_c+\Delta_K)$ to be $\pm \mathcal{E}=\pm \sqrt{i( 1-t_L t_R)\Delta_K}$.

\subsubsection{ Critical properties of the occupied band projector $P_K$ }
The geometric defectiveness of an EP is evident in the divergences observed in the occupied band projector $P$. This divergence occurs because the projector becomes ill-defined when the occupied and valence bands merge. For a 2-component model, $P=\sum_k P_k|k\rangle\langle k|$ is defined (in the biorthogonal basis) as
\begin{equation}
\label{EqPk0}
	\begin{split} 
		P_K =\frac{1}{2} \left(\mathbb{I}-\frac{H(K)}{\mathcal{E}(K)}\right)\ ,\end{split} 
\end{equation} 
where $\mathcal{E}(K)$ is the eigenvalue of the Hamiltonian $H(K)$ (Eq.~\ref{Eqeffectivemodel}). Note that we have analytically continued $P_k\rightarrow P_K$ into the GBZ. 
Ordinarily, away from gap closure points, $P(K)$ projects onto a well-defined occupied band since $\mathcal{E}(K)\neq 0$. But at an EP ($K=K_c$)where $\Delta_K=0$, $ \mathcal{E}(K)= \sqrt{i( 1-t_L t_R)\Delta_K}$ vanishes, rendering $P_K$ singular.

Close to an EP, where $\Delta_K$ is very small, 
\begin{equation}\label{EqPk}
	\begin{split} 
		P_K =\frac{1}{2} \begin{pmatrix}1 & P^{+-}_K\\P^{-+}_K & 1\end{pmatrix}\ ,\quad  {P^{+-}_K}=\frac{1}{{P^{-+}_K}}\ , \end{split} 
\end{equation} 
where
\begin{equation}\label{EqP+}
	P^{+-}_K\approx -\frac{\sqrt{t_L/t_R -1/t^2_R}}{2\sqrt{-i\Delta_K}}\quad ,
\end{equation}

\begin{figure}
	\centering
	\includegraphics[width=\linewidth]{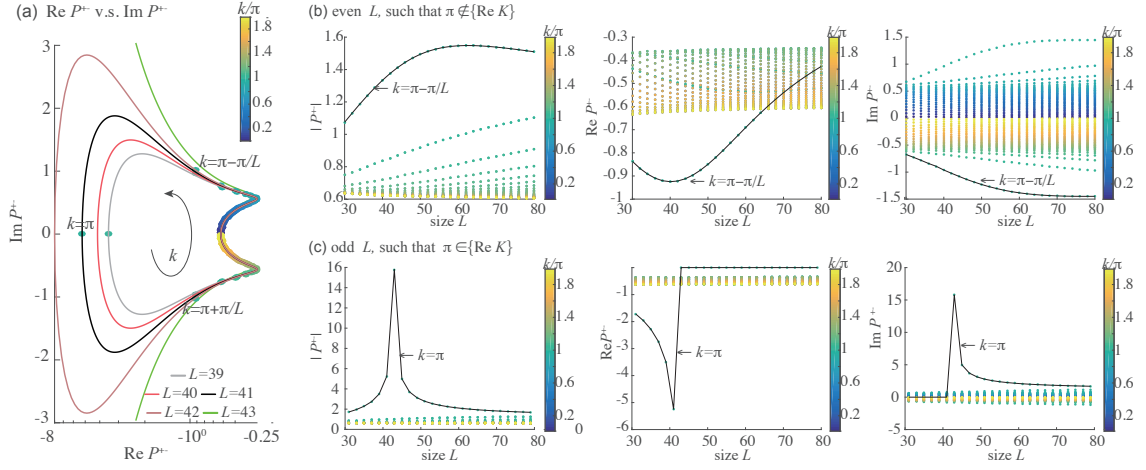}
\caption{\textbf{ Potentially divergent matrix element $P_K^{+-}$ [Eq.~\ref{EqPk}] of our truncated projection operator $P_K$ [Eq.~\ref{EqPk0}] 
in even vs. odd $L$ systems.}
	(a) Complex values of the off-diagonal element $P_{K(k)}^{+-}$, which traces a different loop for each system size $L$ as $k$ cycles over a period $[0,2\pi)$ with discrete points $ (2m-1)\pi/L$, $m=1,2,...,L$ (drawn as solid dots). The $k=\pi$ point exist only for odd $L$, and causes $P_{K(k)}^{+-}$ to diverge to infinity as $L\rightarrow L_c=-\frac{\alpha}{\log t_R}-1=42.79$ [Eq.~\ref{Eqcritical1}], $\alpha= -\log \left|\delta \times \displaystyle\frac{ t_R( t_R- t_L)}{t_L(t_R-1)}\right|$.
	(b, c) The values of $|P_{K(k)}^{+-}|$, Re$(P_{K(k)}^{+-})$ and Im$(P_{K(k)}^{+-})$ as a function of system size $L$, plotted as separate branches for each $k$ value for (b) even and (c) odd $L$. While there exists for even $L$ a $k=\pi-\pi/(L+1)$ branch that departs from the other branches, for odd $L$, the $k=\pi$ branch exhibits a sharp kink for $L=43\approx L_c = 42.79$.
	Parameters are $t_L = 1.2\text{e}^{0.3},t_R = 1.2\text{e}^{-0.3},\delta = 1.6\times 10^{-3}$.}
		\label{picS3}
\end{figure}
with $\Delta_K=K-K_c=k - \pi + i(\frac{\alpha}{L+1}-\frac{\alpha}{L_c+1})$. 
In a finite system, the the quasi-momentum $k=\text{Re}K$ points take the values $2\pi/(L+1),4\pi/(L+1),...$, such that Re$\Delta_K\sim \pi/L$ for even-sized systems, and Re$\Delta_K=0$ for odd-sized systems.
However, the imaginary momentum deviation Im$\Delta_K$ is approximately $\frac{\alpha}{L+1}-\frac{\alpha}{L_c+1}=\frac{\alpha(L_c-L)}{(L+1)(L_c+1)}\sim \alpha(L_c-L)/L^2\approx \alpha(L_c-L)/L_c^2$ near $K_c$, which is much smaller than $1/L$. Since $P^{+-}_K\propto \Delta_K^{-1/2}$, the strength of the singularity depends on both the real and the imaginary parts of $\Delta K$, as illustrated in the complex $P_+$ plot in Fig.~\ref{picS3}a ($L=43$ is the most divergent). To elaborate on the qualitatively different cases of even and odd $L$:

\begin{itemize}
	\item Even $L$, such that min$|\text{Re}\Delta_K|\neq 0$:\\
	\indent Here, the closest momentum point passes approximately within $\pi/L$ of Re$K_c=\pi$, such that 
$|\Delta_K|=\left|\frac{\pi}{L}+i\left(\frac{\alpha}{L_c+1}-\frac{\alpha}{L+1}\right)\right|\approx\frac{\pi}{L}$ for modest system sizes of  $L,\ L_c\ \sim\ 10-10^2$.  The scaling behaviour is dominated by $|\Delta_K|\sim $Re$\Delta_K$-- masking the effect of the imaginary part. As shown in Fig.~\ref{picS3}b, the most divergent contribution to $P^{+-}_K\sim \Delta_K^{-1/2}\sim L^{1/2}$ arises from $k=\pi-\pi/L$, but because $|\Delta_K|$ diverges with $L$ and not $L_c$, it does not really capture the value of $L_c$.

	\item  Odd $L$, such that min$|\text{Re}\Delta_K|= 0$:\\
	\indent Here, $k=\pi$ is visited in the GBZ, and Re$\Delta_K=0$ while $\Delta_K=i\left(\frac{\alpha}{L+1}-\frac{\alpha}{L_c+1}\right)\sim \alpha(L_c-L)/L^2_c$. We hence have $|P^{+-}_K|\sim \Delta_K^{-1/2}\sim L_c/\sqrt{\alpha|L_c-L|}$, which indeed shows up as a divergence at $L\approx L_c$ in the $P^{+-}$ plots in Fig.~\ref{picS3}c. 
\end{itemize}

The fundamental difference between our scaling-induced EPs and usual critical points is that the divergence in the projector does not stem from Re$\Delta_K$, which typically scales like $L^{-1}$, but instead arises from Im$\Delta_K$, which diverges at $L\approx L_c$.

\section{Appendix: Free-fermion entanglement entropy dip due to scaling-induced exceptional points}\label{suppmat2}

In the previous section, we derived the system-size (scaling) dependent generalized Brillouin zone (GBZ) that accounts for the effects of coupling two chains with antagonistic non-Hermitian skin effects (NHSE). This GBZ allows us to approximate our original 4-component system using an effective 2-component Hamiltonian derived from a single chain. A key finding is that, for odd system sizes $L$, the truncated occupied band projector $\bar P$ is expected to diverge at a special $L=L_c$ (in practice, the computed $L_c$ is usually not an integer, and $\bar P$ becomes very large though finite.) 

In this section, we demonstrate how the divergence in the truncated occupied band projector leads to an anomalous dip in the entanglement entropy (EE) scaling of free fermions and how this phenomenon can be analytically estimated and characterized. Unlike typical scenarios where non-Hermiticity is isotropic and quantum correlations are uniformly distributed, resembling those in Hermitian systems, our model shows a distinct behaviour. In conventional non-Hermitian skin effect (NHSE) systems, the Hermiticity can often be effectively removed through a basis transformation, and the resulting hopping asymmetry does not typically alter the EE significantly. However, in our case, the EE scaling deviates from standard area or volume laws and instead exhibits a unique entanglement dip.

For free fermions, the entanglement spectrum can be obtained by truncating the occupied band projector $P$ in real-space. In 1D, we define a real-space partition $[x_L, x_R]$, such that truncated band projector $\bar P$ can be obtained from $P$ via
\begin{equation}\begin{split}\label{2bandP3a}
		& \bar P=  \bar{\mathcal{R}}_{[x_L, x_R] } P  \bar{\mathcal{R}}_{[x_L, x_R] } \ ,\ \\
		& \bar{\mathcal{R}}_{[x_L, x_R] }=\sum^{L}_{x\notin[x_L, x_R] }|x\rangle\langle x|\otimes \mathbb{I}\ ,\\
		&P= \sum_{n,(\text{Re}E_n)<0}|\psi_n^R\rangle \langle \psi_n^L|\ ,
\end{split}\end{equation}
where $\bar{\mathcal{R}}_{[x_L, x_R] }$ is the projector onto sites outside of $[x_L, x_R] $ (note that the index $n$ was expressed as indices $(k,\mu)$ in the main text, such as to emphasize the momentum $k$). For two-component models, the above general expression for $P$ reduces to Eq.~\ref{EqPk0}. 
The eigenvalues of the projection operators can be interpreted as occupancy probabilities. For $P$, they are limited to either 1 and 0. However, when the system is divided in to the $[x_L, x_R]$ and its complement, the truncated occupied matrix $\bar P$ reveals the entanglement between these two regions, resulting in occupancy probabilities (eigenvalues) $p$ away from $1$ and $0$.
The free-fermion entanglement entropy (EE) $S$ quantifies this entanglement, and is given by
	\begin{eqnarray}
	\label{SEE1}
		S(\bar P) &=& -\text{Tr}(\bar P \log \bar P +(\mathbb{I}-\bar P)\log (\mathbb{I}-\bar P)) \notag\\
		&=&-\sum_p (p \log p +(1-p)\log (1-p)) \ .
	\end{eqnarray}

\subsection{Entanglement entropy scaling and dip in the effective 2-component model }
\label{suppmat2a}
To investigate the EE behaviour, we first compute $\bar P$ in its real-space basis, such that $\bar{\mathcal{R}}_{[x_L, x_R] }$ can be implemented as the truncation to a submatrix. Noted with the Fourier Transform $|x,\alpha\rangle =\frac{1}{\sqrt{L}}\sum_k\text{e}^{-ikx}|k,\alpha\rangle$, we can express its real-space matrix elements as 
\begin{equation}\label{eq2cor}
	\begin{split}
		\langle c^{\dagger}_{x_1,\alpha}c_{x_2,\beta}\rangle=	\langle {x_1,\alpha}| \bar P|x_2,\beta\rangle=\left\{\begin{array}{cc}\displaystyle\frac{1}{L}\sum\limits_k P^{\alpha\beta}_K\text{e}^{ik(x_1-x_2)}, &\qquad x_1,x_2\notin[x_L, x_R] \\
			0 &\qquad x_1,x_2\in[x_L, x_R]
		\end{array}\right.\  ,
\end{split}\end{equation} 
where $\alpha,\beta=\pm$ represent the sublattices $A,B$, and $K=k+i\alpha/(L+1)$. Note that even though $P_K$ is evaluated in the GBZ $K$, the fourier transform is still between the real-space lattice and the usual BZ $k$.

\begin{figure}
	\centering
	\includegraphics[width=\linewidth]{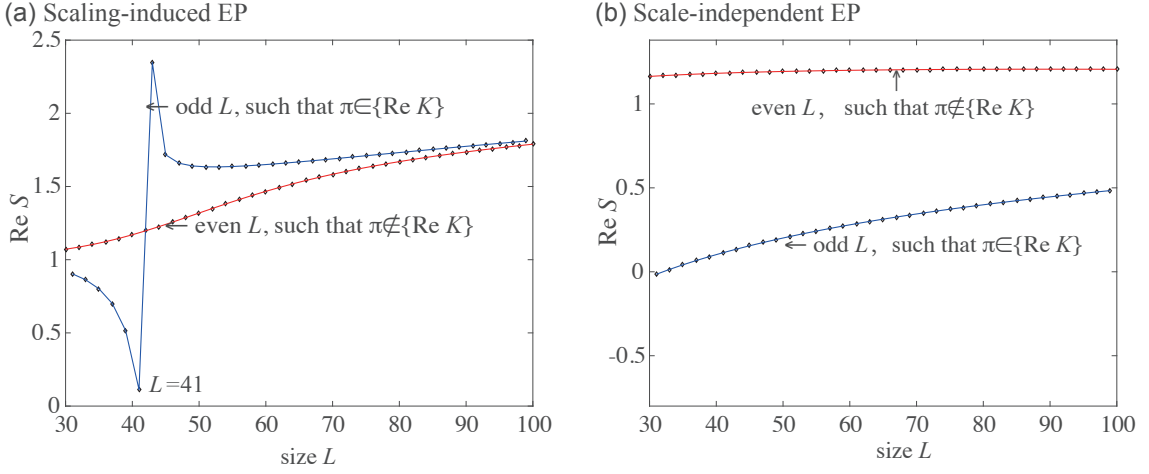}
	\caption{
\label{picS4} 
\textbf{ Comparison of entanglement entropy scaling behaviour with and without scaling-induced exceptional criticality.} 
(a) For our 2-component effective model [Eq.~\ref{Eqeffectivemodel1}] with scale-dependent GBZ given by Im$K = \frac{\alpha}{L+1}$, the entanglement entropy experiences a pronounced dip at $L=41\approx L_c$. This only occurs for odd $L$, where the Re$K=\pi$ point is sampled. 
(b) Had the same model Eq.~\ref{Eqeffectivemodel1} been a physical model with fixed GBZ, rather than an effective model of a $L$-dependent GBZ, there would not have been any entanglement dip. We have fixed Im$K= -\log \left|\delta \times \displaystyle\frac{ t_R( t_R- t_L)}{t_L(t_R-1)}\right|/(L_c+1) =-0.123$, with parameters $t_L = 1.2\text{e}^{0.3},t_R = 1.2\text{e}^{-0.3}, \delta=1.6\times 10^{-3}$. 
	}
\end{figure}

Because of the peculiar scaling properties of our GBZ of our effective 2-band model Eq.~\ref{Eqeffectivemodel}, $\bar P$ behaves very differently for odd and even $L$ [See Fig.~\ref{picS4}]. For odd $L$, we have shown that the discretized momenta in the GBZ contain one point where $Re K =\pi$, i.e., min$|\text{Re}\Delta_K|$ from the EP is zero, where $\Delta_K=i\left(\frac{\alpha}{L+1}-\frac{\alpha}{L_c+1}\right)$. In this case, $P^{+-}_K$ [Eq.~\ref{EqP+}] is divergent as $|L-L_c|^{-1/2}$ near the EP. However, for even $L$, the divergence is only asymptotic with $L^{1/2}$, as explained below Eq.~\ref{EqP+}. In the following, we elaborate on the above:

\begin{figure}
	\centering
	\includegraphics[width=\linewidth]{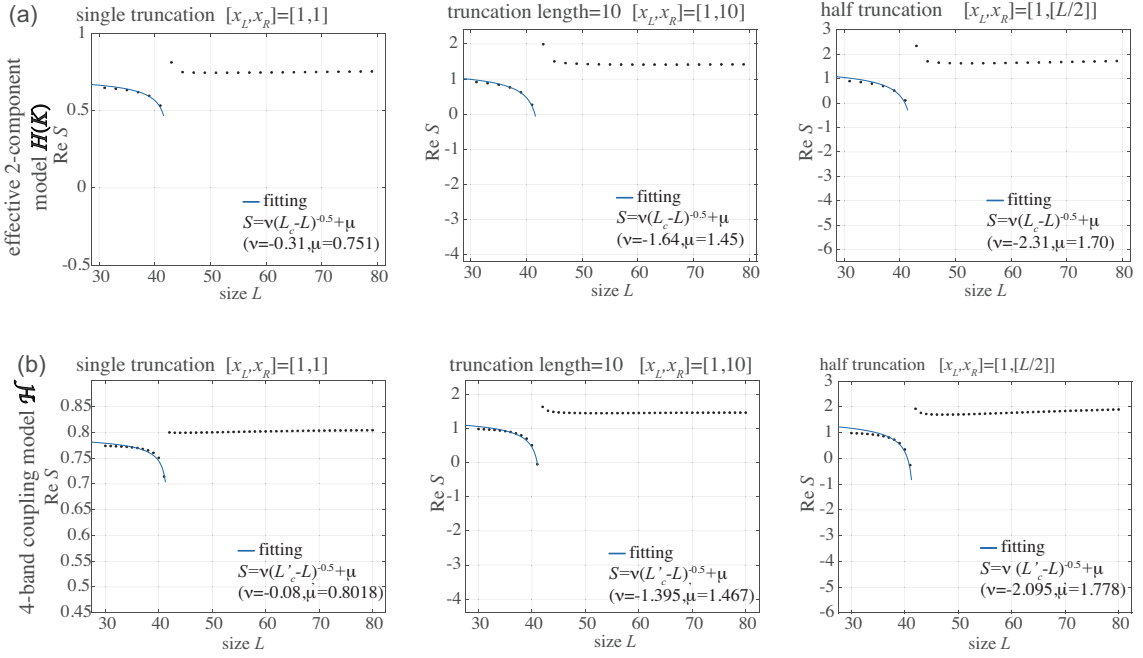}
	\caption{	\label{picS5}
		\textbf{ Entanglement dip with different entanglement cuts.}
		For different truncated regions $[x_L,x_R]$ implemented by the operator ${\bar{\mathcal{R}}}_{[x_L, x_R] }=\sum^{L}_{x\notin[x_L, x_R] }|x\rangle\langle x|\otimes \mathbb{I}$, similar behaviour is observed in the entanglement entropy (EE) of (a) our effective 2-component model $H(K)$ Eq.~\ref{Eqeffectivemodel} and
		(b) its parent 4-component coupled chain
		model $\mathcal H$ Eq.~\ref{EqS1}. For all cases, a prominent entanglement dip occurs, with approximate asymptotic $(L_c-L)^{-0.5}$ behaviour. 
		Parameters are $t_L = 1.2\text{e}^{0.3},t_R = 1.2\text{e}^{-0.3},\delta=1.6\times 10^{-3}$, corresponding to $L_c=42.79$ for (a) and $L'_c=41.8$ for (b). 
		}
\end{figure}
\begin{equation}\begin{split}\label{pp}
		\langle {x_1,+}| \bar P|x_2,-\rangle 
		&\approx \displaystyle\frac{1-(-1)^L}{2L} P^{+-}_{K=\pi+i\frac{\alpha}{L+1}}\text{e}^{i\pi(x_1-x_2)} +\frac{1}{\pi}\int ^{2\pi}_{\pi+\pi/L}P^{+-}_{K=k+i\frac{\alpha}{L+1}} dk.
\end{split}\end{equation}
Here we have split the integral contributions into two parts: the first term is from $\text{Re}\Delta_K=\pi$ and exists only when $L$ is odd, while the second term contains all other momentum point contributions. From the arguments below Eq.~\ref{EqP+}, the first term behaves like:

\begin{equation}\begin{split}\label{pp1}
		 &\displaystyle\frac{1-(-1)^L}{2L} P^{+-}_{K=\pi+i\frac{\alpha}{L+1}}\text{e}^{i\pi(x_1-x_2)}\\
         \approx & -\frac{1-(-1)^L}{2L}\frac{\sqrt{t_L/t_R -1/t^2_R}}{2\sqrt{-i\Delta_K}}\text{e}^{i\pi(x_1-x_2)}\\
		\xrightarrow[L\rightarrow L_c]{\Delta_K=K-K_c,\,\text{Re}\Delta_K=0\quad}&\quad\frac{1-(-1)^L}{2L}\frac{\sqrt{t_Lt_R -1}}{2t_R}\left(\frac{(L+1)(L_c+1)}{\alpha (L_c-L)}\right)^{1/2}\text{e}^{i\pi(x_1-x_2)}\\
		\xrightarrow{L\approx \sqrt{(L_c+1)(L+1)}}&\qquad\quad(1-(-1)^L)\frac{\sqrt{t_Lt_R -1}}{4\sqrt{\alpha}\cdot t_R} (L_c-L)^{-1/2}\text{e}^{i\pi(x_1-x_2)}\ .\\
\end{split}\end{equation}
The second term behaves like 
\begin{equation}\begin{split}\label{pp2}
\frac{1}{\pi}\int ^{2\pi}_{\pi+\pi/L}P^{+-}_{k+i\frac{\alpha}{L+1}} dk
&\approx \frac{1}{\pi}\int^{\pi}_{\frac{\pi}{L}}\frac{\sqrt{t_L/t_R -1/t^2_R}}{2\sqrt{-i\Delta_K}}\text{e}^{i(\pi+\text{Re}\Delta_K)(x_1-x_2)}\text{d} \text{Re}\Delta_K\ \\
				&\propto L^{-1/2}.
				\end{split}\end{equation}
In the above, we have not focused on the spatial $x_1-x_2$ dependence, since we are primarily concerned about the $L$-scaling behaviour. Diagonalizing $\bar P$, it is numerically verified [See Fig.~\ref{picS4}(a)] that due to our scaling-induced EP, odd and even-sized systems behave qualitatively differently. Near the critical size, the EE (computed from the eigenvalues $p$ of $\bar P$ via Eq.~\ref{SEE1}) for odd $L$ exhibits a divergence which we call an \emph{entanglement dip}, while in even-$L$ systems, it changes continuously with size.  

However, this entanglement dip disappears if the system does not possess such a scaling-dependent GBZ. If we were to instead consider Eq.~\ref{Eqeffectivemodel} to be the physical (not effective) model for a non-antagonistic NHSE chain which exhibits a scaling-independent EP point, such that \(K = k + i\frac{\alpha}{L_0 + 1}\) with fixed $L_0$ (instead of $L$),  its EE would no longer show abrupt changes with size, despite still exhibiting odd/even system size effects [See Fig.~\ref{picS4}(b)].\\

\subsection{ Effect of entanglement truncation interval in the effective 2-component model }\label{suppmat2b}

\begin{figure}
	\centering
	\includegraphics[width=\linewidth]{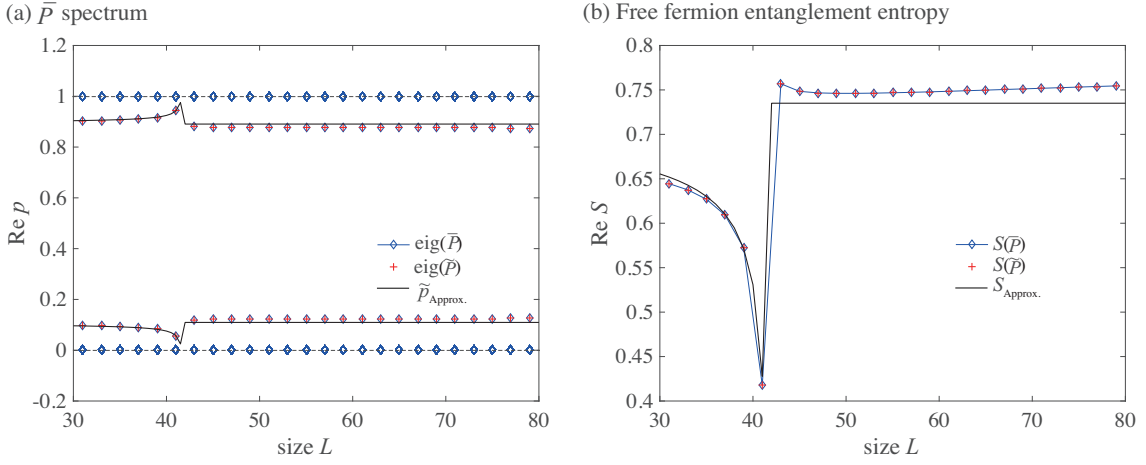}
	\caption{	
\textbf{ Scaling behaviour of the $\bar{P}$ spectrum and the EE of the 2-component model [Eq.~\ref{Eqeffectivemodel}] with single unit-cell truncation.} 
(a) The excellent agreement between the numerical spectra of $\bar{P}$ [Eq.~\ref{2bandP3a}] and its single unit-cell approximation $\tilde{P}$ [Eq.~\ref{spp0}], as well as its analytically approximated spectrum $\tilde{p}_\text{Approx.}$ [Eq.~\ref{spp1}]. Note that $\tilde{P}$ only yields the nontrivial pair of eigenvalues away from 0 and 1, whose kink near $L\approx L_c$ is well-approximated by $\tilde{p}_\text{Approx.}$. 
	(b) The corresponding scaling of the entanglement entropy (EE) $S(\bar{P})$ and $S(\tilde{P})$ from the eigenvalues of
	$\bar{P}$ and $\tilde{P}$ shown in panel (a), along with the analytical result $S_\text{Approx.}$ given by Eq.~\ref{sfit}. The results exhibit good consistency, and in particular all predict the same entanglement dip. 
	Parameters are $t_L = 1.2\text{e}^{0.3},t_R = 1.2\text{e}^{-0.3}, \delta=1.6\times 10^{-3}$. The coefficients $a$ and $b$ in Eq.~\ref{sfit} are calculated as $a = 0.3905$ and $b=0.0456$ with small constants $\mathcal{O}=0.14$, $\mathcal{O}'=-0.035$ [Eq.~\ref{spp1},\ref{sfit}], corresponding to the critical scale being $L_c=42.79$.
	}
	\label{picS6}
\end{figure}

\begin{figure}
	\centering
	\includegraphics[width=\linewidth]{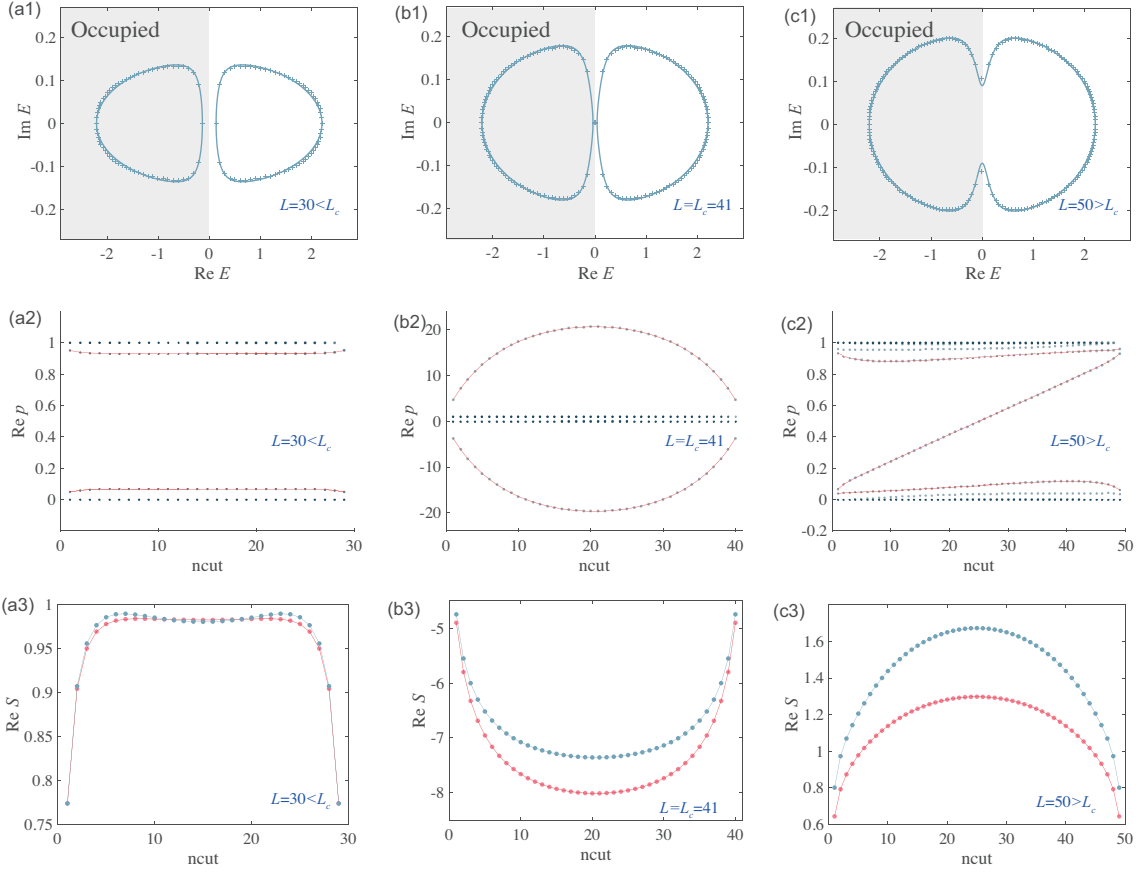}
	\caption{
	\textbf{ Scaling of the energy spectrum, $\bar{P}$ spectrum and entanglement entropy (EE) with different truncation lengths $n_{cut}$ in the 3 regimes: $L<L_c, L=L_c$ and $L>L_c$.}
	(a1-c1) The energy spectrum at different sizes $L$ where the Fermi energy is set at Re$E=0$, identifying the lower energy states in the gray area as the occupied states.
	(a2-c2) The $\bar{P}$ spectrum $p$ with the system divided into two partitions: the truncated region $x\in[1,n_{cut}]$ and non-truncated region $[n_{cut}+1, L]$ which is kept. Eigenvalues that are significantly far from 0 and 1 are joined by a red curve which exhibits qualitatively different behaviour in different regimes.
	(a3-c3) Scaling of the EE with $n_{cut}$. For $L<L_c$, the total EE (blue curve) and the EE contributed only by the red states in (a2-c2) (red curve) are quite similar due to the suppression of the EE by the gap. But for the two other regimes, they differ considerably, albeit still with the same qualitative $n_{cut}$ dependence. Parameters used are $t_L = 1.2\text{e}^{0.3}, t_R = 1.2\text{e}^{-0.3}$, ${\delta=1.6844163\times10^{-3}}$.
	}	
	\label{picS10}
\end{figure}
The entanglement dip at $L\approx L_c$ is universal and occurs independently of the specific entanglement cut region $[x_L,x_R]$, as it is not reliant on the real-space distribution of the matrix elements of $\bar P$. In Fig.~\ref{fig:fig4}(a) for our 2-component model Eq.~\ref{Eqeffectivemodel}, the odd-sized EE is shown to consistently exhibit $(L-L_c)^{-0.5}$ divergence regardless of the truncation region. 

This entanglement dip remains qualitatively similar (as it should be) in the EE of its parent 4-component coupled system [Eq.~\ref{EqS1}], despite the more complex structure of the GBZ and the fact that the real part of the exceptional point (EP) momentum is not strictly fixed at $\pi$. As shown in Fig.\ref{fig:fig4}(b), the entanglement dip is present in both odd and even-sized systems, indicating that it is not intrinsically linked to the positions of the $\text{Re}K$ momentum points. What is fundamental is that the EP lies on a GBZ whose imaginary part of momentum deviation $\Delta_K$ gives rise to a $(L-L_c)^{-0.5}$ divergence in $P$.

\subsubsection{Single unit-cell truncation }

Since the entanglement dip is largely independent of the truncation region, we can make the further analysis by focusing on \emph{single-site} entanglement where only one unit-cell $x_L=x_R$ exists in the untruncated region $[x_L,x_R]=[x_L,x_L]$. For a single unit-cell, the Fourier phase factor $\text{e}^{ik(x_1-x_2)}$ in $\langle {x,\alpha}| P|x,\beta\rangle$ [Eq.~\ref{pp}] disappears, and the single unit-cell truncated projector (we call it $\tilde P$ here to emphasize the single unit cell special case) is
\begin{equation}\label{spp1}
	\begin{split}		
		\tilde P^{+-}&=-\displaystyle\frac{1}{{L}}\sum\limits_k P^{+-}_K \xrightarrow{L\gg1} -\displaystyle\frac{1}{L} P^{+-}_{\pi+i\frac{\alpha}{L+1}} +\frac{1}{2\pi}\int_{k\neq \pi} P^{+-}_{k+i\frac{\alpha}{L+1}} \text{d}k\\
		&=\frac{1}{2L}\frac{\sqrt{t_L -  \text{e}^{\frac{\alpha}{L+1}}}}{\sqrt{t_R -  \text{e}^{-\frac{\alpha}{L+1}}}} +\frac{1}{2\pi}\int_{k\neq \pi} \frac{\sqrt{t_L -  \text{e}^{-ik+\frac{\alpha}{L+1}}}}{\sqrt{t_R -  \text{e}^{ik-\frac{\alpha}{L+1}}}}\text{d}k\\
		&\xrightarrow[t_R -  \exp(\frac{\alpha}{L_c-1}) = 0]{}\frac{\sqrt{t_Lt_R -1}}{2t_R L}\left(\frac{\alpha}{L+1}-\frac{\alpha}{L_c+1}\right)^{-1/2}-\frac{1}{\pi}\sqrt{\frac{-t_L}{t_R}}\mathcal{F}\left(\text{asin}\sqrt{-t_R \text{e}^{-\frac{\alpha}{L+1}}},\frac{1}{t_Rt_L}\right)\ ,
	\end{split}
\end{equation}
where $\mathcal{F}(m,\phi)$ represents the Elliptic integral of the second kind, which varies very weakly with system size $L$ due to the slow $e^{-\frac{\alpha}{L+1}}$ functional dependence, and can be approximated as $\sqrt{(t_L+1)(t_R+1)}/2\pi t_R$. Hence the above is approximated by 
\begin{equation}\begin{split}\label{spp0}
		\tilde P^{+-}&\approx\frac{\sqrt{t_Lt_R -1}}{2\sqrt{\alpha}\cdot t_R} (L-L_c)^{-1/2}+\frac{\sqrt{(t_L+1)(t_R+1)}}{2\pi t_R}\ ,\\
\end{split}\end{equation}
with $\sqrt{(L+1)(L_c+1)}\approx L$ near the entanglement dip. By contrast, the other off-diagonal term $\tilde P^{-+}$
is not sensitive to $L$, as given by
\begin{equation}\begin{split}\label{spp}
		\tilde P^{-+}&= -\displaystyle\frac{1}{{L}}\sum\limits_k P^{-+}_K\approx\frac{ \sqrt{(t_L+1)(t_R+1)}}{2\pi t_L}\ .\\
\end{split}\end{equation}

Hence the occupation probability eigenvalues $\tilde p$ of $\tilde P$ are given by
\begin{small}
	\begin{equation}
		\begin{split} 
			&\tilde{p}_{\text{Approx.}} = \frac1{2}\pm \sqrt{\tilde{P}^{+-}\times\tilde{P}^{-+}}\\
			&=\frac1{2}\pm \left(\displaystyle\frac{1}{4}{\displaystyle\frac{t_R}{t_L}}\sqrt{\displaystyle\frac{t_L/t_R-1/t^2_R}{\alpha}} \frac{1}{\sqrt{L-L_c}}+  \frac{1}{2\pi\sqrt{t_Rt_L}}\sqrt{(t_L+1)(t_R+1)}\right)+\mathcal{O}\ ,
		\end{split} 
	\end{equation}
\end{small}
where the small constant $\mathcal{O}$ is of the order of $10^{-1}$. For entanglement dips that are not too deep (i.e $\tilde p\approx 0$ or equivalently $\tilde p\approx 1$), we can further approximate the EE by
\begin{equation}\label{sfit}
	\begin{split}
		S_{\text{Approx.}}
		&\xrightarrow{|\tilde p| \rightarrow 0 }-2\left(\tilde p\log\tilde p -(1-\tilde p)\tilde p\right)\\
		&=2(b\log(0.5-a)-2ab)(L_c-L)^{-0.5}+0.25+a^2+\mathcal{O}' )\ ,
	\end{split}
\end{equation}
where $a=\frac{1}{2\pi\sqrt{t_Rt_L}}\sqrt{(t_L+1)(t_R+1)}+ \mathcal{O}$, $b=\frac{1}{4}{\frac{t_R}{t_L}}\sqrt{\frac{t_L/t_R-1/t^2_R}{\alpha}}$,
$\mathcal{O}$ and $\mathcal{O}'$ representing small offsets that can be adjusted to mitigate the imperfect approximation [See Fig.~\ref{picS6}(b)].

\subsubsection{Effect of the length of the entanglement truncation interval}

\begin{figure}
	\centering
	\includegraphics[width=\linewidth]{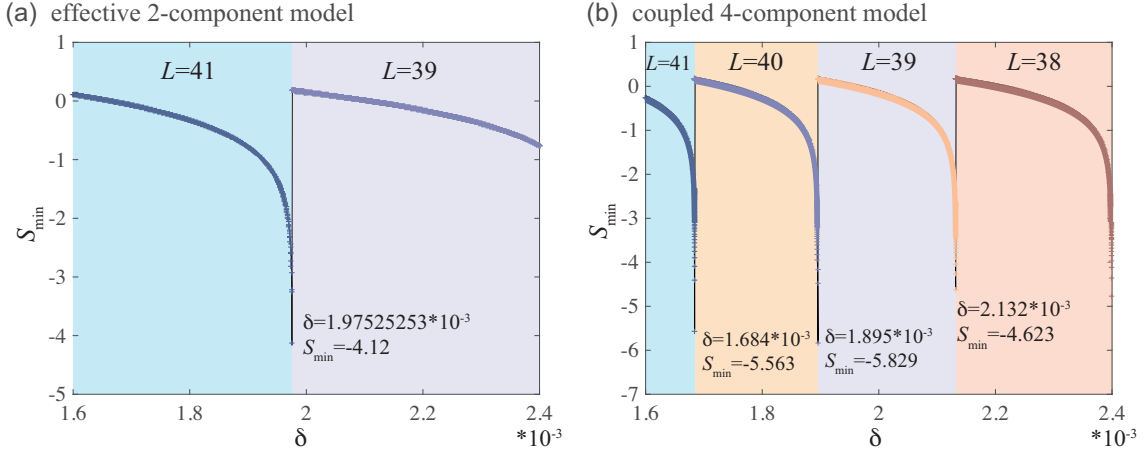}
	\caption{\textbf{ Sensitivity of the entanglement dip minimum $S_\text{min}$ with coupling $\delta$.}
The entanglement dip $S_\text{min}$ can yield negative EE for a wide range of parameters, and can even reach very negative values of $S_\text{min}<-4$ across small parameter windows. Shown are the $S_{\text{min}}$ and the corresponding system chain length $L$ where it is attained, both for our (a) 2-component effective model Eq.~\ref{Eqeffectivemodel} with $\delta$ built into its scale-dependent GBZ and (b) its parent 4-component model Eq.~\ref{EqS1} with $\delta$ being the physical inter-chain coupling. A more complete parameter space plot that shows the dependence of $S_\text{min}$ with $t_L/t_R$ is given in Fig.~3(c) of the main text. Parameters are $t_L = 1.2\text{e}^{0.3}, t_R = 1.2\text{e}^{-0.3}$. }	
	\label{picS7}
\end{figure}

In Hermitian systems, it is well-known that the EE of a cut region $[1,n_{cut}]$ varies strongly like $S\sim \log\left(L\sin \frac{\pi n_{cut}}{L}\right)$, which can be proven with boundary CFT~\cite{calabrese2004entanglement, Cardy:2004hm}. Hence it is prudent to check whether SIEC behaviour is nontrivially influenced by $n_{cut}$. 

In Fig.~\ref{picS10}, it was observed that the effect of $n_{cut}$ qualitatively depends on whether $L<L_c$, $L\approx L_c$ or $L> L_c$. In the first regime $L<L_c$ before the onset of the EE dip, the system is essentially gapped and $n_{cut}$ does not appreciably change either the $\bar P$ spectrum or $S$. However, for the second regime $L\approx L_c$ where the EE dip occurs, the $\bar P$ eigenvalues do become significantly enhanced at $n_{cut}\approx L/2$, lying very far out of $[0,1]$. For the third regime $L>L_c$, the system behaves essentially like an ordinary gapless system, with positive $S$ and $\bar P$ eigenvalues lying within $[0,1]$. One state traverses the $\bar P$ spectral gap between $0$ and $1$, reminiscent of the spectral flow of a topological edge mode.

Interestingly, as shown in the bottom row of Fig.~\ref{picS10}, despite the very different $\bar P$ behaviour at or after the onset of the EE dip (second and third regimes), the EE continues to adhere approximately to the conventional $S\sim \log\left(L\sin \frac{\pi n_{cut}}{L}\right)$ behaviour as $n_{cut}$ varies (for fixed $L$).

\subsection{Controlling the depth of entanglement dips}\label{suppmat2c}

In this section, we explore the relationship between the depth of the entanglement dip (i.e., minumum EE $S_\text{min}$) varying the coupling parameter $\delta$ while keeping other parameters constant. This approach is based on our analytical understanding of how $L_c$ varies with $\delta$ [See Fig.~\ref{picS2}].  Here, we numerically identify parameters that lead to exceptionally negative $S_\text{min}$ values, where the entanglement dip represents drastic departures from usual $\log L$ entanglement scaling.

Although the theoretically predicted $L_c$ varies continuously with $\delta$ [Eq.~\ref{Eqcritical1}], an actual lattice contains only an integer number of unit cells $L$. This prevents us from getting infinitesimally close to $L_c$, where $S$ truly diverges. Away from that, how negative a dip $S$ can reach depends on commensurability considerations, as shown in Fig.~\ref{picS7}. For certain fine-tuned values of $\delta$, the EE through $S_\text{min}$ can dip below $-4$ (such as at $\delta$=1.68441635 $\times 10^{-3}$ or $2.131947\times 10^{-3}$), corresponding to extremely large $p$ eigenvalues. Such dips occur both for our effective two-component model (a) as well as its parent four-component model (b), even though the dip positions are different. But what remains consistent are the shapes and qualitative order of magnitude of the entanglement dips.

\begin{figure}
	\centering
	\includegraphics[width=0.9\linewidth]{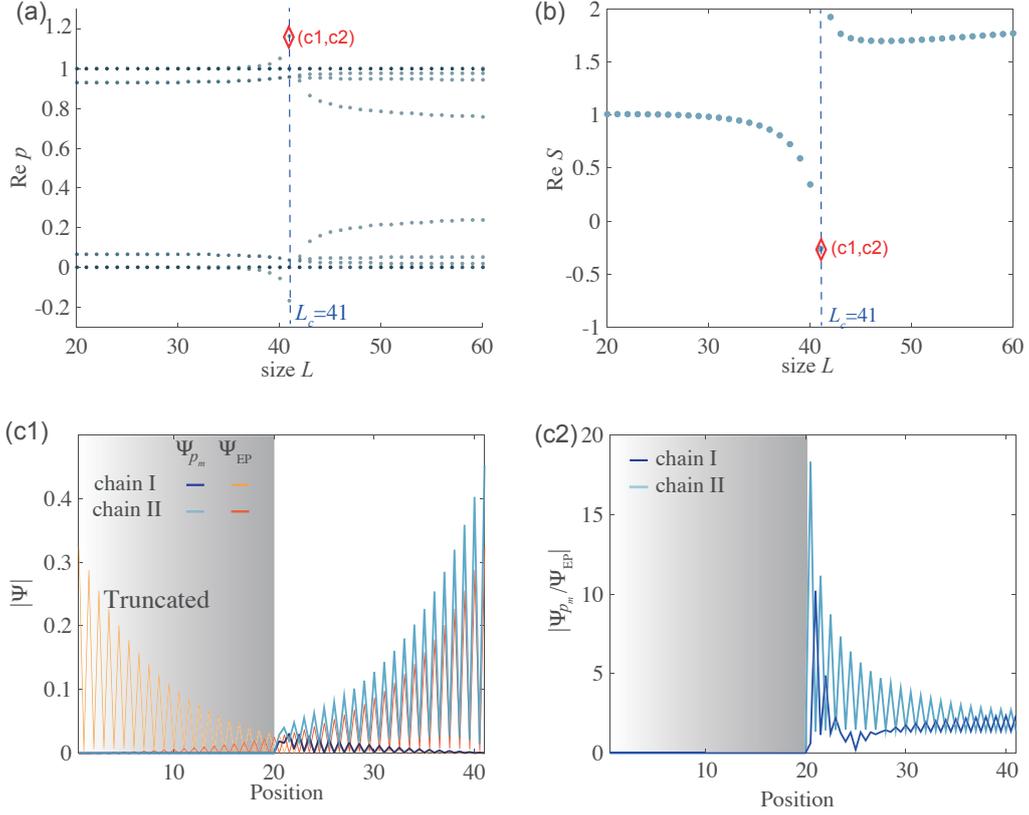}
	\caption{
	\textbf{ Entanglement scaling and $\bar P$ eigenstate profile for our physical 4-component model Eq.~\ref{EqS1}.} 
	(a) $\bar P$ spectrum as a function of system size $L$ at half entanglement truncation. A conventional in-gap spectral branch for critical systems appears for $L>L_c$. 
	(b) The resultant entanglement entropy scaling, with a pronounced entanglement dip before $L\approx L_c$. 
	(c1) The spatial profiles of $\Psi_{p_m}$(the $S=S_\text{min}$ eigenstate of $\bar{P}$, colored red above) and $\Psi_\text{EP}$ (eigenstate of Hamiltonian $\mathcal{H}$) are shown. Since $\bar{P}$ is constructed using the eigenstates of the Hamiltonian, its eigenstate $\Psi_{p_m}$ inherits the skin effect from the Hamiltonian. (c2) To eliminate the skin effect in $\Psi_{p_m}$ and $\Psi_{\text{EP}}$, the profile of $|\Psi_{p_m}/\Psi_{\text{EP}}|$ represents the cumulative wavefunction distribution near the truncation point after excluding the EP eigenstate. Parameters used are $t_L = 1.2\text{e}^{0.3}, t_R = 1.2\text{e}^{-0.3}$, ${\delta=1.6\times10^{-3}}$.
	}
	\label{picS8}
\end{figure}

\begin{figure}
	\centering
	\includegraphics[width=0.9\linewidth]{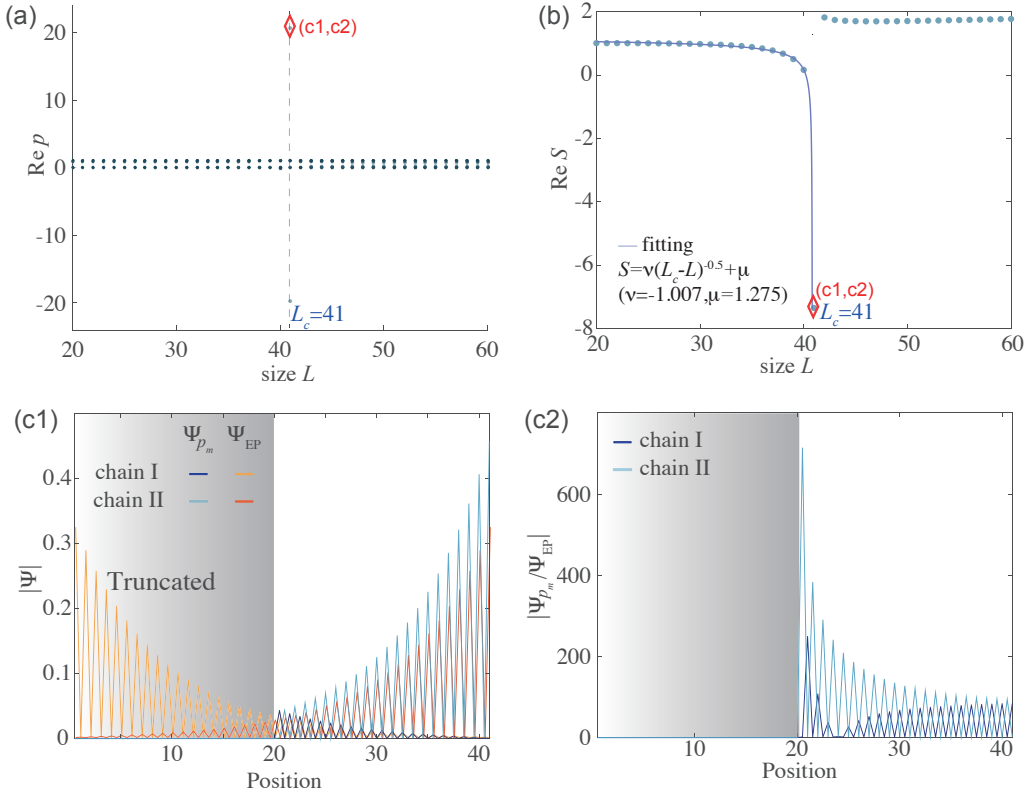}
	\caption{\textbf{ Entanglement scaling and $\bar P$ eigenstate profile for our physical 4-component model Eq.~\ref{EqS1}, for an extremely pronounced entanglement dip.} This figure is similar to Fig.~\ref{picS8}, but with inter-chain coupling tuned to $\delta=1.6844163\times10^{-3}$ such that $L_c = 40.038$ becomes almost an integer, giving rise to very negative $S_\text{min}$. (a) and (b) shows how the $\bar{P}$ spectrum and corresponding entanglement entropy jump dramatically at $L\approx L_c$. 
However, as compared to Fig.~\ref{picS8}(c1,c2), the spatial distribution of $\Psi_{p_m}$ (depicted in (c1) and colored red in (a,b)) and $|\Psi_{p_m}/\Psi_{\text{EP}}|$ in (c2) do not exhibit obvious changes.  
	Parameters are $t_L = 1.2\text{e}^{0.3}, t_R = 1.2\text{e}^{-0.3}$.
}	
	\label{picS9}
\end{figure}

\subsection{Suppressed non-Hermitian skin effect in the entanglement eigenstates }\label{suppmat2d}

Since $P$ is a projector onto the occupied bands, the entanglement eigenstates $\Psi_p$ that satisfy $\bar{P} \Psi_p = p \Psi_p$ can be approximated as linear combinations of the occupied states, which vanish outside the interval $[x_1, x_2]$. Specifically, if the constituent occupied states are all skin-localized due to non-Hermitian skin effect (NHSE), these eigenstates are expected to exhibit an exponentially decaying spatial profile.

However, as demonstrated in Figs.~\ref{picS8} and \ref{picS9}, the entanglement eigenstates $\Psi_p$ exhibit reduced NHSE localization compared to the reference skin localization at the scaling-induced exceptional point (EP). In the physical Hamiltonian, chain I shows a skin effect localized towards the right, whereas chain II shows localization towards the left. This skin localization is evident not only in the eigenstate $\Psi_\text{EP}$ of the physical Hamiltonian but also in the eigenstate $\Psi_{p_m}$ of $\bar{P}$, which corresponds to the most negative eigenvalue $p_m$ and significantly contributes to the entanglement dip.

Though as evident from Figs.~\ref{picS8}(c) and \ref{picS9}(c), the entanglement eigenstate $\Psi_{p_m}$ shows a weaker skin localization than $\Psi_\text{EP}$, despite being approximately made up of skin modes. This can be interpreted as a signature of NHSE suppression in the entanglement eigenstate, and can be qualitatively understood as a consequence of the antagonistic competition between the oppositely directed NHSE chains.

\begin{figure}
	\centering
	\includegraphics[width=\linewidth]{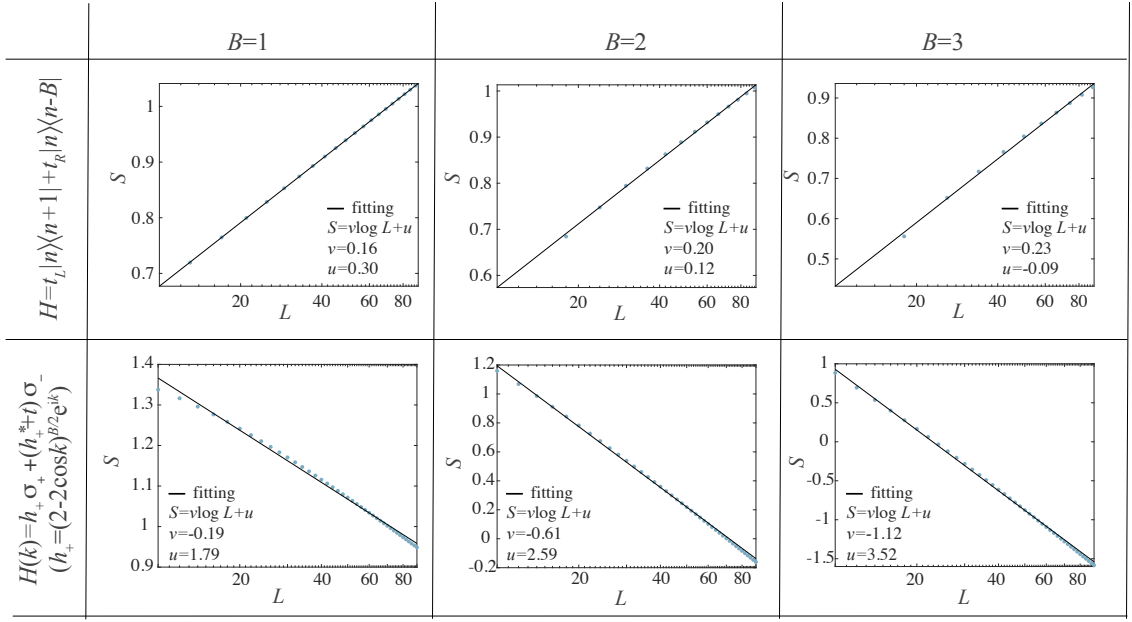}
	\caption{\textbf{ Logarithmic entanglement entropy scaling in generic non-SIEC systems that are critical.}
Shown are examples of the EE scaling of non-Hermitian systems exhibiting the NHSE (top) and exceptional points (bottom), labeled by maximal hopping distance $B$ as defined in the leftmost column. Regardless of form of the model, all cases exhibit $S \sim \log L$ scaling, even through $B>1$ gives rise to irregular generalized Brillouin zones (GBZs) or higher-order EP points. Only other systems exhibiting the SIEC can violate this logarithmic EE scaling. Parameters are $t_L=1.62,t_R=0.89, t=1$.}
	\label{picS12}
\end{figure}

\begin{figure}
	\centering
	\includegraphics[width=1\linewidth]{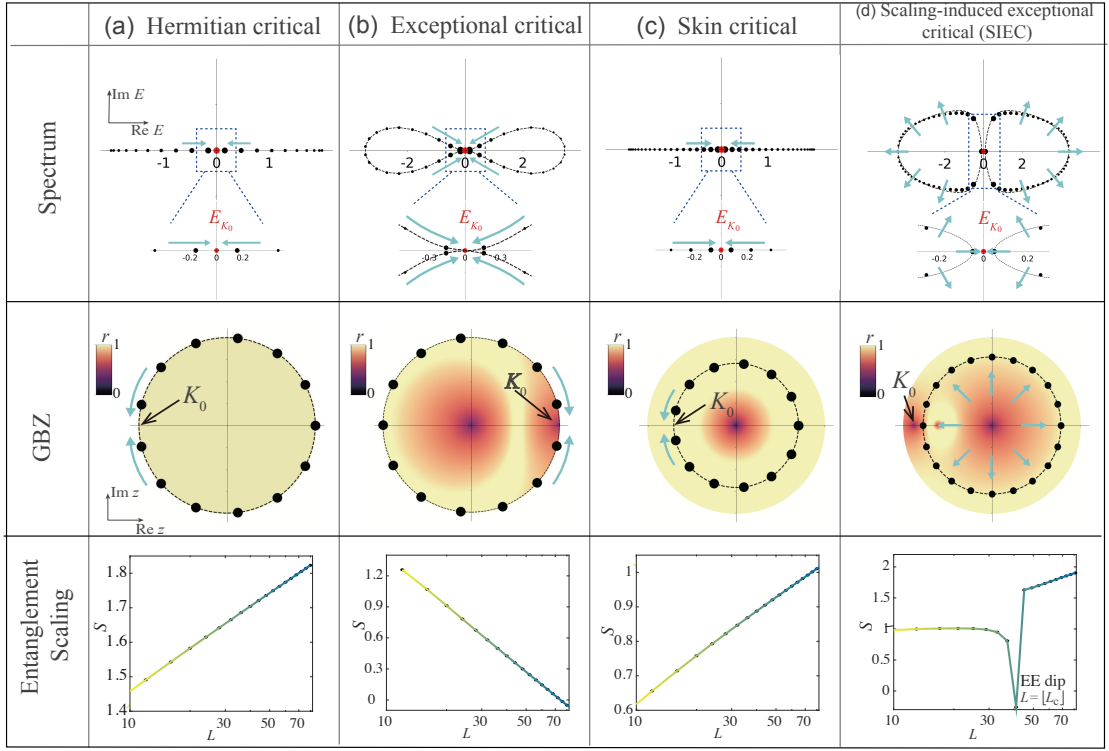}
	\caption{\textbf{ Entanglement scaling behaviour of various types of critical systems, with $S\sim\log L$ scaling violated only in SIEC. }
		(a) For a Hermitian critical point $K_0$ eigenmomentum states (black dots) accumulate near the corresponding energy $E_{K_0}$ (red) as system size $L$ increases (cyan arrows). The nearest approach to $K_0$ hence scales like $\Delta K_\text{min}\sim \pi/L$, leading to logarithmic entanglement entropy scaling, $S \sim \log L$. 
		The Hermitian model used is $H(z) = \left( t_1 + t_2z\right)\sigma_- + \left( t_1 + t_2z^{-1}\right)\sigma_+, z = \text{e}^{ik},k\in \mathbb{R}, t_1=t_2=1.$ 
		(b) Even under the influence of the NHSE, the logarithmic scaling of $S\sim \log L$ persists for a critical system because we still have $\Delta K_\text{min}\sim \pi/L$ in the GBZ $z=\text{e}^{iK (k)}$.
		The non-Hermitian model used is $H(z) = \left( t_R + t_2z\right)\sigma_- + \left( t_L + t_2z^{-1}\right)\sigma_+,  t_L=t_2=1,t_R=0.5$. 
		(c) For a non-Hermitian critical EP $K_0$, the eigenmomentum convergence $\Delta K_\text{min}$ remains proportional to $\pi/L$. Nevertheless, the logarithmic $S \sim -\log L$ exhibits negative scaling, attributable to the divergence of 2-point functions due to geometric defectiveness (vanishing phase rigidity $r\rightarrow 0$). The EP model used is $H(z) = \left( (2 - 2\cos k) z^{-1}\right)\sigma_- + \left( (2 - 2\cos k )z + 1\right)\sigma_+, z = \text{e}^{ik},k\in \mathbb{R}.$ 
		(d) In the case of SIEC, the spectrum and GBZ changes dramatically with $L$, encountering the EP $K_0$ only at a $L=L_c$. This condition induces a characteristic entanglement dip in $S$. The SIEC model used is the same as that in Fig.~1 of the main text.
		}
  \label{picS11}
\end{figure}

\section{Appendix: Comparison with other types of critical scenarios}\label{suppmat3}

To further elaborate on the discussion around Fig.1 of the main text, we present additional examples of NHSE and geometrically defective systems in Fig.\ref{picS12}. These examples also exhibit logarithmic entanglement scaling. Furthermore, Fig.~\ref{picS11} provides a detailed examination of all the cases illustrated in Fig.~1 of the main text.

\section{Appendix: Measuring negative biorthogonal entanglement}\label{suppmat4}
A distinctive feature of non-Hermitian systems is that the left and right eigenstates correspond to eigenstates of different Hamiltonians, specifically $\mathcal{H}$ and $\mathcal{H}^\dagger$. These left and right eigenstates collectively form a biorthogonal basis, which is essential for maintaining the probabilistic interpretation of quantum mechanics. This biorthogonal basis is crucial for defining various physical quantities, including our biorthogonal projector $\bar{P}$ and the entanglement entropy $S$.

However, simultaneously obtaining information from both left and right eigenstates in experimental measurements is usuallt challenging. Here, to address this difficulty, we suggest considering a larger system $\tilde{\mathcal{H}}$ comprising the original system $\mathcal{H}$ and its conjugate $\mathcal{H}^\dagger$. 
These two subsystems are very weakly coupled via a end-to-end coupling~\cite{torma2016physics,vale2021spectroscopic} $\bm \eta =\eta|1\rangle \langle L|+h.c.$ 
that connects the $|1\rangle$ and $|L\rangle$ end sites:
\begin{equation}\label{expri1}
	\begin{split}
		\tilde{\mathcal{H}}=  \begin{pmatrix}\mathcal{H}&\mathbf{\bm \eta} \\ \mathbf{\bm \eta^\dagger} &\mathcal{H}^{{\dagger}} \end{pmatrix}.
	\end{split}
\end{equation}
Importantly, this $\bm \eta$ coupling will not measurably affect the individual subsystems $\mathcal{H}$ and $\mathcal{H}^\dagger$'s eigenenergies and eigenstates, since they already contain equal and opposite NHSE themselves, and hence are no longer subject to net antagonistic NHSE. The purpose of these minuscule couplings $\eta$ is to enable cooperative response between the auxiliary system $\mathcal{H}^{\dagger}$ and the target system $\mathcal{H}$, and their impact on their fundamental characteristics, such as eigenergies and eigenstates, can be considered negligible.\\

The purpose of introducing the auxiliary Hamiltonian as $\mathcal{H}^{{\dagger}}$ is such that the eigenstates of the larger constructed Hamiltonian contains both the left and right eigenvectors of the target Hamiltonian.	
Specifically, for an energy $E$, there are 2 linearly independent degenerate eigenstates $|\tilde{\Psi}_{E,\pm}\rangle$ satisfying
\begin{equation}
	\begin{split}
		\tilde{\mathcal{H}}  |\tilde{\Psi}_{E,\pm}\rangle = E |\tilde{\Psi}_{E,\pm}\rangle,\qquad |\tilde{\Psi}_{E,\pm}\rangle=\begin{pmatrix}|\Psi^R_E\rangle \\ \pm K |\Psi^L_{E}\rangle  \end{pmatrix},
	\end{split}
\end{equation}
where $K$ is complex conjugation operation and $|\Psi^R_E\rangle$ and $|\Psi^L_E\rangle$ represent the eigenstates of the Hamiltonian $\mathcal{H}$ and its hermitian 
conjugate  $\mathcal{H}^{\dagger}$ also with eigenvalue $E$.

\begin{figure}
\centering
\includegraphics[width=0.6\linewidth]{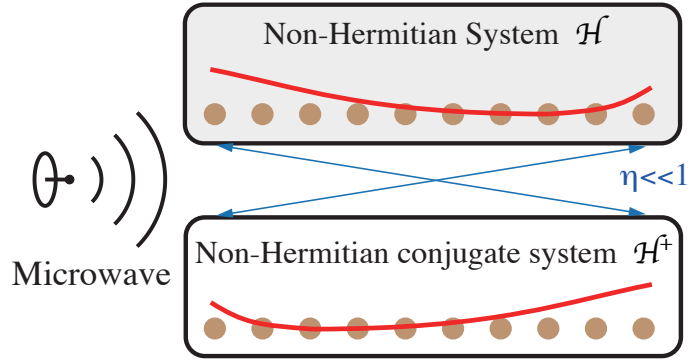}
\caption{ \label{picS13} (a)  A schematic setup in which the non-Hermitian system interacts with the auxiliary system through a weak microwave coupling. (b) A weak coupling $\eta$ are introduced between the endpoints of the $\mathcal{H}$ and $\mathcal{H}^{\dagger}$ systems. 
}
\end{figure}

Below, we show how to connect the biorthogonal normalization factor $\langle \Psi^L_{E} |\Psi^R_{E}\rangle$ of our target system $\mathcal{H}$ to the \emph{physically measurable} non-biorthogonal expectation in terms of the eigenstates of this enlarged Hamiltonian $\tilde{\mathcal{H}}$. We define an operator $\hat\Lambda=\sigma_-\otimes \mathbb{I}\,K$, which is the joint operation of $ \sigma_-\otimes \mathbb{I}$ and the complex conjugate operator $K$. The \emph{non-biorthogonal}, i.e., conventionally measurable expectation of $\hat \Lambda$ in the energy $E$ eigenstate just gives the biorthogonal normalization factor:
\begin{equation}
\begin{split}
	\langle \hat \Lambda \rangle_E=\langle \tilde{\Psi}_{E,\pm} |\hat\Lambda|\tilde{\Psi}_{E,\pm}\rangle={\langle \Psi^L_{E} |\Psi^R_{E}\rangle}.
\end{split}
\end{equation}

Next, we show how the use of $\hat\Lambda$ can connect the non-biorthogonal expectation of an arbitrary operator $\hat A$ with its biorthogonal expectation:
\begin{equation}
\begin{split}
	\sum_{E,\pm}\langle \hat A\hat\Lambda \rangle_E/\langle \hat\Lambda \rangle_E &=\sum_{E,\pm}\frac{1}{\langle \tilde{\Psi}_{E,\pm} |\hat\Lambda|\tilde{\Psi}_{E,\pm}\rangle} \langle\tilde{\Psi}_{E,\pm}| \hat A\hat\Lambda |\tilde{\Psi}_{E,\pm}\rangle\\
	& = \sum_{E,\pm}  \frac{1}{\langle \tilde{\Psi}_{E,\pm} |\hat\Lambda|\tilde{\Psi}_{E,\pm}\rangle} \begin{pmatrix}\langle\Psi^R_E|  & \pm \langle\Psi^L_{E}| K\end{pmatrix} \begin{pmatrix} \textbf{ 0}& \textbf{ 0}\\ \hat A & \textbf{ 0}\end{pmatrix}  K\begin{pmatrix}|\Psi^R_E\rangle \\ \pm K |\Psi^L_{E}\rangle  \end{pmatrix}\\
	& \xrightarrow{K^2=\textbf{ 1}} \sum_{E,\pm}  \frac{1}{\langle \tilde{\Psi}_{E,\pm} |\hat\Lambda|\tilde{\Psi}_{E,\pm}\rangle} \begin{pmatrix}\langle\Psi^R_E|  & \pm \langle\Psi^L_{E}| K\end{pmatrix} \begin{pmatrix} \textbf{ 0}& \textbf{ 0}\\ \hat A & \textbf{ 0}\end{pmatrix}  \begin{pmatrix}K |\Psi^R_E\rangle \\ \pm |\Psi^L_{E}\rangle  \end{pmatrix}\\
	&= \sum_{E,\pm}  \frac{\pm 1 }{\langle \Psi^L_{E} |\Psi^R_{E}\rangle} \begin{pmatrix}\pm \langle\Psi^L_{E}| K\hat A & \textbf{ 0}\end{pmatrix}  \begin{pmatrix}K |\Psi^R_E\rangle \\ \pm |\Psi^L_{E}\rangle  \end{pmatrix}\\
	&=\sum_{E,\pm} \frac{ 1 }{\langle \Psi^L_{E} |\Psi^R_{E}\rangle}  \langle\Psi^L_{E}| K\hat A K\Psi^R_E\rangle\\
	&\xrightarrow{K\hat A=\hat A K}\sum_{E} \frac{ 2 }{\langle \Psi^L_{E} |\Psi^R_{E}\rangle}  \langle\Psi^L_{E}| \hat A|\Psi^R_E\rangle\\
	&=2\text{Tr} (\hat{A}P) = 2\langle A\rangle_{P},
\end{split}
\end{equation}
where $P=\sum\limits_E\displaystyle\frac{1}{\langle\Psi^L_E|\Psi^R_E\rangle} |\Psi^R_E\rangle \langle\Psi^L_E| $ is the biorthogonal projection operator, summed over the occupied energies $E$. Here, the factor of $2$ arises from the doubled  number of ($\pm$) states in $\tilde{\mathcal{H}}$. Although the biorthogonal expectation is arguably not directly measurable when we only have the target system $\mathcal{H}$, the above shows how we can express it in terms of the \emph{physical} expectations $\langle \hat A\hat\Lambda \rangle_E$ and $\langle\hat\Lambda \rangle_E$. In particular, by setting $\hat A=c_{i\alpha}^{\dagger} c_{j\beta} $ , we can obtain the two-point correlation functions $2\text{Tr} (P c_{i\alpha}^{\dagger} c_{j\beta})$, where $i$ and $j$ represent the unit cells and $\alpha$ and $\beta$ denote sub-lattices.\\
 
The 2nd Renyi entropy or purity can be measured based on the expectation values concerning the two-point correlation functions, or equivalently, the distribution of the density matrix.
By evaluating the two-point correlation functions within an entanglement cut region, one can construct the truncated occupied state projector $\bar P$ of the subsystem. The 2nd Renyi entropy \( S_2 \) is then given by
$S_2 = -\log \text{Tr}(\bar P^2)$.
Therefore, by accurately measuring the expectation values corresponding to the correlation functions, we can derive the necessary information to compute the 2nd Renyi entropy.

Below, we provide a brief discussion of how some of the above steps may be implemented in quantum simulator systems. In principle, they are not restricted to any particular platform, but below we shall elaborate mostly in the context of ultracold atomic setups~\cite{li2019observation,ren2022chiral,lapp2019engineering,gou2020tunable,takasu2020pt,ferri2021emerging,rosa2022observing,qin2023non,shen2023proposal}, rather than trapped ions or solid-state spin systems.\\

In cold atom systems, employing either bosonic atoms, such as 87Rb, or fermionic atoms, such as 173Yb or 40K,  radio-frequency (RF) spectroscopy is a widely employed spectral measurement technique~\cite{torma2016physics,vale2021spectroscopic}.  It is based on weakly coupling the system to auxiliary energy levels using radio waves or microwaves, with weak coupling strength achievable by adjusting the laser power ~\cite{cohen1992laser,chu1991laser,muldoon2012control}. General RF spectroscopy as described by~\cite{li2022non,cao2023probing} can measure both the real and imaginary parts of eigenvalues of non-Hermitian systems.
And by inducing momentum transfer with coherent Raman laser beams and combining this with Time-of-Flight and absorption imaging techniques, the momentum distribution and spatial wave function of atoms can be accurately measured~\cite{metcalf1999laser}. For measuring expectation values, there are various techniques specifically geared towards cold atomic experiments~\cite{metcalf1999laser,ketterle1996evaporative,lett1988observation,denschlag2000generating}, such as time-of-flight measurements, absorption spectroscopy, Raman spectroscopy, and interference techniques.

 \SetPicSubDir{ch-bethe}
\SetExpSubDir{ch-bethe}

\chapter{Critical Bethe lattice under non-Hermitian pumping}
\label{ch-bethe}

Continuing from the previous chapters, the non-Hermitian physics, which extends beyond the conventional Hermitian Hilbert space, has revealed rich phenomena such as the NHSE~\cite{PhysRevB.97.121401, PhysRevLett.123.016805, PhysRevB.99.201103, PhysRevLett.124.250402,  PhysRevB.102.205118, PhysRevLett.124.086801,PhysRevLett.125.126402, okuma2021quantum, PhysRevLett.127.066401, PhysRevB.104.L241402, PhysRevB.104.L161106, Li2021, PhysRevB.104.195102, Shen2022, zhang2022review, Shang_2022, Zhang2022, FangHuZhouDing, PhysRevB.107.085426, Zhou2023, Wan_2023, PhysRevLett.131.207201, Okuma_2023, Lin2023, le2023volume, jana2023emerging,shen2023observation, Lei2024}, critical behaviours~\cite{PhysRevLett.113.250401, herviou2019entanglement,guo2021entanglement,tu2022general, PhysRevB.105.L241114,fossati2023symmetry,kawabata2023entanglement,PhysRevB.109.024204, e26030272, Wei2017, PhysRevB.109.024306, PhysRevResearch.6.023081}, and scale-sensitive phase transitions~\cite{li_critical_2020,lee_exceptional_2022}. Among these, the critical non-Hermitian skin effect (cNHSE)~\cite{li2021critical} has drawn particular interest, characterized by eigenenergies and eigenstates exhibiting discontinuous jumps across critical points as the system size changes. This phenomenon unveils a distinct class of discontinuous criticality and holds promising applications in minimal signal sensing and switching. Despite its intriguing nature and potential applications, our current understanding of cNHSE remains largely limited to one-dimensional ladder models, with insufficient insight into the underlying intrinsic physics driving this phenomenon.

In this work, we aim to explore the interplay between cNHSE and the Bethe lattice, a hyperbolic structure commonly used as a foundational model for higher-dimensional studies. Our investigation focuses on how cNHSE manifests across different geometric structures, the robustness of its band properties and transitions, and how its behaviour adapts to varying lattice configurations. This provides insights into the intrinsic nature of cNHSE and its potential extensions to higher-dimensional phenomena.

Moreover, the Bethe lattice is characterized by its hyperbolic geometry and exponentially expanding branching, which amplifies the boundary influence. This topology is frequently used as a framework for studying bulk-boundary correspondence in non-Hermitian physics through mapping to holographic dualities such as AdS/CFT~\cite{PhysRevLett119071602,hahn_observation_2016,nishioka2009holographic}. Its unique boundary impact on the cNHSE could reveal unprecedented bulk-boundary phenomena, as what we will introduce in later sections.

This work reveals that non-Hermitian Bethe lattices exhibit discontinuous cNHSE transitions in both spectra and wavefunctions, similar to conventional cNHSE, but with spectral properties and state localizations unique to Bethe cNHSE. The spectral transitions in Bethe lattices are not characterized by the criticality defined in conformal field theory (CFT)\cite{calabrese2004entanglement, Cardy:2004hm, PhysRevLett.116.026402, Taddia_2016, PhysRevB.95.115122, PhysRevB.96.045140, PhysRevLett.130.250403, PhysRevLett.132.086503, Jones2022} and measured by entanglement entropy\cite{PhysRevLett110100402,Boer_2000,RevModPhys77259}. Instead, they emerge as band properties resulting from the competition between NHSE and the geometric effects of the hyperbolic Bethe lattice. The cNHSE states in Bethe lattices exhibit robust spectral properties and localization patterns determined by boundary conditions. With appropriate boundary design, Bethe cNHSE can display alternating boundary localization across layers, producing distinctive spatial distributions of eigenstates in the bulk that are influenced by the boundary's size. This newly identified bulk-boundary correspondence in wavefunction distribution presents potential applications in signal switching and wave transport control.

\section{The Bethe model}

The one-branch Bethe lattice model is a semi-infinite graph with a specified branching factor $\beta$, which represents the number of neighbours for each non-leaf node. The graph consists of a single root node connected to $(\beta-1)$ nearest-neighbour nodes. Each of these neighbours, in turn, is connected to $(\beta-1)$ new nodes, creating a hierarchical structure. The process continues indefinitely, forming a tree-like graph with an infinite number of levels.
\begin{figure}
\centering
\includegraphics[width=\linewidth]{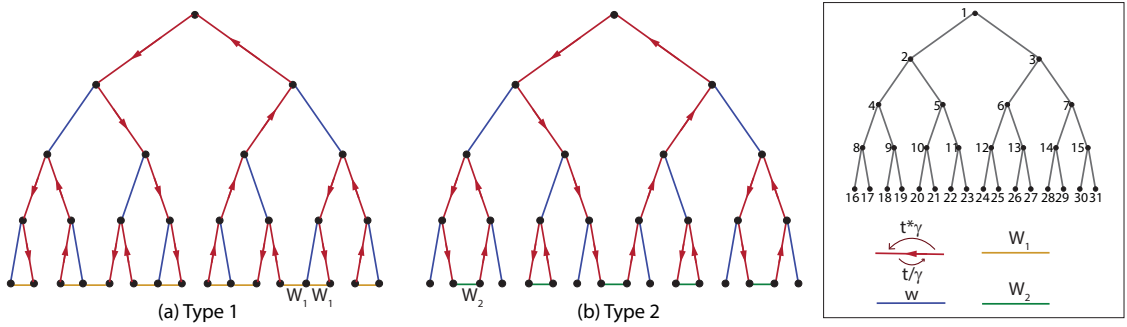}
\caption{\textbf{Non-Hermitian Bethe tree: Two types studied.} A 5-layer Bethe lattice is shown, with the $L^{th}$ ayer containing $L=2^{l-1}$ sites indexed along the layer. Couplings are classified as: (1) Non-Hermitian couplings $t\gamma$, $t/\gamma$ (red), forming $F(l)$ Hatano-Nelson (HN) chains [Eq.~\ref{noHN}] of various lengths, directed from left to right with boundary sites on the Bethe tree's outermost layer. (2) Weak Hermitian couplings $w$, linking adjacent HN chains across layers. (3) Boundary Hermitian couplings $W_1$ (yellow) and $W_2$ (green) closing the open Bethe trees in two distinct ways. The open Bethe tree is given by the Hamiltonian Eq.~\ref{Hodd} and Eq.~\ref{Heven}, while the 2 types of closed Bethe trees is given by Hamiltonian in Eq.~\ref{Htype1} and Eq.~\ref{Htype2}. }
\label{modeltypes}
\end{figure}

In essence, the Bethe lattice can be viewed as a quasi-2D lattice, composed of multiple 1D chains of varying lengths coupled in a controlled manner, as shown in Fig.~\ref{modeltypes}. A 5-layer Bethe lattice is used as an example. The red one-dimensional Hatano-Nelson (HN) chains having coupling strengths $t\gamma, t/\gamma$, are coupled to adjacent chains through weak Hermitian couplings indicated in blue. Generally, shorter chains are located closer to the bottom ``boundary'' layer, whereas longer chains extend deeper into the ``bulk'', corresponding to higher layers. This structure is reminiscent of geodesics in the hyperbolic half-plane, though the actual shapes of the geodesics differ.

In the tight-binding representation, the real-space Hamiltonian is
\begin{equation}
\begin{split}
\hat{H}_{\text{odd}} = &\sum_{i=0}^{m}t_{l}\hat{c}_{3i+1}^{\dagger}\hat{c}_{6i+2} +
t_{r}\hat{c}_{6i+2}^{\dagger}\hat{c}_{3i+1} +
t_{r}\hat{c}_{3i+1}^{\dagger}\hat{c}_{6i+3} + t_{l}\hat{c}_{6i+3}^{\dagger}\hat{c}_{3i+1}
+t_{l}\hat{c}_{3i+2}^{\dagger}\hat{c}_{6i+5}\notag\\ + &t_{r}\hat{c}_{6i+5}^{\dagger}\hat{c}_{3i+2}+ t_{r}\hat{c}_{3i+3}^{\dagger}\hat{c}_{6i+6}+t_{l}\hat{c}_{6i+6}^{\dagger}\hat{c}_{3i+3} 
+w(\hat{c}_{3i+2}^{\dagger}\hat{c}_{6i+4} +
\hat{c}_{3i+3}^{\dagger}\hat{c}_{6i+7} + h.c.),
\end{split}\label{Hodd}
\end{equation}

\begin{figure}
\centering
\includegraphics[width=0.7\linewidth]{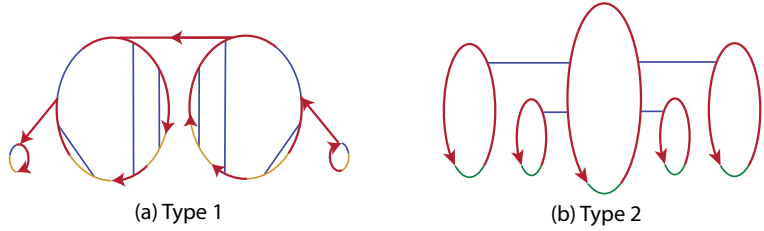}
\caption{\textbf{Simplified illustrations of two types of closed Bethe:} We are interested in the interaction between non-Hermiticity and different quasi-2D structures, specifically the closed non-Hermitian loops within the lattices. With red non-Hermitian couplings dominating over Hermitian couplings of other colors, type 1 and type 2 lattices exhibit distinct patterns of non-Hermitian loops: type 1 lattice has the strong non-Hermitian couplings form directed, discontinuous zigzag paths, with all nested loops primarily composed of weak Hermitian couplings; the type 2 lattice though, has multiple strong HN loops connected in series, forming non-Hermitian loops weakly linked to each other. These two lattice types display distinct characteristics in their band structures.}
\label{cartoon}
\end{figure}
for odd-number-layer lattices, where $\hat{c}_{i}^{\dagger}$($\hat{c}_{i}$) denote the creation(annihilation) operators at the i-th sites, $m=\left \lceil{(2^n-2)/6}\right \rceil-1$ and $l$ being the number of layers; and
\begin{equation}
\begin{split}
\hat{H}_{\text{even}} = \hat{H}_{\text{odd}}
+&\sum_{i=0}^{m}[t_{l}\hat{c}_{3(m+1)+1}^{\dagger}\hat{c}_{6(m+1)+2} +
t_{r}\hat{c}_{6(m+1)+2}^{\dagger}\hat{c}_{3(m+1)+1}\notag\\+&
t_{r}\hat{c}_{3(m+1)+1}^{\dagger}\hat{c}_{6(m+1)+3} + t_{l}\hat{c}_{6(m+1)+3}^{\dagger}\hat{c}_{3(m+1)+1}],
\end{split}\label{Heven}
\end{equation}

for even-number-layers lattice with the same definition of $m$. For each $l$-layers lattice, there existed $F(l)$ number of HN chains of different lengths, with
\begin{align}
F(l)=F(l-1)+2F(l-2).\label{noHN}
\end{align}

Although the red non-Hermitian couplings dominate over the blue inter-chain couplings, the directional pumping in the Bethe tree is complex because the coupled HN chains exhibit oppositely directed skin effects at each pair of connected sites. This interaction creates additional directional pumping pathways (e.g. $21\to 10\to 5\to11\to22$) among the HN chains.

In this work, we focus on the interplay between the coupled non-Hermitian skin effect (cNHSE) and quasi-2D hyperbolic geometry. To explore this, the open Bethe tree lattice is closed at the boundary layer, as shown in Fig.~\ref{modeltypes}, to construct effective lattice loops. Two types of closed Bethe lattices are studied. The type 1 Bethe lattices are closed by inter-chain Hermitian boundary couplings $W_1$ (yellow),forming effective lattices as illustrated in Fig.~\ref{cartoon}a. In this case, the strong HN chains are hardly considered closed, and lattice loops of weak couplings are formed along its zigzag path. The Hamiltonian for this system is given by: 
\begin{align}
\hat{H}_{\text{odd, type 1}} = &\hat{H}_{\text{odd }}
+\sum_{i=0}^{m'}W_1\hat{c}_{2^{n-1}+3i}^{\dagger}\hat{c}_{2^{n-1}+1+3i} +W_1\hat{c}_{2^{n-1}+2+3i}^{\dagger}\hat{c}_{2^{n-1}+3+3i} + h.c.),\\
\hat{H}_{\text{even, type 1}} = &\hat{H}_{\text{even }}
+\sum_{i=0}^{m'-1/3}W_1\hat{c}_{2^{n-1}+3i+1}^{\dagger}\hat{c}_{2^{n-1}+2+3i} +W_1\hat{c}_{2^{n-1}+3i+2}^{\dagger}\hat{c}_{2^{n-1}+3+3i} + h.c.),
\end{align}\label{Htype1}
where $m'=(2^{n-1}-4)/3$.

The type 2 Bethe lattice, in contrast, is closed by the weak intra-chain Hermitian boundary coupling $W_2$ (green) as shown in Fig.~\ref{cartoon}b. In this configuration, all HN chains are individually closed by $W_2$ at their boundaries and connected in series through the weak coupling $w$:
\begin{align}\label{Htype2}
\hat{H}_{\text{odd, type 2}} = &\hat{H}_{\text{odd }}
+\sum_{i=0}^{m'}W_2\hat{c}_{2^{n-1}+1+3i}^{\dagger}\hat{c}_{2^{n-1}+2+3i} + h.c.),\\
\hat{H}_{\text{even, type 2}} = &\hat{H}_{\text{even }}
+\sum_{i=0}^{m'+2/3}W_2\hat{c}_{2^{n-1}+3i}^{\dagger}\hat{c}_{2^{n-1}+1+3i} + h.c.),
\end{align}
with the same $m'$.

These two types of Bethe lattice exhibit different pumping pathways and demonstrate different cNHSE and band properties, as will be shown in later sections.

\section{Size-dependent spectral transition}\label{bethe-secPT}

The critical non-Hermitian skin effect (cNHSE) is a class of criticality where the eigenenergies and eigenstates in the thermodynamic limit jump discontinuously across the critical point as the size of the system varies. This phenomenon was shown by the ansatz model of two oppositely oriented weakly coupled HN chains forming a loop. In the two types of Bethe lattice, we again have the HN chains weakly coupled to each other, but with not only the length of each chain, but also the total number of chains increasing as the size of this two-dimensional lattice increases. The lattice loop is closed by two added Hermitian boundary couplings, $W_1$ in type 1 Bethe lattices closing the inter-chain loops and $W_2$ in type 2 Bethe lattices closing each HN chain.

Holographic cNHSE occurs in these Bethe lattices, with similar real-to-complex eigenenergy transitions but distinct behaviours of the complex eigenenergies.
\begin{figure*}[ht!]
\centering
\includegraphics[width=\linewidth]{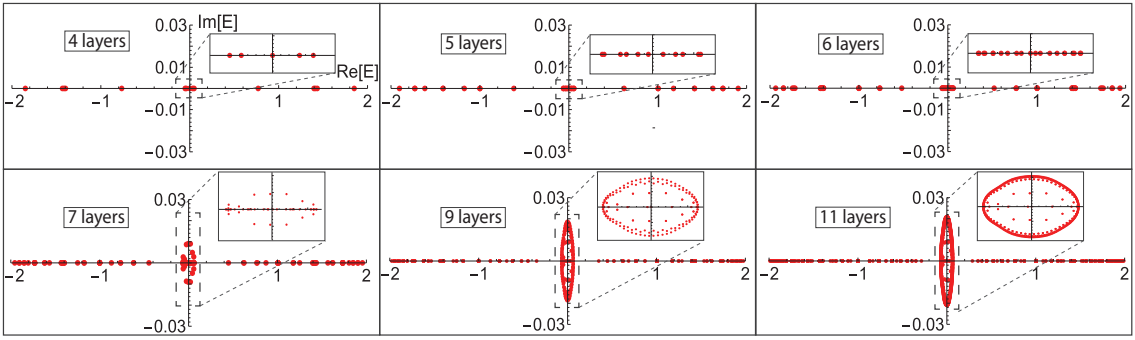}
\caption{\textbf{Gapped critical non-Hermitian spectral transition in type 1 Bethe lattice:} Centralized complex spectral loop in the low-energy band appeared as the number of layer increases up to 7, indicating a discontinuous size-dependent spectral transition named the Bethe critical non-Hermitian skin effect (cNHSE). Further increasing $l$ leads to the appearance of more spectral loops. The multiple complex loops emerge only at low-energy states, separated by the finite gaps from the high-energy regimes, and share the same origin and limiting behaviour for each system size.$t=1$, $\gamma=1.2$ and $w=W_1=0.05$.}
\label{Type1LayerPT}
\end{figure*}

\textit{Type 1 Bethe--}Two intriguing phenomena are observed in type 1 Bethe lattice energy spectra:

Firstly, a Bethe cNHSE is observed. By increasing the number of layers in the system, we observe that the band acquires complex eigenvalues and gradually forms a loop, as illustrated in the example system in Fig.\ref{Type1LayerPT}. As the system further enlarges, more spectral loops appear. Unlike the normal cNHSE observed in ladder structures, our Bethe lattice structure exhibits a rapid transition from real to complex eigenvalues with each additional layer. This rapid transition occurs with only two-site-long chains, a capability that is absent in ladder structures.

Moreover, multiple spectral loops instead of a single spectral loop appeared simultaneously, and their size increases rapidly at first and gradually slow down as they approach the same limit, as evident when comparing the spectra of 9-layer and 11-layer systems. This behaviour contrasts with conventional non-Hermitian systems, where a single complex spectrum typically grows indefinitely. This suggests an intrinsic physical significance of the stabilized complex spectral loop, which highlights a pathway toward achieving stable, controllable, and experimentally feasible non-Hermitian devices.

Secondly, the spectra are separated into three parts with finite gaps in between, with two symmetric high-energy regimes and one low-energy regime. The complex spectral loop only emerge in the low-energy regime, away from the high-energy states which remain real for all sizes of systems, indicating the robustness of its properties.

\begin{figure*}[ht!]
\centering
\includegraphics[width=\linewidth]{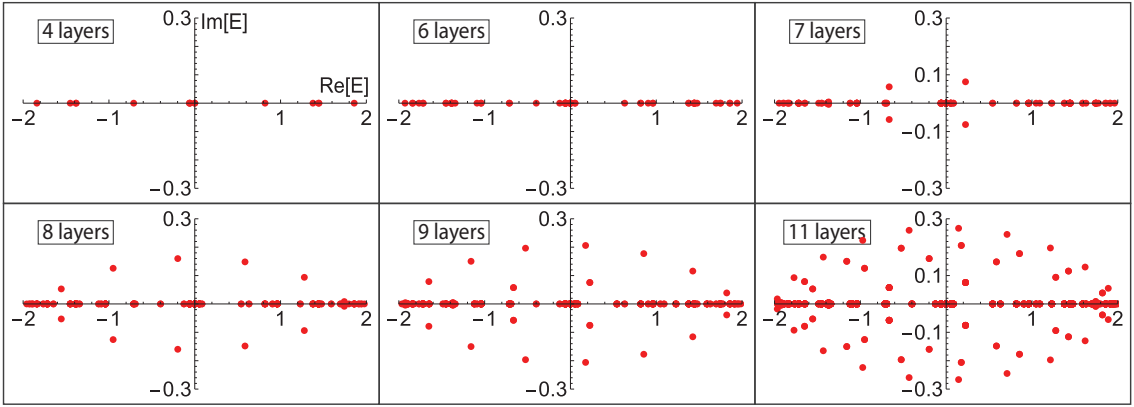}
\caption{\textbf{Gapless critical non-Hermitian spectral transition in type 2 Bethe lattice:} A complex spectral loop spanning the entire real axis emerges when the number of layers reaches 7, signaling a size-dependent spectral transition and the onset of cNHSE. As $l$ increases further, additional spectral loops appear with similar shapes and origins but varying sizes at all system sizes. $t=1$, $\gamma=1.2$ and $w=W_2=0.05$.}
\label{Type2LayerPT}
\end{figure*}

\textit{Type 2 Bethe--}Similar cNHSE is observed in type 2 Bethe lattices, which consist of multiple interconnected closed HN loops. The energy spectra remain real for small systems but become complex when the total number of layers reaches 7, as shown in Fig.~\ref{Type2LayerPT}. However, the strongly non-Hermitian HN loops within the Bethe lattice dominate over the weak inter-couplings, resulting in typical complex spectra once the system transitions to complex eigenenergies. As a result, nearly uniform complex loops emerge and continuously expand without approaching a limiting shape. Additionally, the spectral loops span the entire real line, producing a gapless spectrum.

The cNHSE in Bethe lattices mirrors conventional cNHSE but manifests with multiple complex spectral loops. This arises from the combined contributions of non-Hermitian lattice loops of varying sizes and couplings, and potentially leads to more stable and controllable band structures.

\section{Invariance of entanglement under non-critical spectral transition}

\begin{figure*}[ht!]
\centering
\includegraphics[width=0.8\linewidth]{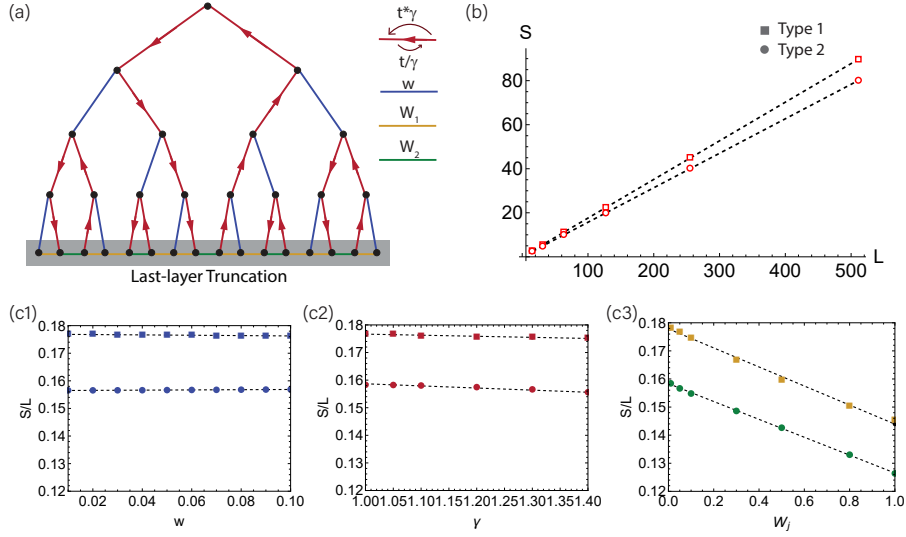}
\caption{\textbf{Area-law entanglement entropy of type 1 and 2 Bethe lattice with last-layer truncation:} (a) 5-layers boundary-truncated Bethe as the lattice model, with the type 1 (type 2) model has $W_2=0$ ($W_1=0$). (b) The numerical entanglement entropy (red dots and squares) forms a linear Area-law relation with the size of system for both models, indicating the absence of conventional critical transition described within the conformal field theory (CFT). (c) The numerical $S/L$ (coloured dots and squares) is shown nearly independent of the bulk coupling $w$, $t_l$ and $t_r$ but linearly related with the boundaries $W_1$ and $W_2$. The boundary sensitivity of the scaling of $S$ characterize the constructed quasi-2D Bethe models. Unless specified in the figure, the parameters are $W_1=0.05$ and $w=0.05$, $t_l=1.2$ and $t_r=\frac{5}{6}$.}
\label{type1_EE}
\end{figure*}
Quantum phase transitions are characterized by CFT through measurement of the entanglement entropy $S$ for the occupied state projection operator $P$ of the Hamiltonian of the system
\begin{align}
P=\sum_{\text{ReE}\leq 0}\mathcal{E}_{i}\ket{\psi_{i}^{R}}\bra{\psi_{i}^{L}},
\end{align}
where $\mathcal{E}_i$ is the $i^{th}$ eigenvalue and $\ket{\psi_{i}^{R}}$ $\left(\bra{\psi_{i}^{L}}\right)$ is the right (left) eigenstate of the Hamiltonian.

For free fermions, the entanglement spectrum can be obtained by truncating the occupied band projector $P$ in real space. In the Bethe lattice, we define a real-space boundary partition $[2^{l-1}, 2^l-1]$, such that the entanglement behaviour between the bulk and the boundary parts of the holographic lattice is studied. The truncated band projector $\bar{P}$ can be obtained from $P$ and the partition projector $\mathcal{R}_{[2^{l-1}, 2^l-1]}$ through
\begin{align}\label{2bandP3}
& \bar{P}= \bar{\mathcal{R}}_{[2^{l-1}, 2^l-1] } P \bar{\mathcal{R}}_{[2^{l-1}, 2^l-1] }, \\
\text{where }& \bar{\mathcal{R}}_{[2^{l-1}, 2^l-1] }=\sum^{L}_{x\notin[2^{l-1}, 2^l-1] }|x\rangle\langle x|\otimes \mathbb{I}.
\end{align}
The occupancy probabilities $p$ are obtained from the eigenvalues of the truncated occupied state projection matrix $\bar P$ and are limited to the range $0\leq p\leq 1$. The correlation between the truncated and untruncated regions of the projection operator space is reflected by this occupied probability $p$.
The free-fermion entanglement entropy $S$ quantifies this entanglement via
\begin{align} \label{SEE}
S(\bar P) &= -\text{Tr}[\bar P \log \bar P +(\mathbb{I}-\bar P)\log (\mathbb{I}-\bar P)\ \notag\\
&=-\sum_p [p \log p +(1-p)\log (1-p)].
\end{align}

We performed simulated calculations of the entanglement entropy for Bethe systems, and the results are as follows:

The free-fermion entanglement entropy $S$ of both type 1 and 2 Bethe lattice, with the boundary layer truncated as demonstrated by Fig. \ref{type1_EE}a, is linearly proportional to the total number of sites $L$ within the system, as indicated by the red dots in Fig. \ref{type1_EE}b. Unlike the exponential relation between the entanglement $S$ and the system's size $L$ in the cNHSE between oppositely orientated weakly coupled HN chains, the entanglement entropy $S$ of the Bethe lattices linearly scales with the boundary length $L$ of the region. The constant proportionality $S/L$ provides information about the uniform density of entanglement along the boundary. Numerically, we observed that $S/L$ remains unchanged with varying non-Hermiticity $\gamma$, as seen in Fig. \ref{type1_EE}c2. The entanglement between the bulk and boundary of the holographic lattice is independent of the skin effect towards the sides of each layer. Fig. \ref{type1_EE}c1-3 further show that the proportionality constant $S/L$ is only affected by the boundary coupling strength rather than the bulk couplings. The similar results of distinct lattice settings in the Bethe structure indicate that the entanglement between bulk and boundary is a structural property inherent to the holographic nature of the lattice and remains stable under non-Hermitian disorder within the bulk.

Concrete exponential entanglement entropy is observed rather than power-law scaling, indicating that the spectral transitions here do not align with conventional criticalities.
\begin{figure*}[ht!]
\centering
\includegraphics[width=\linewidth]{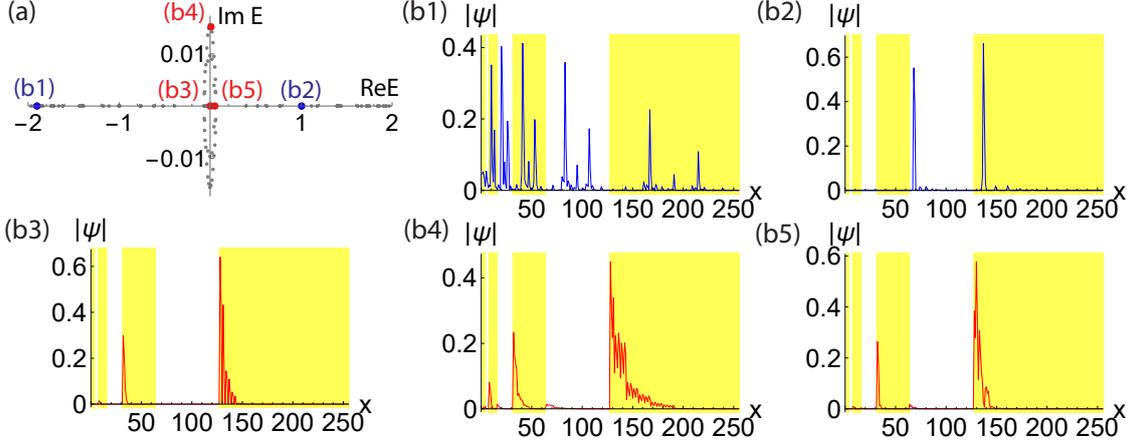}
\caption{\textbf{Competition between geometric effect and NHSE, and the boundary sensitive bulk localization in type 1 gapped states:} (a) Using the 8-layer system from Fig.~\ref{Type1LayerPT} as an example, the low-energy and high-energy bands, separated by finite energy gaps in the spectra of type 1 Bethe lattices, exhibit distinct wavefunction behaviours due to the interplay between non-Hermiticity and geometric localization inherent to the Bethe structure. (b1) States in the high-energy band are dominated by geometric effects, with wavefunctions spreading across all layers of the lattice.
(b2) States in the intermediate high-energy regime are still influenced by layer-wise transport and geometric localization.
(b3-5) States gapped from the above are dominated by the NHSE, exhibiting alternate-layer localization originating from the boundary layer, regardless of the complexity of its energy. The presence of bulk-layer localization of states in the low energy regime is determined by the boundary layer, enabling boundary-sensitive bulk localization. This mechanism suggests potential applications, such as controllable state transport.}
\label{type1_skin}
\end{figure*}

\begin{figure*}[ht!]
\centering
\includegraphics[width=\linewidth]{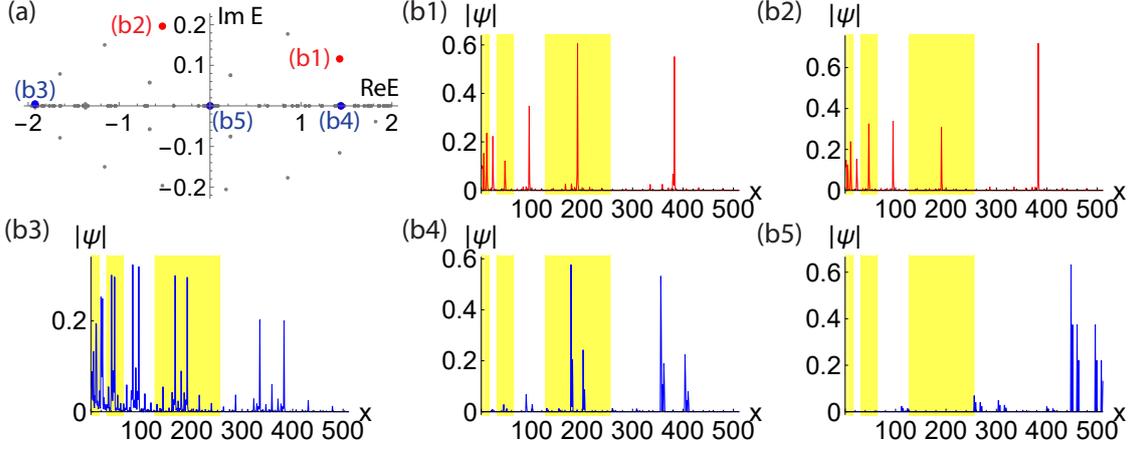}
\caption{\textbf{Competition between geometric effect and NHSE, and the distinct behaviour between real and complex states: } (a) Using the 9-layer system from Fig.~\ref{Type2LayerPT} as an example, the interplay between non-Hermiticity and geometric effects results in different behaviours for complex and real states in type 2 Bethe lattices.
(b1-2) Complex states localization are dominated by the major HN chain, halting transport toward the left and localizing at the left-half sites of the major HN chain of each layer, regardless of whether the states lie in the high- or low-energy regime.
(b3-5) Real energy states are influenced by geometric localization, with state transport throughout the layers dictated by the energy regime.}
\label{type2_skin}
\end{figure*}
\section{Alternating layer confinement alongside boundary localization}
\begin{figure*}[ht!]
\centering
\includegraphics[width=\linewidth]{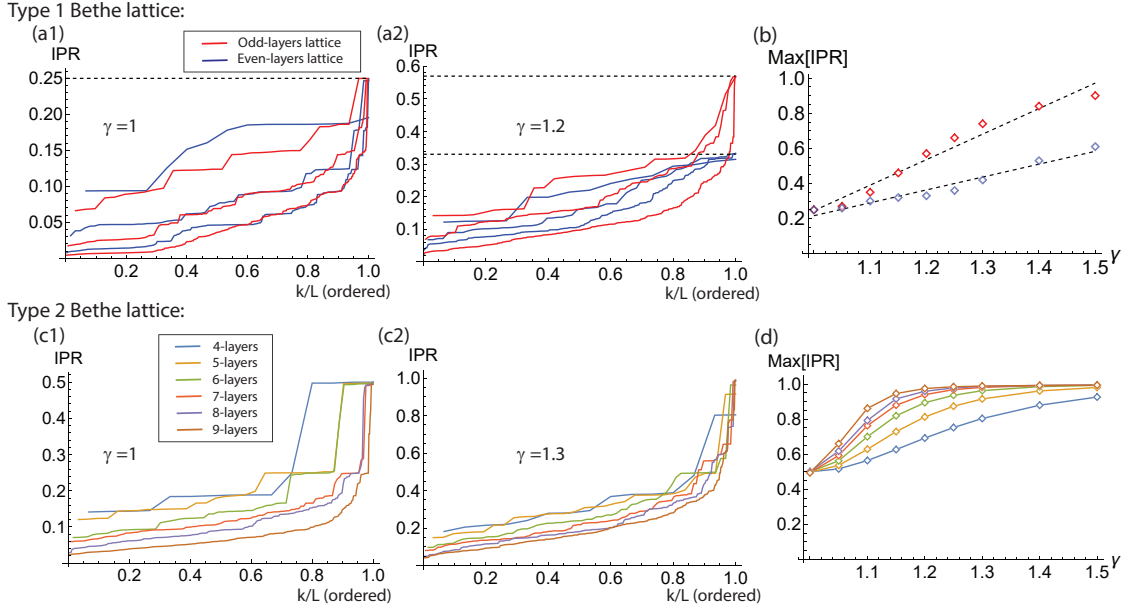}
\caption{\textbf {Inverse participation ratio (IPR) as a measure of localization in type 1 and 2 Hermitian and non-Hermitian Bethe lattices of varying sizes.} (a) Ordered IPR values of the $k^{th}$ eigenstates out of all $L$ with 4 to 9 layers, shown for Hermitian $\gamma=1$ and non-Hermitian $\gamma=1.2$ cases. 
Non-Hermitian type 1 Bethe lattices with even and odd layers exhibit distinct, size-independent maximal IPR, with their dependence on non-Hermiticity illustrated in (b). The odd-even response to non-Hermiticity contributes to alternate localization of low-energy states.
(c) In type 2 Bethe lattices, the maximal IPR is size-dependent, as compared in (d), but no odd-even differences are observed. Increasing $\gamma$ enhances IPR nearly uniformly across all states.}
\label{IPR}
\end{figure*}

The spectral properties of non-Hermitian Bethe lattices in Sec.~\ref{bethe-secPT} is further investigated by studying the localization of states within different phases. 

\textit{Type 1 Bethe--}The gapped spectra consist of two distinct groups of states: low-energy states that exhibit cNHSE as the system size varies, and high-energy states that remain real at all sizes. As shown in Fig.~\ref{type1_skin}, the low-energy states in Fig.~\ref{type1_skin}b3-5 are constrained within a limited energy range for each finite system, displaying stable layerwise localization on alternate layers, counted from the boundary layer. In this regime, NHSE dominates over the geometric effects of the Bethe lattice, causing states to localize on one side of each layer. Meanwhile, the geometric structure drives an alternate-layer localization pattern. That is, adding a layer of sites to the boundary would totally reverse all states' localization within the bulk, giving rise to potential applications of signal transitioning.

In contrast, high-energy states, as shown in Fig.~\ref{type1_skin}b1-2, which do not exhibit cNHSE, are more affected by the geometric structure, leading to non-uniform localization patterns. These states transport more easily across layers, reflecting weaker and less pronounced non-Hermitian directional effects in the high-energy regime.

\textit{Type 2 Bethe--}The gapless spectra of type 2 Bethe lattices exhibits cNHSE across the entire real axis. However, two distinct groups of states with contrasting localization behaviours can still be identified. The complex states in Fig.~\ref{type2_skin}b1-2 after the cNHSE transition have states localized at the longest HN chain situated at the center of each layer. This indicates that the NHSE of the longest closed HN chain dominates over the shorter chains and weakly coupled inter-chain loops for states undergoing cNHSE. In contrast, other states, as shown in Fig.~\ref{type2_skin}b3-5 do not undergo the cNHSE transition. These states are primarily influenced by the geometric structure, allowing them to propagate throughout the layers.

In general, in both types of Bethe lattices, states undergoing cNHSE are stabilized by the interplay between the NHSE and the geometric structure of the Bethe lattice. This results in robust spectra and localization patterns, though the specific behaviour differs between the two lattice types.

We further demonstrate the geometric effect of the Bethe structure and the NHSE on the complete set of wavefunctions through their IPR values. The IPR is defined as follows:
\begin{align}
\text{IPR}=\sum_{x=1}^{L}|\psi_k(x)|^4,
\end{align}
where $x$ is the index of site in the real space and $\psi_k$ is the $k^{th}$ state amplitude with $k=1, 2, ..., L$ and $L$ being the size of the system. For a completely extended state, the wavefunction is spread evenly over all sites and $\psi_k(x)=\frac{1}{\sqrt{L}}$ with a IPR of $\frac{1}{L}$. Similarly, for a fully localized state with the wavefunction localized on a single site $\psi_k(x)=\delta_{x,x_0}$, the IPR would be 1.

 \textit{Type 1 Bethe--}Fig. \ref{IPR}a-b shows the IPR of all states in the example type 1 Bethe lattices with the total number of layers ranging from $l=4$ to $l=9$, with their states ordered in increasing order of the values of IPR. Given that the size of the lattice is $L=2^l-1$, the IPR of all states within the type 1 Bethe lattices is clearly localized even if the systems are Hermitian and driven purely by the geometric effect. This structural localization is bounded by a maximal IPR of $0.25$, which renains independent of lattice size.

When non-Hermiticity is introduced, odd- and even-layer lattices exhibit distinct localization properties. Even-layer lattices, shown by the blue curves, localize more slowly with increasing non-Hermiticity compared to the odd-layer lattices in red. This difference contributes to the alternate layerwise localization pattern.

 \textit{Type 2 Bethe--}The type 2 Bethe lattices exhibit larger IPR and stronger localization in Fig.~\ref{IPR}c compared to type 1 Bethe lattices, with no odd-even difference due to the longest HN chain dominating localization across all layers. As non-Hermiticity increases, localization rapidly approaches an IPR of 1, with larger systems containing more HN loops converging the fastest, as shown in (d). Non-Hermiticity enhances localization in a continuous and uniform manner across all system sizes.

\section{Conclusion}

In this work, we extend the critical non-Hermitian skin effect (cNHSE) to the hyperbolic geometry of Bethe lattices, highlighting size-sensitive spectral transitions where the system maintains a discontinuous shift from real to complex eigenvalues driven by lattice size. Unlike conventioanl cNHSE in ladder structure, not all states within the Bethe lattices undergoes cNHSE, and the comparison between states protected by cNHSE and states otherwise indicates the unusual properties of cNHSE states. In both types of closed Bethe lattices we have studied, the cNHSE states demonstrates robust spectral properties and localization pattern. This further confirms the potential of cNHSE in the applications of perturbation sensing, as the interplay between geometric effects and non-Hermitian skin modes provides enhanced sensitivity to small changes in system parameters.

Moreover, unlike traditional NHSE, which confines states to its boundaries, the critical Bethe lattice of type 1 drives an alternate-layer localization pattern. By fine-tuning lattice parameters, it may be possible to manipulate the localization, enabling precise control over transport and confinement in hyperbolic lattices. This suggests applications in signal switching and wave propagation control. 

Overall, this study demonstrates how cNHSE in Bethe lattices fundamentally diverges from flat models, revealing several new phenomena of Bethe cNHSE and providing a comprehensive analysis to deepen our understanding of cNHSE. The newly discovered phenomena also suggest potential directions for practical applications.

 \SetPicSubDir{ch-concl}

\chapter{Conclusion and Outlook}
\label{ch-concl}

Non-Hermitian physics has rapidly evolved in recent decades, uncovering a range of intriguing phenomena such as the non-Hermitian skin effect (NHSE) and various non-Hermitian phase transitions. These phenomena have revealed numerous unconventional properties, all elucidated by the theoretical framework of the generalized Brillouin zone (GBZ), which also suggests promising applications. This thesis advances our understanding of non-Hermitian physics by offering a detailed analysis of previously unexplored aspects of non-Hermitian band theories and critical phenomena, solely from NHSE or its interplay with other degrees of freedom. The findings include the realness of spectra with broken PT symmetry, the system size-dependent criticality, and the topological phase transitions induced by spatial inhomogeneity.

\section{Conclusion}
\autoref{ch-intro} provides a historical overview of the development of non-Hermitian physics, aiming to help readers understand the foundational concepts and key milestones that have shaped the field. It also situates the current research within the broader historical and scientific context, showing how non-Hermitian physics has evolved over time. Furthermore, it explains why certain concepts, such as the Generalized Brillouin zone (GBZ), are significant by tracing their origins and development. This historical context provides readers with insights into the motivations behind investigating and generalizing these concepts. Additionally, we introduce the evolution of critical phenomena, from everyday occurrences like ice melting to advanced topics such as entanglement entropy scaling, which we frequently use to characterize criticality in our research. This approach aims to equip readers with a comprehensive understanding of why critical phenomena in non-Hermitian systems are of interest and importance.

In \autoref{ch-prel}, we provide a thorough introduction to the technical details and mathematical preliminaries essential for understanding non-Hermitian systems and critical phase transitions. This chapter is crafted to support readers who may be less familiar with these topics, offering foundational knowledge for the subsequent discussions. For example, we outline the key tools necessary for analyzing non-Hermitian systems and explain the construction and application of the Generalized Brillouin zone (GBZ) as a robust framework for describing non-Hermitian inhomogeneous systems. Additionally, we explore critical phase transitions, starting with the basic Ising Model, to demonstrate fundamental critical properties and how to characterize a critical system.

\autoref{ch-realSpec} investigates the appearance of real spectra in non-Hermitian systems, enforced by the skin effect rather than bulk symmetries. The study of the realness of energy spectra has always been a fundamental part of the field of physics and has the potential to open up more directions of study. In this work, we showcase the conditions of the emergence of real spectra, within the broken PT symmetric phase, and provide simple ansatz models with robust open boundary condition (OBC) real spectra, complementing current efforts toward the stable design of non-Hermitian systems.

\autoref{ch-eedip} reports a new class of non-Hermitian critical transitions distinct from known Hermitian or non-Hermitian phase transitions, which are characterized by logarithmic entanglement entropy(EE) scaling with the system size due to the description of criticality in the conformal field theory (CFT). This critical phase transition, named the scaling-induced exceptional criticality (SIEC), exhibits dramatic divergent dips into the super-negativity in their entanglement entropy scaling, strongly violating the current understanding of the EE in the CFT. This critical transition is achieved through multiple differently-oriented weakly-coupled NHSE. Their competition in the skin effect leads to a combined scale-induced critical phase transition and the emergence of a novel type of scaled-induced exceptional points (EPs), associated with an unexpected negative dip in the entanglement entropy. This study also advances the GBZ framework to include the description and explanation of the newly identified scale-dependent energy spectra and critical points.

\autoref{ch-inhomo} introduces spatial inhomogeneity as an additional degree of freedom to conventional non-Hermitian PBC systems, which are characterized by asymmetric but spatially constant hopping strengths. The scaled interplay between inhomogeneity and non-Hermiticity gives rise to a new phase of states in the spectral regimes that emerge from the non-zero difference between maximal and minimal hopping strengths. This phase exhibits numerous unprecedented phenomena, including discontinuous jumps between two branches of position-dependent GBZs and non-trivial topology in 2-component inhomogeneous systems. The phase-space GBZ, defined in a phase space encompassing both momentum and position, is completely constructed to thoroughly describe and analyze the inhomogeneous systems and to successfully explain the emergence and properties of this new phase.

\autoref{ch-bethe} explores the interplay between NHSE and the hyperbolic Bethe lattice, examining the band properties of critical NHSE within two distinct types of Bethe lattices. This study highlights phenomena unique to cNHSE states of Bethe lattices, such as alternate-layer localization acting as a bulk-boundary correspondence. These findings further confirm the potential applications of cNHSE in weak signal sensing and switching, advancing the understanding of the intrinsic nature of cNHSE and paving the way for investigations into cNHSE in higher dimensions.

\section{Outlook}

The study of non-Hermitian systems with spatial inhomogeneity represents a novel and systematically unexplored research direction. Our findings suggest that this approach can, in general, reveal unprecedented topological phenomena, analogous to the early studies of non-Hermitian physics, but on a distinct scale. Just as non-Hermitian physics has unveiled rich phenomena, transformative applications, and fundamental insights, spatial inhomogeneity holds the potential to chart new directions, deepen current understanding, and inspire extensive future research. This thesis establishes a foundational theory for single- and double-band nearest-neighbor hopping lattices in one-dimensional systems. Several intriguing research avenues arise from this work. For instance, in single-band inhomogeneous systems, a complex spectral loop emerges at the center of the open boundary condition (OBC) real-line spectra, with its boundary serving as a transition point between phases with distinct properties. Introducing additional couplings, such as next-nearest-neighbor hopping, alters the originally radial-line-shaped spectra by forming "bubbles," resulting in unconventional band gaps tunable via inhomogeneity. Exploring the topological aspects of inhomogeneous systems also poses an exciting challenge, necessitating the development of new invariants to rigorously define their topology. Extending these concepts to two-dimensional systems promises to uncover even richer topological properties. In summary, the exploration of spatial inhomogeneity in non-Hermitian systems is still at an early stage, with significant potential for future work to uncover new properties and possibly lead to practical applications.

Similarly, systems with size-dependent spectra and critical points, resulting from the interplay of oppositely-directed non-Hermitian effects, have so far only been explored in the simplest model. Extending this research to higher-dimensional or interacting systems could reveal further applications and novel phenomena associated with unconventional scale-dependent criticality.

Finally, the experimental realization of the theoretical models proposed in this thesis represents a promising avenue for future research. The newly introduced systems can generally be addressed using traditional experimental platforms, such as ultracold atomic systems. Spatial inhomogeneity can be implemented through designed potentials, and the negative biorthogonal entanglement can be measured using the method outlined in the Appendix of \autoref{ch-eedip}.

\bookmarksetup{startatroot}
\printbibliography[heading=bibintoc]

\appendix

\end{document}